\documentclass[10pt,twocolumn,floatfix,prb,aps,showpacs,superscriptaddress,longbibliography]{revtex4-2}
\usepackage{graphicx,amsmath,amssymb,color}
\usepackage{multirow}
\usepackage{nicefrac}
\usepackage[titletoc,title]{appendix}
\usepackage{amsmath}
\usepackage{subfigure}
\usepackage{tabularx}
\usepackage{xcolor}
\usepackage{comment}
\usepackage{natbib}
\usepackage{placeins} 
\usepackage{longtable}
\usepackage{booktabs} 
\usepackage{multirow} 
\usepackage[colorlinks,bookmarks=true,citecolor=blue,linkcolor=red,urlcolor=blue]{hyperref}
\usepackage{siunitx}
\usepackage[T1]{fontenc}
\usepackage[utf8]{inputenc}
\usepackage{tabularray}
\usepackage{titlesec}

\makeatletter
\def\@fnsymbol#1{%
  \ifcase#1\relax 
  \or \ensuremath{\dagger}
  \or \ensuremath{*}
  \or \ensuremath{\mathsection}
  \or \ensuremath{\mathparagraph}
  \else\@ctrerr\fi
}
\makeatother

\begin{document}
	
	\title{Topological spin freezing in frustrated quantum materials}
	\author{U. Jena}
	\affiliation{Department of Physics, Indian Institute of Technology Madras, Chennai, 600036, India}
\author{M. Barman}
	\thanks{equal contribution}
    \affiliation{Department of Physics, Indian Institute of Technology Madras, Chennai, 600036, India}
 \author{A. Pradhan}
	\thanks{equal contribution}
    \affiliation{Department of Physics, Indian Institute of Technology Madras, Chennai, 600036, India}
\author{P. Khuntia}
\email{pkhuntia@iitm.ac.in}
\affiliation{Department of Physics, Indian Institute of Technology Madras, Chennai, 600036, India}
\affiliation{Quantum Centre of Excellence for Diamond and Emergent Materials,
Indian Institute of Technology Madras, Chennai, 600036, India}

\date{\today}
\begin{abstract} 
Competing interactions, non-trivial electronic band topology, quantum fluctuations, and the interplay between emergent degrees of freedom in frustrated quantum materials can lead to a myriad of exotic phenomena crucial for addressing recurring themes in contemporary condensed matter while offering immense promise for quantum technologies.
 Glassy dynamics, originally explored in amorphous materials and biological systems, has gained significant attention in quantum condensed matter physics, particularly in the context of collective behavior of electrons, quasiparticle excitations, and exotic topological spin textures. Here, we investigate the manifestation of unconventional glassy spin dynamics in a broad class of frustrated magnets, where the underlying mechanisms of spin freezing exhibit distinct signatures in thermodynamic and microscopic experiments. We identify distinct signatures of topological spin glass behavior in frustrated quantum materials through a comprehensive suite of experimental probes, including thermodynamic measurements, nuclear magnetic resonance, muon spin relaxation spectroscopy, and inelastic neutron scattering. These complementary techniques collectively reveal unconventional spin dynamics, short-range spin correlations, emergent low-energy excitations, and glassy behavior with topological underpinnings, distinguishing this state from conventional spin glasses and disordered magnets. Furthermore, we discuss the role of hydrodynamic spin modes in governing glassy dynamics and the emergence of spin jam states in frustrated lattices that provide a comprehensive framework for understanding the unconventional spin freezing of topological origin, bridging experimental observations with theoretical models. The current review aims at widening our understanding of collective many-body phenomena, stemming from competing interactions, topological defects, collective excitations and their interactions, entanglement, and underlying symmetry, which may aid in the design and investigation of novel quantum  materials to address some of the fundamental questions in contemporary condensed matter, with potential significance in quantum technologies.

\end{abstract}
\maketitle
\tableofcontents
\newpage
\section{Introduction}

The quest for organizing principles in many-body systems provides a route to address many fundamental questions in science. It ranges from the mechanism behind protein folding in biological cells to the emergence of superconductivity in novel materials. At the heart of this progress lie two fundamental concepts: symmetry and topology, each of which governs the phase transitions in matter. Geometry is one of the most fundamental concepts to explain  the underlying symmetry and fundamental laws of nature. The theory of general relativity merges space and time into a single manifold, which was previously thought to be independent of each other~\cite{einstein1916foundation}. On the other hand, quantum mechanics projects the dynamics of a particle in a box onto a spectrum of energy levels~\cite{nakahara2018geometry, tong2012unquantum}. Therefore, the energy spectrum encodes the shape of the manifold in which the particle resides. String theory views a particle as a tiny vibrating loop, and the energy spectrum of this loop in a manifold reveals the equivalence between the large and small compactification scales~\cite{zwiebach2004first}. The effect of geometry also finds a profound implication in magnetism. Particularly in antiferromagnetically interacting Ising spins on a triangular lattice, all the bonds cannot be satisfied simultaneously, which in turn gives rise to magnetic frustration~\cite{PhysRev.79.357, anderson1973resonating}. As a result, the ground state becomes highly degenerate, and exotic states of matter emerge with unconventional spin excitations. Like the quote in a famous article asks, ``Can one hear the
shape of a drum?"~\cite{kac1966can}, can the collective spin excitations of a frustrated spin lattice encode the shape of its underlying free-energy landscape?

Topology is the property of a geometrical object that remains invariant under smooth and continuous deformations such as stretching or bending~\cite{nakahara2018geometry}. In condensed matter physics, the topological perspective provides a powerful tool for describing phases of matter that cannot be distinguished solely by symmetry. The phase transition due to symmetry breaking has a local
order parameter associated with it. However, the topological phases of matter are characterized by a non-local order parameter associated with long-range entanglement~\cite{wen2004quantum,wen1990topological,PhysRevLett.49.405,RevModPhys.82.3045}. The topological phases remain stable under smooth deformations of the Hamiltonian as long as the energy gap does not close. Topological quantum materials host exotic excitations that follow the fractional statistics. 

Unlike conventional order in matter, which is characterized by a local order parameter, topological phases are defined by global features of the many-body wave function—such as Chern numbers, winding numbers, or topological entanglement entropy—that remain stable against smooth deformations of the Hamiltonian~\cite{wen2004quantum,wen1990topological,PhysRevLett.49.405,RevModPhys.82.3045}. The topological state of quantum matter is governed by a synergistic interplay between competing degrees of freedom, collective excitations, the incompatibility of exchange interactions, long-range entanglement, fractional statistics,   and a highly degenerate manifold that transcends the standard paradigm, offering a viable avenue to address fundamental challenges in quantum condensed matter and great potential for next-generation technology~\cite{khatua2023experimental,khuntia2020gapless, basov2017towards,arh2022ising,tokura2017emergent,balents2010spin}. 

The interplay between symmetry, topology, and geometry shapes the fundamental principles that govern how matter organizes itself. The same underlying principle often produces metastable landscapes, where dynamics are trapped and slow relaxation dominates. Metastability refers to an intermediate energetic state, higher than that of the equilibrium state. It is observed across diverse scales, ranging from isomerization to neuroscience. Most emergent properties of matter are in metastable states that span a vast range of scales. For example, diamond, though an inert and hard material, is itself a metastable state. Its conversion to graphite, the true thermodynamic ground state of carbon, is prevented by a large kinetic barrier~\cite{lazicki2021metastability, bovzovic2021unstable}.
 The existence of life itself holds metastability at various levels. For example, protein folding is a metastable phenomenon. Without metastability, proteins might fall into completely collapsed or aggregated forms (like amyloid fibrils), losing their biological functionality. In another instance, the cosmological constant problem~\cite{hawking1984cosmological} suggests that we live in a metastable universe~\cite{de2021glassy}. The study suggests that false vacua can persist in the universe as long-lived metastable states~\cite{PhysRevD.15.2929, PhysRevD.16.1248}. Such false vacua could destabilize the universe if they under go a transition to a true vacuum state, releasing vast amounts of energy that, in turn, can fundamentally change the structure of the universe~\cite{PhysRevD.15.2929, PhysRevD.16.1248, PhysRevD.35.1747, PhysRevD.49.6410}. In condensed matter physics, the concept of false vacuum decay appears in various studies, including ferromagnetic atomic Bose-Einstein condensates~\cite{zenesini2024false}, ultra-cold atom condensation~\cite{braden2018towards}, quantum spin chains~\cite{PhysRevB.104.L201106}, and the destabilization of Kitaev spin liquids~\cite{baskaran2023metastable, PhysRevResearch.6.L022005}. This shows that metastability is a fundamental phenomenon that appears from microscopic quantum systems to the vast scale of the cosmos.

Landau’s Fermi liquid theory has been very successful in explaining the behavior of conventional condensed matter systems that undergo symmetry-breaking phase transitions. It could explain why materials like copper are good conductors of electricity. The Fermi liquid theory describes electrons as quasiparticles that behave almost like free electrons but with renormalized properties due to many-body interactions~\cite{landau1957theory, abrikosov1959theory}. The quasiparticles preserve the fundamental properties such as charge and spin quantum numbers, yet their mass and magnetic moment are renormalized~\cite{coleman2001fermiliquids, coleman2006heavy}. However, attempts to extend this framework to understand high-$T_C$
superconductors and the quantum Hall effect have proven largely unsuccessful. In the presence of strong electronic correlations, the breakdown of coherent quasiparticles occurs, and fractional excitations emerge, signaling the failure of conventional Fermi liquid theory in these systems~\cite{PhysRevLett.35.1779, RevModPhys.56.755, lee1986theories, PhysRevLett.58.2790, PhysRevLett.50.1395, von202040, anderson1973resonating, balents2010spin}. In Mott insulators, strong on-site Coulomb repulsion localizes electrons and opens a charge gap, thus suppressing low-energy charge fluctuations and giving rise to a rich variety of many-body phenomena governed by spin correlations~\cite{ mott2004metal,RevModPhys.40.677}. The parent compounds of cuprate superconductors are Mott insulators, which firmly established the crucial role of strong electron correlations in modern condensed matter physics~\cite{bednorz1986possible, RevModPhys.78.17}. Geometrically frustrated lattices in Mott insulators provide an opportunity to study magnetic phases without conventional symmetry breaking, characterized instead by non-local order parameters~\cite{anderson1973resonating, balents2010spin}. Kagome, triangular, and pyrochlore antiferromagnets with geometric frustration provide platforms for observing exotic quantum spin liquid states (QSLs)~\cite{ 10.1063/1.2186278, KHATUA20231}. In QSLs, excitations like spinons are inherently nonlocal disturbances of the underlying ground state, and a single quasiparticle cannot be created alone but in multiplets~\cite{broholm2020quantum}. The non-local order does not break any spin rotational or translational symmetry, thereby resulting in no Bragg peaks, which makes
it difficult to identify the ground state of QSLs using standard
probes.

\begin{figure*}[t]
\includegraphics[height=307.77475pt, width=520.20218pt]{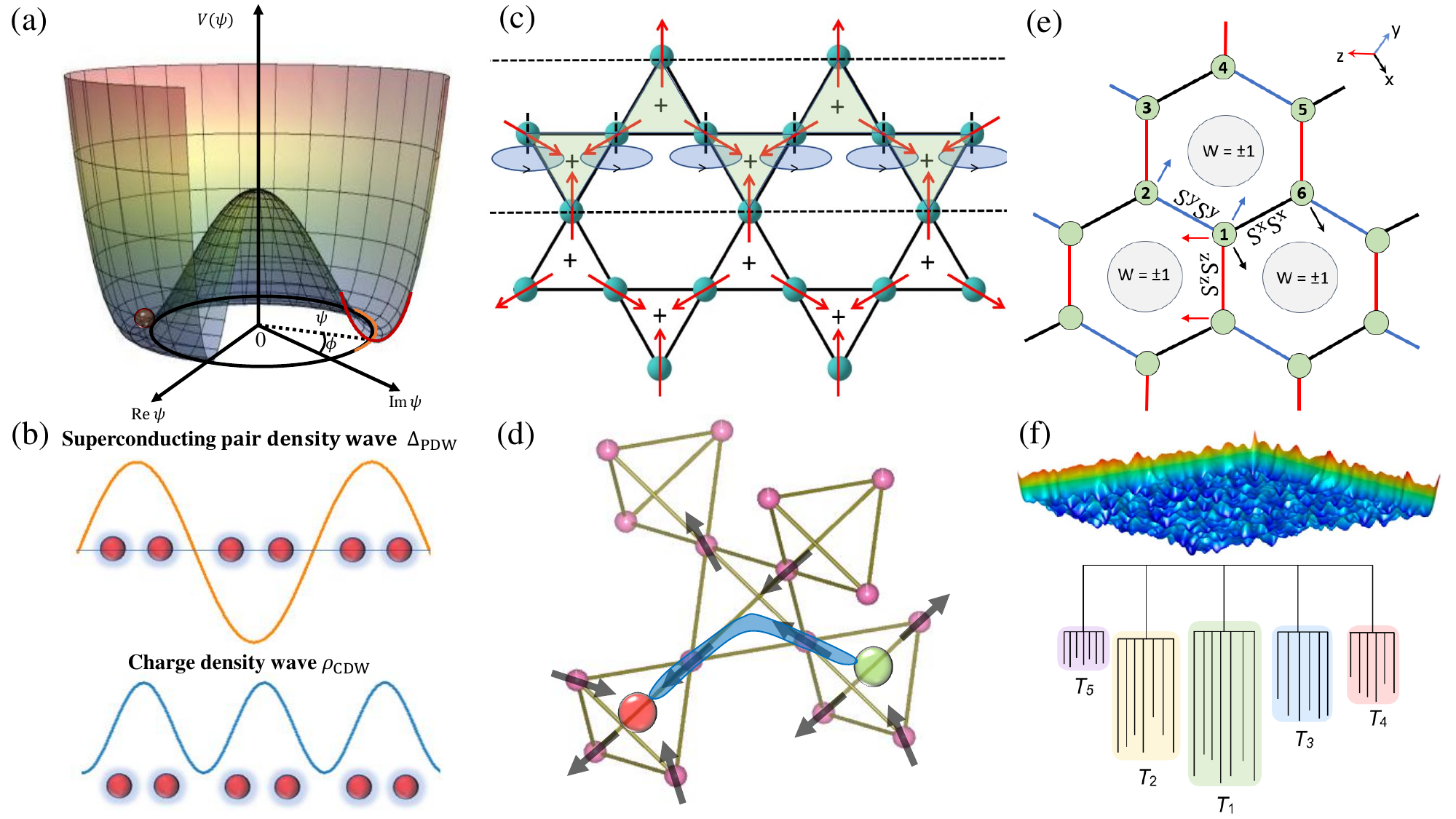}
\caption{\textbf{Symmetry-breaking versus constraint-driven routes to emergent low-energy states in frustrated magnets.} \textbf{(a,b) Symmetry-broken phases.} (a) In systems with spontaneously broken continuous symmetries, the energy landscape often resembles a ``Mexican-hat" profile, where low-energy, gapless excitations (orange path) slide tangentially along the ring of degenerate states, analogous to Goldstone modes. Radial distortions (red path) involve amplitude fluctuations associated with finite energy gaps. (b) Representative examples of ordered states characterized by a nonzero order parameter, such as spin-density-wave and Cooper-pair condensates, illustrating spontaneous breaking of continuous symmetries. \textbf{(c–e) Constraint-driven phases.} (c) Kagome lattice showing the formation of open spin fold excitations in the ground states of $q=0$ spin configurations, illustrating the emergence of spin folds as quasi-localized topological defects.   
        (d) Pyrochlore Ising lattice depicting topological defects as magnetic monopoles connected by Dirac strings, characteristic of spin ice systems. (e) Kitaev honeycomb lattice with bond-dependent Ising interactions, where $S = 1/2$ spins form conserved $\mathbb{Z}_2$ fluxes around each hexagon, representing emergent topological charges. 
        (f) Nonhierarchical energy landscape of a spin jam, characterized by a broad, nearly flat bottom, representing the coexistence of numerous metastable states separated by nonhierarchical barriers, indicative of jammed, non-ergodic dynamics. Source: (f) adapted from~\cite{samarakoon2016aging} with permission from PNAS.
    }
  \label{figure1}
\end{figure*}

Early studies on disordered metals, such as CuMn and AuFe, where Mn or Fe magnetic elements are doped onto nonmagnetic hosts, have revealed unusual spin glass behavior~\cite{PhysRev.106.893,PhysRevB.6.4220,mydosh1993spin}. Below a characteristic temperature $T_g$, the magnetic moments freeze in a random manner without spatial coherence below $T_g$. With time, the system evolves from one metastable state to other by crossing the thermal energy barrier~\cite{edwards1975theory}. The complex nature of spin glasses has attracted extensive study in order to understand the non-equilibrium states of matter. Understanding spin glasses exemplifies the deep curiosity in physics to solve many fundamental problems~\cite{stein2013spin, RevModPhys.58.801}. Whether the spin glass transition is a true phase transition remains ambiguous to date. However, in spin glasses, there can be two fundamental classes of excitations: (i) spin-wave modes and (ii) barrier modes. The spin wave, in general, is the small-amplitude fluctuations around the equilibrium state and the barrier mode excitation involves large-amplitude motions required to overcome the metastable energy barriers~\cite{VILLAIN1980105, PhysRevB.16.2154, jena2025nature, PhysRevB.79.140402}. In canonical spin glasses, a small amount of magnetic elements, such as Mn or Fe, is doped in a nonmagnetic metal host. Here, the doping of magnetic moments is below the percolation threshold. In these systems, the interactions between magnetic moments are mediated by conduction electrons via the long-range RKKY (Ruderman–Kittel–Kasuya–Yosida) interaction~\cite{RevModPhys.58.801,mydosh1993spin,edwards1975theory}. The oscillatory nature of the RKKY interaction accounts for the coupling between magnetic moments to have random exchange signs, resulting in frustration in these metallic systems. In these disordered systems, excitations known as barrier modes are not well-defined because their precise nature remains unclear~\cite{stein2013spin}. Beyond the percolation threshold, where magnetic elements become densely populated, the behavior becomes more complex. Many frustrated magnets exhibit spin-glass-like freezing even with minimal quenched disorder. Here, the weak exchange randomness arising from lattice defects or strain fields can disrupt long-range spatial coherence and stabilize a frozen ground state~\cite{PhysRevLett.98.157201,PhysRevB.89.054433,PhysRevLett.127.017201,PhysRevB.101.024413}. The geometric frustration gives rise to a macroscopically degenerate ground state. Associated with magnetic frustration, the emergent degrees of freedom can arise. In the kagome frustrated antiferromagnet, the spin folding resulting from the degeneracy can give rise to topological constraints~\cite{chandra1993anisotropic,PhysRevB.47.15342}. The folding of spins gives rise to slow spin dynamics. This slowing down of spin dynamics is also evidenced in other frustrated systems due to the gradual slowing down of $Z_2$ fluxes in Kitaev magnets~\cite{PhysRevB.108.165118}, the proliferation of $Z_2$ vortices in triangular antiferromagnets~\cite{kawamura1984phase,kawamura2010chirality}, and the movement of Dirac strings in classical spin ices (see Fig.~\ref{figure1}(b)-(e))~\cite{jaubert2009signature,PhysRevResearch.4.033159}. However, the weakly disordered and disorder-free spin glass may fall into the same category. The theoretical study indicates that a small amount of quenched disorder interacts with intrinsic topological degrees of freedom, enhancing or stabilizing the frozen state~\cite{PhysRevLett.114.247207,PhysRevB.101.024413}. 

In the limit of clean or minimally disordered spin glass Mott insulators, spontaneous symmetry breaking may lead to the emergence of Goldstone modes~\cite{syzranov2022eminuscent, PhysRevB.64.094436, VILLAIN1980105}. For this, the underlying Hamiltonian needs to preserve the rotational symmetry~\cite{PhysRevB.16.2154, fischer1980ferromagnetic}. These low-energy Goldstone modes in frustrated spin glasses are associated with a topological origin that is distinct from those observed in canonical spin glasses~\cite{chandra1993anisotropic, klich2014glassiness,  samarakoon2016aging}. Experimental observations reveal the presence of a ``spin jam'' state in geometrically frustrated systems, where frustration-induced quantum fluctuations and minimal quenched disorder to stabilize a frozen yet complex ground state~\cite{ yang2015spin, samarakoon2016aging, samarakoon2017scaling, jena2025nature}. This highlights how geometric frustration, along with topological constraints, plays a key role in driving unconventional glassy behavior. As we categorize ordered magnets as spin solids and disordered ones as spin liquid, we can classify these frustrated spin glasses as ``spin nematic glasses'' or ``topological spin glasses'', where frustration yields short-range spin correlations driven glassy state plus soft modes from underlying emergent gauge fields. In an analogy to topological spin glasses, a recent study on glasses by Wu et al. showed that the structural defects can be revealed by the topology of the vibrational eigenmodes instead of the spatial structure alone~\cite{wu2023topology,baggioli2023topological}. The findings reveal that plasticity in glasses has a topological origin. The color map in Fig.~\ref{figure1}(f) indicates the spatial distribution of negatively charged topological defects arising from the vibration eigenmodes, and the white crosses mark the location of plastic events. A strong spatial correlation is observed, where plastic events occur in regions with high densities of negative defects. These regions are prone to mechanical instability. Just as negative defects predict plastic events, similarly, the topological defects in frustrated magnets can deform the smooth energy landscape in the proximity of QSLs.

While canonical spin glasses have been widely studied and reviewed over the years~\cite{RevModPhys.58.801,mydosh1978spin}, the idea of topological spin freezing has received far less attention. In this review, we aim to provide a comprehensive overview of unconventional spin freezing of topological origin observed in  promising different magnetic lattices. This review sheds deep insights into some of the key  experimental signatures and the role of external perturbations  that are essential for the design, growth, and investigation of novel quantum materials to realize exotic phenomena with potential technological relevance. For the reader's convenience, we first revisit the phenomena of symmetry breaking, ergodicity, and the canonical spin glass. We then examine the hydrodynamic theory and the spin-jam picture, both of which provide complementary insights into exotic collective excitations in the topological spin freezing state of a broad class of promising frustrated magnets. Next, we review complementary experimental probes, including thermodynamic measurements, as well as microscopic techniques such as nuclear magnetic resonance (NMR), muon spin resonance ($\mu$SR), and inelastic neutron scattering (INS), which reveal both the static and dynamic aspects of topological spin freezing over a broad frequency window. We conclude the review with prospects for topological spin freezing, including technological relevance, future research directions, strategies for the design and investigation of novel frustrated quantum materials to address challenges in contemporary condensed matter and materials science, and potential applications. 

\vspace{-1em}
\section{Theoretical Framework}
\subsection{Symmetry and Ergodicity Breaking in Complex Spin Systems}
 
The phase transition in most condensed matter systems is associated with the spontaneous breaking of continuous symmetry~\cite{anderson2018basic}. The breakdown of a continuous global symmetry gives rise to gapless collective modes known as Nambu–Goldstone modes in conventional magnets~\cite{PhysRevLett.4.380, PhysRev.127.965}. This low-energy excitation, the Nambu–Goldstone mode, traverses the continuous manifold of symmetry-related degenerate ground states and dominates the low-temperature physics of complex many-body systems. A quintessential visualization of this phenomenon can be represented through a ``Mexican hat" potential energy landscape with a continuous ring of minima (Fig.~\ref{figure1}(a))~\cite{weinberg1995quantum}, which captures the essential physics: when the energy is minimized along the brim, the energy cost  of spin fluctuations becomes vanishingly small in that direction and gives rise to Goldstone modes, while radial direction (amplitude) spin fluctuations correspond to gapped and cost finite-energy, often called Anderson-Higgs modes~\cite{PhysRev.130.439,PhysRevLett.13.508}. The Mexican-hat potential as shown in Fig.~\ref{figure1}(a) represents the free-energy landscape of a complex order parameter, $\mathbf{\psi}=|\mathbf{\psi}| e^{i\phi}$, where $|\mathbf{\psi}|$ corresponds to amplitude (Higgs) fluctuations and $\phi$ is the phase (Goldstone) degree of freedom. In condensed matter physics, the Higgs mode emerges in a variety of symmetry breaking states such as superconductors~\cite{PhysRev.130.439,sherman2015higgs,shimano2020higgs}, superfluids~\cite{PhysRevLett.109.010401}, charge density waves (CDW)~\cite{PhysRevB.89.060503,PhysRevLett.122.127001,wulferding2025magnetic}, and uniaxial antiferromagnets~\cite{jain2017higgs,PhysRevLett.119.067201}. Unlike the relativistic version in particle physics, the Higgs mode in condensed matter physics is difficult to detect experimentally. This is because the Higgs mode is a scalar excitation with no spin, charge, or other quantum numbers, so it does not couple linearly to electromagnetic fields, which is why it is difficult to observe using conventional spectroscopic techniques. With advances in experimental techniques, it has become possible to detect the Higgs mode using intense terahertz pulses in superconductors, as well as through nonlinear processes such as Raman scattering or pump–probe spectroscopy~\cite{shimano2020higgs,sherman2015higgs,PhysRevB.101.220507,PhysRevB.100.165131,glier2025non}. In CDW, Fermi surface nesting driven by electron-phonon coupling distorts the lattice and breaks translational symmetry (Fig.~\ref{figure1}(b)), resulting in the opening of a gap at the Fermi level~\cite{RevModPhys.60.1129}. A pseudo-vector-like Higgs mode was identified through quantum pathway interference in Raman scattering in RTe$_3$ (R = La, Gd) CDWs, which carries finite angular momentum arising from inter-orbital electronic coupling~\cite{wang2022axial}. Similarly, in pair density wave (PDW) superconductors, the amplitude of the superconducting pairing $|\mathbf{\psi}|$ oscillates periodically in space so that the spatial average of the superconducting order becomes 
zero~\cite{agterberg2020physics}. This unconventional superconductor hosts two distinct Higgs modes corresponding to in-phase and out-of-phase oscillations of the modulated order parameter, whereas a single Higgs mode exists in a conventional superconductor~\cite{PhysRevB.95.214502}. In certain low-dimensional antiferromagnets near the quantum critical point, the easy-axis anisotropy can create a magnon gap such that the magnon mode lies in higher energies and the Higgs mode remains stable by preventing its decay into spin wave modes~\cite{Su2020StableHiggs}. In frustrated quantum spin ice systems, the Higgs modes are proposed to emerge when the deconfined Coulomb phase transitions to an ordered magnetic state via the condensation of magnetic monopoles~\cite{chang2012higgs,PhysRevLett.124.097203,wulferding2023collective}. Experimentally, this is evident from the disappearance of gapless pinch-point correlations and the appearance of gapped collective modes in neutron spectra. Recent theoretical studies suggest that the Higgs phase behaves like a spin glass in Higgs-gauge theories where global custodial symmetry is spontaneously broken~\cite{PhysRevD.101.054508,PhysRevD.105.034516}.

In general, conventional phases of matter are organized by an order parameter manifold arising from spontaneous symmetry breaking. However, in frustrated magnets, the degeneracy is not merely due to global spin rotation but due to local constraints imposed by lattice geometry and interaction anisotropy, which select particular non-collinear or spin-liquid configurations and often give rise to emergent gauge structures and modified Goldstone spectra~\cite{balents2010spin,PhysRevE.104.034131,PhysRevB.101.024413,khuntia2020gapless,takagi2019concept}. As a result, the phase space of a frustrated spin system is fragmented into distinct regions which are separated by topological defects~\cite{lhotel2020fragmentation,PhysRevB.71.064421,canals2016fragmentation,sorokin2019topological,castelnovo2008magnetic}. These defects are vortices, $Z_2$ fluxes, spin folds, or monopoles, and their physical nature depends on the underlying symmetry and dimensionality (Fig.~\ref{figure1}(c-e)). In such systems, the energy landscape evolves from a simple Mexican-hat form to a complex, nonhierarchical structure (Fig.~\ref{figure1}(f)) with many degenerate states. In Fig.~\ref{figure1}, we have schematically illustrated the progression from symmetry-breaking phase to constrained manifold in various frustrated spin systems and eventually to non-hierarchical energy landscape of topological spin glasses whose details are addressed in subsequent sections. The above studies exemplify how emergent order arising from a change in the material's underlying topology or symmetry can give rise to new collective modes.

In conventional ferromagnets, the rotational symmetry SO(3) is broken spontaneously. As a result, a low-energy excitation is produced, known as magnons, which manifest as the Goldstone modes of the ordered state. The dynamics of these modes can be modeled through the precessional motion of spins and the Fourier-transformed spin field $\mathbf{S}_{\mathbf{k}}(t) = \int d^3 r \, \mathbf{S}(\mathbf{r}, t) e^{i \mathbf{k} \cdot \mathbf{r}}$,  which evolves with a well-defined frequency spectrum. The coherent excitations are directly probed by neutron scattering, which reveals sharp and dispersive modes that are characteristic of magnetically ordered systems. In contrast, topological spin glasses challenge this conventional characterization. Instead of having sharp Bragg peaks in neutron scattering, they show broad, continuum-like excitations. These systems possess a small fraction of frozen moments embedded within a predominantly dynamic spin background~\cite{PhysRevLett.127.017201, PhysRevLett.114.247207, samarakoon2016aging}. Topological spin glasses show unconventional low-energy excitations without long-range magnetic order, as revealed by specific heat, INS, and NMR. Experimentally finding these unconventional excitations motivates deeper theoretical investigation with appropriate models~\cite{RevModPhys.58.801, samarakoon2016aging, chandra1993anisotropic}. 

Halperin and Saslow (HS) provided an early framework that states that spin glasses can host collective low-energy modes, analogous to Goldstone modes, even without conventional magnetic ordering~\cite{PhysRevB.16.2154}. These Goldstone modes originate from the frozen but spatially random local magnetization of the system, which possesses a finite spin stiffness and supports long-wavelength excitations associated with symmetry breaking~\cite{PhysRevB.16.2154}. Unlike magnons in regular ordered magnets, the eigenvectors of collective excitations in spin glasses are not simple plane waves; instead, they are spatially disordered superpositions of Fourier components. The long-wavelength averaging behavior exactly mirrors that of the acoustic phonons in structural glasses, where collective vibrational coherence survives in atomic-level disorder~\cite{VILLAIN1980105}. Models such as the
isotropic Mattis spin glass confirm that the frequency of
these excitations scales linearly with wavevector \(k\), which is consistent with
the HS framework~\cite{mattis1976solvable}. The HS theory generalizes the Goldstone theorem to disordered glassy systems, where the necessity of broken symmetry survives even in the absence of long-range order. These modes in spin glasses are called HS modes, which encode not only the topology of the energy landscape~\cite{chandra1993anisotropic, klich2014glassiness} but also the emergent rigidity in glassy states~\cite{PhysRevB.16.2154}.

Understanding the link between symmetry breaking and spin glass behavior necessitates revisiting Anderson’s notion of generalized rigidity~\cite{anderson2018basic, anderson1997concepts}. The rigidity explains how a force applied at one point in a solid that propagates coherently across the entire system, which extends over distances many orders of magnitude larger than the range of microscopic interactions. This also explains why solids resist shear, why magnets exhibit collective spin alignment that results in long-range ordered magnetic phases, and why supercurrents flow persistently without dissipation.

Though the glassy systems do not exhibit conventional order, the nature and origin of rigidity remain elusive~\cite{stein2013spin}, and this long-standing puzzle invites revisiting in the context of frustrated magnets, where unconventional spin glass behavior is observed in the presence of minimal disorder or in disorder-free conditions. This unconventional spin glass behavior in a disorder-free system gives a hint at hidden and topological form of rigidity that does not rely on crystalline order~\cite{syzranov2022eminuscent, PhysRevB.47.15342, klich2014glassiness}. 

While the symmetry breaking concept provides us with an understanding of the emergent phases of matter, the dynamical counterpart of these phases is governed by the principle of ergodicity. The assumption deeply rooted in equilibrium statistical mechanics, the concept underlying ergodicity asserts that the time averages are equivalent to the ensemble averages~\cite{aleksandr1949mathematical}. This means that if one waits a sufficient amount of time, a system will explore all accessible microstates consistent with its energy. This ergodicity assumption fails dramatically in spin glass below the freezing temperature $T_g$~\cite{palmer1982broken}. The spin glass system becomes trapped in a subset of configuration space in the energy landscape, thus it is unable to fully explore its phase space within experimentally relevant timescales-the ergodicity is broken~\cite{RevModPhys.58.801}. This ergodicity breakdown arises due to a complex free energy landscape, which consists of enormous metastable minima separated by finite energy barriers~\cite{mezard1987spin}. As a result, the system exhibits aging and memory effects and loses self-averaging~\cite{PhysRevB.38.373}. Above the freezing temperature $T_g$, thermal fluctuations are sufficient to overcome the energy barrier, and the system behaves as a paramagnet, thereby restoring ergodicity. Below $T_g$, the system becomes arrested in one of the metastable minima and the relaxation time of the system diverges so that the macroscopic observables no longer reflect equilibrium properties but instead probe a frozen snapshot of the system’s phase space~\cite{RevModPhys.58.801}. The phenomenon of ergodicity breaking in spin glasses reflects a broader class of non-ergodic phenomena in quantum many-body systems, where interactions among spins fail to ensure thermalization and instead confine the dynamics to restricted regions of Hilbert space—setting the system apart from diverse settings such as many-body localization and quantum scars in Rydberg atom arrays~\cite{palmer1982broken,nandkishore2015many, basko2006metal,bluvstein2021controlling}.
As disorder and frustration inhibit thermal relaxation, the spin glass does not explore its phase space fully and thereby stabilizing emergent non-ergodic phases. The non-ergodic phases challenge the conventional thermodynamic framework and demand new approaches to elaborate their out-of-equilibrium dynamics and statistical behaviors. In conventional second-order phase transitions, such as the paramagnet to ferromagnet transition at the Curie temperature \(T_C\), the free energy can be expressed in the Landau paradigm as:~\cite{ginzburg2009theory}
\begin{equation}
F = F_0 + a(T) M^2 + b M^4 + \dots,
\end{equation}
where \(a(T) \sim (T - T_\text{C})\), \(F_0,~b > 0\) and \(M\) is magnetization as the order parameter of the system with time-reversal symmetry. Below Curie temperature \(T_\text{C}\), the system undergoes spontaneous symmetry breaking and selects one of two degenerate minima at \(M = \pm M_0\), which leads to macroscopic magnetization aligned either up or down. Although this selection of the system divides it into two disjoint subspaces in phase space, ergodicity is still preserved within each subspace, where the system fully explores microstates consistent with the chosen magnetization ~\cite{venkataraman2012beyond}.  Unlike ferromagnets, the spin glass states do not possess a clear symmetry-related phase space structure and do not partition the phase space into well-defined ergodic components. Instead, the system becomes dynamically confined within a local basin in configuration space, unable to explore the full ensemble even over experimental timescales~\cite{mezard1987spin,RevModPhys.58.801}. 

The breakdown of ergodicity does not happen only in disordered magnets. Instead, there occurs a broader class of non-equilibrating many-body phenomena, including many-body localized phases \cite{nandkishore2015many}, quantum many-body scars \cite{turner2018weak}, and discrete time crystals \cite{PhysRevLett.116.250401}, in which each state is stabilized by hidden constraints that prevent conventional thermalization. Rather than focusing solely on ergodicity breaking, a unified perspective can consider how field theory and statistical mechanics encode such constraints. For instance, the emergence of topological defects in symmetry-broken phases gives rise to the rugged free-energy landscapes of spin glasses~\cite{zurek1985cosmological,peskin2018introduction}.
When such a system is cooled rapidly, distinct regions of phase space may fall into different minima, resulting in topological defects such as domain walls, vortices, or monopoles whose nature depends on the symmetry and dimensionality (Fig.~\ref{figure1}). Analogous to spin glasses, below the freezing temperature $T_g$ during cooling, the system breaks into many metastable configurations exponentially, which can be interpreted as the emergence of ``defect-like" structures in configuration space.  These structures are not topological in real space, but they are topologically protected in phase space, where the system cannot transition between two states without crossing large free energy barriers. A bridge between glassy dynamics and field-theoretic descriptions of symmetry breaking hints at the deep structure behind the rugged free energy landscapes, which is responsible for ergodicity breaking in the same way. The glassiness in disordered magnets, defect formation in field theories, or localization in quantum matter can be viewed through a common lens of constrained dynamics and fragmented phase space.

\begin{figure}[t]
		\begin{center}
			\includegraphics[height=193.80319pt, width=244.79468pt]{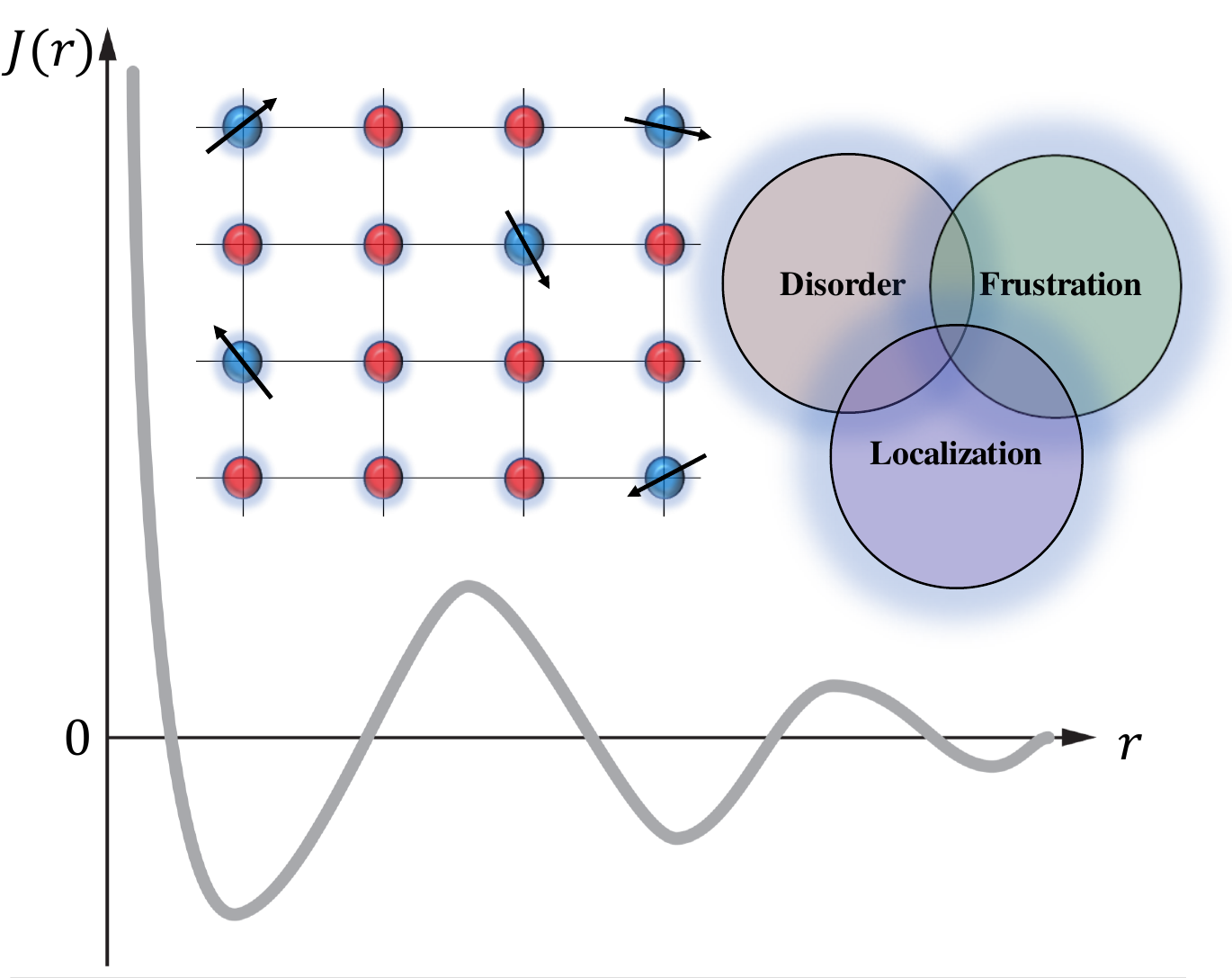}
			\caption{\textbf{The Ruderman–Kittel–Kasuya–Yosida (RKKY) interaction}. The RKKY interaction oscillates in sign and magnitude as a function of interatomic distance, leading to competing ferromagnetic and antiferromagnetic couplings between dilute magnetic moments. These randomly distributed interactions generate magnetic frustration. The inset shows a lattice with dilute magnetic atoms embedded in a nonmagnetic matrix, illustrating the presence of quenched disorder and spatial inhomogeneity. The interplay between magnetic frustration, disorder, and localization gives rise to spin freezing in metallic spin glasses. Adapted from the ref.~\cite{stein2013spin} with permission.}
			\label{fig: RKKY}
		\end{center}
	\end{figure}
    
\subsection{Canonical Spin Glass}

Complexity in nature is an inevitable phenomenon that prevails everywhere. The general definition of complexity is challenging to articulate, although key features include nonlinearity, randomness, localization, and hierarchy. Most of the emergent phenomena that occur around us involve complexity. Therefore, understanding the nature of complexity is crucial to comprehending the fundamental laws of nature. For more than half a century, the study of canonical spin glass has revealed the complexity of the disordered systems in condensed matter physics ~\cite{RevModPhys.58.801}, providing valuable insights with significant applications across the disciplines, such as neural networks~\cite{dotsenko1995introduction}, computer science, and environmental science~\cite{stein2013spin}. Spin glass physics originated from early studies of magnetic properties and resonance effects in diluted magnetic alloys. In canonical spin glasses such as AuFe or CuMn, a small fraction (less than a few percent) of magnetic elements like Fe or Mn resides on a non-magnetic substrate~\cite{PhysRev.106.893,arrott1965neutron,gonser1965magnetic,PhysRevB.1.349, PhysRevB.6.4220}. The Kondo effect is a phenomenon observed in these alloys at low concentrations of magnetic impurities, where the conduction electrons form cluster singlets with localized magnetic moments, which is reflected in the electrical resistance~\cite{kondo1964resistance}. As the concentration of magnetic impurities increases beyond a certain threshold, the system undergoes a transition into a spin glass state. 

In canonical spin glasses below the freezing temperature $T_g$, each spin develops a nonzero local magnetization (\(m_i = \langle S_i \rangle \neq 0\)), even though the total magnetization of the sample remains zero (\(M = \sum_i m_i / N = 0\), where \(N\) is the total number of moments. The localized moments interact via the long-range RKKY interaction~\cite{PhysRev.96.99,kasuya1956theory,PhysRev.106.893}. The RKKY interaction is an indirect exchange coupling mediated by conduction electrons that become spin-polarized through their interaction with localized magnetic moments. Its interaction strength \(J(r)\) varies with distance \(r\) as   \(J(r) \propto \cos(2k_\text{F} r + \varphi) / (k_\text{F} r)^3\) where  \(k_\text{F}\) is the Fermi wave vector and \(\varphi\) is a phase factor (Fig. \ref{fig: RKKY}). The RKKY interaction requires itinerant electrons to mediate the exchange coupling, so it is highly unlikely to occur in Mott insulator spin glasses, where strong electron correlations localize electrons and suppress conduction. However, in some Mott insulators, similar indirect exchange interactions can occur via other mobile excitations. For example, in Mott insulator spin glass SrLaGaRuO$_6$ ~\cite{yatsuzuka2022spin}, RKKY-like interactions can arise that are mediated by mobile excitonic states such as $J_\text{eff}=1$ excitons rather than conduction electrons. Conversely, insulating spin glasses such as Eu\(_x\)Sr\(_{1-x}\)S and Eu\(_x\)Sr\(_{1-x}\)Te, exhibit frustration due to competing ferromagnetic and antiferromagnetic interactions among nearest and next nearest neighbors ~\cite{tholence1979spin,maletta1982magnetic,kobler2001impact}.

Here, we discuss the early models proposed for understanding the mechanism behind the canonical spin glass. In 1975, Edwards and Anderson (EA) proposed a short-range Ising Hamiltonian \cite{edwards1975theory}: $\mathcal{H}_\text{EA} = \sum_{\langle i,j \rangle} J_{ij} S_{i} S_{j}$, where \( S_{i} = \pm 1 \) represents spin variables, and \( J_{ij} \) denotes the exchange interaction strength between spins \( S_i \) and \( S_j \). In the EA model, interactions are limited to nearest neighbors in a regular lattice, resulting in \( O(N) \) coupling terms for a system of \( N \) spins. The quenched disorder is introduced by random exchange parameters \( J_{ij} \), which follow a Gaussian distribution with zero mean and variance \( J^2 \):
$ P(J_{ij}) = \frac{1}{\sqrt{2\pi J^2}} \exp{\left(-\frac{J_{ij}^{2}}{2J^2}\right)}$. The low-temperature spin-glass phase can be distinguished from the high-temperature paramagnetic state using an order parameter \( q \). This order parameter is defined as the time-dependent correlation of a spin \( S_i \) at time \(t=0\) with itself at a later time \( t \): $ q = \frac{1}{N} \sum_{i=1}^{N} \left< S_i(0) \cdot S_i(t) \right>$. $q$ measures how much a spin ``remembers" its initial orientation or how much overlap to its previous spin configuration as time evolves. This memory in the spin-glass phase persists even over longer times t. As a result, the system gives a nonzero value of $q$, while in the high temperature paramagnetic phase, the system loses this memory over time and $q$ value falls to zero. The EA order parameter in spin glasses reveals temporal ordering, which manifests through broken ergodicity and slow relaxation dynamics, in contrast to the conventional magnet with long-range order. Palmer expanded this framework to non-equilibrium systems by introducing the notion of broken ergodicity~\cite{palmer1982broken}, in which the system becomes dynamically confined within metastable regions of phase space. The EA order parameter ($q_{\rm EA}$) measures over the long-time behavior of local spin correlations in a quenched disordered sample and can be written as: $
q_{\rm EA} \;=\; \frac{1}{N}\sum_{i=1}^N \big[\,\langle S_i\rangle^2\,\big]_{J} \;=\; \frac{1}{N}\sum_{i=1}^N \Big[\,\lim_{t\to\infty}\langle S_i(0)S_i(t)\rangle\,\Big]_{J}$, 
where $\langle\cdot\rangle$ denotes the thermal average and $[\cdot]_J$ is the average over different disorder realizations. At absolute zero temperature, \( q_{EA} \) is 1, indicating completely frozen spins. While above the $T_g$ in the classical Ising spin glass, it drops to zero. Between \( 0 < T < T_g \), \( q_{EA} \) takes a finite value, reflecting partial freezing of spin orientations. The zero-field solution of the EA model predicts a Curie-Weiss-like temperature dependence of the magnetic susceptibility: \( \chi(T) = \frac{1-q}{T} \), assuming \( k_B = 1 \). This susceptibility shows a cusp at $T_g$~\cite{nagaoka2012anderson}, signaling the onset of the spin glass transition. Specific heat calculations within the model suggest a singularity near $T_g$, but experiments on spin glasses usually show a broad maximum in specific heat around \(T_g\). This discrepancy could arise because thermal measurements are often performed over timescales too short for the spins to fully equilibrate in spin glasses~\cite{stein1989spin}. Crucially, the EA model predicts no finite-temperature spin glass transition in one- and two-dimensional systems with short-range interactions~\cite{RevModPhys.58.801,PhysRevB.64.180404}. In three dimensions, the existence of a true thermodynamic spin glass transition remains a subject of ongoing debate, although numerical evidence supports the phase transition~\cite{PhysRevB.32.7384,PhysRevLett.54.928,PhysRevB.53.R484}.  In the limit of infinite spatial dimensions, the EA model maps onto the Sherrington-Kirkpatrick (SK) model, which clearly exhibits a finite-temperature spin glass transition~\cite{PhysRevLett.35.1792}.  Moreover, a finite-temperature transition has been reported in four-dimensional systems~\cite{PhysRevB.32.7384, reed1978high, PhysRevB.53.R484}. The theoretical calculations on EA model predict that the lower critical dimension for the existence of a spin glass phase is $D_c=2.5$~\cite{maiorano2018support}.

The canonical spin-glass phase is characterized by a highly complex landscape of many metastable states that are not equivalent to each other. This unique form of order does not correspond to conventional spatial symmetry breaking but is instead revealed through a detailed distribution of overlaps between different spin configurations~\cite{PhysRevLett.43.1754,parisi1980order}. While each individual spin behaves as a quantum object with fluctuations, the macroscopic glassy state arises due to collective dynamics that become well-defined only in the thermodynamic limit. In the case of spin glasses, it is rather the temporal ordering that is meaningful, describing slow relaxation and aging phenomena in these systems~\cite{RevModPhys.58.801, mydosh1993spin}. Spatial symmetries are spontaneously broken, and spins align steadily in space via dipole moments in the case of conventional magnetic ordering. The spin glass can then be seen as a many-body phenomenon that cannot be explained by its individual components, echoing the concept of `More is Different'~\cite{anderson1972more}. In quantum spin glasses, modeled by Hamiltonians such as $
\mathcal{H} = \sum_{ij} J_{ij} \hat{S}_i^z \hat{S}_j^z + \Gamma \sum_i \hat{S}_i^x,$ the competition between static disorder ($J_{ij}$) and quantum fluctuations ($\Gamma$) leads to quantum tunneling between various metastable states~\cite{rieger2005quantum}. One of the main challenges in investigating canonical spin glasses is understanding the nature of their low-energy excitations. It was proposed in the spin wave theory of these disordered metals that the Goldstone-like modes may survive even in the presence of strong disorder~\cite{PhysRevB.16.2154,dzyaloshinsk1978concept,takayama1978contribution,PhysRevB.21.3708}. However, their direct experimental observation has so far remained elusive~\cite{VILLAIN1980105}. Dzyaloshinskii and Volovik further stressed that the coupling to barrier modes conceals such excitations~\cite{dzyaloshinsk1978concept}. Still, locally rotational symmetry breaking may give rise to the localized collective modes~\cite{takayama1978contribution}. Although spatial inhomogeneity tends to localize the spin waves and limit their mobility~\cite{PhysRev.109.1492}, several extended spin wave modes still persist.

\subsection{Emergent Glassiness in Frustrated Quantum Magnets}

The key features of glassy dynamics are slow relaxation, memory effects, and ergodicity breaking, which have conventionally been linked to quenched disorder. The complexity gives rise to an energy landscape full of metastable states. However, recent studies provide many examples in which glassy behavior arises from intrinsic factors such as geometrical constraints or kinetic limitations that prevent the system from fully accessing its available Hilbert space~\cite{PhysRevLett.94.040402, garrahan2011kinetically, klich2014glassiness}. Before discussing topological spin glasses in frustrated magnets, it is helpful to consider examples where the nature of dynamical arrest occurs. Fracton phases of matter provide a remarkable example where the motion of individual excitations is not restricted due to energy barriers, because of subsystem symmetries and emergent conservation laws~\cite{pretko2020fracton, PhysRevB.101.174204}. Nontrivial topological sectors can lead to disconnected regions in configuration space, making transitions between them dynamically forbidden or exponentially slow. For example, loop constraints in quantum dimer models or emergent gauge charges in spin liquids can effectively pin the system into a glassy state~\cite{castelnovo2009topological}. This reveals a broader principle: glassy dynamics can be an emergent, stable property of constrained many-body systems, where the complexity of the energy landscape arises not from external disorder but from the internal architecture of interactions and symmetries. Many disorder-free glassy systems exhibit a fundamental characteristic: a macroscopically degenerate manifold of low-energy states, which usually arises from geometric or topological frustration~\cite{chandra1993anisotropic}. The energy landscape in these systems consists of many local minima, which keeps the system in metastable states and hinders relaxation. A notable instance is the quantum model on an octahedral spin lattice, wherein frustration combined with kinetic constraints in the Hilbert space, results in dynamically arrested behavior, despite the Hamiltonian's translational invariance~\cite{PhysRevLett.94.040402}. The number of low-energy states that can be reached in both frustrated magnetic systems and constrained quantum models grows exponentially with the size of the system,  \( \mathcal{N}(E \lesssim \epsilon) \sim e^{\alpha N} \). However, quantum tunneling among these states is exponentially terminated, which makes the timescales very different and stops equilibration with accessible experimental or numerical timescales.

Another important mechanism of non-ergodicity is the appearance of local integrals of motion, which break up the Hilbert space into different sectors.  This phenomenon represents a variant of disorder-free localization, similar to many-body localization. It has been shown in constrained quantum lattice models that a proliferation of dynamical symmetries limits the system's evolution~\cite{PhysRevLett.118.266601, PhysRevB.95.155133}. The effective Hamiltonian is compatible with a large set of commuting operators \(\{Q_i\}\), each satisfying \([\mathcal{H}, Q_i] = 0\) for all \(i\). These constraints create an exponentially large number of disconnected subspaces, which makes the dynamics non-ergodic and results in an initial-state-dependent relaxation, exhibiting memory effects similar to those seen in structural glasses.

A system near a quantum critical point (QCP) can exhibit glass-like behavior due to the profound sensitivity of their many-body ground states, even without quenched disorder. This phenomenon can be heuristically connected to the Anderson's orthogonality catastrophe in gapless Fermi systems, demonstrating how many low-energy modes induce ground states highly sensitive to local perturbations~\cite{PhysRevLett.18.1049}. In the language of ground-state fidelity, the overlap between ground states at parameters 
$g$ and $g+\delta g$, $
F = \left| \langle \psi(g) \,|\, \psi(g+\delta g) \rangle \right|$, measures how sensitive the many-body wavefunction is to infinitesimal perturbations. For small $\delta g$, this overlap is controlled by the fidelity susceptibility $\chi_F$,  such that $
F \approx \exp\!\left[-\tfrac{1}{2}\,\chi_F\, \delta g^2 \right]$. The fidelity susceptibility $\chi_F$ shows a strong dependence on system size and scales universally with the critical exponents near a QCP. Even very small changes to the parameters can cause nonperturbative reorganizations of the many-body wavefunction, which breaks adiabatic continuity~\cite{PhysRevLett.99.095701,gu2010fidelity,RevModPhys.83.863}. That enhanced sensitivity can make local changes more powerful and favor long-lived nonthermal behavior, which provides a clean route toward dynamically arrested relaxation. Kinematic constraints, conserved quantities, and topology can break up the Hilbert space into many subspaces that are not well connected with each other, and can lead to disorder-free glassiness~\cite{PhysRevX.10.011047,PhysRevB.101.174204,moudgalya2022quantum}. In this case, local moves can't connect different sectors well, which engender relaxation and memory effects happen slowly, even when the energy density is finite. In numerous physical systems, especially frustrated magnets exhibiting limited local dynamics, proximity to criticality (which increases fragility) in conjunction with these constraints can yield resilient, disorder-free glassy dynamics~\cite{PhysRevLett.94.040402}. These results depend upon various fundamental mechanisms. A QCP by itself does not cause universal glassiness or fragmentation of Hilbert space, instead, the resulting glassy dynamics are determined by the particular microscopic constraints and conservation laws inherent to the system.

Topologically ordered systems host certain local excitations that are intrinsically immobile due to global conservation laws. Mainly in fracton models, quasiparticles may only move collectively or along restricted subspaces, which leads to severely suppressing dynamics~\cite{PhysRevB.101.174204, PhysRevB.103.235133, grover2014quantum}. These kinematic constraints do not arise from energetic barriers of the energy landscape but from the inherent topological structure, which gives the system ultra-slow, glassy dynamics. The dynamical glass phase unable to be resolved through any local operator, which is a key point of a stable dynamical glass phase emerging from topological order~\cite{pretko2020fracton}.
All of these diverse mechanisms reveal that emergent glassiness can originate from deep structural features of the system’s Hilbert space and dynamical constraints, rather than from extrinsic disorder. In frustrated spin glasses, self-generated complexity that destabilizes thermalization leading to non-ergodic phases in either generated by frustration induced degeneracy, topological constraints, local integrals of motion or critical sensitivity. The study of glassiness without disorder prompts a reevaluation of the conventional taxonomy of phases and suggests a unifying framework for ergodicity breaking, rooted in the geometry and topology of the quantum many-body wavefunction. The physics of spin freezing in highly frustrated quantum materials is quite intricate owing to topological order, a complex energy landscape,  massive ground state degeneracy, unavoidable disorder, and non-trivial low-energy excitations that are manifested in thermodynamics and microscopic experiments such as \(\mu\)SR, NMR, and INS. The HS  hydrodynamic framework and spin jam theory offer simple prototype models to elucidate some of the intricacies in frustrated magnetic materials~\cite{PhysRevB.16.2154,PhysRevB.79.140402,jena2025nature}.

\subsection{Hydrodynamic Formalism of Spin Glasses} 

Halperin--Saslow (HS) theory proposed a hydrodynamic model for low-energy excitations in disordered magnets, where the local spin configurations break continuous rotational symmetry, even in the absence of long-range magnetic order~\cite{PhysRevB.16.2154, PhysRev.188.898}. In this framework, a frozen spin configuration with finite spin stiffness provides collective support for the emergence of long-wavelength modes analogous to Goldstone excitations. These modes exhibit a linear dispersion $\omega \sim k$, leading to gapless excitations that arise below the freezing temperature. A key consequence is a characteristic power-law contribution to thermodynamic responses, such as $C_m \sim T^{D}$ for linear modes, yielding $C_m \sim T^{2}$ in two-dimensional systems~\cite{PhysRevB.79.140402}.

Including an anisotropy term in the macroscopic free energy lifts the degeneracy of the spin-wave spectrum: the transverse modes develop a finite energy gap at $k=0$, , and the longitudinal mode can hybridize with the transverse modes, with the degree of mixing determined by the relative orientation of the remanent magnetization, the anisotropy axis, and the applied magnetic field~\cite{PhysRevB.22.1174}. For a simple uniaxial anisotropy of strength $K$, Saslow showed that the gap magnitude scales as
$\Delta \sim \gamma \sqrt{\frac{K}{\chi}},$ so that even weak anisotropy shifts the resonance away from zero frequency. Henley et al. further demonstrated that for Dzyaloshinskii--Moriya--type randomness the effective anisotropy takes the form of a unidirectional ``$\cos\theta$'' term, which leads to three coupled modes at $k=0$ whose frequencies depend sensitively on the angle between the remanent magnetization, the anisotropy axis, and the applied field~\cite{PhysRevB.25.5849}. The eigenvalue problem showed that in some high-symmetry setups, one transverse mode remains gapless while the others acquire finite frequencies. While for general orientations, all three modes become intermixed and exhibit gapped excitation spectra~\cite{PhysRevB.22.1174}. The anisotropy does not just change the spin wave velocity. It also changes the low-energy dynamics in a way that creates finite-frequency resonances and angle-dependent spectral weight that can be measured with ESR or neutron scattering. Andreev et al. \cite{andreev1976macroscopic} and Ginzburg \cite{ginzburg1978microscopic} studied the existence of collective excitations in spin glasses and providede the dynamics of small angular fluctuations around a disordered, frozen spin configuration. Their approach goes beyond macroscopic hydrodynamics but links the low-energy excitations to a microscopic model of randomly interacting spins. The time evolution of spin precession equations, where each spin \( \boldsymbol{S}_i \) is considered as a vector of substantial magnitude \( S \) in the classical limit:
\begin{equation}
\frac{d\boldsymbol{S}_i}{dt} = \sum_j J_{ij} \, \boldsymbol{S}_i \times \boldsymbol{S}_j,
\end{equation}
where \( J_{ij} \) represents the coupling exchange interaction between spins located at \( i \) and \( j \).  Beginning with a stable and quenched spin arrangement  \( \boldsymbol{S}_i^{(0)} \), the dynamics of low-energy excitations can be examined by incorporating infinitesimal angular perturbations \( \boldsymbol{\varphi}_i \) which describe local rotations written as:
\begin{equation}
\boldsymbol{S}_i = \boldsymbol{S}_i^{(0)} + 2 \boldsymbol{\varphi}_i \times \boldsymbol{S}_i^{(0)}.
\end{equation}
Assuming a stable and quenched spin configuration \( \boldsymbol{S}_i^{(0)} \), one considers small deviations \( \boldsymbol{\varphi}_i \) as local angular rotations of the form:
\begin{equation}
\boldsymbol{S}_i = \boldsymbol{S}_i^{(0)} + 2 \boldsymbol{\varphi}_i \times \boldsymbol{S}_i^{(0)}.
\end{equation}
After linearizing the equations of motion with respect to these angular variables, we obtain a second-order differential system:
\begin{equation}
\frac{d^2 \boldsymbol{\varphi}_i}{dt^2} = - \sum_j \Lambda_{ij} \, \boldsymbol{\varphi}_j,
\end{equation}
where \( \Lambda_{ij} \) is a stiffness matrix that depends on the spin configuration and the disorder in \( J_{ij} \). In the long-wavelength limit, the solutions are modulated plane waves that follow a linear dispersion relation of the form $\omega_k = c |k|$, where \( c \) is the spin-wave velocity, which comes from microscopic parameters as $c^2 = \frac{\rho}{\langle \chi \rangle},$ with \( \rho \) being a positive-definite coefficient that relates to the distribution of torques in the system and \( \langle \chi \rangle \) being the average local magnetic susceptibility. These soft excitations arise from the rotational symmetry of the spin configuration, despite the absence of global magnetic order. The presence of such modes in disordered magnetic systems can still leads to coherent, propagating excitations, which complements the microscopic picture of the hydrodynamic theory, offering a first-principles derivation of similar Goldstone-like modes in spin glasses~\cite{ginzburg1978microscopic}. The hydrodynamic framework establishes the essential continuum field theory for the unique topological spin glass and topological order with non-trivial low-energy modes, which directly explains the experimental observation of unusual low-frequency magnetic noise, unconventional thermodynamic properties, and extremely slow spin dynamics.

\subsection{Spin Jam Formalism}

A ``spin jam" is a distinct class of glassy states found in densely populated frustrated magnets, where frustration and quantum effects play a crucial role in stabilizing this state. Unlike canonical spin glasses, where disorder plays a significant role in inducing glassy behavior, spin jams emerge from a
highly ordered, crystalline environment~\cite{yang2015spin}. This state is particularly relevant in transition metal-based frustrated oxides, where the nearest neighbors interact via the Heisenberg interaction. The energy landscape of a spin jam state features a rugged but nearly flat bottom with large degenerate states corresponding to different locally collinear configurations~\cite{PhysRevB.99.054416}. To understand the energy landscape, which is influenced by quantum fluctuations, the Holstein–Primakoff representation of the Heisenberg Hamiltonian: $\mathcal{H} = J \sum_{\langle i, j \rangle} \mathbf{S}_i \cdot \mathbf{S}_j$ to calculate the spin wave. By expanding the square roots to the lowest order in \(1/S\), the Hamiltonian is approximated as a quadratic Holstein–Primakoff Hamiltonian~\cite{klich2014glassiness}:
\[
\mathcal{H}_{HP} = E_{MF} + JS \sum_{\langle i, j \rangle} \left[ h_{(1),ij} a_i^\dagger a_j + h_{(2),ij} a_i^\dagger a_j^\dagger + \text{h.c.}\right].
\]
Here $E_{MF}$ is the mean field energy of the spin configuration. The free energy contributions, including thermal corrections and zero-point energy, are calculated from the following hamiltonian :
\[
F = E_{MF} - k_B T \sum_{k, a} \log \left(1 - e^{\hbar\omega_{k,a} / k_B T} \right) + \sum_{k, a} \frac{\hbar\omega_{k,a}}{2}
\],
where the sum is over the Brillouin zone and $\alpha$ is band index. The second term is the thermal energy corrections, and the third term refers to the zero point energy. For a system in the  long-range ordered states, the quadratic Hamiltonian is diagonalized in momentum space,  while for non-ordered states, a real-space treatment is employed, where symplectic diagonalization is involved to obtain ground state properties. The emergence of low-energy modes, referred as the HS modes, gives a response in the low-temperature thermodynamic behavior. The specific heat \( C_v \) and the imaginary part of the dynamic susceptibility \( \chi''(\omega) \) are described by, $
C_v \propto T^2
$ and $
\chi''(\omega) \propto \omega$~\cite{PhysRevB.79.140402}. These behaviors directly highlight the linearly dispersive nature of HS modes, and their contribution stabilizes the spin jam state, which is in sharp contrast with conventional spin glasses. The freezing temperature \( T_\mathrm{g} \) for the spin jam state is governed by the interplay of the quantum energy cost \( E_{SW} \) to flip a set of spins and the temperature-dependent correlation length \( \xi(T) \) of the system. The relation between \( T_\mathrm{g} \) and \( \xi(T) \) can be expressed as  $T_g \propto \mathcal{F}(E_{SW}, \xi(T))$, where  \( \mathcal{F} \) is a function that captures the dependence of \( T_\mathrm{g} \) on quantum energy scales and system correlations. This formalism allows us to study how \( T_\mathrm{g} \) changes with nonmagnetic doping, which is expected to reduce \( T_\mathrm{g} \) as the distance between the nonmagnetic impurities becomes comparable to \( \xi(T) \)~\cite{klich2014glassiness}.

In frustrated magnets, the underlying spin-interacting Hamiltonian shapes the topology of the energy landscape. In a spin solid, the energy landscape has a global minimum, and the system falls into an ordered state as a result of magnetic phase transition below the critical temperature. However, more complex landscapes with numerous local minima result in the system being trapped in metastable states, which give rise to disordered, glassy behavior. To understand the glassy behavior, the idea of rugged energy landscapes is invoked and extends beyond magnetic systems to other materials like structural glasses, polymers, and neural networks~\cite{dotsenko1995introduction}. Canonical spin glasses such as  CuMn and AuFe have a rugged energy landscape arising from random exchange interactions governed by the RKKY interaction, with the freezing temperature \( T_g \) being comparable to the Curie-Weiss temperature \( |\Theta_{\text{CW}}|\). In contrast, in the spin jam state like SrCr\(_{9p}\)Ga\(_{12-9p}\)O\(_{19}\) (SCGO), the freezing temperature is significantly lower than \( |\Theta_{\text{CW}}| \), which indicates fundamentally different energy landscape, shaped by both frustration and quantum effects, distinguishing spin jam behavior from conventional spin glasses. Both canonical spin glasses and
spin jams exhibit aging and memory effects,
but they differ characteristically. Spin glasses like CuMn show pronounced memory effects
even with relatively short waiting times below $T_g$. In
contrast, spin jams, including material like SCGO, show
more subdued memory effects and slower aging, indicating a flatter and less hierarchical energy landscape
with more uniform ruggedness. A detailed discussion of these distinctions and their physical implications is presented in later sections of this paper.

\section{Experimental Signatures of Topological Spin Freezing}
The experimental realization, detection, and elucidation of non-trivial spin freezing of topological origin in highly frustrated quantum magnets is highly significant and topical, as it shows distinct experimental signatures associated with competing exchange interactions, disorder, and topological excitations that are robust against weak perturbations. Understanding of topological spin freezing, which often coexists with dynamic and cooperative quantum states, is crucial to gain insights into novel quantum states. Probing topological spin freezing via an array of complementary experimental techniques, including thermodynamics, NMR, INS, and $\mu$SR aims to provide deep insights into topological spin freezing characteristics and associated exotic excitations, such as unconventional spin textures and exotic low-energy excitations over a wide range of time scales in frustrated magnets, is pivotal in this context. This sets an attractive platform to address some of the fundamental questions in contemporary condensed matter and potential applications in technologies related to high-density storage and computation.

\subsection{Thermodynamic Characteristics}
\begin{figure*}[t]
\includegraphics[height=236.02365pt, width=408.00156pt]{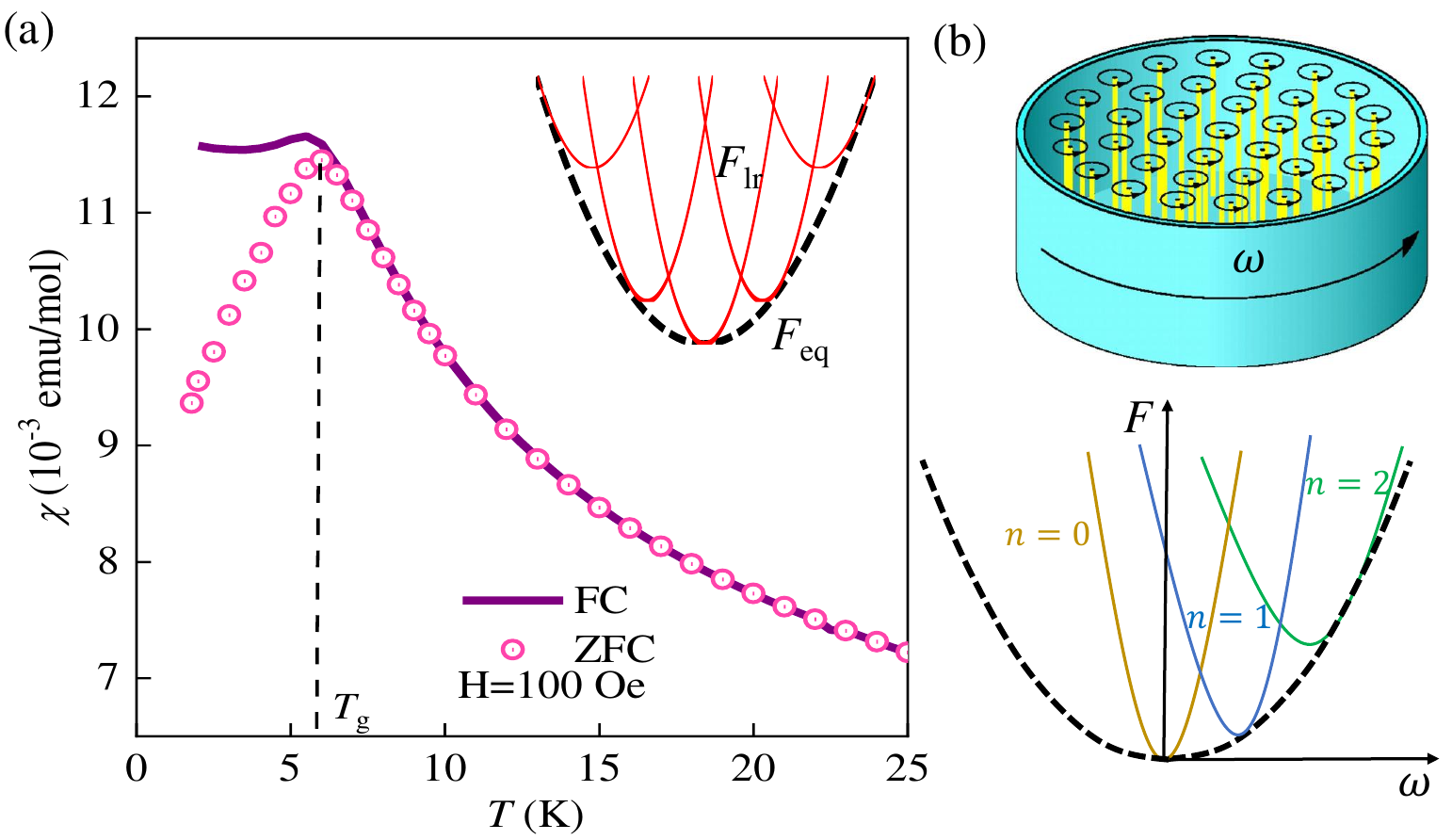}
  \caption{\textbf{Manifestations of broken ergodicity in diverse quantum systems.} 
(a) ZFC–FC bifurcation in the magnetic susceptibility of a disordered Kitaev magnet Li$_2$RhO$_3$ which exhibits topological spin freezing. The inset illustrates a rugged free-energy landscape comprising a multitude of metastable minima (solid lines), representative of frozen yet fluctuating spin configurations. The field-cooled path (dashed line) reflects traversal through a convex envelope of these states, highlighting the absence of full thermal equilibration below the freezing temperature. (b) Free-energy branches in rotating superfluid $^3$He (bottom), where discrete vortex states appear as local minima in angular momentum space (top). The piecewise equilibrium curve reveals first-order transitions as the rotation rate is tuned, signaling metastability and quantization. This analogy emphasizes that broken ergodicity—here governed by topological quantization and conservation laws—can arise even in clean, well-defined quantum fluids. Source:(a) adapted from \cite{PhysRevB.96.094432} with permission from APS ,(b) adapted from \cite{lounasmaa1999vortices,nagaoka2012anderson} with permission from PNAS. }
  \label{Compare_fig 2}
\end{figure*}

The freezing phenomenon can be examined by the temperature dependence of magnetic susceptibility that shows bifurcation measured following zero field cooled (ZFC) and field cooled (FC) protocols below the spin freezing temperature $T_g$ (Fig.~\ref{Compare_fig 2} (a)), analogous to canonical spin glass behavior. In the ZFC mode, susceptibility shows a linear response ($\chi_\text{lr}$) due to thermal fluctuations in a metastable state with a long relaxation time. Below the freezing temperature, the spin state is stable, but the magnetic susceptibility increases asymptotically with temperature. In the FC mode, as the material cools under an applied magnetic field, it approaches one of the lowest free energy states, with susceptibility $\chi_\text{eq}$ representing the average of all ensembles ~\cite{nagaoka2012anderson}. Susceptibilities in ZFC and FC modes can be expressed as:
$
\chi_\text{eq} = \left(\frac{\partial^2 F_\text{eq}}{\partial M^2}\right)^{-1} , \quad \chi_\text{lr} = \left(\frac{\partial^2 F}{\partial M^2}\right)^{-1}
$. Here, $\chi_\text{eq}$ corresponds to the equilibrium free energy $F_\text{eq}$, and $\chi_\text{lr}$ represents the response associated with the metastable free energy $F$. Since the second derivative of free energy $F$ measures curvature, $\chi_\text{eq} > \chi_\text{lr}$. This deviation from linear response arises because in FC conditions the system allows exploration of the entire phase space, whereas in ZFC conditions it explores only a subset, a phenomenon known as broken ergodicity characterized by long relaxation times~\cite{palmer1982broken}. The DC magnetic susceptibility carries information about the equilibrium state of the spin glass with static order parameters. Since the spin glass state is a non-equilibrium state that undergoes continuous relaxation, the AC susceptibility is an appropriate probe for capturing its dynamic nature. The frequency-dependent ac susceptibility ($\chi_\text{AC}=\chi^\prime+i\chi^{\prime\prime}$), measured
below the spin freezing temperature $T_\text{g}$, offers crucial insight
into the complex dynamical behavior of spin glass
systems. 

The real component of it, $\chi'$ corresponds to the in-phase response of the magnetization to an oscillating field and therefore observes reversible and non dissipative processes such as spin alignment and domain wall motion.
However, in glassy magnetic systems, the rapid growth of relaxation time near $T_g$ observed in  $\chi'$  indirectly evidences the establishment of irreversible dynamics manifested by the frequency dependent characteristic cusp. As the probing frequency increases, the observed peak in $\chi'$ shifts towards higher temperatures.
Because at lower applied frequencies the measurement effectively probes slower spin degrees of freedom; hence it allows more spins to equilibrate and freezing occurs at lower temperatures, whereas at higher frequencies only the highly fluctuating spins contribute to the signals hence freezing occurs at relatively higher temperatures~\cite{mydosh2015spin}.
This frequency-dependent characteristic is widely considered an essential criterion to distinguish the glassy magnetic state from the conventional long-range magnetic ordered state. On the other hand, the imaginary part $\chi''$ corresponds to the out-of-phase component and energy loss within the system, displaying a corresponding peak that becomes more prominent at lower frequencies. The frequency dependence of both components of AC susceptibility near $T_g$ is an essential signature of spin glass materials~\cite{mydosh1993spin,binder1986spin}.

A true transition of the spin-glass state can be seen in the measurement of nonlinear magnetic susceptibility, which becomes divergent at $T_g$~\cite{RevModPhys.58.801,mydosh1993spin}. In the spin glass state, the orientation of magnetic moments remains static but randomly distributed, so the spatial average of their pairwise correlations is nearly zero i.e., \( \langle \mathbf{S}_i \cdot \mathbf{S}_j \rangle \approx 0 \)~\cite{PhysRevLett.56.1601}. The corresponding spin-glass susceptibility is defined as \( \chi_{\text{SG}} = \frac{1}{N} \sum_{i,j} \left[ \langle \mathbf{S}_i \cdot \mathbf{S}_j \rangle^2 \right]_{\text{av}} \), which remains finite and positive in the frozen phase and grows strongly as the system approaches the glass transition. This growth signals the development of long-ranged frozen correlations without conventional spin-rotational symmetry breaking, indicating a genuine thermodynamic transition into a glassy state. Experimentally, this critical behavior is not accessed directly through \(\chi_{\text{SG}}\) but through the nonlinear magnetic susceptibility \(\chi_{\text{nl}}\), which quantifies the cubic response of magnetization to an external field. Expanding the magnetization as \( M = a_1 H - a_3 H^3 + \dots \), the coefficient \( a_3 \) (or equivalently \(\chi_{\text{nl}}\)) serves as the measurable analogue of \(\chi_{\text{SG}}\)~\cite{PhysRevLett.64.2070}. According to mean field theory, \(a_3\) shows a sharp enhancement as \(T\to T_g\) which is consistent with the growth of correlated spin domains. This divergence in non-linear susceptibility is of fundamental importance. Whereas a cusp in the linear susceptibility can arise from superparamagnetic blocking or dynamical cluster freezing, phenomena lacking a diverging correlation length. The enhancement of \(a_3\) uniquely signifies the establishment of a true spin-glass transition governed by cooperative freezing of spins.

An insightful analogy to understand the intriguing energy landscape of spin glasses can be considered the behavior of the superfluid $^3$He in a bucket rotating with angular velocity $\omega$, leading to the formation of quantized vortices~\cite{lounasmaa1999vortices}. The free energy of the system $F$ is plotted against angular velocity and it exhibits a denumerable set of metastable valleys~\cite{nagaoka2012anderson} (Fig.~\ref{Compare_fig 2}(b)). Each valley corresponds to an integer number of vortices and is separated from its neighbors by energy barriers associated with vortex nucleation. Each valley refers to a conserved topological quantity called the vortex number, which marks distinct regions in the configuration space of the system. As the angular velocity $\omega$ changes, the system undergoes a sequence of  first order phase transitions leading to abrupt changes in the vortex number, causing discontinuous changes in the angular momentum of the superfluid. This multivalley landscape provides an insightful intermediate case between two extremes: (i) a rotating bucket of superfluid helium, where the central valley corresponds to zero vortex number and a set of valleys corresponds to higher order vortex numbers. Here, the valleys correspond to a discrete number of vortices. (ii) Spin glass, where the distribution of valleys is continuous. A small variation in the angular frequency of the rotating He bucket gives rise to successive first-order phase transitions. Labelling a conserved number for Ising or Heisenberg spin glass to account for the first-order phase transition between continuous valleys still remains an open problem.

Mathematically, the analogy can be formalized by comparing the energy landscapes:
\begin{itemize}
    \item[$\ast$] In superfluid $^3$He, the free energy can be modeled as $ F_n(\omega) = F_0 + \frac{1}{2} I (\omega - \omega_n)^2,$
    where $\omega_n \propto n$ is the angular velocity at which the $n$-vortex state becomes stable, and $I$ is the moment of inertia.
    \item[$\ast$] In spin glasses, the free energy landscape $F[\{\vec{S}_i\}]$ is highly rugged, with minima separated by energy barriers whose topology or cohomology class is not yet well-defined but may encode nonlocal information.
\end{itemize}

This analogy reinforces the broader theme of this review: topological spin glasses may be understood not just in terms of energy barriers and metastability but through an emergent structure of topologically distinct valleys~\cite{chandra1993anisotropic}, possibly governed by conserved or quasi-conserved quantities yet to be identified.

In the specific heat ($C_p$) measurements of unconventional spin glasses, a broad maximum is often detected in the vicinity of the spin freezing temperature \(T_g\), indicating  short-range spin correlations rather than a sharp phase transition. This broad maximum depicts the gradual nature of spin freezing process, where the spins become trapped in a disordered, non-equilibrium state with a wide distribution of energy barriers. In contrast to conventional magnetic phase transitions, which anticipate a pronounced divergence in \(C_p\), spin glasses display a smeared-out anomaly originating from frustration and competing interactions. At temperature below \(T_g\), the specific heat exhibits a linear response with temperature ($C_m\propto T$), which is a hallmark of spin glass behavior as observed in canonical systems such as CuMn, AuFe and PtMn~\cite{PhysRevB.20.368,mirza1985magnetic,kimishima1977specific}. However, the lack of sharp anomaly near \(T_g\) in $C_p$ raises questions about whether the glass transition in canonical spin glass constitutes a genuine phase transition \cite{stein2013spin}. 
The class of antiferromagnetically interacting Mott insulators such as Na$_4$Ir$_3$O$_8$~\cite{PhysRevLett.99.137207}, SrCr$_{9p}$Ga$_{12-9p}$O$_{19}$~\cite{ramirez1990strong}, NiGa$_2$S$_4$~\cite{nakatsuji2005spin} and CoAl$_2$O$_4$~\cite{PhysRevB.88.174415} show peculiar $T^2$ dependence below $T_g$, suggesting unconventional low-energy excitations characteristic of these frustrated magnets. In 2D frustrated spin-lattices, the quadratic temperature dependence of magnetic specific heat ($C_m\propto T^{2}$) can be described by the mean-field excitations `spaghetti mode', characterized by a length scale $L_0$ extending upto an order of 10$^2$ number of spins. It assumes a finite spin-stiffness constant within the spin texture of length scale $L_0$. For a conventional antiferromagnetically ordered system, the rigidity in terms of spin-stiffness constant can be derived as: $\mathcal{D}_0^2= [3\sqrt{3}\zeta(3)/2\pi](ak_B\theta_\text{CW}/\hbar)^2/\ln(2S+1)$~\cite{nakatsuji2005spin}. Here, $\theta_{CW}$ refers to the energy scale of the exchange interaction and $\mathcal{D}_0$ is the spin stiffness associated with long-range ordering in the absence of any magnetic frustration. However, in the presence of magnetic frustration, the spin stiffness is reduced (softened), leading to a suppression of the spin-wave rigidity constant $\mathcal{D}$. The spin stiffness can be quantitatively estimated from the low-temperature magnetic specific heat, which is described by the expression: $\frac{C_m}{R} = \left[\frac{3\sqrt{3}\,\zeta(3)}{2\pi}\right]\left(\frac{a k_B T}{\hbar \mathcal{D}}\right)^2 - \left(\frac{\sqrt{3}\pi}{2}\right)\left(\frac{a}{L_0}\right)^2$
, valid in the temperature regime $T < T_g \ll |\theta_{CW}|$.~\cite{nakatsuji2005spin}

The anomalous features of unconventional spin freezing can be systematically described within the HS framework, which treats the long-wavelength spin fluctuations of glassy magnets in terms of hydrodynamic modes, analogous to transport phenomena in fluids. In this approach, the low-temperature behavior of magnetic susceptibility and magnetic specific heat serves as key diagnostics of the frozen state. The zero-field-cooled magnetic susceptibility $\chi_{m}$ (in the limit $T \rightarrow 0$) together with the scaling of the magnetic specific heat, $C_{m}/T^{2}$, below  $T_{g}$, reveal the presence of two intrinsic energy scales, $E_{1}$ and $E_{2}$. The first energy scale, \(E_1\), is expressed as:
\begin{equation}
    E_1 = \frac{2g^2\mu_B^2 S(S+1) N_A}{z \chi_m} = k_B \tau_1,
\end{equation}
where \(N_A\) is Avogadro's number, \(z\) is the coordination number. This relationship links the susceptibility to \(E_1\), which is associated with the characteristic energy related to the exchange interactions of the magnetic material. The second energy scale, \(E_2\), can be derived from the specific heat \(C_m\) using the relation~\cite{PhysRevB.79.140402, jena2025nature}:
\begin{equation}
    \frac{C_m}{N_A \nu} = \frac{3 \zeta(3)}{\pi d} \frac{k_B^3 T^2}{\hbar^2} \sum_i \left( \frac{1}{c_i^2}\right) - \frac{3 k_B \pi}{L_0^2 d},
\end{equation}
where \(\nu\) is the volume of the unit cell, \(d\) is the spacing between decoupled spin-lattice layers, \(a\) is the bond length between nearest magnetic moments, and \(L_0\) is the characteristic length scale to which the Goldstone modes are well defined. By neglecting the second term and assuming three Goldstone modes, the expression simplifies to:
\begin{equation}
    \frac{C_m}{N_A \nu} = \frac{3 \zeta(3)}{\pi d} \frac{k_B^3 T^2}{\hbar^2} \frac{n_p}{c^2}.
\end{equation}
Substituting \(c = \gamma \left( \frac{\rho_s}{\chi_m} \right)^{1/2}\), leads to:
\begin{equation}
    \frac{\rho_s}{N_A \nu} = \frac{3 n_p \zeta(3)}{\pi d} \frac{k_B^3}{g^2 \mu_B^2} \frac{\chi_m}{C_m / T^2}.
\end{equation}
For in-plane interactions, this results in the energy scale \(E_2\):
\begin{equation}
    E_2 = \frac{\rho_s d}{N_A \nu} = \frac{3 n_p \zeta(3)}{\pi} \frac{k_B^3}{g^2 \mu_B^2} \frac{\chi_m}{C_m / T^2} = k_B \tau_2,
\end{equation}
where \(n_p\) is the number of degenerate hydrodynamic modes. These energy scales, \(E_1\) and \(E_2\), are linked to the exchange interaction energy and spin stiffness, which correspond to distinct temperature scales, the Curie-Weiss temperature \(|\theta_{CW}|\) and the spin freezing temperature \(T_g\), respectively.

Interestingly, the hyperkagome spin lattice Na$_4$Ir$_3$O$_8$ exhibits a quadratic specific heat dependence below 30 K, which remains field independent up to 12 T ~\cite{PhysRevLett.99.137207}. However, despite its 3D structure, the possible reduction in the effective dimensionality of magnetic excitations from 3D to 2D has not been systematically investigated in this material. In contrast, the pyrochlore Y$_2$Mo$_2$O$_7$ exhibits anomalous glassy behavior near 22 K~\cite{greedan1986spin}. In this case, local displacements of O$^{2-}$ ions relative to their ideal crystallographic positions lead to a distribution of Mo–O–Mo bond angles. This, in turn, modifies the superexchange pathways and introduces a distribution of exchange interactions, thereby creating intrinsic interaction disorder~\cite{PhysRevLett.124.087201}. This variation in exchange interactions ($J$) between Mo-Mo pairs intrinsically contributes to the glassy nature, which transforms a flat energy landscape into a rugged one \cite{PhysRevLett.124.087201, yang2015spin}.  Notably, it exhibits a quadratic specific heat dependency on temperature, characteristic of two-dimensional magnon excitations. This observation suggests that orbital degeneracy may effectively reduce the dimensionality of magnetic interactions from three to two \cite{PhysRevB.89.054433, PhysRevLett.124.087201}.

The presence of broad peak in the magnetic specific heat per temperature ($C_m/T$) near $T_g$ suggests the onset of short range spin correlations in these frustrated magnetic systems. This behavior is quite different from isotropic  unfrustrated or weakly frustrated systems where a sharp $\lambda$ type peak is observed, that corresponds to long-range magnetic order. This is suppressed by a critical magnetic field \(H_C(0) \sim k_BT_\text{broad}/g\mu_B\sqrt{S(S+1)}\) \cite{ PhysRevLett.84.2957,nakatsuji2005spin}. However, the presence of broad peak in $C_m/T$ even at higher magnetic fields (\(\geq H_C(0)\)) in the frustrated magnetic systems suggests that it is not associated with long-range magnetic order rather with robust short-range spin correlations, which is a characteristic of the unconventional spin glass state~\cite{PhysRevB.96.094432,PhysRevLett.84.2957}. These short-range spin correlations encounter higher magnetic fields, underscoring the intriguing spin dynamics near the spin glass transitions and unique magnetic correlations  in such systems. These broad maxima underscore the inherent stability of the spin glass state against external perturbations, reflecting pronounced spin fluctuations and spin correlations. Such behavior is further supported by microscopic experiments, discussed in the subsequent sections. In frustrated magnets, the magnetic specific heat exhibits a field-independent $T^n$  ($n=2$ for 2D and $n=3$ for 3D frustrated magnetic systems) behavior below the broad maximum. This behavior can be understood through several interconnected mechanisms as discussed in the following paragraphs: 

1. \textbf{Hydrodynamic Theory and Quasi Two-Dimensional Spin Glasses}: In quasi two dimentional materials exhibiting spin glass behavior, the hydrodynamic theory plays an essential role in elucidating the $T^2$ dependence of magnetic specific heat which persists even under the application of magnetic fields. This theory describes three hydrodynamic modes such as \(\omega_+\), \(\omega_-\), and \(\omega_0\) whose degeneracy is lifted by the magnetic field, resulting in distinct frequency modes~\cite{PhysRevB.16.2154, PhysRevB.79.140402}. Below the spin glass transtion temperature, the quadratic polarization mode \(\omega_-\) compensates for the gapped mode \(\omega_+\), preserving the \( T^2 \) specific heat behavior despite the external field~\cite{PhysRevB.79.140402,jena2025nature}.

2. \textbf{Symmetry Reduction in Noncollinear Antiferromagnets}: In noncollinear antiferromagnetic ordered state, the robustness of the \( T^2 \) specific heat below $T_N$ against applied magnetic fields arises from the reduction of symmetry from \( O(3) \) to \( O(2) \) in the Hamiltonian. This reduction in symmetry leads to a decrease in the number of Goldstone modes from three to one. However, an ``accidental" degeneracy persists, unrelated to the \( O(2) \) symmetry, similar to the zero-field case which leads to a pseudo Goldstone mode. It helps in preserving the $T^2$ dependence of specific heat even under the application of magnetic field. Such modes, resulting from an approximate or weakly broken continuous symmetry, manifest as unexpectedly low-energy excitations across a wide range of systems, from quantum chromodynamics~\cite{PhysRevLett.29.1698, weinberg1995quantum} to high-$T_\text{C}$ superconductors~\cite{PhysRevLett.128.141601} and quantum magnets~\cite{PhysRevLett.121.237201, PhysRevResearch.2.043023}. A well-known example is the generation of pions due to the chiral symmetry breaking which is associated with the pseudo-Goldstone modes. ~\cite{PhysRevLett.4.380, ecker2007chiral, kocic1993universal}. In such cases, the continuous symmetry is weakly and explicitly broken by a perturbation such as a weak applied magnetic field, leading to the opening of a finite energy gap in the Goldstone mode and hence it becomes massive. This scenario is in sharp contrast with systems where local (gauge) symmetry is broken, such as superconductors, and the Goldstone modes do not manifest
as physical excitations. Here, the Goldstone modes are absorbed by the gauge fields through the Anderson-Higgs mechanism leading to a massive gauge bosons~\cite{PhysRev.130.439, PhysRevLett.13.508}. In superconductivity, this corresponds to the acquisition of a finite penetration depth by the electromagnetic field reflecting the Meissner effect and the opening of an energy gap in the electronic excitation spectrum~\cite{PhysRev.130.439, leggett2006quantum}. 

3. \textbf{Short-Range Spin Correlations and Spin Singlets}: Another important contributing factor is the coexistence of glassy behavior with the formation of short lived spin singlets near the spin glass transition temperature $T_g$ ~\cite{PhysRevLett.73.3306,PhysRevLett.84.2957, PhysRevLett.127.157204}. These singlets show a distinctive kind of spin correlation, which appears as an undecouplable Gaussian-shaped relaxation profile in $\mu$SR asymmetry. The lifetime of these singlets is significantly shorter than the muon lifetime, suggesting the ephemeral nature of these correlations\cite{PhysRevLett.127.157204}. The interplay among thermal fluctuations, fluctuation stiffness, and non-analytic temperature dependence gives rise to distinct quantum behavior and thermodynamic signatures that support the quadratic specific heat behavior even under an applied magnetic field. The scenarios outlined above address the field-independent behavior of the power-law exponent \(n\) in the magnetic specific heat (\(C_m \propto T^{n}\)). Where \(n=2\) for two-dimensional spin glasses and \(n=3\) for three-dimensional networks such as hyperhoneycomb spin glasses. Overall, the combined interplay of hydrodynamic modes, symmetry reduction, accidental degeneracy, and short-range spin correlations contributes to this robust behavior.

\subsubsection{Two peak nature of specific heat}
\begin{figure}[t]
		\begin{center}
			\includegraphics[height=193.80673pt, width=244.8032pt]{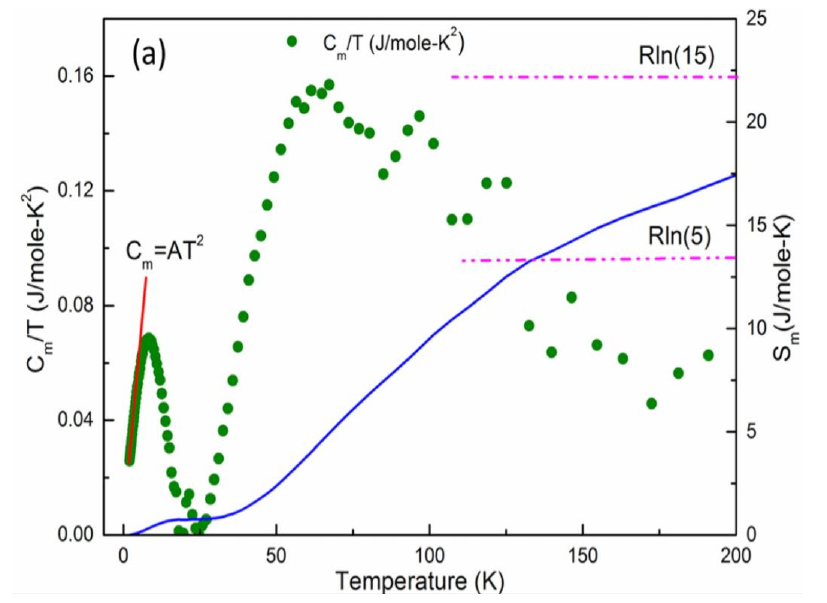}
			\caption{\textbf{Double-peak structure in the specific heat as a signature of complex spin freezing}. The specific heat divided by temperature of the frustrated spinel FeAl$_2$Se$_4$ as a function of temperature exhibits a double-peak structure, indicative of two distinct correlation regimes. Adapted from the ref.~\cite{PhysRevB.99.054421} with permission from APS }
			\label{fig: DoublePeakHC}
		\end{center}
	\end{figure}

Experimental results indicate that there appear two peaks in the specific heat data for some topological spin glasses. The theoretical findings suggest that the dual-peak nature in specific heat could be due to the presence of two energy scales associated with two different types of spin correlations. This two peak nature of specific heat is evident in triangular lattice NiGa\(_2\)S\(_4\), van der Waals chalcogenides Mn$_2$Ga$_2$S$_5$, and spinel FeAl\(_2\)Se\(_4\), etc (see Fig. \ref{fig: DoublePeakHC} and Fig. \ref{fig: NiGa2S4 Chi, Cm, NMR, Neutron}(b))~\cite{shen2022spin,PhysRevB.99.054421, nakatsuji2005spin}. The high-$T$ peak that is observed around the energy scale of exchange energy is due to the onset of quadrupolar spin correlations, which is the spin nematic behavior, and the peak that appears near $T_g$ is due to the dipolar spin correlation. The essential physics arises from the existence of biquadratic exchange interaction, expressed as \((\mathbf{S}_i\cdot\mathbf{S}_j)^2\), which exhibits non trivial behavior for \(S\geq1\). Unlike spin \(S=1/2\) systems, where the quadrupolar moments are absent, high spin \(S >1/2\) can host a quadrupolar tensor defined as 
\begin{equation}
Q_{i}^{\mu\nu}=\tfrac{1}{2}\langle S_i^\mu S_i^\nu+S_i^\nu S_i^\mu\rangle-\tfrac{1}{3}S(S+1)\delta^{\mu\nu},
\end{equation}
which encodes the fluctuation of higher-order moments beyond conventional moments. These degrees of freedom are directly coupled to the biquadratic term, producing an additional channel of correlations that grow at temperatures of the order of the exchange scale \(J\). Microscopically, such terms arise from multi-orbital Hubbard picture or via spin-lattice coupling in these materials.~\cite{PhysRevB.74.092406,PhysRevB.74.092406}.

Theoretical studies~\cite{PhysRevB.79.214436,tsunetsugu2006spin} have shown that when the system is close to the quantum critical point, separating the quadrupolar (spin-nematic) phase from the conventional magnetic phase, two distinct temperature scales typically arise. First, the intermediate scale \(T_Q\) marks the quadrupolar correlations,  and the second is a much lower scale \(T^*\)($\sim T_\text{broad}$) corresponding  to the regime where dipolar correlations dominate. The first temperature scale manifests as the high-temperature peak in \(C_p(T)\), while the latter produces a broad maximum at lower temperatures associated with slow spin dynamics, glassiness or collective excitations such as HS modes. These features do not correspond to true phase transitions in two dimensions which is consistent with the Mermin-Wagner theorem. Instead they represent crossovers governed by the different symmetry regions of the Hilbert space.

In NiGa\(_2\)S\(_4\), the combined study of neutron and NMR experiments confirms that quadrupolar spin fluctuations start to appear well above the freezing temperature \(T_g\)~\cite{PhysRevB.77.054429}. This behavior is consistent with the two peak nature observed in \(C_p\) as shown in Fig.~\ref{fig: NiGa2S4 Chi, Cm, NMR, Neutron}(b). This makes an essential difference between \(S=\tfrac{1}{2}\) and \(S\geq 1\) based topological spin glasses. Systems with \(S\geq 1 \) possess a sufficiently rich Hilbert space to sustain both quadrupolar and dipolar correlations, which leads to multiple entropy-releasing processes, resulting in a characteristic double peak in the specific heat.

\subsubsection{Contrasting nature of specific heat in canonical and topological spin glasses}
Specific-heat measurements on canonical spin glasses show a linear temperature dependence ($C_m\propto T$) near $T_g$. This linear-in-\(T\) behavior can be established by using the  Anderson-Varma-Halperin (AVH) mechanism~\cite{anderson1972anomalous}. In this formalism, low-energy excitations emerge from localized two-level systems (TLS) associated with the collective nature of spin freezing. Here, the system can fluctuate between two nearly degenerate spin configurations that are separated by a small energy barrier. The quantum tunneling between these two metastable states captures the dynamics of spin glasses. As the temperature is lowered down to $T_g$, the density of states arises primarily from the quantum tunneling among these metastable states with minimal energy difference (\( \Delta E \approx 0 \)). For a statistical distribution of localized TLS, the specific heat is given by  $
C_{TLS} = \frac{\pi^2}{6} k_B^2 T n(0), $ where \( n(0) \) represents the density of states. This scenario is predicted to be a universal feature of 'glassy' systems~\cite{anderson1972anomalous}.
The Hamiltonian is given : $
\mathcal{H} = \begin{pmatrix} E_1 & \Delta \\ \Delta & E_2 \end{pmatrix}$, with \( \Delta \propto e^{-V / 2\hbar} \), where \( V \) is the energy barrier. Here, the energies of two metastable states ($E_1$ and $E_2$) are nearly degenerate, and for the energy barriers are sufficiently small ($\Delta E\sim k_\text{B}T$), the tunneling gives rise to zero-point fluctuations that yield the observed $C_m\propto T$ behavior.

The AVH framework assumes a wide distribution of local magnetic fields, \(H_\text{loc}\) which arises from random orientation of magnetic moments present in the system. The distribution of the field facilitates the quantum tunnelling among nearly degenerate spin configurations, giving rise to characteristic low-energy excitations in canonical spin glasses. In contrast, topological spin glasses possess an energy landscape shaped not by random disorder but by significant frustration and quantum fluctuations. As a result, the spins are not trapped in random directions and local magnetic fields remain broadly distributed rather than forming a sharp peak at \(H_{\text{loc}}=0\) which is a defining feature of the canonical systems. The aftermath of this effect is observed in the distinct power law of specific heat which shows a unique density of states associated with collective HS modes~\cite{PhysRevB.16.2154,PhysRevB.79.140402,jena2025nature}. In systems with partial magnetic dilution, where both spin-jam and canonical spin-glass behaviors coexist, the total magnetic specific heat can be modeled as a combination of two components: $
C = f\,C_{\text{HS}} + (1 - f)\,C_{\text{TL}},$
where \( C_{\text{HS}} \) corresponds to contributions from the HS collective modes and \( C_{\text{TL}} \) represents tunneling-like (TL) excitations of localized spin clusters. The factor \( f \) quantifies the relative fraction of the spin-jam component and depends sensitively on the degree of disorder and magnetic dilution~\cite{PhysRevB.109.104420,jena2025nature}. When the magnetic concentration increases beyond the percolation threshold, the collective HS excitations start to dominate over the localized two-level modes. This marks a crossover from a canonical to a topological spin-glass regime, which is clearly observed in the specific heat behavior. The low-temperature specific heat evolves from a linear dependence $C\sim T$ to a power-law form $C\sim T^n$ with $1 < n\leq2$. In the case of $n=2$, typical for two-dimensional frustrated spin glasses, the excitations originate purely from collective Goldstone modes. This crossover is clearly observed in the doped quasi-two-dimensional spin glass 
Ba$_2$Sn$_2$ZnCr$_{7p}$Ga$_{10-7p}$O$_{22}$. For $p \geq 0.93$, the specific heat follows a $T^2$ dependence indicative of HS mode excitations. With increasing  non-magnetic (Ga$^{3+}$) doping, the low-temperature specific heat below $T_g$ shows a hybrid behavior. It contains contributions from both the quadratic temperature-dependent term (\(C_m\propto T^2\)) associated with the HS modes and the linear-in-T component(\(C_m\propto T\)), which corresponds to localized two-level excitations~\cite{PhysRevB.109.104420}. This coexistence signifies the gradual transition from a regime dominated by a collective HS mode to a localized spin glass-like state as the amount of non-magnetic dilution increases.

Zero-point entropy is defined as the residual entropy at absolute zero temperature. In conventional spin glasses, the randomness often leads to a frozen state with a finite residual entropy. Theoretical calculations predict zero-point entropies of about \( S_0 \approx 1.66~\text{J mol}^{-1}\text{K}^{-1} \) for Ising spins and \( S_0 \approx 4.30~\text{J mol}^{-1}\text{K}^{-1} \) for XY systems~\cite{edwards1980ground,tanaka1980ground}. These values are consistent with experimental observations in dilute dipolar spin glasses such as \(\text{LiHo}_x\text{Y}_{1-x}\text{F}_4\)~\cite{PhysRevLett.98.037203}. Unlike canonical spin glass, which has hierarchical energy landscapes shaped by random disorder, the spin jam or topological spin glass exhibits a flat but rugged energy landscape that originates from strong spin correlations and quantum fluctuations. In spin jam systems, the zero point entropy expressed as \( S_{\text{SJ}} \) arises from collective frustration rather than random disorder. Thermodynamic measurements on the quasi-two-dimensional frustrated kagome system Ba\(_2\)Sn\(_2\)ZnCr\(_{7p}\)Ga\(_{10-7p}\)O\(_{22}\) reveal a continuous crossover between these two regimes. For high spin density (\( p \geq 0.93 \)), the magnetic specific heat follows a \( T^2 \) law, characteristic of HS modes, while increased nonmagnetic dilution introduces an additional linear-in-\(T\) contribution from localized two-level excitations~\cite{PhysRevB.109.104420}. From specific heat the total magnetic entropy can be estimated as, $
S(T) = S_0 + \Delta S(T) = S_0 + \int_{T_{\text{base}}}^{T} \frac{C_{\text{mag}}}{T}\, dT,$  
where \( S_0 \) is the zero-point entropy. For \( p = 0.98 \), the experimental observation indicates that about 55\% of the total magnetic entropy remains unreleased at the lowest measured temperature, corresponding to \( S_0 \approx 6.3~\text{J mol}_{\text{Cr}}^{-1}\text{K}^{-1} \)~\cite{PhysRevB.109.104420}. 

In conclusion, in a densely frustrated spin glass, the freezing mechanism is primarily dominated by quantum fluctuations and the short-range spin correlation, which gives rise to Goldstone modes. However, in proximity to or below the percolation threshold, the random occupation of spin moments opens a path to the canonical spin glass nature, as the spins are not well connected in this region. Here, the excitations are not well defined, and the localized two-level excitations dominate the low-temperature behavior.

\subsection{ Signatures of Topological Spin Freezing via Local Probes}

\subsubsection{Nuclear magnetic resonance }
\begin{figure*}[t]
\includegraphics[height=404.99358pt, width=510.0pt]{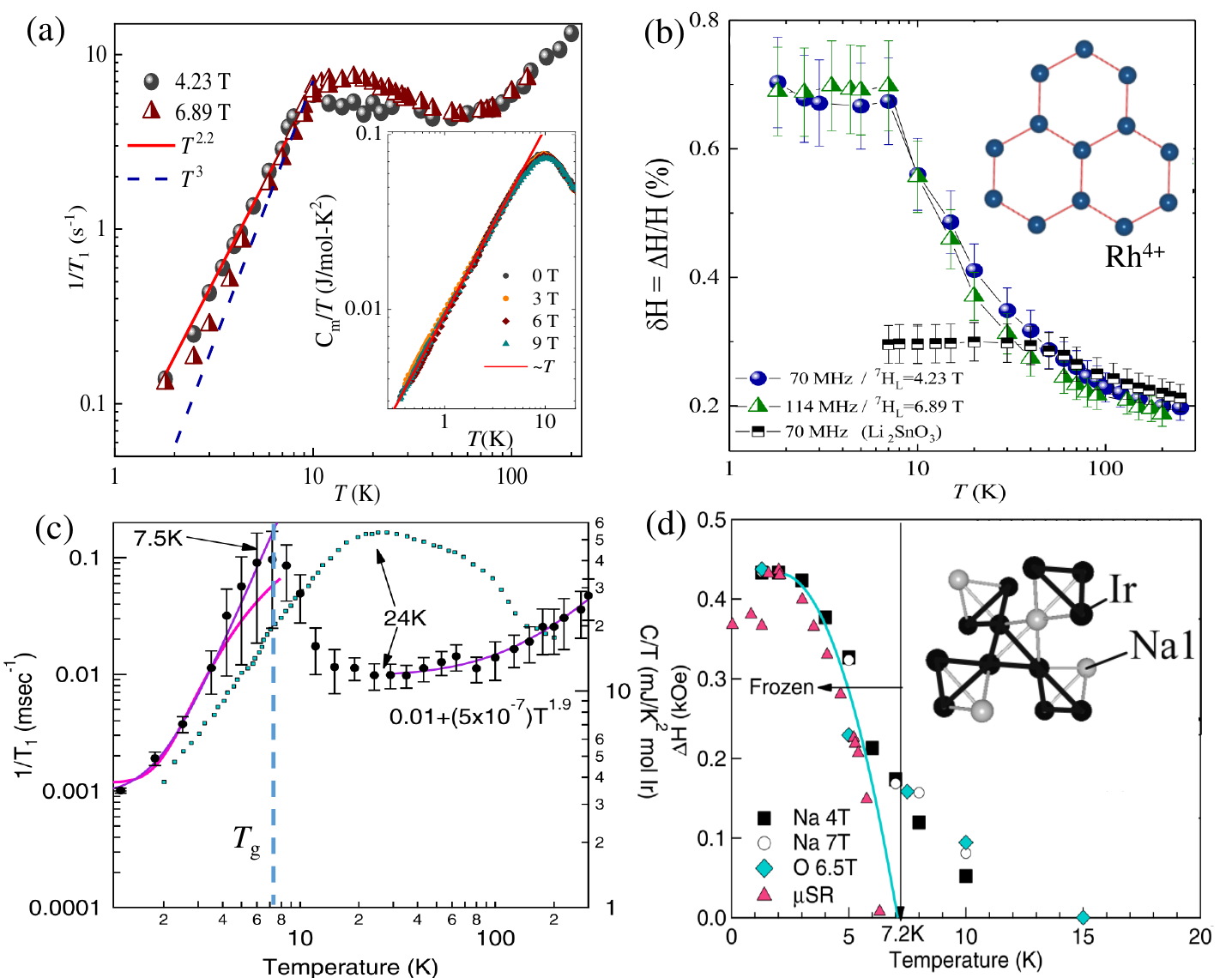}
  \caption{\textbf{NMR and thermodynamic signatures of slow spin dynamics and spin freezing in Kitaev and spin-orbit coupled frustrated magnets.} (a) Temperature dependence of the spin-lattice relaxation rate below $T_g$, \(1/T_1\) for Li\(_2\)RhO\(_3\), measured under applied magnetic fields of 4.23~T and 6.89~T. A broad maximum and subsequent rapid drop in \(1/T_1\) signify the onset of spin freezing. Inset: Field-independent magnetic specific heat \(C_{m}/T\) vs $T$ data up to 9~T, highlighting the robust low-energy spin dynamics characteristic of topological spin freezing in Kitaev magnets. 
(b) Relative NMR line width \(\delta H = \Delta H/H\) at 70 and 114~MHz as a function of temperature, showing significant broadening in Li\(_2\)RhO\(_3\) compared to the nonmagnetic analog Li\(_2\)SnO\(_3\), indicative of internal field distribution associated with spin freezing. 
(c) \(^{23}\)Na spin-lattice relaxation rate \(1/T_1\) (left axis, black circles) and specific heat \(C/T\) (right axis, blue squares) for Na\(_4\)Ir\(_3\)O\(_8\), plotted against temperature. Power-law fits (purple lines) illustrate the nontrivial scaling behavior expected from HS-type low-energy excitations. 
(d) Gaussian full width at half maximum (FWHM) broadening for \(^{23}\)Na at 4 and 7~T in Na\(_4\)Ir\(_3\)O\(_8\), plotted as a function of temperature below 16~K. The broadening reflects the emergence of static spin textures in the hyperkagome lattice as temperature decreases. Source: (a,b) adapted from \cite{PhysRevB.96.094432} with permission from APS, (c,d) adapted from \cite{PhysRevLett.99.137207, PhysRevLett.115.047201} with permission from APS.}
\label{nmr_material}
\end{figure*}

Nuclear magnetic resonance (NMR) occupies a unique position among microscopic techniques  employed to examine the  magnetic and electronic properties of frustrated magnets. Its effectiveness stems from the interaction between nuclear magnetic moments and local fields produced by surrounding electrons, allowing us to probe the fine details of charge and spin fluctuations at the atomic scale. Each nuclear isotope is characterized by a nuclear spin quantum number $I$ and gyromagnetic ratio $\gamma_n$, allowing the resonance condition to be selectively tuned by target nuclei such as $^1$H, $^{13}$C, or $^{31}$P. This capability provides exceptional site selectivity and sensitivity to explore the microscopic spin dynamics and local environments of the complex materials~\cite{fukushima2018experimental}. Notably, $^1$H stands out for its large gyromagnetic ratio \((\gamma/2\pi = 42.576~\mathrm{MHz/T})\)~\cite{fukushima2018experimental}. The particular strength of NMR lies in its ability to probe both the static and dynamic properties of spin systems. The experimentally observed Knight shift $K$ originates from the polarization of electron spins surrounding the nucleus through hyperfine coupling and hence provides a direct link between the local magnetic field and the bulk magnetic susceptibility. The NMR linewidth arises from the local field distribution within the material.
In topological spin glass phases, the study of temperature dependent NMR spectra reveals the crossover from a high-temperature paramagnetic state to a low-temperature, correlated and frozen state. Typically, we see a sharp resonance in the NMR lineshape in the paramagnetic phase. However, when the system is cooled to the spin-freezing temperature $T_g$, the shape of the lines gradually broadens and becomes asymmetric~\cite{PhysRevB.96.094432}. This crossover in the NMR lineshape reflects the emergence of static local magnetic fields and the slowing down of spin dynamics below \(T_g\). We can observe the development of  magnetic correlation and features associated with emergent spin excitations by investigating the NMR shift, the linewidth, and the relaxation rate. At the heart of these observations, the hyperfine interactions play a vital role, which couples nuclear and electronic spins through both contact and dipolar channels~\cite{abragam1961principles,khuntia2019novel}. The hyperfine Hamiltonian,
\begin{equation}
\begin{split}
\mathcal{H}_{\text{hf}} = \frac{8\pi}{3} g_I \mu_B \gamma_n \hbar\, \mathbf{I}\cdot\mathbf{S(r)}
\\+ g_I \mu_B \gamma_n \hbar\, \mathbf{I} \cdot \left( \frac{\mathbf{S}}{r^3}
- \frac{3\mathbf{r}\left(\mathbf{S}\cdot\mathbf{r}\right)}{r^5} \right)
\end{split}
\end{equation}
encompasses both a contact term and a dipolar term. The contact term accounts for the Knight shift and tracks intrinsic spin susceptibility, whereas the dipolar interaction accounts for anisotropic NMR line broadening.

In topological spin glasses, the HS modes are of particular interest because they remain gapless and long-wavelength fluctuations even when spin freezing sets in ~\cite{PhysRevB.16.2154,huang2024emergent}. Experimentally, the Knight shift ($K$) offers a direct insight into the spin susceptibility through the hyperfine coupling constant $A_\text{hf}:
K = \frac{H_{\text{loc}}}{H_0} = A_\text{hf}\,\chi(q=0,\omega=0).$
Variations in the Knight shift ($K$) with temperature capture how the electronic system reorganizes during the glassy transition. At the same time the NMR linewidth (\(\Delta H\)) increases sharply below \(T_g\), indicating the emergence of static (or quasi static) local fields and broad distribution of magnetic moments of owing electron spin ~\cite{PhysRevLett.115.047201}. The relative linewidth, defined as $\delta H = \Delta H / H$, is a useful parameter to quantify the glassy transition. A sudden increase in $\delta H$ indicates the onset of spin freezing in canonical spin glasses, while its saturation at low temperatures suggests the presence of fluctuating regions that persist even at the lowest temperatures. The magnetic moment derived from the saturated value of NMR at low temperatures is typically very small compared to that expected for a spin system, which suggests frustration-induced fluctuations in topological spin glass materials. This behavior distinguishes the topological spin glass from its canonical counterpart, in which the linewidth continues to broaden as the freezing progresses~\cite{PhysRevB.76.054452}.
The dynamical window of NMR is primarily determined  by the $1/T_1$, which governs how the nuclear spins exchange energy with their electronic environments. By investigating its temperature and field dependence, one can gain insight into the signatures of collective spin modes and underlying spin dynamics of the systems. In general, the relaxation rate is sensitive to the low-energy spectrum of the material:
$\frac{1}{T_1} \propto \sum_q |A_\text{hf}(q)|^2 \frac{\chi^{\prime\prime}_m(q,\omega)}{\omega}$,
where $A_\text{hf}(q)$ is the wave vector-dependent hyperfine coupling, and $\chi^{\prime\prime}_m(q,\omega)$ is the imaginary part of the dynamic spin susceptibility~\cite{abragam1961principles}. For collective HS modes, this susceptibility often takes the form~\cite{fischer1979electrical}: $
\chi^{\prime\prime}_m(q,\omega) = \frac{\omega\,\chi_m\, D_s q^2}{2}
\Bigg[ \frac{1}{(\omega - cq)^2 + (D_s q^2)^2} + \frac{1}{(\omega + cq)^2 + (D_s q^2)^2} \Bigg],$
with $D_s$ the spin-diffusion constant and $c$ the spin wave velocity. In the limit where $D_s\rightarrow0$ the susceptibility simplifies to $\chi^{\prime\prime}_m(q,\omega) = \frac{\pi}{2} \omega\chi_m \left[ \delta(\omega - cq) + \delta(\omega + cq) \right].$ This reflects sharply peaked excitations, corresponding to gapless, propagating spin waves. Both the NMR relaxation rate \(1/T_1\) and the magnetic specific heat coefficient \(C_m/T\) reflect the density of states (DOS) of low-energy spin excitations as \(C_m/T \propto \overline{N(E_F)}\) and \(1/T_1 \propto T \overline{N(E_F)}^2\), which leads to \(1/T_1 \propto T^3\). So for two-dimensional magnets with linearly dispersing antiferromagnetic spin waves, where \(\overline{N(E_F)} \propto T\) leads to a power law behavior in spin-lattice relaxation rate as: \(1/T_1 \propto T^3\). In 2D triangular topological spin glass NiGa$_2$S$_4$ a $T^3$ dependence of $1/T_1$ is observed below 1 K. The power-law temperature dependence in $1/T_1$ follows a broad hump near $T_g$ complementary to the broad peak observed in $C_m$, which marks the onset of short-range spin correlations. In both the 2D Kitaev spin glass $\mathrm{Li_2RhO_3}$ and the 3D hyperkagome material $\mathrm{Na_4Ir_3O_8}$, the $1/T_1$ exhibits a maximum near $T_g$, followed by a sharp decrease and eventually a power law decay at the lowest temperatures~\cite{PhysRevLett.115.047201,PhysRevB.96.094432}. The experimental observation reveals that, below $T_g$, $1/T_1$ follows a temperature dependence as $T^n$, with $n$ approximately $2.2$ in $\mathrm{Li_2RhO_3}$ and nearly $3.8$ in $\mathrm{Na_4Ir_3O_8}$. The HS theory predicts $n=3$ for 2D systems, associated with gapless hydrodynamic spin modes. However, the deviation of this value from ideal is not surprising because in real materials, unquenched quantum fluctuations, spin-orbit coupling, or subtle exchange anisotropies shift the low-temperature exponent~\cite{jena2025nature,PhysRevB.78.094403,RevModPhys.88.041002}. The reduced value in 2D suggests a rich landscape of low energy states that remain accessible even in freezing states, while the enhanced value in 3D frustrated  spin glasses hints at partial gapping or anisotropic stiffness~\cite{PhysRevLett.115.047201,VILLAIN1980105}.

  As for the NMR linewidth, its detailed temperature and field dependence contains valuable clues about the microscopic freezing process. In both $\mathrm{Li_2RhO_3}$ and $\mathrm{Na_4Ir_3O_8}$, the linewidth $\Delta H$ increases upon cooling, and it saturates for $T < T_{\text{broad}}$, a trend that closely tracks features in the specific heat. The saturated linewidth at low temperatures suggests the presence of long-lived but not truly static moments: only a fraction of spins appears to be frozen, while a dynamic background still persists. The fact that $\Delta H$ below $T_g$ shows minimal dependence on the applied magnetic field. In the hyperkagome material \(\mathrm{Na_4Ir_3O_8}\), for example, \(\Delta H\) remains nearly unchanged between 4~T and 7~T for \(T < 7\)~K (Fig.~\ref{nmr_material}(d))~\cite{PhysRevLett.115.047201}. A similar field-independent broadening is also found in the Kitaev compound \(\mathrm{Li_2RhO_3}\) between 4.23~T and 6.89~T below \(T_g = 6\)~K~\cite{PhysRevB.96.094432}. This insensitivity to external fields complements the field-independent nature of specific heat in topological spin glasses.

Overall, NMR provides an exceptional microscopic view of the intricate interplay between topology and disorder in highly frustrated spin glasses. It provides access to local magnetic correlations, exotic dynamics, and emergent low-energy excitations. Complementary measurements such as $\mu$SR, AC susceptibility and inelastic neutron scattering are essential for a complete classification of topological spin glasses.

\subsubsection{Neutron scattering}
Neutrons are electrically neutral subatomic particles that carry a magnetic moment due to their spin. The de Broglie wavelength of neutrons is comparable to the distances between atoms in solids, making them a regular probe in condensed matter research. Neutrons can deeply penetrate matter because they interact weakly with matter, enabling the measurement of the bulk properties. Another feature is that the neutron probe is sensitive to isotopic effects as they interact with the nuclei. In a magnetic material the neutron spin moment couples to the magnetic field created by the electronic environment. Overall, neutron scattering has become a regular probe used for studying the structural and spin dynamical behavior in frustrated magnets. In elastic neutron scattering, we observe the scattered neutron with zero energy loss ($E_i=E_f$). For a long-range ordered magnetic state, elastic neutron scattering sees the Bragg peaks. In a disordered spin system, the magnetic scattering becomes diffusive because the periodicity is lost. This diffusive magnetic scattering shows the onset of short-range spin correlations. We can measure the spatial extent of these correlations by looking at the shape and width of the diffuse signal. In INS, we measure the energy change in neutrons after they scatter. Here $E_i\neq E_f$ and the change in energy is $\hbar\omega=E_i-E_f$.  The probed neutrons can create or annihilate excitations in a magnetic material, such as spin waves (magnons), spinons, or more exotic modes. By measuring the change in energy, one can map the spectrum of these collective phenomena. This is done using energy-sensitive tools like monochromators, time-of-flight spectrometers, or detectors that can precisely track changes in neutron energy. Because of this, inelastic neutron scattering provides valuable insight into the dynamic behavior of materials. This allows us to extract information on the spatial correlation and the temporary changes of spin moments.

The dynamical structure factor $S(Q,\omega)$ is the key observable in INS measurements, characterizing spatial and temporal spin correlations.. The momentum transfer is defined as: $Q=|\mathbf{k}_i - \mathbf{k}_f|$, where $\mathbf{k}_i$ and $\mathbf{k}_f$ are the incident and scattered neutron wavevectors, respectively. In terms of the experimentally defined scattering angle $2\theta$, the magnitude of $Q$ is given by: $Q = \sqrt{k_i^2 + k_f^2 - 2 k_i k_f \cos(2\theta)}$.
In the special case of elastic scattering ($k_i = k_f = 2\pi/\lambda$), this reduces to 
$Q = \frac{4\pi}{\lambda} \sin\theta,$
where $\theta$ is half of the scattering angle. In general, the structure factor $S(\mathbf{Q},\omega)$ represents the spherically averaged scattering function
of the powder sample and it is given as~\cite{lee1996spin}:
\begin{equation}
\begin{split}
S(Q,\omega) = \int \frac{d\Omega_{\hat Q}}{4\pi} \frac{1}{2} \sum_{\alpha,\beta}
(\delta_{\alpha\beta} - \hat{Q}_\alpha \hat{Q}_\beta) |F(Q)|^2 \frac{(g\mu_B)^2}{2\pi\hbar} \\
\times \int dt\, e^{i\omega t} \frac{1}{N} \sum_{\mathbf{R},\mathbf{R}'} \langle S_{\mathbf{R}}^\alpha(t) S_{\mathbf{R}'}^\beta(0) \rangle
e^{-i\mathbf{Q} \cdot (\mathbf{R} - \mathbf{R}')}
\end{split}
\end{equation}
By integrating over all energy transfers, one obtains the static structure factor $S(\mathbf{Q})=\int S(\mathbf{Q},\omega)d\omega$. The static structure factor represents the Fourier transform of the pair correlation function \( G(r) \), ), which expresses the probability of finding two spin moments separated by a distance \( r \). Consequently, the structure factor can be written as $
S(\mathbf{Q}) = \int G(\mathbf{r}) e^{i\mathbf{Q} \cdot \mathbf{r}} \, dr$. This quantity provides a direct link between real-space spin correlations and the intensity distribution observed in reciprocal space~\cite{boothroyd2020principles}.

Neutron scattering stands out as a remarkably versatile tool for probing magnetic order and dynamics. At the heart of this approach lies the double-differential cross-section,$
\frac{d^2\sigma}{d\Omega\,d\omega},$
which tells us how likely neutrons are to scatter in a particular direction while exchanging energy with the material. The cross-section is directly related to the experimentally observed dynamic structure factor $S(Q, \omega)$, via: $\frac{d^2\sigma}{d\Omega\,d\omega} \propto |F(Q)|^2 S(Q, \omega)$,
where $F(Q)$ is the magnetic form factor that carries the information of the spatial distribution of
magnetic moments and their orientation relative to the
scattering vector $Q$. A closer examination of $S(Q, \omega)$ probes the momentum- and energy-resolved map of spin correlations and thus plays a pivotal role in revealing both
the nature of the ground state and the spectrum of low energy excitations~\cite{lovesey1984theory, PhysRevLett.86.1335,han2012fractionalized}. For instance, in quantum spin liquids the lack of any sharp features in $S(Q, \omega)$ infers the absence of conventional long-range magnetic order, while broad, diffuse continua indicate the presence of fractionalized excitations such as spinons~\cite{broholm2020quantum,han2012fractionalized,banerjee2017neutron}. In such cases the excitation often involves two spinons each carrying individual wave vectors $q_1$ and $q_2$ such that their combined momentum equals the total momentum transfer $\mathbf{Q}$. These two spinon excitations form a broad continuum in the excitation spectrum which fills the area between the dispersion of single spinon and conventional magnons. Whereas in conventional magnets exhibiting long-range order, the low-energy spectrum is dominated by sharp spin-1 magnon modes. Therefore, measuring the dynamical structure factor, $S(Q, \omega)$, using inelastic neutron scattering provides a powerful means to distinguish between the static ordered phase and the more subtle emergent excitations characteristic of exotic quantum phases.

The correlation length \(\xi\) is another important quantity that can be found from neutron scattering, which measures how far magnetic order or correlations persist spatially in a magnetic material. Generally, \(\xi\) can be calculated by examining the width of the static structural factor \(S(Q)\). For systems where the short-range correlation dominates the static structure factor can be approximated as $
S(Q) \sim \frac{1}{1 + (Q\xi)^2}$. This Lorentzian-type form signifies that when the peak in the reciprocal space broadens, the correlation in the real space diminishes. The correlation length is inversely related to the full width at half maximum (FWHM) of the peak:($
\xi = \frac{1}{\text{FWHM}}$)~\cite{boothroyd2020principles}. Therefore, a sharp peak corresponds to long-range correlation, and a broader peak indicates the more localized and short-range magnetic correlation. In topological spin glasses, this length scale \(\xi\) typically extends only a few lattice spacings~\cite{PhysRevB.79.140402} and reflects how frustration inhibits the formation of long-range magnetic order. Whereas the stiffness length \(L_0\)\  is the distance over which collective, long-wavelength excitations, like spin waves or coherent spin fluctuations, can propagate collectively and coherently. Therefore, in short, for topological spin glasses, \(\xi\) captures the frozen, short-range correlations that arise from frustration, whereas, \(L_0\) generally a larger scale is for the coherent propagation of collective magnetic excitations, reflecting the distinct static and dynamic length scales in magnetic materials.

In addition to static correlation, INS maps the imaginary part of the dynamic susceptibility, $\chi^{\prime\prime}(Q, \omega)$, which measures how spins respond to time-dependent perturbations. The connection to the dynamic structure factor is given by the fluctuation-dissipation theorem : $
S(Q, \omega) = \frac{1}{\pi} \chi^{\prime\prime}(Q, \omega)\left[1+n(\omega)\right],$
with $n(\omega)$ representing the Bose occupation factor~\cite{lovesey1984theory,squires1996introduction}. In canonical spin glasses, the complicated, rugged energy landscape traps spin excitations and leads to localized, slow relaxations. This often yields a simple Lorentzian form:
\begin{equation}
\chi^{\prime\prime}(Q, \omega) \sim \chi^{\prime\prime}(Q, 0) \frac{\Gamma_L \hbar \omega}{(\hbar \omega)^2 + \Gamma_L^2},
\end{equation}
where $\Gamma_L$ sets the energy scale for these processes~\cite{kofu2021spin}. 

On the other hand, in frustrated magnets like SrCr$_{9p}$Ga$_{12-9p}$O$_{19}$, the landscape is much richer, blending features from both localized and collective excitations. It is common to find a functional form combining Lorentzian and arctangent contributions~\cite{yang2015spin, PhysRevLett.127.017201}:
\begin{equation}
\chi^{\prime\prime}(Q, \omega) \sim \alpha \left[ \frac{\omega}{\omega^2 + \Gamma_L^2} \right] + \tan^{-1} \left(\frac{\omega}{\Gamma_{\text{min}}} \right),
\end{equation}
where the arctangent is associated with the influence of long-wavelength and collective modes. It is important to understand this distinction in order to separate quantum topological phenomena from classical glassy relaxation~\cite{jena2025nature}. Consider the doped square-lattice antiferromagnet Sr$_2$CuTe$_{0.5}$W$_{0.5}$O$_6$ a spin-$\frac{1}{2}$ system showing glassy behavior below $T_g \approx 1.7$~K. Here, experiments find a linear $\hbar\omega$ dependence in $\chi^{\prime\prime}$ for $\hbar\omega < k_B T_g$, and a magnetic specific heat scaling as $C_m \sim T^2$~\cite{PhysRevLett.127.017201,PhysRevB.98.054422}. This points to the persistence of gapless, Goldstone-like spin excitations in a state that lacks global order but still breaks some continuous symmetries. The freezing mechanism here seems quantum in origin, deeply tied to frustration and gauge constraints~\cite{PhysRevB.101.024413}. With added disorder or dilution, systems like SCGO($p$) increasingly show broad, Lorentzian spectral weight, a clear mark of classical spin glass relaxation taking over from quantum coherence~\cite{yang2015spin}.

The elastic magnetic structure factor, $S_{\text{mag}}(Q)$, indeed provides information on spatial correlations in spin glass systems. In canonical spin glasses, it is typically broad and nearly featureless, reflecting the dominance of finite clusters and the absence of long-range magnetic order. In topological spin glasses, $S_{\text{mag}}(Q)$ often produces diffuse or moderately structured features, indicating well-defined short-range correlations. Quasi-elastic scattering, observed as spectral broadening at low energies, points to slow dynamics and a wide spread in relaxation times. $S_{\text{mag}}(Q)$ also helps distinguish interaction regimes---the appearance of broad peaks for RKKY-type systems is a classic fingerprint.

Tracking the evolution of $S_{\text{mag}}(Q)$ below $T_g$ can quantify the freezing transition, especially by focusing on the integrated low-energy spectral weight:
\begin{equation}
\tilde{S}(T) = \int_{Q_{\text{min}}}^{Q_{\text{max}}} \int_{\omega_{\text{min}}}^{\omega_{\text{max}}} \frac{S(Q, \hbar\omega)}{|f(Q)|^2} d\omega dQ,
\end{equation}
A sharp drop in $\tilde{S}(T)$ signals the freezing of spin fluctuations, mirrored by the emergence of a finite staggered moment $\langle S \rangle$ even without global order~\cite{PhysRevLett.98.107204,PhysRevLett.127.017201,PhysRevLett.65.3173}.

The static structure factor, correlation length, and dynamic susceptibility all highlight why neutron scattering is  an advantageous instrument to explore the frustrated magnets. These observables allow capturing both the spatial arrangements and temporal evolution of spins. In addition, these observables allow to distinguish between canonical spin glass and topological spin glass~\cite{VILLAIN1980105}.

For the discerning readers, juxtaposing $S(Q, \omega)$ with complementary probes such as thermodynamic and NMR measurements yields coherent evidence for topological glassiness. The absence of Bragg peaks below $T_g$, along with broad and overdamped low-energy excitations, is very different from the well-defined spin wave modes observed in conventional magnets. The time-resolved experiments can capture the aging phenomena where the magnetic dynamics evolve over experimental timescales, direct experimental signatures of glassy behavior. When analyzed using theoretical frameworks such as the Sherrington–Kirkpatrick or Halperin–Saslow models, neutron scattering provides significant insights into the influence of quantum coherence, frustration, and topological constraints on spin dynamics.

 \begin{figure*}[t]
\includegraphics[height=285.589pt, width=510.0pt]{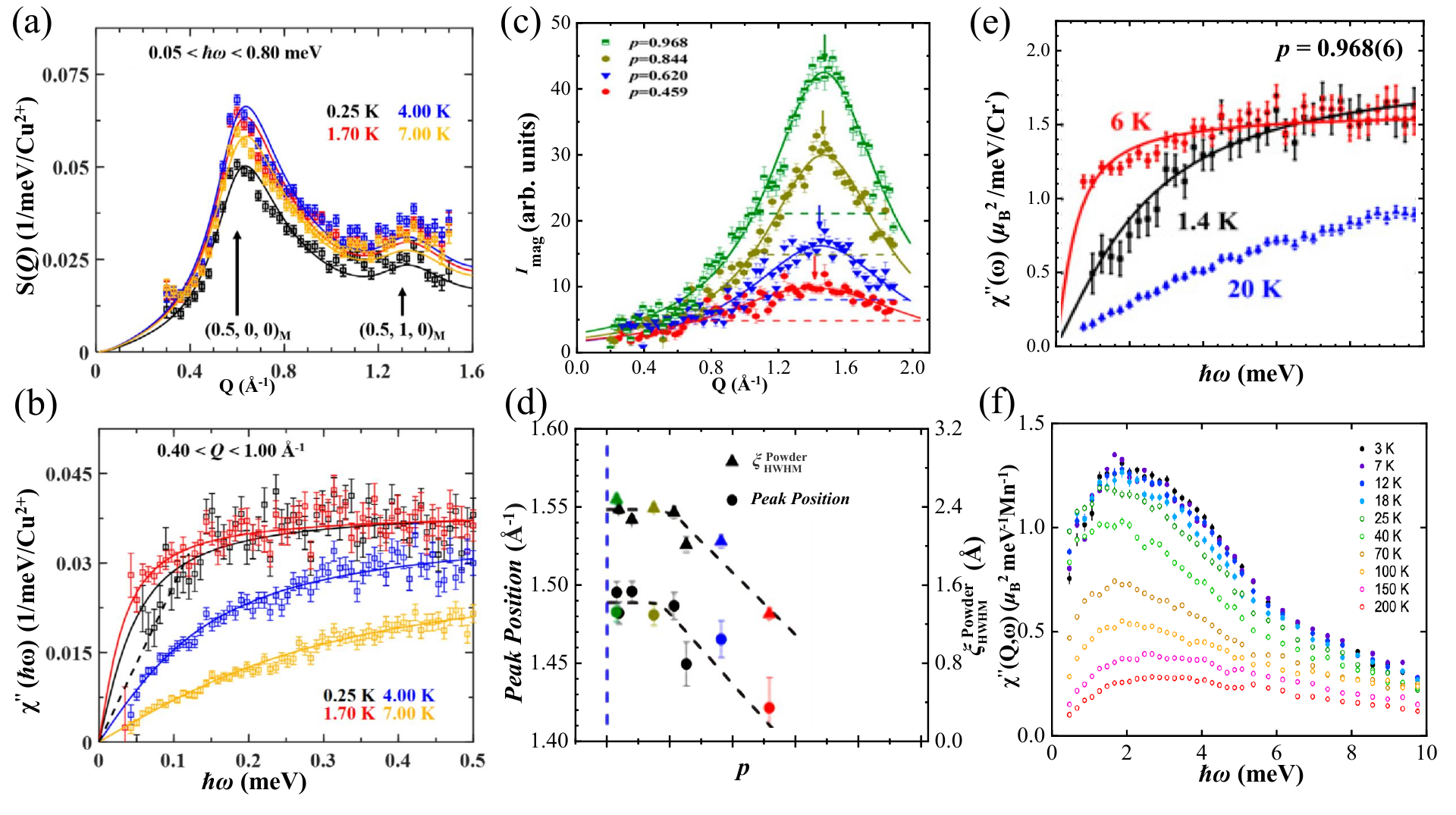}
  \caption{\textbf{Inelastic neutron scattering reveals topological spin freezing in frustrated quantum materials.} (a) Momentum dependence of low-energy magnetic fluctuations $S(Q)$ in a doped square lattice antiferromagnet Sr$_2$CuTe$_{0.5}$W$_{0.5}$O$_6$, integrated over the energy range 0.05–0.80 meV, measured at various temperatures (0.25 K, 1.70 K, 4.00 K, and 7.00 K). The integrated intensities reveal temperature-dependent evolution in spectral weight and broadening, with smooth fits shown as colored curves. (b) Imaginary part of the dynamic susceptibility Sr$_2$CuTe$_{0.5}$W$_{0.5}$O$_6$, $\chi''(\hbar\omega)$ in, obtained via detailed balance correction from energy-integrated scattering $S(\hbar\omega)$ averaged over the momentum window $Q = [0.4, 1.0]\,\text{\AA}^{-1}$. A near-linear behavior in $\chi''(\hbar\omega)$ is seen at low temperatures up to approximately 0.15 meV, indicative of overdamped, gapless excitations. (c) Elastic magnetic scattering intensity of SCGOp as a function of momentum $Q$, measured at 1.4 K (or 0.27 K for $p = 0.459$). Nonmagnetic background subtraction was performed using data collected at 20 K. The profiles are well-described by Lorentzian fits, with dashed lines indicating the extracted peak widths and positions and the evolution of the static spin correlation length and corresponding peak positions as a function of dilution parameter $p$, revealing how structural disorder modulates magnetic correlations(d). (e) Temperature-dependent behavior of $\chi''(\omega)$ for $p=0.968(6)$ in kagome material SrCr$_{9p}$Ga$_{12-9p}$O$_{19}$, derived from inelastic neutron scattering at low energies. The observed suppression of spectral weight with increasing temperature reflects thermally activated damping of slow fluctuations. (f) Energy- and momentum-resolved magnetic susceptibility $\chi''(Q, \omega)$ integrated over $Q = [1.5, 2.1]\,\text{\AA}^{-1}$, measured using incident energies of 5.3, 9.0, and 18.7 meV. Data are shown for comparison with the canonical spin glass Cu$_{1-x}$Mn$_x$ ($x = 0.034$, $T_g = 20.2$ K), illustrating a broad continuum of excitations characteristic of glassy spin dynamics. Source: (a,b) adapted from \cite{PhysRevLett.127.017201} with permission from APS, (c,d,e) adapted from \cite{yang2015spin} with permission from PNAS, (f) adapted from \cite{PhysRevResearch.6.013006} with permission from APS. }
  \label{fig: 4 INS}
\end{figure*}

\subsubsection{\texorpdfstring{$\mu$}~\text{SR experiments}}

Muon spin relaxation ($\mu$SR) is a powerful microscopic probe to study the magnetic properties of frustrated systems. The muon is an elementary particle with a spin of $\frac{1}{2}$ and a mass that is about 207 times that of the electron~\cite{uemura1994spin,scheck1978muon,blundell2021muon}. It is short-lived, with an average lifetime of 2.2 $\mu$s, and possesses a magnetic moment characterized by a gyromagnetic ratio of $\gamma_\mu = 2\pi \times 135.5$ MHz T$^{-1}$~\cite{blundell1999spin}. In condensed matter studies, the positive muon ($\mu^+$) is particularly useful because it interacts directly with the electronic environment. The high-intensity muon beam particles are created when high-energy proton beams are fired into a target. This event gives rise to the production of pions, which are again short-lived particles with an average lifetime of 26 ns, and they further decay into a muon via the reaction $\pi^+\rightarrow \mu^+ + \nu_{\mu}$. Neutrinos have a very intrinsic feature of negative helicity, where their spins are antiparallel to their momentum. As the $\pi^+$ has zero spin, in the rest frame, the muon should follow the same trend as the neutrino, i.e, the muon must have a negative helicity. So, the pion decay gives rise to 100\% spin-polarized muons, which is a major advantage of the $\mu$SR technique over other spectroscopy methods. The decay muons have an energy of 4.1 MeV, which is implanted into the material, and they undergo a number of events, such as the ionization of atoms, muonium formation via $e^-$ capture, and inelastic collisions, all of which occur in a very short time. The above events involve only Coulombic interactions, so the polarization of muons remains 100\%. Finally, following a sequence of successive  electron capture and loss reactions, the muon comes to rest and occupies an interstitial site with an energy of a few eV, which is suitable for probing low-energy many-body phenomena in condensed matter. The implanted muon undergoes decay into a positron via $\mu^+\rightarrow e^+ + \nu^-_{\mu}+\nu_e$.

\begin{figure*}[t]
\includegraphics[height=384.04645pt, width=510.0pt]{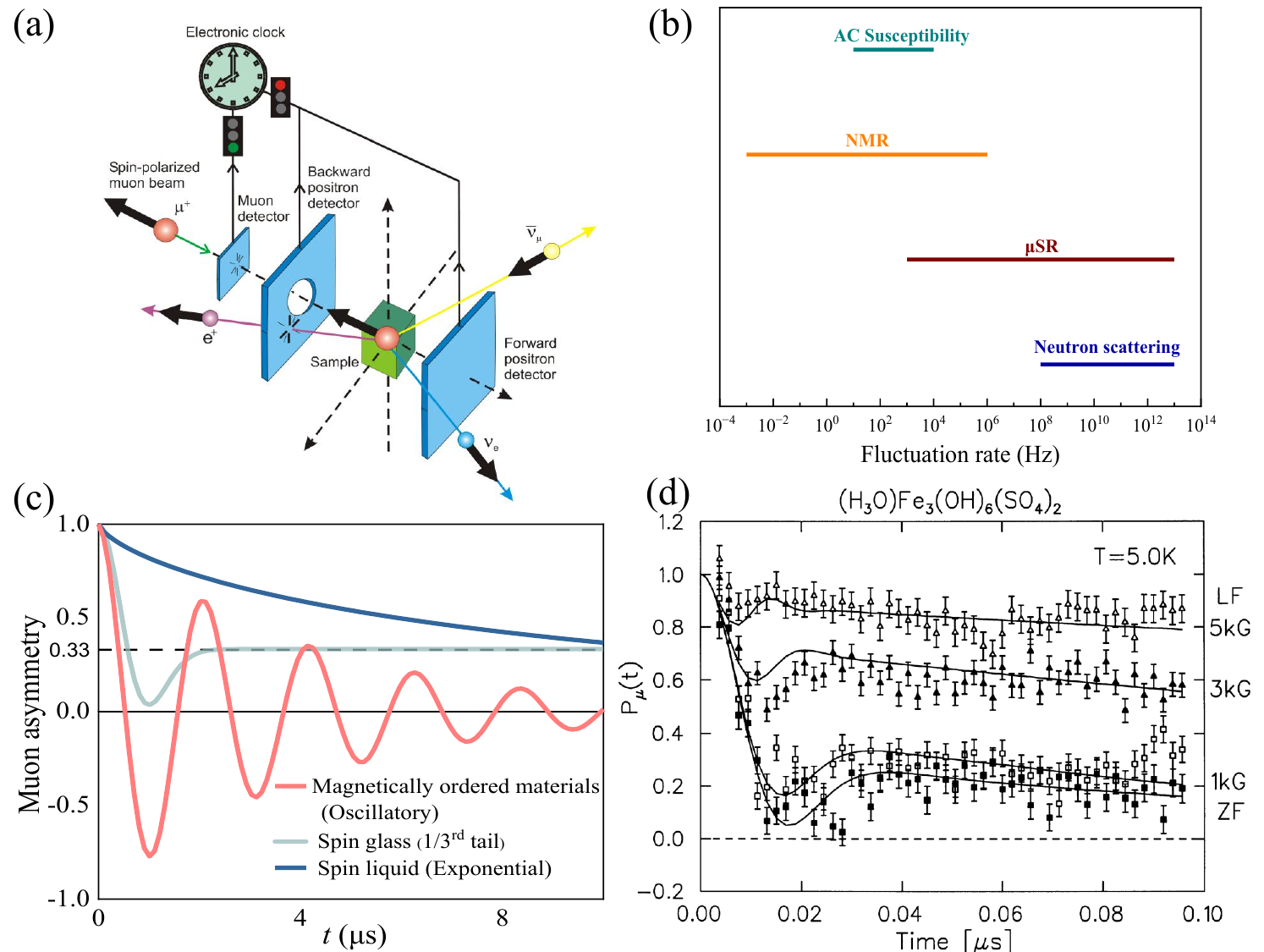}
  \caption{ \textbf{ $\mu$SR as a local probe to track novel ground states and low-energy excitations in frustrated magnets.} (a) Spin-polarized muons are introduced into the sample, where their spin dynamics are tracked over time. This is achieved by detecting the time-dependent asymmetry A(t) in the emission of positrons, which are preferentially emitted along the direction of the muon spin at the moment of decay.
 (b) Schematic comparison of the frequency windows probed by various experimental techniques commonly employed to study spin dynamics in frustrated magnets, including spin glasses. Each technique—neutron scattering, $\mu$SR, NMR, and a.c.\ susceptibility—accesses a distinct range of electron spin fluctuation frequencies, spanning over several orders of magnitude. $\mu$SR is particularly sensitive to fast local fluctuations (KHz to GHz) and shares an interface with other  complementary techniques as depicted, while a.c.\ susceptibility captures the slow collective dynamics (Hz to kHz), making their overlap narrow yet valuable in diagnosing glassy behavior. NMR bridges these regimes, providing site-specific insights into the spin environment. This comparative plot emphasizes the complementarity of these probes in mapping out the full dynamical landscape of emergent glassy or topologically constrained spin systems. (c) Time evolution of muon asymmetry in magnetically ordered, spin glass, and spin liquid materials. (d) Muon depolarization in hydronium iron jarosite H$_3$OFe$_3$(OH)$_6$(SO$_4$)$_2$ at 5 K, conducted in zero field and with applied longitudinal fields up to 5 kG. Source: (a) adapted from \cite{sonier2002muon} with permission, (b) adapted from \cite{le2011muon} with permission, (d) adapted from \cite{harrison2000musr} with permission from APS.}
  \label{musr}
\end{figure*}

Fully polarized muons act as an ideal probe to investigate the local magnetic environments in solids. Once the muons are implanted into the material, they come to rest at  interstitial sites, where they experience the local magnetic field due to unpaired spin moments. The local magnetic field in a magnetic
material probed by µSR is dipolar in nature. The core principle behind the $\mu$SR technique is that the muon spin will precess around the local magnetic field direction at a frequency directly proportional to the field strength. The precession frequency, $\omega _\mu$, of the muon spin is directly related to the local field $\mathbf{B}_{\text{local}}$ by the relation: $ \omega_\mu = \gamma_{\mu} \left|\mathbf{B}_{\text{local}}\right|
$, where $\gamma_{\mu}$ is the gyromagnetic ratio of the muon, which is $2\pi\times $135.53 MHz/T~\cite{blundell2004muon}. The precession results in periodic oscillations of the muon spin polarization along the beam direction, which can be monitored through the detection of the emitted positrons from the muon decay. The typical muon lifetime is approximately 2.2 $\mu$s, during which the muon interacts with its local environment of the sample under study and subsequently decays via weak interactions, emitting a positron preferentially aligned with its spin. Fig. \ref{musr}(a) presents a schematic of a typical $\mu$SR experimental arrangement. In this setup, muons are implanted into the sample under investigation, where muons decay into positrons after a lifetime. In forward and backward directions of the the muon, positron detectors are placed which detects positrons coming from the sample. The positron counts recorded by the both detectors are used to construct the asymmetry function A(t), which is directly proportional to the muon spin polarization function P(t). By analyzing the time evolution of P(t), one can extract information about the static and dynamic magnetic properties of the material~\cite{le2011muon,blundell2021muon,hillier2022muon,khatua2023experimental,nuccio2014muon}.

$\mu$SR is a uniquely powerful and complementary technique for exploring subtle magnetic and electronic ground states. The central key of $\mu$SR experiment is the precession or relaxation of the muon spin which bears close resemblance to NMR and electron spin resonance (ESR)~\cite{blundell1999spin,yaouanc2011muon,blundell2004muon}. Since these local probes track spin dynamics, the essential distinction lies in their initial polarization of the probe spins. While muon beams can be produced with nearly $100\%$ polarization, the nuclear or electron spins in NMR and ESR are weakly polarized under typical laboratory conditions. As a result, NMR and ESR require the application of radio-frequency (MHz range) or microwave (GHz range) photons to produce signal. In contrast, $\mu$SR does not rely on such resonant driving; instead, it is inherently a nonresonant technique where muons' natural precession and decay provide the necessary signal.  NMR requires NMR-activated nuclei with suitable magnetic moments and often suffers from weak signal-to-noise ratios in insulating or dilute systems, while $\mu$SR does not depend on isotopic abundance. The implanted muon acts as a very sensitive local probe of the internal magnetic field, it can detect tiny static or fluctuating moments (down to $\sim 10^{-3}\,\mu_B$), often below the resolution of neutron diffraction. In contrast to neutron scattering, which requires large single crystals and provides information averaged over reciprocal space, $\mu$SR can be performed on small amounts of polycrystalline or even powdered samples, offering microscopic, site-specific insight into the local environment.  NMR is sensitive to very low-frequency dynamics  and neutron scattering accesses much higher frequencies  (Fig. \ref{musr}(b)), while $\mu$SR  bridges the intermediate-frequency regime between fast electronic processes and slower nuclear spin dynamics, making it particularly well suited for studying spin-glass freezing, magnetic order, spin dynamics, superconducting vortex states, and quantum fluctuations in frustrated magnets~\cite{lacroix2011introduction,amato1997heavy}.
It is important to note that the muon probe is not completely passive, as it carries a positive charge and can perturb the electronic and magnetic environment of the host lattice locally. It is proven from various studies that the implanted muon typically induces a local structural distortion, thereby modifying exchange pathways and leading to a local magnetic perturbation~\cite{lancaster2018quantum,huddart2022mufinder}. Especially in the Pr$^{3+}$ pyrochlore Pr$_2$B$_2$O$_7$ (B$=$Sn, Zr, and Hf), the implanted muons induce an anisotropic distortion field that splits the non-Kramers doublet ground state of Pr$^{3+}$ moment~\cite{PhysRevLett.114.017602}. As calculated, the splitting is of several meV, which leads to an enhancement of the hyperfine coupling of Pr nuclear moments close to the muon site and results in a static distribution of magnetic fields observed in $\mu$SR experiments. As a result, it shows an increase in static relaxation rates and rms field widths on cooling, and it reflects the muon-induced effects rather than the behavior of the quantum spin ice in Pr$_2$B$_2$O$_7$~\cite{PhysRevLett.114.017602}. Muon polarization carried out in ZF configuration at 50 mK on ZnCu$_3$(OH)$_6$Cl$_2$ shows damped oscillatory behavior; however, this is not due to spin freezing. It is attributed to the formation of a muon–hydroxyl (OH–$\mu$) complex and the dipolar interaction
between the nucleus and the muon moments gives rise to the damped oscillatory behavior~\cite{PhysRevLett.98.077204}. Substituting H-atom with deuterium, i.e., ZnCu$_3$(OD)$_6$Cl$_2$, there appears an exponentially decaying function in ZF polarization without any oscillatory nature. In the deuterium version, the oscillation frequency is too small such that it escapes the muon time window. Upon applying a weak LF field of 8 mT, we see that the relaxation is largely suppressed and easily decoupled from the ZF polarization. Importantly, no additional relaxation or loss of oscillatory behavior is observed down to 50 mK, suggesting that the Herbetsmithite is in a dynamic state. Therefore, special care should be taken into account while interpreting $\mu$SR data—particularly in highly frustrated or low-dimensional spin systems. It is noteworthy that intrinsic electronic and spin degrees of freedom respond on timescales much faster than the muon precession. So, except for some special cases as discussed in the above examples, $\mu$SR probes a fully relaxed local equilibrium configuration in the presence of the muon~\cite{blundell2021muon}, rather than a transient perturbation. In this sense, the measured local fields by $\mu$SR experiment correspond to a renormalized magnetic environment.

The $\mu$SR technique can be performed in three different geometries to investigate the time-evolution polarization function of the measured sample. A common geometry is the transverse-field(TF-$\mu$SR) method, where the external magnetic field is set perpendicular to the initial polarized muon. In this configuration, the muon precesses around the applied field and eventually depolarizes, which is detected through the positron detector. The other two geometries are longitudinal-field(LF-$\mu$SR) method, where the external magnetic field is set parallel to the initial muon polarization, and the Zero-field(ZF-$\mu$SR) method, where the muon probes with its own polarized spin and without any external magnetic field. In this method, the muon precesses around the intrinsic magnetic field and provides insight into the local magnetic environment. All of these three methods are widely utilized to track the evolution of magnetic correlation, spin fluctuations, low-energy excitations, and magnetic ordering, which is inaccessible through other bulk magnetization techniques~\cite{nuccio2014muon,blundell2021muon}.

The normalized muon spin polarization function, $P_{\alpha}(t)$, with $\alpha = x, y, z$, represents the projection of the ensemble-averaged muon spin along a chosen axis. The $P_{\alpha}(t)$ originates from the muon spin orientation after interaction with the local magnetic field, $B_{\mathrm{loc}}$, which exerts a torque on the muon’s magnetic moment. The local field generally consists of dipolar contributions due to neighboring nuclear spins, and the electron spin accounts for the contact hyperfine term. For a static local field inclined at an angle $\theta$ relative to the initial muon spin, the muon polarization precesses around the field direction, and the longitudinal component of the polarization follows a characteristic time dependence: $P_{z}(t) = \cos^{2}\theta + \sin^{2}\theta \cos\!\left( \gamma_{\mu} B_{\mathrm{loc}} t \right)$.
For a polycrystalline sample where the angle $\theta$ is uniformly distributed, the average muon polarization function typically follows the form:

\[
P_{z}(t) = \frac{1}{3} + \frac{2}{3} \cos(\omega_{\mu} t),
\]

The oscillations in muon polarization function (Fig. \ref{musr}(c)) indicate that the material hosts magnetically long-range ordered state, where the magnetic fields are static and uniform\cite{de1997muon,blundell1999spin, nuccio2014muon}.
In frustrated spin glass materials, the local magnetic moments are disordered and exhibit slow fluctuations. Topological spin freezing in frustrated magnets, caused by competing exchange interactions, is marked by persistent spin dynamics linked to residual spin fluctuations, topologically non-trivial spin textures, and a wide range of local fields due to uneven magnetic correlations, which leads to a lower glass temperature. $\mu$SR is especially good at examining these fluctuations because it can reveal both static and dynamic behaviors of the local fields. In spin glasses, the muon spin relaxation displays unique signatures that vary with the timescale and dynamics of the local fields. A wide range of local fields results in a specific Kubo-Toyabe function:

\[
P_{z}(t) = \frac{1}{3} + \frac{2}{3} \left( 1 - \Delta^2 t^2 \right) \exp \left( -\frac{1}{2} \Delta^2 t^2 \right),
\]
where $\Delta$ is the size of the local field distribution.This function illustrates the relaxation of muon polarization due to local field inhomogeneities. The absence of coherent oscillations in spin glasses, coupled with the distinctive ``1/3 tail'' (Fig. \ref{musr}(c)) observed at extended time intervals, indicates that the spin configurations are either frozen or disordered. This is common for unusual spin-glass materials, such as (H$_{3}$O)Fe$_{3}$(OH)$_{6}$(SO$_{4}$)$_{2}$, which is shown in Fig. \ref{musr}(d) \cite{le2011muon,blundell2021muon, PhysRevB.96.094432}.

 In systems with rapidly fluctuating local fields, such as those found in spin liquids, the muon polarization decays exponentially rather than oscillating(Fig. \ref{musr}(c)) This can be described by a stretched exponential function\cite{le2011muon}:
\[P_{z}(t) = \exp \left[ - (\lambda t)^{\beta} \right],\] where $\lambda$ is the muon spin relaxation rate, and $\beta$ is the stretching exponent, which reflects the distribution of relaxation rates in the system. The stretched exponential decay is commonly observed in spin liquids implying a distribution of relaxation times, where the local magnetic moments fluctuate on a timescale that is fast compared to the muon precession frequency, $\nu \gg \frac{\gamma_{\mu}}{2\pi} B_{\text{local}}$.
The relaxation rate $\lambda$ can be related to the fluctuation rate $\nu$ and B$_{loc}$ of the local magnetic fields of electronic origin at the muon site via the Redfield relation\cite{slichter2013principles,le2011muon}:
\[\lambda = \frac{2 \gamma_{\mu}^2 B_{\text{local}}^2 \nu}{\nu^2 + (\gamma_{\mu} B_{\text{o}})^2}.\]

The muon relaxation rate provides detailed insight into the spin dynamics of the system under study. The temperature dependence $\lambda$ increases upon cooling and shows a plateau at low temperatures, which indicates persistence of spin dynamics at low temperatures. The plateau type nature is key signature of the spin-liquid ground state~\cite{le2011muon,blundell2021muon}.

\begin{table*}[t]
\centering
\caption{Fundamental differences between canonical and topological spin glasses.}
\label{tab:spin_glass_comparison}

\renewcommand{\arraystretch}{1.95}
\setlength{\tabcolsep}{6pt}

\begin{tblr}{
colspec={ Q[5cm] Q[5cm] Q[6cm]}
}
\hline\hline

\textbf{Feature} & \textbf{Canonical Spin Glass} & \textbf{Topological Spin Glass} \\

\hline

Origin of frustration 
& Randomness and disorder (e.g.\ dilute alloys)~\cite{RevModPhys.58.801,stein2013spin,mydosh1978spin}
& Frustration due to the geometry of the spin lattice or competing exchange interactions with quantum fluctuations, even near the clean limit~\cite{klich2014glassiness,syzranov2022eminuscent,yang2015spin,PhysRevB.101.024413} \\

Driving mechanism 
& Random RKKY interactions with varying sign~\cite{RevModPhys.58.801} 
& Lifting of degeneracy via order-by-quantum-fluctuations~\cite{klich2014glassiness,PhysRevB.101.024413} \\

Topology of energy landscape 
& Hierarchical rugged funnel with multiple energy scales~\cite{mezard1988spin} 
& Nonhierarchical, wide, nearly flat but rough energy landscape~\cite{samarakoon2016aging} \\

Energy scale hierarchy 
& Strong hierarchy (ultrametric-like)~\cite{RevModPhys.58.765} 
& Weak hierarchy (uniform branching) \\

Freezing temperature $T_g$ 
& Comparable to $\theta_{\mathrm{CW}}$, i.e., $T_g\sim|\theta_\text{CW}|$ so that the slowing down of spin fluctuations occur in a narrower region $T_g\leq T\leq 2T_g$~\cite{PhysRevB.31.546}
& Much lower than $\theta_{\mathrm{CW}}$ ($T_g<<|\theta_\text{CW}|$), so it provides a wide temperature range for slowing down of spin fluctuations~\cite{PhysRevLett.73.3306} \\

Aging effect 
& Strong; appears even for short waiting times 
& Weak at short times; develops slowly~\cite{samarakoon2016aging} \\
Topological character 
& Absent 
& Present (topological constraints on spin dynamics) \\

Spin stiffness 
& No true rigidity~\cite{PhysRevLett.85.3017} 
& Finite effective stiffness supporting collective modes~\cite{PhysRevB.16.2154,reed1979spin} \\

Spin dynamics 
& Localized cluster relaxation 
& Collective, long-wavelength excitations \\

Low-energy excitations 
& Two-level systems and barrier hopping~\cite{anderson1972anomalous,VILLAIN1980105} 
& Gapless collective modes (Halperin--Saslow type) \\

Sensitivity to impurities 
& Strong 
& Weak (robust near the clean limit) \\

DC susceptibility (ZFC–FC splitting) 
&Strong bifurcation below $T_g$; sharp onset 
& Present but often broader and smoother onset \\

Specific heat, $C_p$ 
& $\sim T$~\cite{anderson1972anomalous} 
& $\sim T^{D}$, where $D$ is the effective dimensionality~\cite{PhysRevB.16.2154,PhysRevB.79.140402}\\

Neutron scattering 
& Broad, featureless magnetic scattering 
& Finite-$Q$ peaks indicating short-range correlations \\

Dynamic susceptibility, $\chi^{\prime\prime}$ 
& Lorentzian (localized relaxation) 
& Broad spectrum with hydrodynamic contributions \\

Correlation length 
& Very short, no clear spatial structure 
& Finite short-range correlations with measurable length scale \\

NMR Spectroscopy
&  NMR line broadens significantly below $T_g$, Non-exponential magnetization recovery as $M(t)\sim \exp [-(t/T_1)^{\beta}]$, $\beta<1$~\cite{PhysRevLett.40.250} 
& NMR relaxation rate,  $1/T_1\propto T^D$ and the linewidth $\Delta H$ saturates below $T_g$ with field independent nature~\cite{PhysRevB.96.094432,PhysRevLett.115.047201,PhysRevB.79.140402}\\

$\mu$SR measurement& $1/3$ tail recovery in ZF~\cite{blundell1999spin}&$1/3$ tail recovery in ZF or undecouplable Gaussian relaxation in LF polarization due to formation short-lived singlets with no $1/3$ tail recovery~\cite{PhysRevLett.73.3306}\\

Representative systems 
& CuMn, AuFe alloys with distribution of magnetic moments are less dense (below threshold limit) in real space
& SCGO($p$), NiGa$_2$S$_4$, Li$_2$RhO$_3$ with high density of magnetic moments near to clean limit \\

\hline\hline
\end{tblr}
\end{table*}

\section{Candidate Quantum Materials for Topological Spin Freezing}
Here, we present a few representative 2D and 3D frustrated quantum magnets that are potential contenders for hosting topological spin freezing. We provide common experimental signatures through the lens of complementary thermodynamic, NMR, $\mu$SR, and INS experiments that are compared with suitable theoretical models to draw a generic framework applicable to a large class of frustrated quantum materials. A comparative summary of the experimental signatures and theoretical features distinguishing canonical and topological spin glasses is presented in Table~\ref{tab:spin_glass_comparison}.

\subsection{Kagome lattice antiferromagnets}
The kagome lattice is a two-dimensional network of corner-sharing triangles where each vertex is connected to four nearest neighbors. The two-dimensional nature, along with the low coordination number, makes it an ideal host for many unconventional magnetic and electronic behaviors~\cite{mekata2003kagome, husimi1950statistics, syozi1951statistics, PhysRevLett.65.3173, balents2010spin}. In metallic kagome lattices, spin-orbit coupling brings about electronic correlations that give rise to unconventional superconductivity, the anomalous Hall effect, magnetic skyrmions, Berry curvature-driven transport, and Weyl fermions~\cite{RevModPhys.88.041002, yan2011spin, PhysRevLett.98.107204}. In many cases, the observed physical properties of kagome compounds agree closely with theoretical expectations, reflecting the simplicity yet profound richness of the underlying spin-lattice. From an electronic viewpoint, kagome magnets are known for hosting a flat band along with a Dirac band in their band structure. The flat band in the kagome lattice arises from destructive interference between electron wave functions on the lattice. This behavior can be well described by a simple tight binding model where electrons hop between nearest-neighbor sites, given by the Hamiltonian: $\mathcal{H} = -t \sum_{\langle ij \rangle, s} \left( c_{i,s}^\dagger c_{j,s} + c_{j,s}^\dagger c_{i,s} \right),$ where $t$ is the hopping amplitude and $c_{i,s}^\dagger$ ($c_{i,s}$ ) are electron creation (annihilation) operators for spin  s at site i~\cite{mielke1991ferromagnetic}. The solution to this simple model yields a flat band at $E=-2t$ across the Brillouin zone, which is a direct manifestation of the underlying geometry of the kagome lattice. The cancellation of hopping amplitudes in the eigenstates results from interference conditions imposed by the lattice symmetry~\cite{mielke1991ferromagnetic, leykam2018artificial}. 

Flat bands are far from a mathematical curiosity: they dramatically enhance electron–electron interactions, often leading to correlation-driven ground states such as Mott insulators, unconventional superconductors, and fractional quantum Hall–like phases~\cite{yin2022topological}. Alongside these, kagome materials frequently display Dirac-like dispersions at special symmetry points in the Brillouin zone, where energy varies linearly with momentum. These Dirac cones host topologically protected edge states that are resilient to weak disorder, establishing kagome metals as fertile ground for realizing quantum spin Hall effects and Weyl semimetals~\cite{kang2020dirac, PhysRevLett.106.236802}. The coexistence of flat bands and Dirac cones within the same framework instigates a striking interplay between localization and itinerancy.

Apart from its electronic properties, the unconventional spin freezing in kagome magnets with minimal structural disorder has attracted considerable attention. A few promising examples include the magnetoplumbite SrCr\(_{9p}\)Ga\(_{12-9p}\)O\(_{19}\) (SCGO)~\cite{PhysRevLett.64.2070} and the spinel-type Ba\(_2\)Sn\(_2\)ZnCr\(_{7p}\)Ga\(_{10-7p}\)O\(_{22}\) (BSZCGO)~\cite{PhysRevLett.86.894}. These Cr$^{3+}$-based Kagome–triangular slab (pyrochlore-derived) magnets show spin freezing even with weak quenched disorder. The unconventional spin freezing is the result of geometrical frustration and the peculiarity of kagome structure. Here, the intrinsic chirality plays an important role. For classical Heisenberg spins on the kagome lattice, the system favors a 120° spin arrangement. Each triangular plaquette satisfies the constraint $\sum_{i \in \triangle} \mathbf{S}_i = 0$. This $\sum_{i \in \triangle} \mathbf{S}_i = 0$ constraint supports a manifold of highly degenerate coplanar ground states characterized by continuous spin-folding zero modes. These modes correspond to collective deformations of the spin configuration that preserve the local 120° constraint on each triangle. Such spin folds can either be open, extending across the lattice (Fig.~\ref{figure1}(c), or closed, forming local loop-like excitations commonly referred to as weathervane modes. The open one is associated with  uniform-chirality state called the $q = 0$ state (Fig.~\ref{figure1}(b)), and the closed loop has staggered chirality with a  $\sqrt{3} \times \sqrt{3}$ structure. The scalar chirality on a triangle is defined as $
\gamma = \frac{2}{3\sqrt{3}}\left( \mathbf{S}_1 \times \mathbf{S}_2 + \mathbf{S}_2 \times \mathbf{S}_3 + \mathbf{S}_3 \times \mathbf{S}_1 \right),$ serving as a way to measure the topological winding and handedness of spins on that triangular motif~\cite{chandra1993anisotropic, PhysRevB.47.15342}.

The introduction of an infinitesimal XY anisotropy in frustrated kagome magnets introduces a new length scale, $l_0$, beyond which out-of-plane spin fluctuations acquire a finite energy cost, effectively suppressing the long-wavelength renormalization of spin stiffness~\cite{chandra1993anisotropic}. This anisotropy energetically favors coplanar spin configurations and converts what would otherwise be a crossover behavior into a genuine Kosterlitz-Thouless (KT) transition at a temperature \( T_{\mathrm{KT}} \). In this regime, topological defects become bound below \( T_{\mathrm{KT}} \), while remaining unbound above a higher crossover scale \( T_\chi \), leading to a complex thermodynamic landscape. Even exponentially small amounts of anisotropy are sufficient to induce this transition, characterized by a transformation from a defect-rich regime at high temperatures to a low-temperature phase with an exponentially small density of defects. The KT phase transition sees a divergence in spin correlation length below the transition temperature~\cite{pokrovsky1979properties}. In the kagome lattice, there appear to be two types of defects. One is the line defects, which are open loops with uniform chirality (see Fig.~\ref{figure1}(b)), while the other is closed loops with alternating chirality. Along these line or loop defects, spins can rotate freely without incurring any energy. These defects are a natural consequence of geometrical frustration in the kagome structure. The conventional KT transition involves the 360° vortex–antivortex pairs~\cite{kosterlitz1973ordering}. In the kagome case, however, the lattice geometry and frustration enable defects with lower winding angles of 120° vortices. Comparatively, this carries a smaller gradient field that the full 360$^{\circ}$. As it costs less energy to produce these fractional defects than to create a full $360^{\circ}$ vortex, the transition temperature is reduced. The theoretical upper bound, $
T_{\text{KT}} \approx \frac{|\theta_{\text{CW}}|}{48}$
arises from calculations of these defect energies and matches well with what is observed in real kagome spin glasses~\cite{PhysRevB.64.094436}. The predicted KT-transition temperature in the hydronium and deuterium jarosite matches well with the glass transition temperature $T_g$.

A few representative kagome magnets exhibiting unconventional glassy dynamics is presented in Table~\ref{material_name_part1} and the specific heat with quadratic $T^2-$ dependence is depicted in the Fig.~\ref{fig:figure1}(a). Kagome-derived frustrated magnets SrCr$_{8.82}$Ga$_{3.18}$O$_{19}$ and Ba$_2$Sn$_2$Ga$_3$ZnCr$_7$O$_{22}$ are highly frustrated, and this, in turn gives rise to collective low-energy excitations below the glass transition temperature with a renormalized spin stiffness as predicted by the hydrodynamic theory of spin glasses ~\cite{PhysRevLett.84.2957, PhysRevLett.86.894}.

\begin{figure}[t]
		\begin{center}
			\includegraphics[height=240.00139pt, width=237.09279pt]{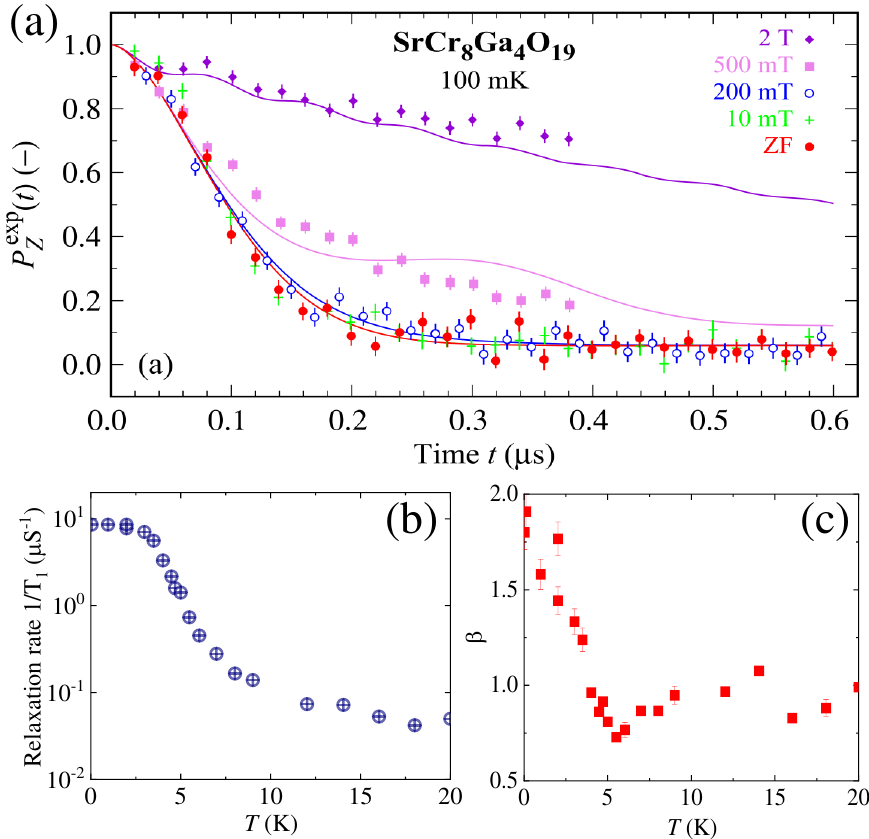}
			\caption{\textbf{Longitudinal field-dependent muon polarization in kagome lattices.} (a) LF muon polarization for kagome lattice SrCr$_8$Ga$_4$O$_{19}$ measured at 100 mK. (a) Muon spin relaxation rate $1/T_1$ recorded in an applied field of 10 mT and fitted with a power-law exponential form of $e^{{-(t/T_1)}^{\beta}}$. The corresponding coefficient $\beta$ is given in (c) as a function of temperature. Adapted from refs.~\cite{PhysRevLett.73.3306,PhysRevB.105.L241104}.}
			\label{kagome_muSR}
		\end{center}
	\end{figure}

The kagome spin lattice retains the memory of past defect movements. The defect dynamics become even more complex due to the presence of non‑Abelian disclinations. In this case, the homotopy group that determines how defects interact is non‑Abelian, which makes the system path‑dependent~\cite{chandra1993anisotropic}. To capture this complex development, one observes the chirality overlap function defined as: $ \gamma_{\text{overlap}} = \frac{1}{N} \sum_{\langle i,j \rangle} \vec{\gamma}_i \cdot \vec{\gamma}_j,$. It measures how the spatial correlation between chiralities $\vec{\gamma}_i$ and $\vec{\gamma}_j$ of neighboring triangular plaquettes is associated in a coplanar spin configuration~\cite{PhysRevB.47.15342}. These chiral degrees of freedom arise as collective fluctuations of a higher order that are beyond the harmonic spin‑wave theory and emerges within the semiclassical manifold of the kagome antiferromagnet. Non‑Abelian disclinations put strong constraints on defect dynamics, thereby suppressing thermal activation by slowing down the relaxation processes. The higher‑order spin‑wave fluctuations can stabilize staggered chiral states in kagome antiferromagnets by pushing the system into local minima of the energy landscape~\cite{PhysRevLett.69.832}. These emergent chiral states are characterized by nontrivial scalar chirality and are relevant to topological spin glass behavior in kagome magnets. Here, the spin freezing not a mere manifestation of random orientation of spin dipoles; instead, it is the chiral order that drives the freezing phenomenon. However, the complete description of the global energy landscape is still unresolved. The self‑energy corrections from anharmonic fluctuations are complex and depend on the wavevector, which makes it hard to state whether the chiral state is the true ground state or merely a metastable one~\cite{chandra1993anisotropic}.  
This intricate energy landscape could be responsible for the partial ordering or glassy freezing seen in kagome systems like SCGO and similar compounds. 

An interesting result is observed in the $\mu$SR data of SrCr$_8$Ga$_4$O$_{19}$ (SCGO with $p=8/9)$ spin glass with $T_g=3.5$ K~\cite{PhysRevLett.73.3306}. The zero field $\mu$SR asymmetry function doesn't show 1/3 recovery in zero field data down to 100 mK. In case of static magnetic fields, the $\mu$SR relaxation function should exhibit Gaussian damping and recovery to 1/3 initial symmetry. Conversely, a dynamic field leads to exponential or stretched-exponential relaxation without such recovery. For SrCr$_8$Ga$_4$O$_{19}$, the muon polarization function $P^\text{exp}_z(t)$ is well fitted with $P^\text{exp}_z(t)=\exp[-(t/T_1)^\beta]$ (Fig.~\ref{kagome_muSR}(a)). The relaxation rate $1/T_1$ showed an increase by over 2 orders of magnitude while cooling from 20 K to 100 mK, and it saturates below $T_g$. This marks a gradual slowing of Cr$^{3+}$ spin fluctuations over a wide temperature range, whereas in the canonical spin glass it is narrower. The line shape below $T_g$ apparently approaches a Gaussian shape with $\beta=2$ as $T\rightarrow0$. This Gaussian shape is expected for a dense static random field near the muon site, and a small field is enough to cause a decoupling of the LF $P^\text{exp}_z(t)$. However, the observed LF $\mu$SR spectra, as shown in Fig.\ref{kagome_muSR}(a), show no significant field dependence upto 200 mT. The observed ``undecoublable Gaussian'' suggests the formation of short-lived singlet pairs (RVB-like correlations). Furthermore, the formation of short-lived singlets suppresses the local magnetic field for most of the time, and a small fraction of unpaired magnetic moments intermittently generate strong local fields at the muon site, which produces an unexpected ``undecoublable Gaussian'' line that mimics a static dense field distribution. The slowing of spin dynamics over a broad temperature range, along with undecoupled lines, suggests that the low-energy state is mainly driven by quantum fluctuations, and the freezing is seemingly spin-liquid-driven glassiness. The undecouplable spectral lines can have another origin too. For example, when considering the approximate pyrochlore structure of SrCr$_8$Ga$_4$O$_{19}$ by connecting the sandwiched triangular layer to the nearby kagome layers, we obtain a near pyrochlore structure. Like in pyrochlore ZnCr$_2$O$_4$, the Cr moments form a composite structure of hexagonal spin loops, and each loop is characterized by a ``director''. The undecoupled nature of the muon spin relaxation response may be due to the slow, large-amplitude fluctuations of these directors~\cite{lee2002emergent}.

Herbertsmithite ZnCu$_3$(OH)$_6$Cl$_2$ is a nearly perfect realization of the $S = 1/2$ kagome Heisenberg antiferromagnet~\cite{PhysRevLett.98.077204,khuntia2020gapless}. Despite the large Curie-Weiss temperature $\theta_\text{CW}\approx-300$ K, it does not order down to sub-Kelvin temperatures~\cite{PhysRevLett.98.107204}.  However, when a moderate magnetic field is applied, the QSL state becomes unstable. It then transitions into a frozen spin‑solid state below a characteristic temperature \( T_c(H) \), which rises with increasing field strength~\cite{PhysRevLett.107.237201}. In this region, the spin fluctuations are exponentially suppressed and a spin gap of about $\Delta \approx 2.3\,k_B\,T_c$ opens up. 
Static magnetic moments also appear, with a magnitude of roughly $0.1\,\mu_B$ — much smaller than the full moment expected for a Cu$^{2+}$ spin-$1/2$. This strong reduction highlights the effect of quantum renormalization~\cite{PhysRevLett.107.237201}. Interestingly, the transition shows no critical divergence in the spin‑lattice relaxation rate. Instead, the NMR spectra have Gaussian line shapes, pointing to a static yet spatially disordered state. The Dzyaloshinskii–Moriya (DM) interactions in this material is estimated to be around $D \sim 0.04J - 0.08J$~\cite{zorko2008dzyaloshinsky,el2010electron} and places Herbertsmithite near a quantum critical point at $D_c \sim 0.1J$~\cite{PhysRevB.78.140405}. At this point, the system can transition from a QSL to an ordered phase. The applied magnetic field, along with the DM interactions, drives the system across this QCP. This is supported by the scaling relation $T_c \propto (B - B_c)^\phi$, with $B_c \approx 1.55$ T and $\phi \approx 0.6$, indicating that the freezing arises from field‑induced quantum criticality~\cite{PhysRevLett.107.237201}. The emergent freezing arises from the intrinsic many-body effects, and it challenges the realization of a true QSL.


\begin{table*}[t]
\centering
\renewcommand{\arraystretch}{1.5}
\caption{Magnetic and thermodynamic parameters of frustrated quantum magnets exhibiting unconventional spin-freezing behavior (Part I).}
\begin{tblr}{
colspec={ |Q|Q|Q[1.4cm]|Q|Q|Q|Q[4cm]|}
}

\hline
\textbf{Material [ref.]} & \textbf{Lattice} & \textbf{Effective Spin} & $T_g$ (K) & $\theta_\text{CW}$ (K) & $f=\frac{|\theta_\text{CW}|}{T_g}$ & \textbf{Topological characteristics}\\
\hline

(H$_3$O)Fe$_3$(SO$_4$)$_2$(OH)$_6$~\cite{majzlan2004thermodynamic} & Kagome & 5/2 & 16.5 & -833 & 50.5 & \SetCell{r=7}Local constraint $\sum_{\triangle}\vec{S}=0$; emergent divergence-free field; loop (weathervane/spin fold) modes; fractionalized excitations in quantum limit \\
(D$_3$O)Fe$_3$(SO$_4$)$_2$(OD)$_6$~\cite{wills1998magnetic} & Kagome & 5/2 & 13.8 & -700 & 50.7 & \\
Cd$_2$Cu$_3$(OH)$_6$(SO$_4$)$_2\cdot$4H$_2$O\cite{fujihala2014unconventional} & Kagome & 1/2 & 5 & -40.5 & 8.1 & \\
BCGO ($p=0.902$)~\cite{yang2016glassy} & Kagome & 3/2 & 5 & -695 & 139 & \\
SrCr$_{8.82}$Ga$_{3.18}$O$_{19}$~\cite{PhysRevLett.84.2957} & Kagome & 3/2 & 3.5 & -500 & 143 & \\
Ba$_2$Sn$_2$Ga$_3$ZnCr$_7$O$_{22}$~\cite{hagemann2001geometric}  & Kagome & 3/2 & 1.5 & -312 & 208 & \\
Mn$_3$C$_6$S$_6$~\cite{murphy2021exchange} & Kagome & 5/2 & 12 & -253 & 21 & \\

\hline

Mn$_2$Ga$_2$S$_5$~\cite{shen2022spin} & Triangular & 5/2 & 12 & -260 & 22 & \SetCell{r=9}No strict local constraint; frustration-driven 120$^\circ$ manifold; topological defects (Z$_2$ vortices); chirality as emergent degree of freedom; defect-mediated transitions \\
NiGa$_2$S$_4$~\cite{nambu2006coherent} & Triangular & 1 & 8 & -80 & 10 & \\
FeGa$_2$S$_4$~\cite{PhysRevLett.99.157203} & Triangular & 2 & 16 & -160 & 10 & \\
Ni$_{0.7}$Zn$_{0.3}$Ga$_2$S$_4$~\cite{nambu2006coherent} & Triangular & 1 & 4.5 & -57 & 12.7 & \\
CuCrSSe~\cite{tewari2024signature} & Triangular & 3/2 & 18 & -81 & 4.5 & \\
Cu$_2$(OH)$_3$(C$_7$H$_{15}$COO)\cite{girctu2000glassiness} & Triangular & 1/2 & 20 & -140 & 7 & \\
FeAl$_2$Se$_4$\cite{PhysRevB.99.054421} & Triangular & 2 & 14 & -200 & 14 & \\
YbZnGaO$_4$\cite{PhysRevLett.120.087201} & Triangular & 1/2 & 0.1 & -2.46 & 24.6 & \\
NaRuO$_2$\cite{ortiz2023quantum} & Triangular & 1/2 & 1.5 & -- & -- & \\

\hline

Sr$_2$CuTe$_{0.5}$W$_{0.5}$O$_6$\cite{PhysRevLett.127.017201} & Square & 1/2 & 1.7 & -71 & 41.8 & Valence-bond-glass state with disorder-induced spin-liquid correlations \\

\hline

Li$_2$RhO$_3$~\cite{PhysRevB.96.094432} & Kitaev & 1/2 & 6 & -60 & 10 & \SetCell{r=6}Exact local integrals of motion ($Z_2$ flux); fractionalization into Majorana fermions + gauge field; non-Abelian anyons (with field) \\
Ru$_{0.9}$Cr$_{0.1}$Cl$_3$~\cite{PhysRevB.99.214410} & Kitaev & -- & 3.1 & -85 & 28 & \\
Na$_2$Ir$_{0.89}$Ti$_{0.11}$O$_3$~\cite{PhysRevB.89.241102} & Kitaev & 1/2 & 5 & -55 & 11 & \\
Na$_2$Ir$_{0.9}$Ru$_{0.1}$O$_3$~\cite{PhysRevB.92.134412} & Kitaev & 1/2 & 5.6 & -127 & 23 & \\
Li$_2$Ir$_{0.91}$Ti$_{0.09}$O$_3$~\cite{PhysRevB.89.241102} & Kitaev & 1/2 & 3.5 & -33 & 9 & \\
Li$_2$Ir$_{0.78}$Ti$_{0.22}$O$_3$~\cite{PhysRevB.89.241102} & Kitaev & 1/2 & 2 & -33 & 16 & \\

\hline
\end{tblr}
\label{material_name_part1}
\end{table*}

\begin{table*}[t]
\centering
\renewcommand{\arraystretch}{1.5}
\caption{Magnetic and thermodynamic parameters of frustrated quantum magnets exhibiting unconventional spin-freezing behavior (Part II, continued).}
\begin{tblr}{
colspec={ |Q|Q|Q[1.5cm]|Q|Q|Q|Q[4cm]|}
}
\hline
\textbf{Material [ref.]} & \textbf{Lattice} & \textbf{Effective Spin} & $T_g$ (K) & $\theta_\text{CW}$ (K) & $f=\frac{|\theta_\text{CW}|}{T_g}$ & \textbf{Topological characteristics}\\
\hline

$\beta$-ZnIrO$_3$~\cite{PhysRevMaterials.6.L021401} & Hyperhoneycomb & 1/2 & 12 & 45 & 3.75 & 3D extension of Kitaev physics; loop-like $Z_2$ flux excitations\\
$\beta$-MgIrO$_3$~\cite{zubtsovskii2022topotactic} & Hyperhoneycomb & 1/2 & 23 & 62 & 2.7 & \\

\hline

Y$_2$Mo$_2$O$_7$~\cite{greedan1986spin} & Pyrochlore & 1 & 22 & -200 & 9 & \SetCell{r=10}Strong local constraint $\sum_{\text{tetra}}\vec{S}=0$; emergent Coulomb phase; deconfined magnetic monopoles; Dirac strings; pinch-point correlations; emergent hexagonal loop excitations\\
Lu$_2$Mo$_2$O$_7$~\cite{PhysRevLett.113.117201} & Pyrochlore & 1 & 16 & -160 & 10 & \\
Y$_2$Ru$_2$O$_7$~\cite{ito2001nature,PhysRevB.74.104425} & Pyrochlore & 1 & 76 & -1250 & 16.5 & \\
Y$_{1.75}$La$_{0.25}$Ru$_2$O$_7$~\cite{chatterjee2026tunable} & Pyrochlore & 1 & 69 & -893 & 12.8 & \\
Y$_2$Ir$_2$O$_7$~\cite{taira2001magnetic,kumar2016nonequilibrium} & Pyrochlore & 1/2 & 150 & -386 & 2.5 & \\
NaCaNi$_2$F$_7$~\cite{PhysRevB.92.014406} & Pyrochlore & 1 & 3.6 & -129 & 36 & \\
NaCaCo$_2$F$_7$~\cite{PhysRevB.89.214401,PhysRevB.93.014433} & Pyrochlore & 1 & 2.4 & -140 & 56 & \\
CsNiCrF$_6$~\cite{xvxt-whns} & Pyrochlore & -- & 2.3 & -60 & 26 & \\
ZnCr$_{1.6}$Ga$_{0.4}$O$_4$~\cite{PhysRevLett.110.017203,fiorani1985antiferromagnetic} & Pyrochlore & 3/2 & 2.4 & -115 & 48 & \\
Zn$_{0.95}$Cd$_{0.05}$Cr$_2$O$_4$~\cite{PhysRevB.64.024408,PhysRevB.65.220406} & Pyrochlore & 3/2 & 10 & -500 & 50 & \\
\hline

Na$_4$Ir$_3$O$_8$~\cite{PhysRevLett.99.137207} & Hyperkagome & 1/2 & 6 & -650 & 108 & 3D corner-sharing network; loop constraints; quantum spin liquid candidate; extended fractionalization \\
Gd$_5$Ga$_5$O$_{12}$~\cite{PhysRevLett.74.2379} & Hyperkagome & 7/2 & 0.14 & -2 & 14.3 & \\

\hline
KSrFe$_2$(PO$_4$)$_3$~\cite{boya2022signatures} & Trillium & 5/2 & 3.5 & -70 & 20 & non-centrosymmetric cubic symmetry and structural chirality 
\\

\hline
(Fe,Ga)$_2$TiO$_5$~\cite{li2023evolution} & Pseudobrookite & 5/2 & 55 & -900 & 16 & Competing exchange paths; frustration without well-defined topological constraint \\

\hline
Ca$_2$LaRuO$_6$~\cite{ann2025electronic} & double-perovskite & 5/2 & 13 & -137 & 10.5  & spin-orbit coupled state; valence bond-glass formation (Mo-variant) \\
Sr$_2$CaReO$_6$~\cite{PhysRevB.65.144413} & double-perovskite & 1/2 & 14 & -443 & 31.6 &  \\
Ba$_2$YMoO$_6$~\cite{PhysRevLett.104.177202,PhysRevB.82.094424} & double-perovskite & 1/2 & 1.3 & -160 & 123 &  \\

\hline
\end{tblr}
\label{material_name_part3}
\end{table*}

\begin{figure*}[t]
\includegraphics[height=139.8955pt, width=510.0pt]{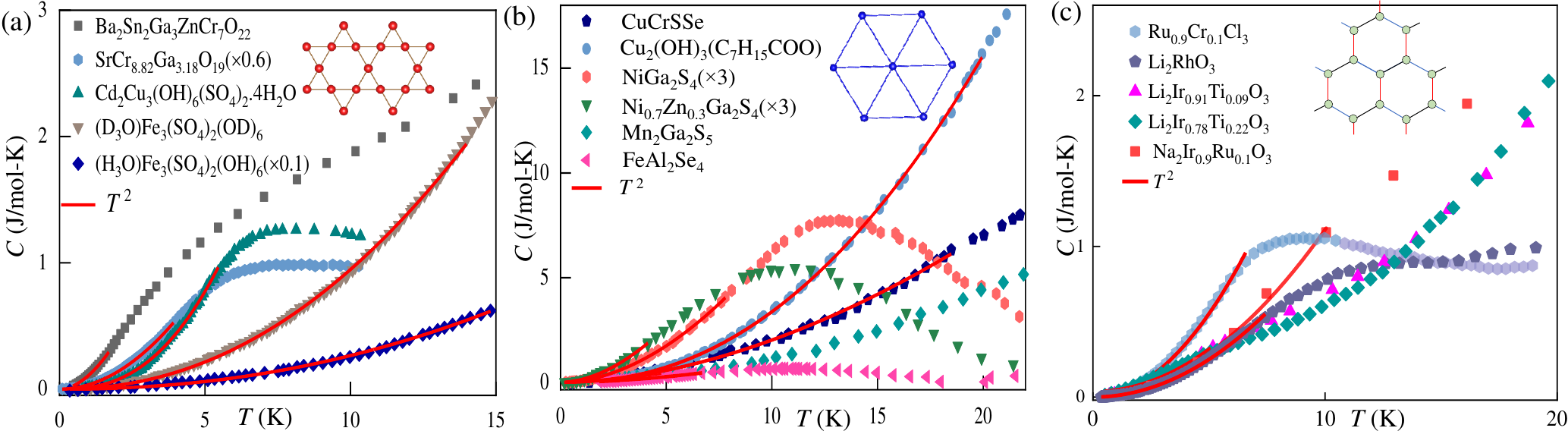}
  \caption{\textbf{Temperature dependence of the magnetic specific heat in frustrated spin glasses} Temperature dependence of some of the frustrated spin glasses (a) kagome, (b) triangular, and (c) honeycomb spin lattices. The low-temperature regime (below $T_{g}$) is analyzed using spin stiffness fitting, providing insights into the nature of the ground state and the role of frustration in governing low-energy excitations, as discussed in the text. The data presented here are adapted from the references listed in Tables I and II.}
  \label{fig:figure1}
\end{figure*}

In summary, the intrinsic spin freezing phenomenon in frustrated kagome magnets arises from the interplay between emergent defects and the non-Abelian nature of their dynamics. In these systems, XY anisotropy, complex defects, and chiral order produce spin relaxation which are dominated by the braiding of defect configurations. The combined effect of XY anisotropy, non-Abelian defects, and complex chiral order leads to this intrinsic glassiness. The spin relaxation is mainly dominated by the non-trivial braiding of defect configurations rather than simple thermal excitations. To illustrate the role of topological defects in glassy behavior, let us consider an example from a non-magnetic system. Such as the recent study on crystallization of a spherical surface shows that topological constraints can drive the system into a glassy state~\cite{guerra2018freezing}. The topological constraint imposed by geometry on a spherical surface requires the existence of at least twelve five-coordinated disclinations~\cite{caspar1962physical}. Grain-boundary scars are frequently formed by additional defects to further alleviate the stress caused by curvature~\cite{guerra2018freezing}. These inevitable defects remain mobile and tend to cluster in certain regions, leading to persistent structural and dynamical heterogeneity. Even though such systems are not magnetic, they show the universality of topological defects in forming glassy states.

\subsection{Triangular lattice antiferromagnets} 
\begin{figure*}[t]
\includegraphics[height=358.01799pt, width=510.0pt]{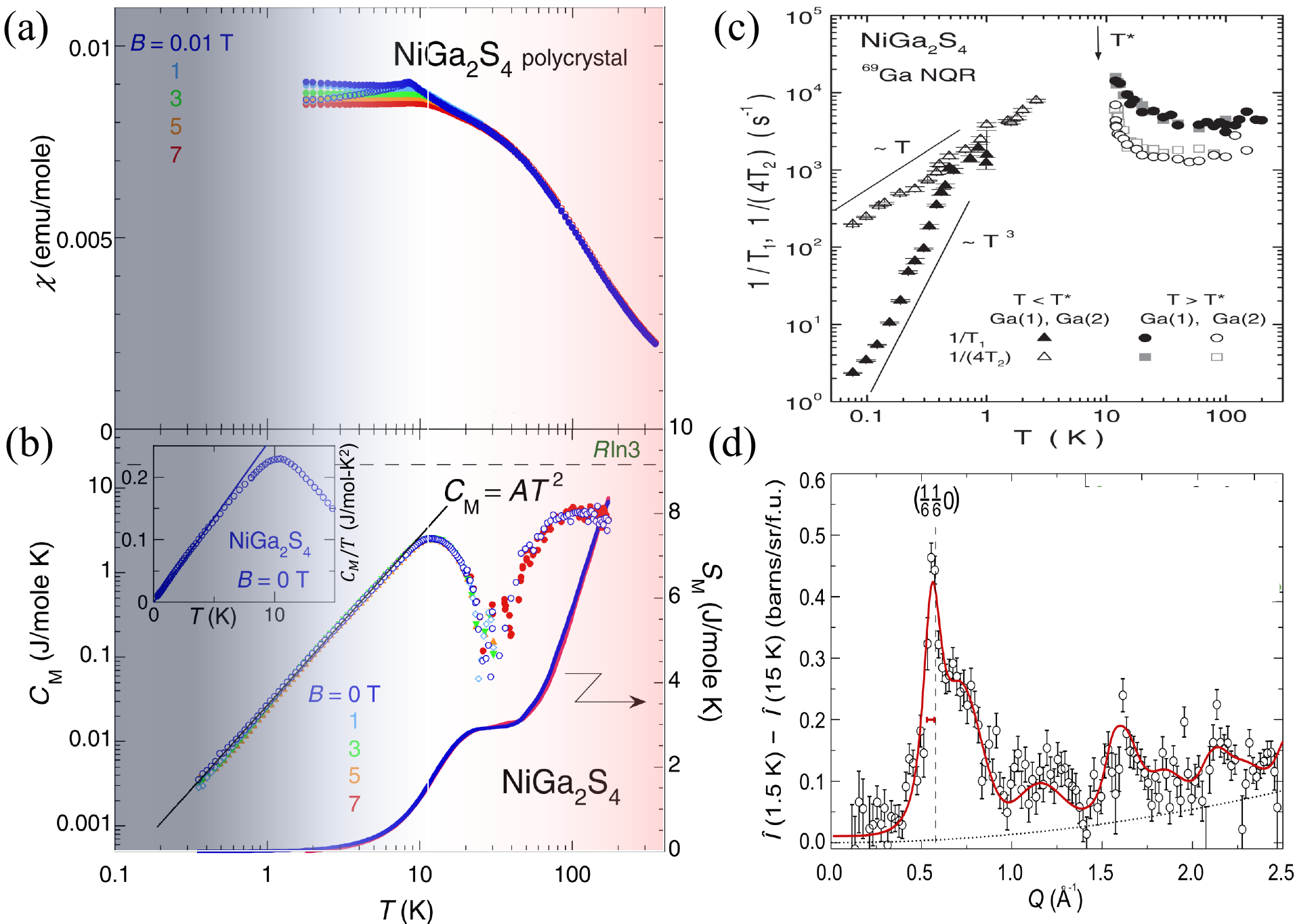}
  \caption{ \textbf{Unconventional spin freezing in frustrated triangular lattice antiferromagnet} (a) Temperature dependence of DC magnetic susceptibility measured under various applied fields. 
    (b) Magnetic specific heat ($C_M$, left axis) of polycrystalline NiGa$_2$S$_4$ under different magnetic fields, plotted on a full logarithmic scale. The temperature dependence of entropy ($S_M$, right axis)  at 0 T (blue) and 7 T (red). Inset: $C_M/T$ versus temperature at zero field. The solid line indicates the quadratic behavior of CM. (c) Nuclear quadrupole resonance relaxation rates (spin-lattice relaxation rate,$1/T_1$ and spin-lattice relaxation rate, $1/4T_2$) as functions of temperature for $^{69}$Ga(1) and $^{69}$Ga$(2)$ sites. (d) Elastic neutron scattering intensity difference between 1.5 K and 15 K, with the red horizontal bar denoting the FWHM resolution width.The red solid line depicts the  calculated pattern fir the quasi-2D structure and $\sim Q^2$ term (dashed) to account for nuclear incoherent scattering. Source:(a,b,d) adapted from \cite{nakatsuji2005spin} with permission from AAAS, (c) adapted from \cite{PhysRevB.77.054429} with permission from APS.}
  \label{fig: NiGa2S4 Chi, Cm, NMR, Neutron}
\end{figure*}

Two-dimensional triangular lattice antiferromagnets are
formed by edge-sharing triangular motifs with magnetic
moments at the vertices of the triangle, with each magnetic moment is connected to six neighbors. Triangular lattice is the simplest prototype to demonstrate geometric frustration. The presence of geometric frustration in the antiferromagnetic interaction of spin moments leads to substantial degeneracy. The strong quantum fluctuations that arise from magnetic frustration prevent long-range magnetic ordering even at absolute zero temperature. Thereby, this can pave the way for the realization of a diverse array of quantum phenomena in the frustrated spin lattice. G.H. Wannier's pioneering work on the statistical behavior of the two-dimensional Ising antiferromagnetic triangular lattice attracted early interest in the study of frustrated spin networks~\cite{PhysRev.79.357}. For nearest-neighbor Heisenberg interactions, the classical ground state adopts on a 120° coplanar configuration and long-range  order can survive even at $T = 0$ K~\cite{chubukov1994large,PhysRevLett.69.2590,PhysRevLett.68.1762}. However, the next-nearest-neighbor interaction, anisotropic exchange interaction, and structural disorder can induce exotic  many-body phenomena in frustrated triangular lattice antiferromagnets~\cite{arh2022ising,khatua2023experimental,balents2010spin}. It has been suggested that quantum fluctuations can stabilize a resonating valence bond state in spin-$1/2$ Heisenberg triangular antiferromagnets~\cite{anderson1973resonating}.

The finite-temperature behavior of the triangular antiferromagnet has possible instabilities toward topological defect binding to promote anomalous spin freezing~\cite{nakatsuji2005spin}. In particular, theoretical and numerical studies have proven that the triangular lattice supports $Z_2$ vortices, which are topological point defects of the vector spin chirality that arise from the nontrivial homotopy of the 120$^\circ$ spin structure. These excitations originate from the noncollinear 120$^\circ$ spin configuration and can unbind upon heating, thereby producing slow spin dynamics and possible finite-$T$ phase transitions. Such a phase transition does not necessarily exhibit a sharp anomaly in any thermodynamic quantity. The $Z_2$ vortices originate from the topology of the noncollinear order parameter. The 120$^\circ$ state is special only up to a global spin rotation and a global spin reflection. Consequently, the associated order-parameter manifold is $\text{SO}(3)$, associated with an oriented triad of spin directions. Its fundamental group, $\pi_1(\text{SO}(3)) = \mathbb{Z}_2,$ hosts stable point defects, recognized as $Z_2$ vortices~\cite{PhysRevB.97.195115}. In classical Heisenberg triangular antiferromagnets, a finite-temperature transition associated with the binding and unbinding of these vortices has been proposed and supported by Monte Carlo simulations~\cite{kawamura1984phase,kawamura2010z}. Importantly, this transition is not equivalent to the Berezinskii–
Kosterlitz–Thouless (BKT) transition of the XY model. More precisely, the BKT transition in systems possessing a continuous U(1) symmetry takes place with topological excitations: the integer vortices are characterized by the winding of the spin angle $\theta$~\cite{kosterlitz1973ordering}. In such a scenario, the BKT transition, at some finite characteristic temperature $T_\text{BKT}$, is believed to be related to the vortex-antivortex pairing. This, in turn, leads to the propagation of the long-wavelength spin-wave excitations due to algebraic spin-correlation decay. Here, the spin wave propagates coherently through the system. Contrasting this, in the triangular-lattice Heisenberg antiferromagnet, the binding of $Z_2$ vortices gives rise to topological ordering. The spin-wave propagates only within the coherent domains of the spin-gel state. This leads to slow and overdamped dynamics rather than coherent, long-range spin-wave propagation~\cite{kawamura2010z}.

A canonical example of triangular frustrated spin glass is the $S=1$ NiGa$_2$S$_4$~\cite{nakatsuji2005spin}. Despite the large antiferromagnetic Curie-Weiss temperature ($\theta_\text{CW}=-80$ K), it does not show any long-range magnetic ordering down to sub-Kelvin temperatures. Instead, it exhibits short-range spin correlations
with incommensurate wavevectors and finite correlation
lengths below the spin glass transition $T_g\sim 8$ K~\cite{nakatsuji2007coherent, PhysRevB.78.220403}. The bulk thermodynamic measurements provide evidence for the occurrence of anomalous spin freezing in novel frustrated magnets. The observed magnetic specific heat exhibits two broad peaks at different temperature ranges (see Fig.~\ref{fig: NiGa2S4 Chi, Cm, NMR, Neutron}(b)). The temperature region below the low-temperature broad peak exhibits a $T^2$ power-law behavior. This indicates the presence of Goldstone modes emerging in this two-dimensional triangular lattice without any long-range magnetic ordering~\cite{PhysRevB.78.180404, tsunetsugu2006spin}. The NMR spin lattice relaxation rate shows $1/T_1 \propto T^3$ at temperatures below 1 K as shown in Fig. \ref{fig: NiGa2S4 Chi, Cm, NMR, Neutron}(c) \cite{PhysRevB.77.054429}. Both the NMR and ESR studies further reveal the anomalous line broadening and slowing down of spin dynamics below $T_g$, which suggests the presence of some exotic low energy excitations in this Ni-based triangular lattice~\cite{PhysRevB.79.140402}. The spin freezing observed in NiGa$_2$S$_4$ has been suggested as a phenomenon of topological origin. This phenomenon is thought to result from the pairing of $Z_2$ vortices inherent in the frustrated Heisenberg triangular lattice~\cite{nakatsuji2010novel,PhysRevB.78.180404}. Neutron scattering performed at 1.5 K suggests that the intra-plane spatial spin correlations extend up to seven lattice spacings, and in comparison, the interplane correlation is very weak (Fig. \ref{fig: NiGa2S4 Chi, Cm, NMR, Neutron}(d)). An alternative perspective on the unconventional spin freezing behavior can be obtained from the role of quadrupolar or spin-nematic correlations, which appear naturally in the case of $S=1$ Ni$^{2+}$ moments. The spin nematic states emerge from the dominant contribution of higher-rank spin correlations, which give rise to long-range order of spin tensors. This is in contrast to the conventional long-range ordering due to spin dipole moments~\cite{tsunetsugu2006spin,PhysRevB.79.214436}. It thus accounts for the absence of magnetic Bragg peaks in neutron scattering below $T_g$ while low-energy spin excitations along with finite susceptibility survive. In these cases, freezing results from the slowing down of the nematic directors that give rise to a hidden order with slow, critical-like fluctuations in the low-temperature limit. A valuable comparison can be made with the isostructural compound FeGa$_2$S$_4$ ($S = 2$). The $\mu$SR experiments show an onset of quasistatic magnetism at $T\approx 31$ K, well
above a lower freezing temperature $T_g\approx$ 16 K~\cite{PhysRevB.86.064435}. Both the NiGa$_2$S$_4$ and FeGa$_2$S$_4$ exhibit field-independent specific heat with a two-peak nature in $C_m/T$. The high-temperature peak is observed at 60 K, and the lower one is visible at 10 K~\cite{PhysRevLett.99.157203}. Interestingly,  triangular lattice Mn$_2$Ga$_2$S$_5$ shows a spin-glass-like freezing near the temperature 12 K, where the magnetic specific heat follows a quadratic power-law behavior (Fig.~\ref{fig:figure1}(b)) and the ac susceptibility reveals strong frequency dependence~\cite{shen2022spin}. The above study highlights unconventional spin freezing in triangular lattices, and some representative candidate materials with their unique characteristic features are presented in Table~\ref{material_name_part1} and the specific heat with quadratic $T^2-$ dependence is depicted in the Fig.~\ref{fig:figure1}(b).

Now moving towards the 4$f$ triangular frustrated glassy magnets, YbZnGaO$_4$ and its sister compounds exhibit anomalous glassy dynamics at low temperatures below the characteristic exchange energy scale~\cite{PhysRevLett.122.137201, PhysRevLett.120.087201, huang2024emergent, li2020kosterlitz}. The specific heat shows a broad anomaly at a temperature slightly above the glass transition temperature. Spin freezing associated with this 4$f$ oxide is quite different from that of transition metal-based Mott insulating spin glasses. Because the 4$f$ orbitals are highly localized as compared to their 3$d$ counterparts, the effective exchange energy is typically much smaller than that of the 3$d$ frustrated magnets. In 4f magnetic oxides, the low-energy magnetic properties are largely shaped by single-ion anisotropy, which originates from the interplay of spin-orbit coupling and splitting of multiplets due to the crystalline electric field (CEF)~\cite{arh2022ising,khatua2024classical,PhysRevB.111.094408,PhysRevB.102.054428}. The ground state is often a CEF doublet or singlet with a strong anisotropic $g$-tensor, and the low-energy excitations are mainly the transition between CEF levels rather than gapless modes. The anisotropy present in 4$f$-based frustrated spin glass can gap out the transverse spin-wave modes~\cite{PhysRevB.16.2154}. The nature of low-energy excitations in the spin glass state of YbZnGaO$_4$ is mainly governed by the localized excitations. Such behavior underlines the distinct nature of the spin glass state in this material.
Localized modes, which are also observed in 2D amorphous glasses, indicate regions of structural irregularities where plastic deformation is likely to occur~\cite{wu2023topology}. Those modes usually correspond to topological defects~\cite{baggioli2023topological} and serve as a precursor of shear transformation upon mechanical stress~\cite{wu2023topology}. Another example is the triangular lattice spin glass TmMgGaO$_4$, which contains Tm$^{3+}$ ions with an electronic configuration of 4$f^{12}$ and a total angular momentum $J = 6$~\cite{PhysRevX.10.011007,shen2019intertwined}. These ions, due to crystal-electric-field splitting, form a non-Kramers ground state doublet that is well separated from higher energy levels by approximately 400 K~\cite{hu2020evidence}, and the material can be treated as low spin system at low temperatures. This compound, as a quantum Ising magnet, shows that quenched disorder coupled with interlayer coupling in the triangular lattice Ising antiferromagnet transforms the emergent 2D BKT phase into a U(1) gauge glass. This, in turn, gives rise to an emergent HS mode~\cite{huang2024emergent}. The weak quenched disorder favors the emergence of the HS mode~\cite{PhysRevB.101.024413}, while the ferromagnetic interlayer coupling makes the 2D gauge glass become a 3D gauge glass in which the HS mode survives~\cite{huang2024emergent}.

\subsection{Square lattice antiferromagnets}

The square lattice, a fundamental geometry of bipartite latice in condensed matter physics, hosts a variety of fascinating quantum states, including high-$T_\text{C}$ superconductors, spin liquids, and non-collinear magnetic ordering, etc. Lars Onsager's pioneering work on the ferromagnetic Ising square lattice showed that even in a 2D system~\cite{PhysRev.65.117}, a finite-temperature phase transition from a disordered to an ordered magnetic state is possible. 

Another fascinating property in a square lattice is the physics of superconductivity in many copper oxide materials. Some cuprate oxides materials such as (La, Sr)$_2$CuO$_4$ and YBa$_2$Cu$_3$O$_7$, where Cu$^{2+}$ ($S=1/2$) ion constitute a square planar coordination with oxygen atoms, exhibit strong covalent bonding between copper and oxygen through the hybridization of 3$d$ and 2$p$ orbitals. These CuO$_2$ planes play a vital role in realizing superconductivity, where Cu-O bonds in these planes form high covalency. The key requirement for superconductivity to form an electron pair comes from the multiple oxidation states of copper atoms(Cu$^{+}$, Cu$^{2+}$, Cu$^{3+}$) in the Cu-O bond~\cite{o2022electron}.

Although the cuprate lattices described in terms of perovskite-related structures, the conducive electronic structure elucidates from the square CuO$_2$ planes. The chemical bonds of Cu-O contributing to the states near the Fermi level are anti-bonding in nature, which contributes to the superconductive states. As a two-dimensional geometry, the cuprate square lattices lead antiferromagnetic spin fluctuations, which strongly renormalize the ground-state properties. The CuO$_2$ planes create unconventional pairing channels, mostly the $d$-wave pairing symmetry observed in high $T_\text{C}$ cuprate superconductors~\cite{RevModPhys.78.17,keimer2015quantum}. Beyond this superconductivity, the square lattice is also important for exploring frustration-induced quantum states such as quantum spin liquid and spin glass. The nearest and next-nearest neighbor antiferromagnetic couplings ($J_{1}-J_{2}$model) in square lattice introduce competing exchange pathways, which are capable of destabilizing conventional Néel order and give a complex magnetic ground state~\cite{schulz1996magnetic,PhysRevLett.97.117204}. 

The square lattice offers a versatile platform to study the transition between these phases. For instance, tuning the square lattice material by controlled external stimuli such as doping, hydrostatic pressure can drive the system from a superconducting state to a spin-liquid-like state or a spin-glass phase. The coexistence of spin-glass and spin-liquid states in the Li-doped quasi-two-dimensional square lattice systems such as La$_2$Cu$_{1-x}$Li$_x$O$_4$ reveals the subtle interplay between disorder and quantum fluctuations~\cite{PhysRevB.72.184401}. The interconnection between all of these phases suggests that spin fluctuations in the square plane and frustration play a crucial role in hosting these phenomena. The spin freezing phenomena in this doped cuprate superconductor suggest that glassy spin dynamics may underlie the disruption of long-range coherence, coexistance with fast spin fluctuations, offering a microscopic window on how magnetic disorder breaks superconductivity and the emergence of unconventional pairing mechanisms~\cite{PhysRevLett.83.604,PhysRevB.62.9148,PhysRevB.69.014424}.

In spin-$\frac{1}{2}$ square lattice Heisenberg antiferromagnets, the interplay between nearest-neighbor ($J_1$) and next-nearest-neighbor ($J_2$) interactions leads to exchange frustration. Within the range $ 0.4 \leqslant \frac{J_2}{J_1} \leqslant  0.6$, a quantum disordered ground state can be realized~\cite{PhysRevB.98.054422}. The square lattice Sr$_2$CuTeO$_6$ with spin $S=1/2$ shows antiferromagnetic long-range ordering at 29 K. Here, the magnetic interaction is mainly due to nearest-neighbor exchange $J_1 = -7.18$ meV and a weaker next-nearest-neighbor interaction ($J_2 = -0.21$ meV)~\cite{PhysRevLett.117.237203}. Substituting Te$^{6+}$ at W$^{6+}$ in the square-lattice Sr$_2$CuWO$_6$ results in columnar Néel ordering, with a lower Néel temperature ($T_N$) of 24 K. This columnar ordering arises due to magnetic interactions characterized by $J_1 = -1.2$ meV and $J_2 = -9.5$ meV~\cite{PhysRevB.94.064411, vasala2014magnetic}. This indicates that the magnetic ground state is heavily influenced by the cation composition. The W$^{6+}$ is a $d^{0}$ cation, while Te$^{6+}$ is a $d^{10}$ cation. By varying the proportions of these ions in the compound Sr$_2$CuTe$_x$W$_{1-x}$O$_6$, the magnetic ground state of the square Cu$^{2+}$ lattice can be tuned accordingly~\cite{mustonen2018spin}. For a 50:50 \% composition of Te$^{6+}$ and W$^{6+}$ in Sr$_2$CuTe$_{0.5}$W$_{0.5}$O$_6$, no magnetic ordering is observed down to sub-Kelvin temperatures despite a strong antiferromagnetic interaction as indicated by $\theta_\text{CW}=-71$ K~\cite{PhysRevB.105.184410,mustonen2018spin,PhysRevB.98.054422,uematsu2017randomness}. Specific heat measurements show a change from linear to quadratic behaviour below 7 K and 1.5 K, respectively. This means that the material goes from being a spin liquid to a frozen state because of local freezing effects caused by the random distribution of Te and W ions~\cite{PhysRevB.98.054422}.

\begin{figure}[t]
		\begin{center}			
        \includegraphics[height=229.49759pt, width=244.80055pt]{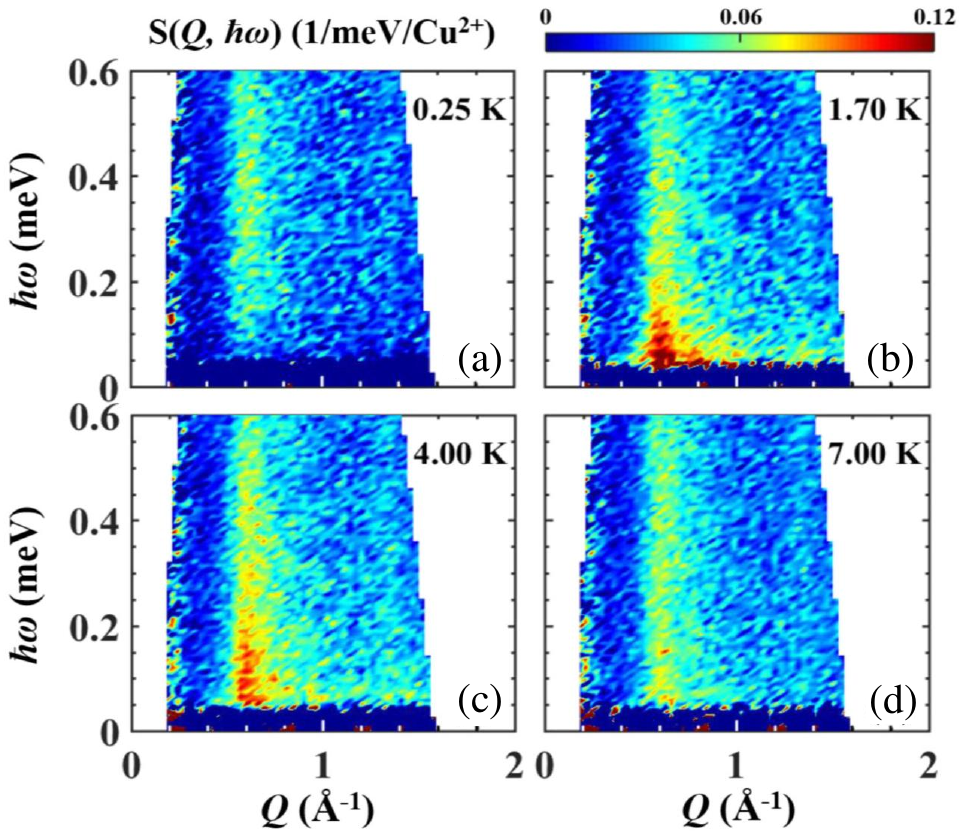}
			\caption{\textbf{Inelastic neutron scattering probing novel quantum states in frustrated magnets}.Color contour maps illustrate the low-energy spin fluctuations in Sr$_2$CuTe$_{0.5}$W$_{0.5}$O$_6$, obtained via inelastic neutron scattering with an incident energy of 1.55 meV. Measurements were conducted at temperatures of 0.25 K, 1.70 K, 4.00 K, and 7.00 K. Source: adapted from \cite{PhysRevLett.127.017201} with permission from APS.}
			\label{square neutron}
		\end{center}
	\end{figure}

Neutron scattering experiments on Sr$_2$CuTe$_{0.5}$W$_{0.5}$O$_6$ confirmed the transition from a dynamic state to a weakly frozen state at temperatures below 1.7 K~\cite{PhysRevLett.127.017201}. The imaginary part of the dynamic susceptibility, $\chi^{\prime\prime}$, derived from the neutron scattering cross section S(Q, $\hbar\omega$), demonstrates a linear relationship with temperature for energy transfers where $\hbar\omega<k_BT_g$. The existence of weak frozen moments indicates the equilibrium between quantum fluctuations and static spin correlations influenced by bond randomness. The dynamic structure factor $S(Q, \hbar \omega)$, examined at temperatures of 7 K, 4 K, 1.7 K, and 0.25 K, as illustrated in Fig. \ref{square neutron}, reveals gapless excitations near $Q \approx 0.6$ \AA$^{-1}$ and maintains this gaplessness down to $\hbar \omega \approx 0.05$ meV, which corresponds to the antiferromagnetic wave vector of (1/2,0,0) and reflects the system's intrinsic magnetic structure. As the temperature drops from 7 K to 1.7 K, the intensity of $S(Q, \hbar \omega)$ rises above 0.3 meV, and the spectral weight moves to lower energies. It is surprising that when the temperature drops from 1.7~K to 0.25~K, $S(Q, \hbar \omega)$ weakens below 0.2 meV. This shows that spin freezing occurs as the temperature approaches closer to the critical point. Together with this result with linear energy dependence of the dynamic susceptibility $\chi''(\hbar \omega)$ suggests that a significant portion of the spectral weight with approximately 99.4$\%$ of the total spectral weight being fluctuating rather than static. the specific heat measurement shows a quadratic power-law dependence on temperature. These results emphasize that quantum fluctuations and partial spin freezing coexist, giving rise to rich and unconventional spin dynamics.

\begin{figure}[t]
		\begin{center}
			\includegraphics[height=367.21576pt, width=239.70473pt]{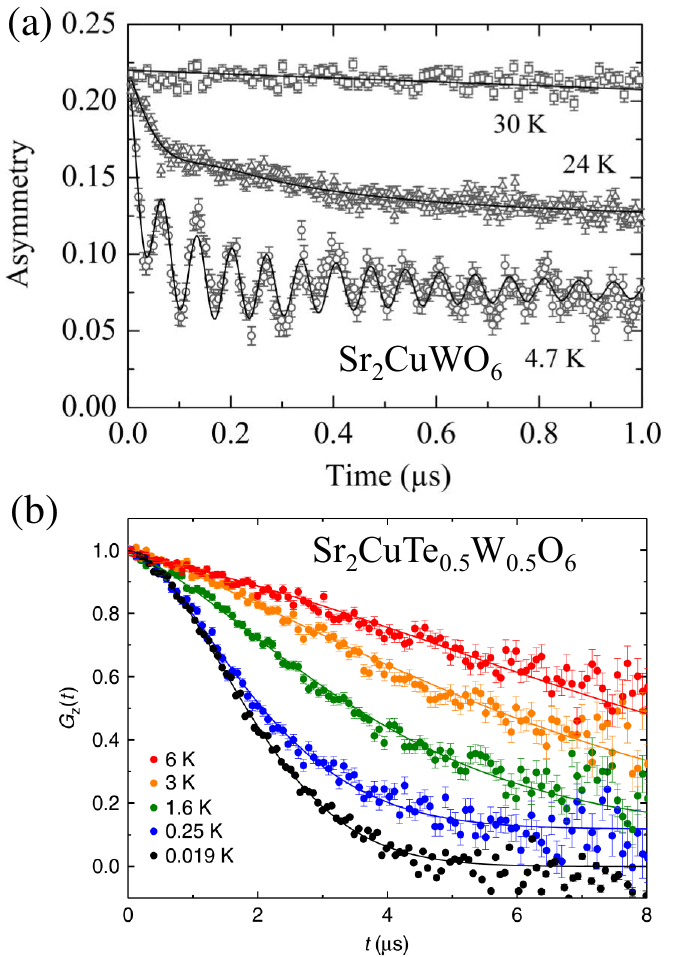}
			\caption{\textbf{$\mu$SR as a local probe to track intriguing quantum states and excitations in frustrated magnets}. (a) Zero-field muon spin relaxation asymmetry as function of time in Sr$_2$CuWO$_6$.(b)Sr$_2$CuTe$_{0.5}$W$_{0.5}$O$_6$ recorded at various temperatures, illustrating the evolution of intriguing ground states and associated spin dynamics with decreasing temperature. Source: (a) adapted from \cite{PhysRevB.89.134419} with permission from APS, (b) adapted from \cite{mustonen2018spin} with permission from NPG.}
			\label{Square_Muon}
		\end{center}
	\end{figure}

Muon spectroscopy shows that Sr$_2$CuWO$_6$ and Sr$_2$CuTe$_{0.5}$W$_{0.5}$O$_6$ have different magnetic properties. In Sr$_2$CuWO$_6$, long-range Néel order manifests at 24 K, as indicated by oscillatory asymmetry in the ZF-$\mu$SR spectra~\cite{PhysRevB.89.134419}. Above this transition temperature at 30 K (Fig.~\ref{Square_Muon}(a)), the spectra do not show any oscillations, which means that the state is paramagnetic. Below $T_N$ at 4.7 K, the oscillatory signal remained strong, indicating that the ordered phase is stable and the internal field is well-defined. The 1/3 asymptotic behaviour of the depolarization curve below $T_N$ shows that the sample is polycrystalline. One-third of the implanted muon spins align with the local field and do not precess, while the other two-thirds experience transverse components that cause depolarization~\cite{le2011muon}.In the case of Sr$_2$CuTe$_{0.5}$W$_{0.5}$O$_6$, the ZF-$\mu$SR spectra show a different pattern of behavior~\cite{mustonen2018spin}. The time-dependent polarisation function, $G_z(t)$, which is directly linked to the asymmetry, does not exhibit oscillatory behavior or the typical recovery of the asymmetry to 1/3 in zero field measurements down to 19 mK as shown in Fig.~\ref{Square_Muon}(b)~\cite{PhysRevB.89.134419}. As shown in Fig.~\ref{Square_Muon}(b), the depolarization increases as the temperature decreases, indicating a broadening of the internal field distribution at lower temperatures. Fitting the depolarization function using a power-law form 
\( G_z(t) = \exp[-(\lambda t)^{\beta}] \) reveals that above 2 K, 
\(\beta\) is close to 1. However, as the temperature decreases, 
\(\beta\) increases toward 1.8~\cite{mustonen2018spin}. In topological spin glasses, 
a Gaussian-like relaxation with \(\beta \approx 2\) is characteristic, 
as observed in the spin glass SrCr$_8$Ga$_4$O$_{19}$. The relaxation rate \(\lambda\) increases at low temperatures, which is consistent with the picture that local field fluctuations become slower. The absence of a 1/3 tail in $\mu$SR in Sr$_2$CuTe$_{0.5}$W$_{0.5}$O$_6$ compound can be ascribed to its topological spin-glass like nature, since 1\% of spins are frozen as estimated from neutron scattering measurements~\cite{PhysRevLett.127.017201}. Neutron scattering experiments reveal that spin-wave like excitations of dynamic spin correlations extended over the entire magnetic bandwidth even in the presence of partial spin freezing~\cite{PhysRevLett.127.017201}. This observation suggests that the existence of spin wave excitations defies standard paradigm, suggesting to an unconventional interplay between frozen and dynamic spin components.

The observation of low-energy, linearly dispersing excitations and a corresponding linear dependence of the dynamic susceptibility in Sr$_2$CuTe$_{0.5}$ W$_{0.5}$O$_6$ suggests the presence of gapless modes associated with finite spin stiffness in a partially frozen state. In Sr$_2$CuTe$_{0.5}$W$_{0.5}$O$_6$, the interplay between quantum fluctuations (due to $S=1/2$ for Cu$^{2+}$) and bond disorder leads to a crossover regime in which both thermodynamic and microscopic probes reveal signatures of collective, glassy spin dynamics with quasi-linear excitation spectra.

\subsection{Kitaev magnets}

\begin{figure}[t]
		\begin{center}
			\includegraphics[width=0.47\textwidth,height=0.73\textwidth]{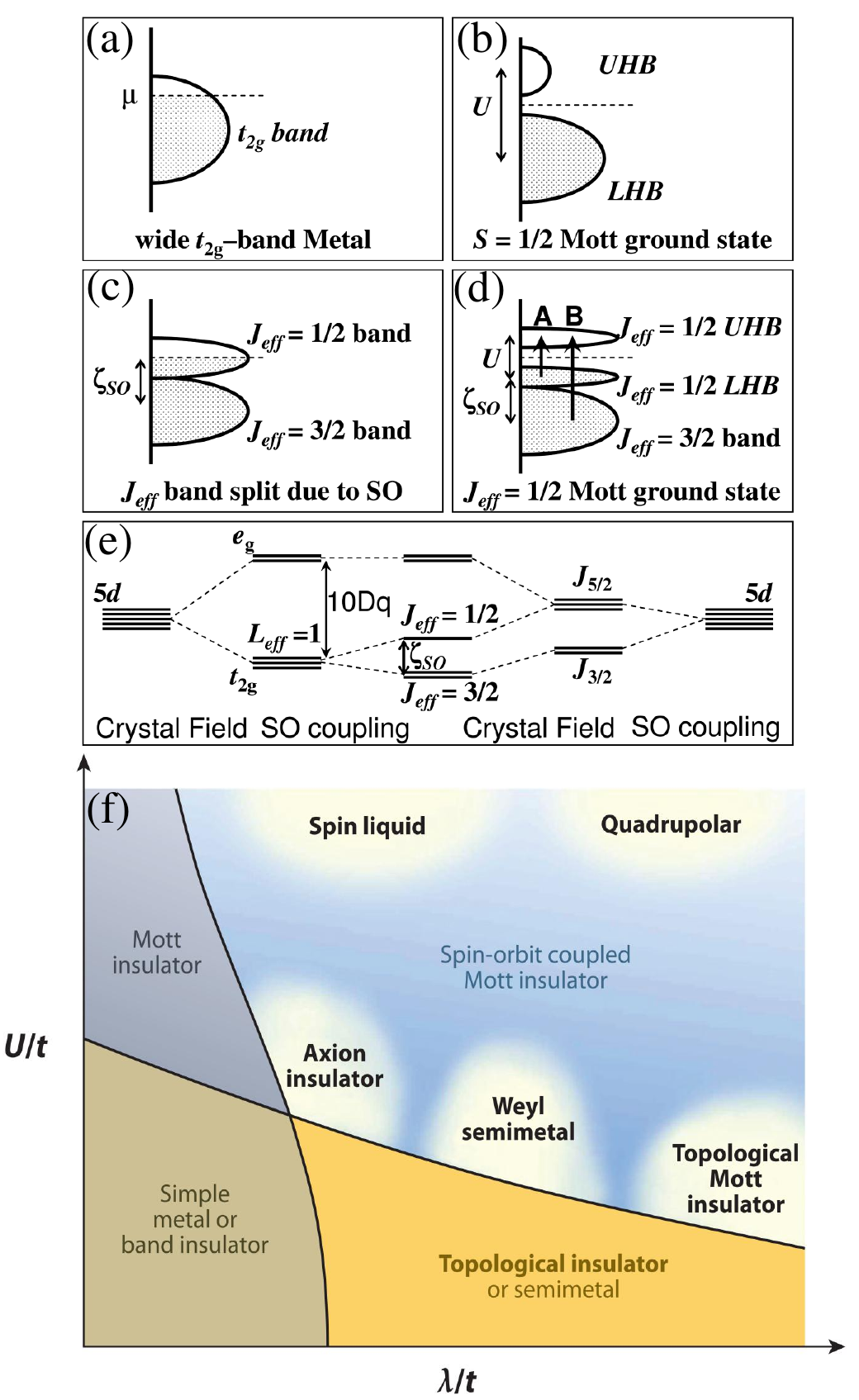}
			\caption{\textbf{Novel quantum states in frustrated 4d and 5d magnets.} Illustration of the energy level structure for a t$_{2g}^5$ band under varying interactions: (a) in the absence of both SOC and Coulomb repulsion U, where the system remains in a simple crystal-field-split state; (b) in the presence of a significantly large U but without SOC, leading to enhanced electron localization; (c) with SOC included but neglecting U, resulting in spin-orbital entangled states; and (d) incorporating both SOC and U, where their competition dictates the final electronic structure. Optical transitions, denoted as A and B, are indicated by arrows. (e) Energy level modifications induced by the interplay of crystal field effects and SOC. (f) A qualitative phase diagram mapping the impact of electron correlation strength (U/t) and SOC strength ($\lambda$/t), illustrating emergent electronic and magnetic phases in strongly correlated frustrated quantum materials. Source:(a-e) adapted from \cite{PhysRevLett.101.076402} with permission, (f) adapted from \cite{witczak2014correlated} with permission.}
			\label{KitaevPhase}
		\end{center}
	\end{figure}

The honeycomb lattice is well recognized for its geometric efficiency and mechanical stability in both natural and engineered systems~\cite{darwin1859origins}. The two-dimensional nature and low connectivity of the honeycomb lattice make it a model system in condensed matter physics, which can host a number of electronic and magnetic properties that offer an ideal platform to test theoretical conjectures. Because the antiferromagnetic honeycomb lattice is bipartite, it can support long-range magnetic order. However, recent studies of the spin-orbit-coupled Kitaev honeycomb lattice have revealed the presence of exotic Majorana fermions without any conventional spatial magnetic ordering. The Majorana fermions are significant both for advancing our understanding of fundamental quantum matter and for realizing fault-tolerant quantum computation~\cite{takagi2019concept, trebst2022kitaev,PhysRevResearch.6.L022005,sahasrabudhe2024chiral,sahasrabudhe2020high}. A significant change in physical properties occurs when moving from 3$d$ to 4$d$ and 5$d$ transition-metal compounds. Compared to 3$d$ transition metals, the heavier transition metals have spatially extended orbits which get hybridized with the ligand $p$ orbitals. This in turn broadens the electronic bandwidth in 4$d$ and 5$d$ transition-metal compounds~\cite{cao2021physics, cox1998exotic}. The electronic correlations, spin–orbit coupling, and crystal field effects all play important roles in driving novel ground states in these materials~\cite{RevModPhys.70.1039, cao2013frontiers,trebst2022kitaev}. Due to the extended $d$ orbital, the physical properties are highly susceptible to their surroundings. Because the competing energy scales are comparable in magnitude, even small changes in these energy parameters can significantly alter the magnetic ground state (Fig.~\ref{KitaevPhase}).

In the 3$d$-transition metal oxides, the Coulombic interaction is strong enough that the energy states are well localized. In an octahedral environment, for example, the crystal field splits the triply degenerate $t_{2g}$ level from the doubly degenerate $e_g$ levels. This energy level splitting is roughly around 0.1-1 eV, and the spin-orbit coupling is about two orders of magnitude lower than this~\cite{cao2013frontiers}. So in 3$d$-oxides, the orbital effects are quenched, and the dynamic behavior is mostly governed by spin-only Hamiltonian. However, with an increase in atomic number $Z$ in the periodic table, the orbital contribution becomes prominent. In 4$d$ and 5$d$ transition-metal oxides, the ground state manifests from a subtle competition among three principal interactions: crystal-field splitting, spin–orbit coupling (SOC), and electron correlations (Coulomb repulsion $U$, Hund’s coupling). The crystal field imposed by an octahedral ligand splits the $d$-orbitals into a lower-energy $t_{2g}$ manifold and higher-energy $e_g$ states. Because 5$d$ orbitals are more spatially extended, the bandwidths are large, and in many Ir-based compounds the nominal electron count leaves the $t_{2g}$ manifold partially filled (e.g.\ $t_{2g}^5$)(Fig. \ref{KitaevPhase}). However, when electron–electron interactions become appreciable, the on-site Coulomb repulsion $U$ suppresses charge fluctuations and favors localization, steering the system toward a Mott-insulating regime (as seen in many 3$d$ oxides). In 5$d$ systems, this localization tendency is mitigated by the generally larger bandwidth, yet even a moderate $U$ can act on narrowed orbital subbands. When SOC becomes prominent, it changes the $t_{2g}$ manifold: the effective coupling of orbital and spin angular momenta splits the $t_{2g}$ energy level which set into a lower-energy $J_{\mathrm{eff}} = 3/2$ quartet and a higher-energy $J_{\mathrm{eff}} = 1/2$ doublet. In a $t_{2g}^5$ system, the $J_{\mathrm{eff}} = 3/2$ states are fully filled, leaving the half-filled $J_{\mathrm{eff}} = 1/2$ band susceptible to gap opening by correlation effects. This narrowing of the $J_{\mathrm{eff}} = 1/2$ band makes it easier for even moderate Coulomb repulsion to stabilize an insulating state—this is the concept of a “relativistic Mott insulator,” famously demonstrated in Sr$_2$IrO$_4$.

\begin{figure}[t]
		\begin{center}
			\includegraphics[height=101.99959pt, width=239.69852pt]{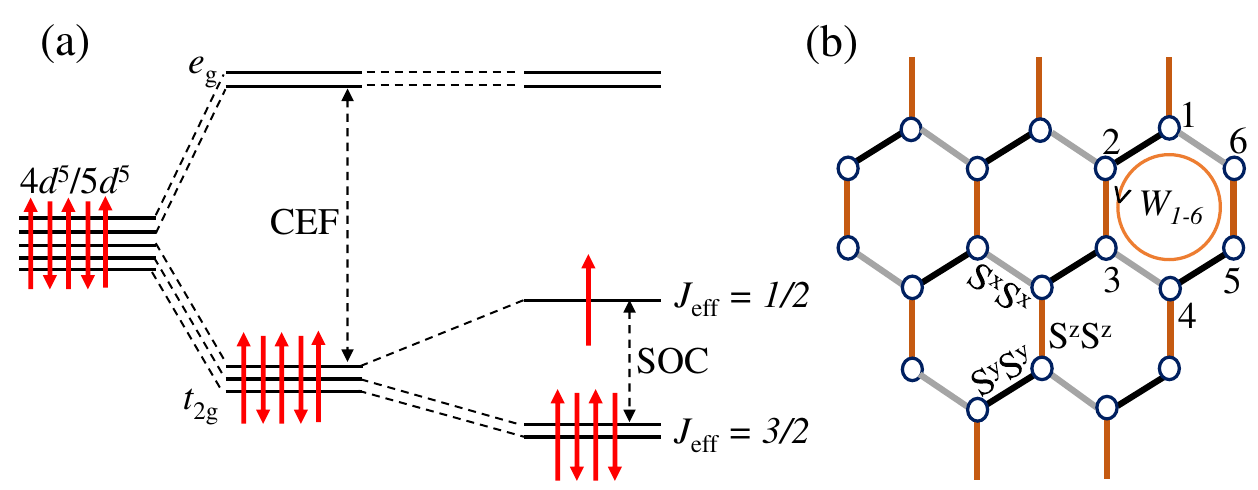}
			\caption{\textbf{ Exotic states in Frustrated  Kitaev quantum magnet.}(a) A schematic representation of a frustrated Kitaev magnet decorated on a honeycomb lattice with bond-dependent anisotropic exchange interactions (b) The combined effect of crystal electric field (CEF) and spin-orbit coupling in the $4d^5$ and $5d^5$ based quantum magnets realises a low energy $J_\text{eff}=1/2$ state potential to host Kitaev physics.}
			\label{kitaevsop}
		\end{center}
	\end{figure}

The understanding of rigorous mathematical platform for spin-orbit coupled QSL. The Kitaev model is an exactly solvable Kitaev Hamiltonian, which is defined on a two-dimensional honeycomb lattice with spin-$\frac{1}{2}$ moments located at each site.
The Hamiltonian is expressed as: $
\mathcal{H} = -K_x \sum_{\langle i,j \rangle_x} S_i^x S_j^x - K_y \sum_{\langle i,j \rangle_y} S_i^y S_j^y - K_z \sum_{\langle i,j \rangle_z} S_i^z S_j^z$, where $K_x$, $K_y$, and $K_z$ are the exchange couplings along the $x$-, $y$-, and $z$-bonds, respectively, and $S_i^\gamma$ represents the spin operator at site $i$ along the $\gamma$-axis. The bond-dependent nature of these interactions leads to a highly anisotropic Ising-type exchange that induces exchange frustration, preventing the system from ordering magnetically and stabilizes a QSL state in the absence of an external magnetic field. This solvability is achieved by mapping the spin operators onto Majorana fermions, a process that introduces a set of conserved quantities associated with each plaquette (hexagon) of the lattice. These conserved quantities, known as flux operators, are defined as: $
W_p = \prod_{\langle i,j \rangle \in p} \sigma_i^\gamma \sigma_j^\gamma$, where the product runs over the six bonds of a hexagon $p$, and $\sigma_i^\gamma$ denotes the Pauli matrix corresponding to the spin at site $i$ along the $\gamma$-direction (Fig. \ref{kitaevsop}). The eigenvalues of $W_p$ are $\pm 1$, corresponding to different flux configurations that are invariant under the dynamics governed by the Kitaev Hamiltonian. The entire system can thus be described in terms of these fluxes and the itinerant Majorana fermions that hop between lattice sites.

When an external magnetic field $\mathbf{h} = (h_x, h_y, h_z)$ is applied, the Kitaev model becomes non-integrable, and the first-order perturbation effectively renormalizes the Kitaev couplings~\cite{kitaev2006anyons}:
\begin{equation}
\mathcal{H}_{\text{eff}}^{(1)} = -\sum_{\gamma} (K_\gamma + \delta K_\gamma) \sum_{\langle ij \rangle_\gamma} S_i^\gamma S_j^\gamma, \quad \text{with} \quad \delta K_\gamma \propto h_\gamma.
\end{equation}
However, the topological transformation arises only at third order, where a chiral term of the form
\begin{equation}
\mathcal{H}_{\text{eff}}^{(3)} = -\kappa \sum_{\langle\langle\langle i,j,k \rangle\rangle\rangle} S_i^\alpha S_j^\beta S_k^\gamma, \quad \kappa \propto \frac{h_x h_y h_z}{\Delta^2},
\end{equation}
emerges. Here $\alpha$, $\beta$, $\gamma$ cycle over $x$, $y$, $z$, and $\Delta$ is the flux gap.

Kitaev honeycomb magnets are characterized by spin-orbit-driven interactions that are highly anisotropic and bond-dependent, specifically between $J_\text{eff} = 1/2$ degrees of freedom. The Kitaev model predicts the emergence of a QSL state as a result of bond frustration. Inherently, it gives rise to deconfined fractional excitations which are known as Majorana fermions~\cite{takagi2019concept, kitaev2006anyons}. One of the most intriguing features of the Kitaev QSL is its ability to host non-Abelian anyons. This remarkable property broadens our understanding of unconventional quantum matter and opens promising avenues for topological quantum computation, where non-Abelian anyons serve as a natural basis for fault-tolerant qubits.
\\
Recent studies on some Kitaev magnets have shown glassy behavior at low temperatures.  For example, the most celebrated Kitaev material, $\alpha$-RuCl$_3$, has revealed glass-like behavior under the application of an intermediate magnetic field, evidenced by the non-linear susceptibility measurements~\cite{holleis2021anomalous}. Zero-temperature density-matrix renormalization group calculations in the intermediate field regime indicate that the emergence of glassy dynamics originates from the slowing down of \( Z_2 \) fluxes in proximity to the \( U(1) \) spin-liquid phase \cite{PhysRevB.108.165118}. In integrable systems, the large number of conserved quantities constrains the system’s dynamics, preventing the initial state from sampling the full Hilbert space and thereby giving rise to weak ergodicity breaking in Kitaev materials~\cite{serbyn2021quantum}. In the honeycomb Kitaev model, these conserved quantities correspond to flux operators defined around each hexagonal plaquette. Each flux operator is represented as a product of six spin components, written as \( W_{1\text{-}6} = 2^6 S_1^z S_2^x S_3^y S_4^z S_5^x S_6^y \) (see Fig.~\ref{kitaevsop}(b))~\cite{PhysRevLett.117.037209}. This constrained dynamics is not unique to two-dimensional systems. Recent investigations of one-dimensional spin-1 Kitaev chains have revealed that the Hilbert space becomes fragmented into unequal, disconnected subspaces, further illustrating how local conservation laws can profoundly restrict ergodicity~\cite{PhysRevResearch.4.013103}. In the real materials, the Heisenberg interaction dominates over the Kitaev exchange. A recent study predict that a honeycomb isotropic Heisenberg magnet can harbour Kitaev spin liquid as a metastable state~\cite{baskaran2023metastable}. The slow decay and nontrivial dynamics of these metastable phases may naturally produce glassy behavior.
\\
Recent investigations in frustrated Kitaev spin glass magnets have revealed intriguing low-temperature spin freezing behavior~\cite{PhysRevB.96.094432,PhysRevB.89.241102, PhysRevB.99.214410, PhysRevB.92.134412}. This behavior highlights characteristic features reminiscent of the hydrodynamic modes with gapless excitations, as proposed for other frustrated spin glass materials~\cite{jena2025nature}. In these Kitaev magnets, specific heat measurements at low temperatures have demonstrated the presence of a linearly dispersing mode. The magnetic specific heat follows a $T^2$ dependence below the freezing temperature (see Fig.~\ref{fig:figure1}(c)). In the $J_\text{eff} = 1/2$ Kitaev spin glass Li$_2$RhO$_3$, the specific heat remains field-independent up to an applied magnetic field of 9 T below the glass transition temperature $T_g$, demonstrating the robustness of the low-temperature state~\cite{PhysRevB.96.094432}. The NMR study of this material is discussed exclusively in the NMR probe section of this review, where it provides clear evidence for the existence of topological spin freezing. A recently reported Kitaev material Ag$_3$LiIr$_{1.4}$Ru$_{0.6}$O$_6$ with $\theta_\text{CW}\sim -160$ K shows long-range magnetic order around 20~K with a possible quantum spin-liquid-like phase below 10~K~\cite{xfm3-px8f}. The ZFC--FC bifurcation in magnetic susceptibility, together with the absence of a sharp $\lambda$-type anomaly in the specific heat, indicates the presence of frozen magnetism rather than a true long-range ordered state. The $ T^2$-power-law behavior and field-independent magnetic specific heat up to an applied magnetic field of 7 T mimic the behavior of an unconventional spin glass, as observed in Li$_2$RhO$_3$. Furthermore, the $\mu$SR and $^7$Li NMR results support the freezing behavior. The $\mu$SR measurements show a partial loss of asymmetry and a temperature-independent relaxation rate below 10~K. This implies the coexistence of static and dynamic components with persistent low-energy spin dynamics. Similarly, the incomplete wipeout of the $^7$Li NMR intensity and the power-law dependence of $1/T_1$ indicate the dynamics of novel low-energy excitations in the frozen spin background. whereas, the parent compounds Ag$_3$LiIr$_2$O$_6$ and Ag$_3$LiRu$_2$O$_6$ show different magnetic behaviors. The ground state of $J_\text{eff}=1/2$ Ag$_3$LiIr$_{2}$O$_6$ is a long-range antiferromagnetic state which is evident from the clear oscillating nature of the asymmetry function in $\mu$SR~\cite{PhysRevB.103.094427,PhysRevB.104.115106}. For the $S=1$ Ag$_3$LiRu$_{2}$O$_6$, the low temperature behavior is reported to be a spin glass below 5 K~\cite{PhysRevB.99.054417}, but the spectroscopic study along with theoretical calculations, predicts the ground state as a spin-orbit entangled singlet state with $J_\text{eff}=0$~\cite{PhysRevResearch.4.043079}. So the anomalous spin freezing nature of 
Ag$_3$LiIr$_{1.4}$Ru$_{0.6}$O$_6$ could be due to the bond randomness caused by the presence of disorder between Ir$^{4+}$ ($J_{\text{eff}}=1/2$) and Ru$^{4+}$ ($S=1$) ions on a honeycomb lattice. The specific data of some of the promising Kitaev topological spin glass materials are presented in Fig. \ref{fig:figure1}(c), showing the quadratic temperature dependence, which is adapted from the ref. \cite{bastien2019spin,manni2014effect,mehlawat2015fragile,PhysRevB.96.094432}
\\

In quantum magnets with dominant ferromagnetic interactions, the low-temperature spin dynamics can also have clear fingerprints in thermodynamic quantities. In particular, an undamped linear spin-wave mode in three dimensions gives rise to a characteristic \( T^3 \) contribution to the magnetic specific heat at temperatures below a crossover or freezing scale \( T_g \)~\cite{fischer1979electrical}.  Such behavior has been reported in the hyperhoneycomb (three-dimensional analog of Kitaev honeycomb magnet) iridates \(\beta\)-ZnIrO\(_3\) and \(\beta\)-MgIrO\(_3\), where nominal Ir\(^{4+}\) ions carry effective pseudospin-\(1/2\) (\( J_{\mathrm{eff}} = 1/2 \)) moments and where Curie–Weiss fits suggest dominant ferromagnetic coupling with \(\theta_{\mathrm{CW}}\) of order 40–60 K in these systems~\cite{PhysRevMaterials.6.L021401,zubtsovskii2022topotactic, haraguchi2023monoclinic}. In these lattices, a nontrivial \( Z_2 \) flux operator can be defined around a ten-site loop (a conserved quantity in the pure Kitaev limit), which plays a key role in mediating low-energy excitations and constraining fluctuations~\cite{PhysRevB.89.045117}. Experimentally, both \(\beta\)-ZnIrO\(_3\) and \(\beta\)-MgIrO\(_3\) show weak anomalies in bulk thermodynamic measurements around \( T_g \sim 12\,\mathrm{K} \) and \( 22\,\mathrm{K} \), respectively—interpreted as spin-freezing or crossover phenomena~\cite{PhysRevMaterials.6.L021401, zubtsovskii2022topotactic}. Below those temperatures, the magnetic specific heat is well described by a leading \( C_{\mathrm{mag}} \propto T^3 \) term, consistent with the presence of an undamped, linearly dispersing magnon branch. This contrasts with conventional ferromagnets, where long-range order often leads to quadratic dispersion (and hence \( C_{\mathrm{mag}} \propto T^{3/2} \) in 3D). The persistence of the \( T^3 \) dependence even under magnetic fields (up to 5 T in \(\beta\)-ZnIrO\(_3\)) suggests a robustness of the linear magnon modes. The robustness is similar to that of its two-dimensional Kitaev spin glasses, as discussed above. Moreover, since these systems lie close to competing quantum phases, perturbations such as magnetic field, pressure, or strain may push them across crossovers or transitions, making the interplay between spin waves, freezing phenomena, and frustration a rich arena for exploring emergent quantum states.
\\
\subsection{Pyrochlore lattice}
\begin{figure}[t]
		\begin{center}
			\includegraphics[height=255.00139pt, width=260.09279pt]{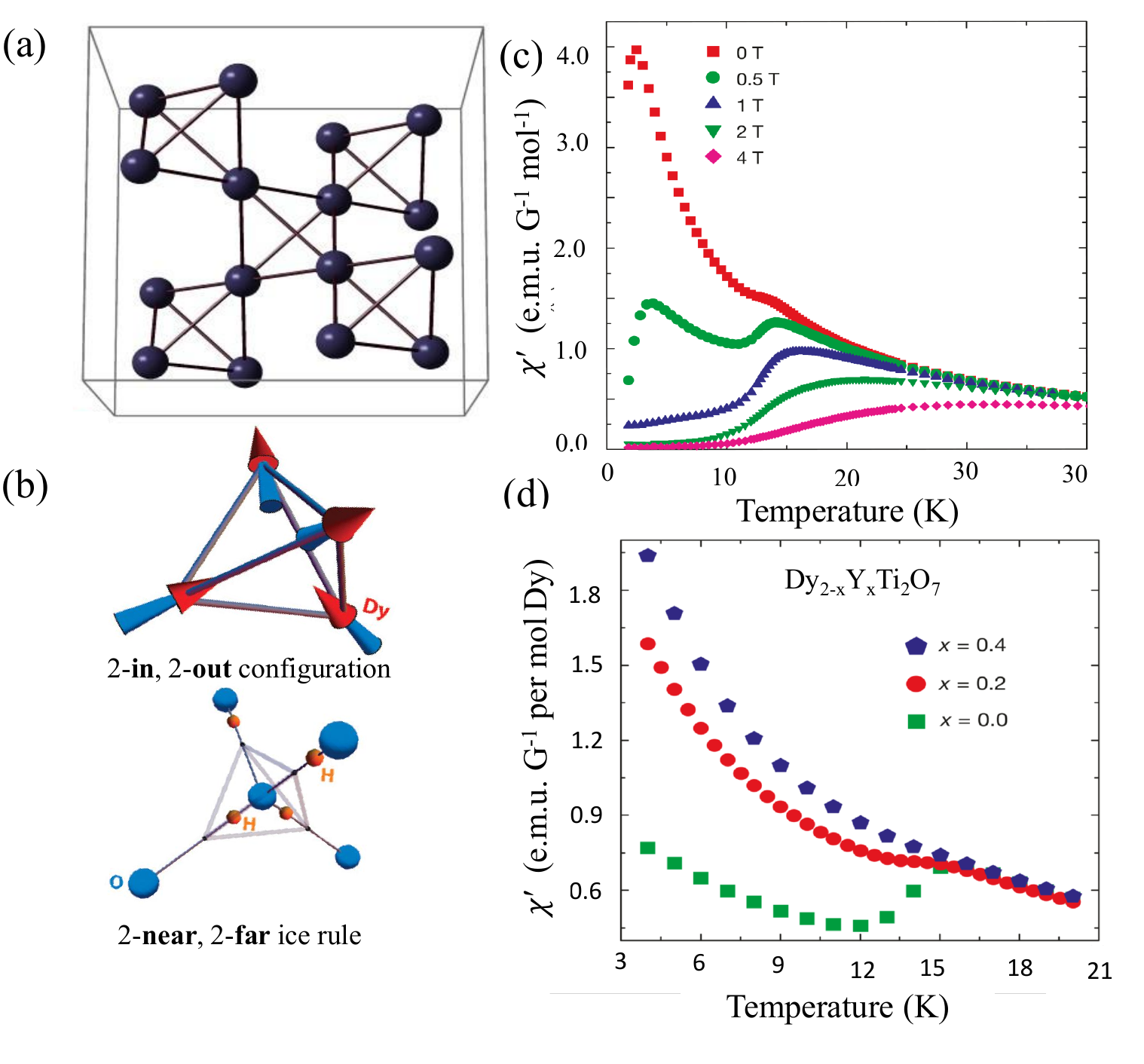}
			\caption{\textbf{Emblematic frustrated pyrochlore lattice.}(a) A schematic representation of a pyrochlore lattice structure (b) Diagram depicting the correspondence between spin ice (upper panel) and water ice (lower panel). In the spin ice structure, a spin directed outward (inward) represents a hydrogen atom displaced away from (toward) the central oxygen atom of the tetrahedron (c) Real part of the a.c. susceptibility of Dy$_2$Ti$_2$O$_7$ as a function of temperature in various dc fields at 100 Hz. The magnetic field increases the freezing temperature, contrary to the behavior in typical spin glasses (d) Real part of the a.c. susceptibility for Dy$_{2-x}$Y$_x$Ti$_2$O$_7$ $(0\leqslant x \leqslant 0.4)$as a function of temperature at 500 Hz, measured without a d.c. field. Substituting magnetic Dy$^{3+}$ with non-magnetic Y$^{3+}$ ions weakens the freezing transition, highlighting the role of spin-spin interactions in the process. Source:(b)adapted from \cite{castelnovo2012spin} with permission from AR, (c,d) adapted from \cite{snyder2001spin} with permission from NPG.}
			\label{pyro}
		\end{center}
	\end{figure}

The pyrochlore lattice is a three-dimensional network of corner-sharing tetrahedra, representing one of the most interesting geometrically frustrated structures in condensed matter. In the pyrochlore lattice, each magnetic ion (or vortex) is connected to six nearest neighbors (Fig.~\ref{pyro}(a)), which leads to a high degree of magnetic frustration. The general formula for the pyrochlore oxides is A$_{2}$B$_2$O$_7$, where the A site is typically occupied by rare-earth or large ions and the B site hosts transition metal ions~\cite{subramanian1983oxide,PhysRev.102.1008,RevModPhys.82.53}. The intricate geometry and the competing degrees of freedom in the pyrochlore lattice give rise to a highly degenerate ground state. The macroscopic degeneracy is the key to the variety of emergent phenomena observed in pyrochlore lattices, such as spin liquid~\cite{RevModPhys.82.53,PhysRevLett.80.2929,PhysRevB.61.1149}, spin ice associated with magnetic monopole excitations~\cite{PhysRevLett.79.2554,ramirez1999zero,castelnovo2008magnetic}, spin glass~\cite{PhysRevLett.124.087201,PhysRevLett.107.047204} and anomalous Hall effect~\cite{taguchi2001spin}. Early investigation of spin ice materials such as Dy$_2$Ti$_2$O$_7$~\cite{ramirez1999zero, morris2009dirac} and Ho$_2$Ti$_2$O$_7$~\cite{bramwell1998frustration, fennell2009magnetic}, revealed the intriguing possibility of hosting magnetic monopoles. This has drawn considerable attention as the magnetic monopole emerges as a collective quasiparticle excitation, reflecting the hypothetical elementary magnetic charges that high-energy physicists had long sought but never observed in real system~\cite{dirac1931quantised,t1974magnetic}. In the rare-earth pyrochlore spin ice materials, the magnetic moments are constrained along the local Ising axes~\cite{snyder2001spin, PhysRevLett.95.217201}. The dynamics of these Ising spins are mostly governed by the nearest-neighbor exchange strength and the long-range dipolar interactions~\cite{castelnovo2008magnetic}. In spin ice materials, the magnetic moments (Ising spins) on the pyrochlore lattice interact ferromagnetically but are constrained by strong local anisotropy to align along the ⟨111⟩ axes connecting the centers of neighboring tetrahedra. As a result, the ground state configurations obey the so-called “ice rule”~\cite{ramirez1999zero}, analogous to the proton disorder in water ice(Fig. \ref{pyro}(b))~\cite{pauling1935structure}. Here, each tetrahedral vertex has two spins pointing inward and two pointing outward ( “2-in, 2-out”  configuration). This local constraint results in a macroscopically degenerate ground state and extensive residual entropy~\cite{bramwell2001spin}. In spin ice, flipping a single spin breaks the local “two-in, two-out” ice rule, effectively creating a pair of defects that behave like magnetic charges of opposite sign. This quasiparticle is referred to as a magnetic monopole, which moves through the lattice as neighboring spins flip, forming a current of magnetic charge. Here we can pause for a moment and think about an in-depth question~\cite{wen2003quantum, wen2019choreographed}: are the particles we observe in nature truly elementary, or could they emerge from the collective behavior of many interacting degrees of freedom ?

However, the experimental study has shown that the potential spin ice candidate Dy\(_2\)Ti\(_2\)O\(_7\) shows spin glass behavior. But the glassy nature of this pyrochlore fundamentally differs from that of canonical spin glass. The slowing down of spin dynamics could be due to the presence of topological defects which are magnetic monopoles. The freezing behavior reflects an entropic balance: as the temperature is lowered, the creation and motion of monopoles are increasingly hindered, resulting in a dynamical slowdown and eventual arrest of the spin system~\cite{snyder2001spin,PhysRevLett.104.107201,PhysRevLett.114.247207}. The glassy nature of spin ice exhibits a relatively narrow distribution of relaxation times, which  contrasts with the broad distribution range found in canonical spin glasses~\cite{snyder2001spin}. A distinct trend of glassy behavior is observed with the application of a magnetic field. With an increase in magnetic field, the real part of the ac susceptibility $\chi'$ marks a suppression below the freezing temperature \(T_g\) (Fig.~\ref{pyro}(c)). This anomaly is a direct consequence of the cooperative dynamics within the ice-rule-constrained manifold. When an external magnetic field is applied, it polarizes the background spin configuration, increasing the energy barrier for monopole motion and thereby stabilizing the frozen state. Conversely, doping Dy\(_2\)Ti\(_2\)O\(_7\) with just 20\% of non-magnetic Y\(^{3+}\) ions at Dy$^{3+}$ suppresses glassiness, as reflected by an enhancement of \(\chi'\) (Fig.~\ref{pyro}(d)). This sensitivity arises because the energy scales governing interactions in 4$f$-based spin ices are intrinsically small due to the weak exchange on the order of 1–2 K. These 4$f$ electrons are subject to large crystal electric field splittings ($\sim$300 K), which isolate the Ising doublets but reduce the overall energy landscape for collective behavior. As a result, even weak dilution disrupts the delicate network of frustrated tetrahedra, diminishing the topological constraints and enabling faster spin dynamics. From a field-theoretical standpoint, the low-energy excitations in spin ice are governed by emergent U(1) gauge fields, and the excitations, mapped onto Goldstone modes in the presence of broken symmetry, are highly susceptible to perturbations. In the regime of weak exchange disorder, characterized by a disorder strength \(\Delta\) (\(0 < \Delta \ll J\)), a modified hydrodynamic theory predicts that these emergent gauge fields interact with the disordered spin background, leading to a renormalized velocity \(c \sim a\Delta\) for the collective excitations, where \(a\) is the spin lattice spacing~\cite{PhysRevB.101.024413}. This disorder-induced reduction in excitation velocity increases the density of low-energy states, thereby enhancing the magnetic specific heat \(C_m\) as the system approaches the glassy state. Thus, slow spin dynamics, arising not from conventional random disorder but from the constrained topological manifold and emergent gauge structure, define a fundamentally different route to spin freezing in spin ices like Dy\(_2\)Ti\(_2\)O\(_7\).

In contrast to 4$f$-based pyrochlores, where the glassy state arises from the emergence of exotic quasi-particle excitations and a corresponding gauge field, the glassy behavior observed in 3$d$ and 4$d$ transition metal pyrochlores, such as Y$_2$Mo$_2$O$_7$\cite{greedan1986spin, yang2015spin}, NaCaNi$_2$F$_7$\cite{PhysRevB.92.014406, cai2018musr}, and Lu$_2$Mo$_2$O$_7$~\cite{PhysRevLett.113.117201}, is instead primarily governed by a combination of strong geometric frustration and antiferromagnetic exchange randomness, which conspire to produce a manifold of nearly degenerate states and low-energy collective fluctuations. Table \ref{material_name_part3} presents some of the transition metal based frustrated spin glasses and the associated thermodynamic parameters.  Despite the three-dimensional nature of the pyrochlore lattice, the specific heat in this class of spin glasses shows $T^2$ power-law behavior, which is close to that observed in 2D antiferromagnetic spin waves. The apparent reduction in dimensionality, from the expected three-dimensional to an effective two-dimensional spin-wave modes, may originate from strong spin–orbit coupling~\cite{PhysRevLett.93.156407,yang2015spin}, which can anisotropically constrain magnetic interactions and effectively lower their dimensionality. The Mo$^{4+}$ based $S=1$ frustrated spin glass Y$_2$Mo$_2$O$_7$  has effective $\theta_\text{CW}$ =$-$200 K with $T_g=22 $ K~\cite{greedan1986spin}. However, the sister compound Lu$_2$Mo$_2$O$_7$, which also shows spin-glass behavior, has an effective $\theta_\text{CW}$ of $-$160 K with $ T_g=16$ K~\cite{clark2013oxygen}. Despite the substitution of Y$^{3+}$ with Lu$^{3+}$, the nature of spin freezing qualitatively remains the same. In diffuse magnetic neutron diffraction for both Y$_2$Mo$_2$O$_7$ and Lu$_2$Mo$_2$O$_7$, broad elastic magnetic peaks are observed at $Q \approx 0.4 \, \text{\AA}^{-1}$~\cite{PhysRevLett.83.211} and $0.6 \, \text{\AA}^{-1}$~\cite{PhysRevLett.113.117201}, respectively.  These differences in $Q$ values suggest that Y$_2$Mo$_2$O$_7$ has longer-range magnetic correlations compared to Lu$_2$Mo$_2$O$_7$. Nitriding the Lu$_2$Mo$_2$O$_7$ spin glass oxide, resulting in Mo$^{5+}$ $S=1/2$ Lu$_2$Mo$_2$O$_5$N$_2$ magnet. The $\mu$SR measurements confirm the lack of long-range magnetic ordering or spin freezing down to 300 mK in Lu$_2$Mo$_2$O$_5$N$_2$~\cite{PhysRevB.107.024407}. This can be interpreted as a crossover from a spin glass to a spin liquid state, where the quantum fluctuations are amplified due to the spin-1/2 magnetic moment. The linear-$T$ dependence of specific heat and the absence of a broad magnetic scattering peak signal the emergence of a spinon Fermi surface~\cite{PhysRevLett.113.117201}.  Despite the introduction of bond randomness due to oxynitridation, the glassy state is completely destabilized in Lu$_2$Mo$_2$O$_5$N$_2$, suggesting the presence of a dynamic ground state resulting from the interplay between quantum fluctuations and exchange randomness. In the Ru-based Y$_2$Ru$_2$O$_7$, the thermodynamic probes reveal a distinct form of spin freezing. It shows ZFC-FC bifurcation in magnetization without any trace of hysteresis~\cite{taira1999magnetic} and the magnetic specific heat shows a sharp anomaly at $T_g=76$ K~\cite{ito2001nature}. This indicates that the phase transition is of second order, and the neutron diffraction data show a nearly long-range magnetic order with an estimated ruthenium magnetic moment of approximately 1.36 $\mu_B$ \cite{ito2001nature}. A similar glassy behavior is observed in Y$_2$Ir$_2$O$_7$~\cite{taira2001magnetic,kumar2016nonequilibrium}, indicative of a conventional spin-freezing transition that occurs despite the strong magnetic frustration inherent to its pyrochlore geometry.

\begin{figure}[t]
		\begin{center}
			\includegraphics[height=200.00139pt, width=245.09279pt]{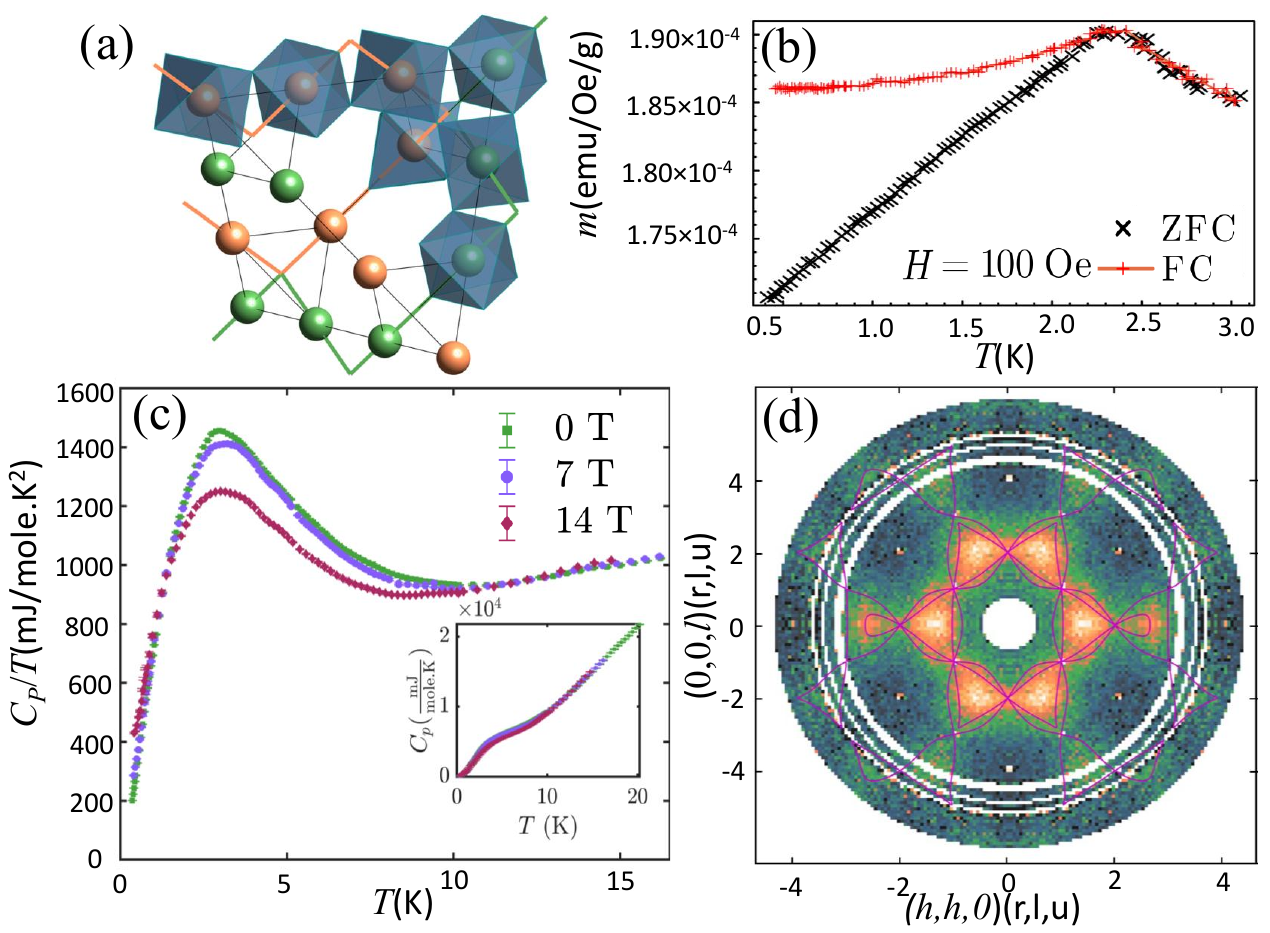}
			\caption{\textbf{Frustration-driven coexistence of liquid-like correlations and spin freezing in the pyrochlore CsNiCrF$_6$.}(a) Schematic of the pyrochlore lattice formed by a mixed occupancy of Ni$^{2+}$ and Cr$^{3+}$ ions on a network of corner-sharing tetrahedra, highlighting the geometrical frustration and underlying charge-ice constraints. (b) DC magnetic susceptibility measured at 100 Oe showing ZFC-FC bifurcation at $T_g=2.3$ K. (c) Specific heat measured under various magnetic fields (applied along the (1,1,1) direction), exhibiting a broad maximum around $T_g$ consistent with the absence of sharp long-range ordering and the presence of fluctuating, correlated degrees of freedom.(d) Experimental magnetic diffuse neutron scattering measured at 1.5 K, showing characteristic pinch-point features in the structure factor $S(\textbf{Q})$. (a) and (d) are adapted from ref.~\cite{fennell2019multiple}, (b) and (c) are adapted from ref.~\cite{xvxt-whns}.}
			\label{CsNiCrF6}
		\end{center}
	\end{figure}

The pyrochlore fluoride CsNiCrF$_6$ provides a compelling example of a frustrated magnet where correlated disorder, emergent gauge structure, and unconventional freezing coexist~\cite{fennell2019multiple,xvxt-whns}. Here,  Ni$^{2+}$ and Cr$^{3+}$ cations form a pyrochlore lattice and the cation distribution is not random; instead, it follows an Anderson-type constraint where each tetrahedron contains two Ni$^{2+}$ and two Cr$^{3+}$ ions, as shown in Fig.~\ref{CsNiCrF6}(a). This ``charge-ice'' rule puts a constraint on the local conservation law that is equivalent to divergence-free condition on an emergent flux field, $\nabla \cdot \mathbf{B} = 0$. The constraint places CsNiCrF$_6$ in the class of Coulomb phases, and the dynamical behavior of the system is not governed by the conventional symmetry-breaking state, but rather by an emergent gauge structure. The observed pinch points in the diffuse neutron scattering structure factor $S(\mathbf{Q})$ (Fig.~\ref{CsNiCrF6}(d)) further confirm the presence of the Coulomb phase~\cite{fennell2019multiple}. The significant consequence of the "charge-ice'' constraint is that the cation configuration transforms into a fully packed loop model on the diamond lattice, with each loop consisting of ions of the same species, i.e., either the loop of Ni$^{2+}$ or Cr$^{3+}$ ions. A power-law distribution of loop lengths provides a topological backbone for both structural and magnetic correlations. INS carried out at low energy ($\hbar\omega<2$ meV) yields a structure factor with broadened pinch-point features, indicating that the system remains in the Coulomb phase. Although there is no long-range magnetic order at low temperatures, bulk thermodynamic measurements reveal a freezing transition at $T_f \approx 2.3$~K~\cite{xvxt-whns}. The transition is marked by the ZFC-FC splitting observed in the measurement of DC susceptibility, as shown in Fig.~\ref{CsNiCrF6}(b).  The specific heat data, as illustrated in Fig.~\ref{CsNiCrF6}(c), exhibits broad maxima below 10 K, with the maximum obtained at $T_g$. It is also observed that the field evolution of $T_g$ follows an opposite trend to that of the canonical spin glass. The observed unconventional spin freezing may be due to a fluctuation-driven instability toward a spin-nematic state induced by weak interloop couplings~\cite{xvxt-whns}. In proximity to this instability, the spin can have a pronounced slowing down, as the relaxation processes require the coordinated rearrangements of spin loop structures rather than any local spin flip transition. This provides a natural mechanism for slowing spin dynamics in CsNiCrF$_6$. In a broader perspective, CsNiCrF$_6$ exemplifies a fundamentally different route to spin freezing where the phase space is fragmented by topological constraints due to correlated disorder between  Ni$^{2+}$ and Cr$^{3+}$ moments than random disorder.

In the $3d-$fluoride based pyrochlore NaCaNi$_2$F$_7$, neutron scattering and thermodynamic measurements reveal an unusual hierarchy of magnetic regimes~\cite{plumb2019continuum}. A QSL-like dynamic state exists around the dominant exchange scale, and ultimately, a frozen state is reached below ~3.6 K. 
Remarkably, nearly 90\% of the low-temperature spectral weight appears in the inelastic scattering, far exceeding semiclassical expectations for an $S = 1$ system and pointing to the persistence of strong quantum fluctuations. Below the $T_g$, the specific heat follows a power-law behavior as $C_m\propto T^{2.2}$. The sub-quadratic nature of $C_m$ indicates the presence of linearly dispersing bosonic excitations with an effective two-dimensional constraint despite the underlying cubic geometry. The reduction in dimensionality is attributed to frustration-induced soft nodal lines~\cite{bergman2007order} that restrict the low-energy spin fluctuations akin to topological constraints in phase space~\cite{kawamura1998universality,henley2010coulomb}. Although the system eventually freezes at low temperature, it nonetheless reflects the existence of quantum fluctuations within a highly entangled manifold. Thus, NaCaNi$_2$F$_7$ pyrochlore illustrates how glassiness can emerge as the terminal fate of a quantum liquid, resulting from a combined effect of geometric frustration and topology.
  
  Thus, pyrochlore-based frustrated magnets provide a paradigmatic model for realizing exotic physical phenomena. The interplay between competing interactions, disorder, and emergent electromagnetic responses may facilitate the establishment of a theoretical framework and a deeper understanding of many-body quantum phenomena, such as spin liquids, topological spin freezing, and spin ice, in a broad class of frustrated 3D magnets.

\subsection{Other promising frustrated spin-lattices}

\begin{figure*}[t]
\includegraphics[height=398.35463pt, width=510.00493pt]{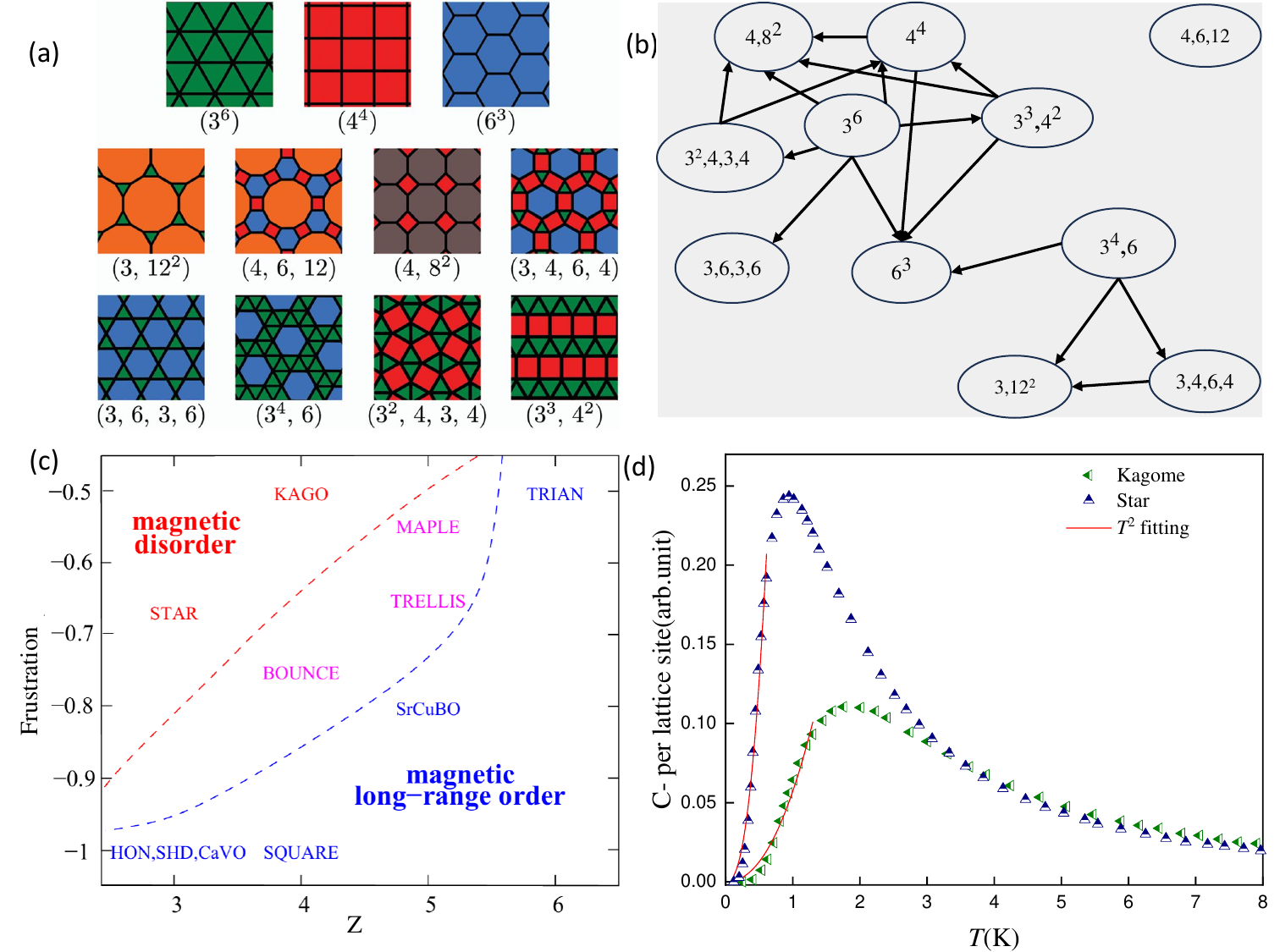}
  \caption{\textbf{Archimedean lattices are promising contenders for frustrated spin glass physics} (a) All eleven Archimedean lattices, also known as uniform tilings, consist of regular polygons arranged such that each vertex has an identical configuration of surrounding polygons which are labeled by some sequential rules (b) The relationships (arrows) between all eleven Archimedean lattices, where one tiling transforms into another by selectively removing edges (bonds) and applying suitable distortions (c) A schematic representation of semiclassical magnetic order and quantum magnetic disorder in Archimedean lattices within a parameter space defined by frustration (defined by ref \cite{kobe1995frustration}) and coordination number z. Here the notation of Archimedean lattices is used as $(3^{6}$:TRIAN),$(4^{4}$:SQUARE),$(6^{3}$:HON), (($3^{4}.6$:MAPLE);$(3^{3}.4^{2}$:TRELLIS);($3^{2}.4.3.4$:SrCuBO);(3.4.6.4:BOUNCE);(3.6.3.6:KAGO);$(3.12^{2}:$STAR);(4.6.12:SHD);$(4.8^{2}$:CaVO)  (d) The specific heat per lattice site for the star($3,12^{2}$) and Kagomé lattices as a function of temperature digitized the simulated data. Source: (a)adapted from \cite{de2019topological} with permission from RSC, (c)adapted from \cite{farnell2014quantum} with permission from APS, (d)adapted from \cite{zhang2005thermodynamic} with permission from World Scientific.}
  \label{Archi}
\end{figure*}

An Archimedean lattice is in general, a uniform tiling two-dimensional arrangement of regular polygons where all vertices share the same topologically identical and they frequently form the foundational magnetic structures of insulating, quasi-2D quantum magnetic materials \cite{grunbaum1987tilings,shrock1997lower,farnell2018interplay}.  The Archimedean lattice is generic to most of the highly frustrated lattices, including kagome, honeycomb, and triangular ( Fig.~\ref{Archi}(a)), which are potentially important for demonstrating novel electronic, magnetic, and superconducting states of fundamental appeal and technological relevance. Insights into the structure- property relationships, non-trivial electronic bands,  ground state degeneracies, and topological phases in these novel lattices are crucial for designing artificial lattice architectures, including cold atoms and artificial spin ice, where the tunability of structure and associated properties is seamless compared to their natural counterparts. The name Archimedean lattice comes from the discoverer Archimedes. In the Pappus(fl. AD 300-350) collection, he described that Archimedes had discovered thirteen ``solids" with faces composed of regular polygons of more than one type \cite{field1997rediscovering}. Later Kepler showed that there exist only 11 distinct Archimedean lattices \cite{Yu2015IsingAO}. These 11 Archimedean lattice patterns are significant not only in mathematics \cite{conrad1971essential,martinez1973archimedean} but also in materials science, as many of these lattices correspond to natural material systems. Archimedean lattices are visually captivating and can be found in various forms of art and architecture. These uniform patterns are appearing in everything from household ceramic tile and bowl designs and basket weaves (with ``kagomé" referring to a weave pattern in Japanese) to the atomic structures of materials\cite{farnell2018interplay}. The Archimedean lattice has different notations that might be labeled by the sequence of polygons that makes a complete circuit surround each vertex in every direction, because all vertices share the same topological environment, which can be expressed in the mathematical notation for an Archimedean lattice, $ \Delta=\prod_{i}P_{i}^{a_{i}}$ \cite{shrock1997lower}. Where the mathematical notation $P_{i}^{a_{i}}$ refers that the regular polygon $P_{i}$ occurs continuously $a_{i}$ times, sometime it can occur noncontinuously. Noncontiguously, since the starting point is irrelevant, the symbol remains unchanged under cyclic permutations. For future reference, when a polygon 
$P_{i}$ appears multiple times in a noncontinuous way within the product, we will represent 
$a_{i}$  as the sum of all $a_{i}$  values corresponding to each occurrence of $P_{i}$ in the product. These eleven Archimedean latices in the form of sequences are, 
\begin{equation}
\begin{aligned}
\{\Delta\}=\{(3^{6});(4^{4});(6^{3});(3^{4}.6);(3^{3}.4^{2});(3^{2}.4.3.4);\\ (3.4.6.4);(3.6.3.6);(3.12^{2});(4.6.12);(4.8^{2})\}
\end{aligned}
\end{equation}

Among these lattices, three are homopolygonal, or monohedral, meaning they consist of only one type of regular polygon:\\$(3^{6}$:Triangular),$(4^{4}$:square),$(6^{3}$:Honeycomb).The remaining eight lattices are hetero-polygonal (($3^{4}.6$:maple leaf);$(3^{3}.4^{2}$:trellis);($3^{2}.4.3.4$:SrCuBO);(3.4.6.4:bounce);\\(3.6.3.6:kagome);$(3.12^{2}:$star);(4.6.12:SHD);$(4.8^{2}$:CaVO), meaning they involve tilings with more than one type of regular polygon~\cite{shrock1997lower}.

Each edge of the Archimedean lattices narrates an exchange interaction (or bond) $J$, which interacts between two spins sitting at the vertices. These bonds ideally assume a uniform strength spread over the lattices. However, in real materials the exchange interaction between two spins often varies in magnitude due to lattice distortions, anisotropies, chemical substitutions, or pressure-induced changes to the lattices. The variation of bond strength can effect the magnetic ground state and associated low-energy excitations of the system. Thus, it has become both an experimental and a theoretical interest to explore how the underlying lattice and hence the ground state change upon eliminating and joining such a bond within the broader Archimedean framework. A particularly interesting scenario ariese when a specific lattice among 11 Archimedean lattices can be continuously transformed into another one by tuning some specific bonds toward zero strength ($J$→0). For instance, the square lattices can be obtained from the triangular lattice by selectively removing the diagonal term, reducing the connectivity in the triangular lattice. In the same way, more complex transformations are possible. The honeycomb lattice can be derived from the triangular lattice,
while the other lattices such as  the maple-leaf, trel-
lis (bounce), and star lattices can be obtained by removing or modifying particular bond~\cite{richter2004quantum}. A schematic diagram of relationships among the 11 Archimedean lattices is shown in Fig. \ref{Archi}(b), where the arrows illustrate the transformation one lattice into another~\cite{richter2004quantum}. The family of Archimedean lattices provide considered a rich platform for exploring quantum many-body phenomena, where quantum fluctuations dominate over classical ordering.
In general, 2D quantum magnets develop magnetic ordering with nearest-neighbor isotropic Heisenberg interactions at finite temperatures, unless suppressed by frustration or competing interactions. In order to investigate the exotic magnetic behavior of the 11 distinct 2D Archimedean lattices,  it is necessary to shed insight into the combined effect of lattice topology, spin coordination number ($z$), quantum fluctuations, the interplay between nearest-neighbor and next-nearest-neighbor interactions, and exchange interactions~\cite{kobe1995frustration}. Building on this framework, Farnell et al. provide a comprehensive analysis of the Archimedean lattices, summarizing in Fig.\ref{Archi} (c) \cite{farnell2014quantum} how these competing factors influence the emergence of different magnetic phases. Semiclassical magnetic long-range ordered states : which may be collinear or non-collinear—are observed in bipartite lattices such as the square, honeycomb, CaVO, and SHD lattices. Magnetically disordered states: often described as cooperative quantum paramagnets or spin liquids, emerge in highly frustrated lattices such as the triangular, kagome, and star lattices. Intermediate or amorphous-like phases: referred to as amorphous lattices, are found in more complex geometries like the trellis, bounce, and maple-leaf lattices, where competing interactions neither lead to conventional order nor a fully disordered state.

However, recent experimental studies have highlighted exceptions and refinements to this classification. For instance, the square-lattice compound Sr$_{2}$CuTe$_{0.5}$W$_{0.5}$O$_{6}$, though bipartite and expected to support long-range order, exhibits a highly frustrated spin-glass state instead \cite{hu2021freezing}. In this material, Cu$^{2+}$ ($S=1/2$) spins constitute a square lattice, but the random distribution of Te and W ions leads to disorder-induced frustration. In addition to long-range order, the square lattices can exibit disorder type ground states like spin glass\cite{mustonen2018spin,farnell2014quantum}. Similarly, in honeycomb-lattice systems such as those hosting Kitaev interactions, magnetic behavior deviates from the phase diagram due to the bond-dependent anisotropic Ising-like couplings. The Kitaev model, which was originally introduced as a solvable quantum spin liquid model, predicts ground states with fractionalized excitations and nontrivial topological order. In real materials such as $\alpha-$RuCl$_{3}$ doped with 1$\%$ Cr or Na$_{2}$IrO$_{3}$ doped with Ti and Ru show glassy behavior, accompanied by a low-temperature specific heat scaling as $C \propto T^{2}$, indicative of linear dispersive modes, rather than classical magnetic order~\cite{bastien2019spin,mehlawat2015fragile,jena2025nature}. These findings suggest that the simple classification into ordered, disordered, or intermediate phases must be expanded to include the effects of bond disorder, doping, and anisotropies in real materials, all of which can lead to unexpected behaviors like spin glassiness or quantum paramagnetism in systems previously proposed to host long-range order.

We analyze the simulated specific heat per lattice site for the star and kagome lattices as a function of temperature, shown in Fig. \ref{Archi}(d) \cite{zhang2005thermodynamic}. By fitting the  specific heat data as presented in ref.\cite{zhang2005thermodynamic} using the quadratic equation $C=aT^{n}$, we determine that $n\approx 2.17$ for the kagome lattice and $n\approx 2.09$ for the star lattice. The observed power-law behaviors in the specific heat is indicative of low-energy, linearly dispersing collective excitations, which naturally arise in highly frustrated magnetic systems with an extensive manifold of nearly degenerate states. Such scaling is due to the presence of soft modes associated with fluctuating spin textures and residual correlations, even in the absence of conventional long-range magnetic order.

Most Archimedean lattices contain triangular motifs hosting antiferromagnetic interaction, which inherently introduce geometric frustration with a highly degenerate manifold that can lead to exotic states such as QSL or  spin  glass  in this class of materials. In the preceding sections of this review, we have already explored spin freezing behavior in triangular and kagome lattice geometries. Spin freezing is a defining signature of frustrated magnetic systems, where the intricate lattice connectivity enforces local constraints and defects that restrict spin dynamics, leading to an inability to establish QSL as well as conventional magnetic order even at extremely low temperatures.

\begin{figure}[t]
		\begin{center}
			\includegraphics[height=193.79901pt, width=239.6988pt]{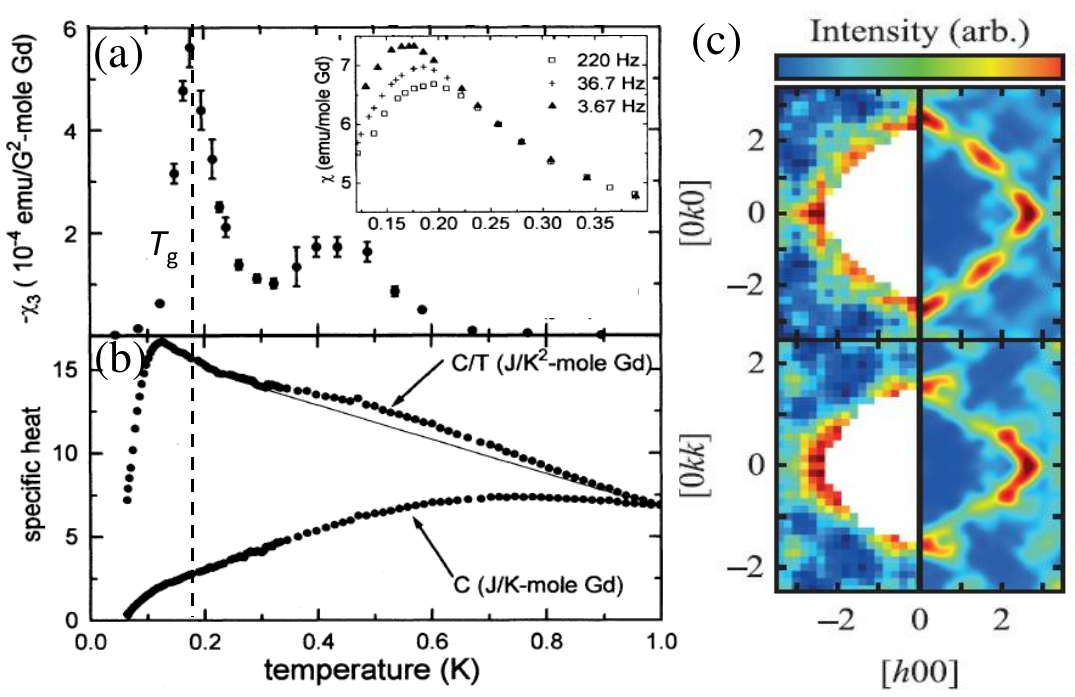}
			\caption{\textbf{Non-trivial phenomena  in 3D garnet.} (a) Temperature dependence of the nonlinear susceptibility $\chi_3(T)$. The inset of the panel shows the frequency dependence of the imaginary part of the linear susceptibility. (b) The $C/T$ versus $T$ in a 3D hyperkagome-like spin lattice Gd$_3$Ga$_5$O$_{12}$ shows a broad maximum below the glass transition. (c) Magnetic diffuse scattering in the single crystals of Gd$_3$Ga$_5$O$_{12}$ in two reciprocal-space planes: the upper panel shows the $(hk0)$ plane and the lower panel shows the $(hkk)$ plane. The left side presents the experimental neutron-scattering data, while the right side shows the corresponding reverse Monte Carlo refinement, with the white regions in the experimental data indicating areas without coverage. Source: adapted from refs.~\cite{PhysRevLett.74.2379,paddison2015hidden} with permission from APS and AAAS.
}
			\label{Gd3Ga5O12}
		\end{center}
	\end{figure}

Another frustrated spin system is the hyperkagome Na$_{4}$Ir$_{3}$O$_{8}$ with effective spin $J_\text{eff}= 1/2$, showing ZFC-FC bifurcation around 6 K in spite of having a large dynamic region with Curie-Weiss temperature $|\theta_{CW}|\sim 650$ K~\cite{okamoto2007spin}. This is a derived spinel from spinel oxides AB$_{2}$O$_{4}$ with B-site occupied by Ir and Na due to cation ordering and can be rewritten as (Na$_{1.5}$)$_1$(Ir$_{3/4}$, Na$_{1/4}$)$_{2}$O$_{4}$. The appearance of the glassy phase in this low-spin system is enhanced by the incomplete occupancy of Na$^{1+}$ ion in the A-site resulting in random Ir$^{4+}$ spin exchange interaction path. The specific heat shows a  broad peak around temperature 30 K with no ordering down to 500 mK. The $\mu$SR results indicate the onset of spin freezing below 6 K, attributed to the behavior of bulk spins rather than a dilute impurity contribution~\cite{PhysRevLett.113.247601}. The $\mu$SR results confirm the existence of quasistatic, short-range magnetic correlations localized within a single unit cell length~\cite{PhysRevLett.113.247601}. The reduction of spin fluctuation below freezing temperature is further evident from the NMR spin-lattice relaxation rate $1/T_1$~\cite{PhysRevLett.115.047201}. 

In a spin-orbit coupled system, degeneracy can be lifted, leading to significant lattice distortion known as the Jahn-Teller effect. A recent study on the frustrated pyrochlore lattice A$_2$Mo$_2$O$_7$ (A = Y, Dy, Ho, Tb) reports a glass transition at temperatures around 20 K without any quenched disorder~\cite{PhysRevLett.124.087201}. NMR, X-ray absorption spectroscopy, and $\mu$SR results reveal Mo-Mo pair distortion in the Mo-pyrochlore network in a two-in, two-out manner~\cite{PhysRevLett.118.067201}. The relative displacement of O$^{2-}$ atoms results in a distribution of Mo-O-Mo angles, altering the exchange interactions and introducing interaction disorder. This distribution of exchange interactions $J$ among Mo-Mo pairs intrinsically contributes to the glassy nature by transforming a flat energy landscape into a rugged one. Given the critical role of spin-orbit coupling in driving this glassy behavior, it can be designated as a spin-orbital glass. Remarkably, it exhibits a quadratic specific heat dependency $C_m \sim T^{2}$, characteristic of 2D magnon excitation. This suggests that orbital degeneracy may reduce the dimensionality of magnetic interactions from three to two.

A paradigmatic example of emergent complexity in geometrically frustrated magnets is found in gadolinium gallium garnet (Gd$_3$Ga$_5$O$_{12}$, or GGG), where spin freezing occurs in a site-ordered chemically clean lattice. The glass transition occurs around the temperature 0.14 K~\cite{PhysRevLett.74.2379} which is well below the characteristic antiferromagnetic interaction scale $\theta_{\rm CW} \approx -2$\,K~\cite{wolf1962magnetic}. The suppression of conventional magnetic ordering is due to the high degree of magnetic frustration inherent to its hyperkagome network of Gd$^{3+}$ ions. The complexity in GGG is evident from its magnetic diffuse scattering shown in Fig.~\ref{Gd3Ga5O12}(c). The development of short-range spin correlations beyond the next-nearest neighbor is evident from the anisotropic pattern of the diffused scattering~\cite{paddison2015hidden}. Comparing it with the reverse Monte Carlo (RMC) refinements confirms that the spin correlations originate from the coherent dynamics of finite spin textures. 

The combined study of neutron scattering and RMC analyses revealed that the intermediate-temperature spin-liquid state contains a form of hidden long-range order which cannot be described within the dipolar framework. Instead of multipolar ordering consists of ten-spin loops, where the spin orientations conspiringly form into a higher-rank tensor configuration. These configurations are characterized by a vector-like ``director'' that behaves analogously to orientational order in nematic liquid crystals~\cite{paddison2015hidden}.  This ``hidden multipole crystal'' preserves the crystallographic symmetry but exhibits nontrivial spatial correlations of the loop directors, which identified in thermodynamic measurement were in the form of a broad anomaly in specific heat around 0.5\,K together with a pronounced peak in the nonlinear magnetic susceptibility(see Fig.~\ref{Gd3Ga5O12}(a)). It reinforces the interpretation that GGG undergoes a freezing transition not of conventional dipolar order, but of emergent loop degrees of freedom. The experimental observation of frequency dependence of the ac susceptibility, together with the absence of a sharp anomaly in the specific heat indicates unconventional freezing behavior (Fig.~\ref{Gd3Ga5O12}(b)). The inelastic neutron scattering reveals the multipolar organization associated with low-energy spin dynamics, which suggests that there are present gapless hydrodynamic modes~\cite{PhysRevB.16.2154}. The low-energy excitations of gapless hydrodynamic modes are governed not by long-range dipolar order, but by loop director fields, potentially a new class of modes rooted in multipolar freezing rather than conventional magnetic order. Unlike 3$d$-based canonical spin glasses, GGG exemplifies intrinsic, frustration-driven glassiness arising in the limit of negligible quenched disorder.

The observed behavior can be revised in the framework where frustration alone, coupled with anisotropy and multipolar degrees of freedom, stabilizes glassy states with both local liquid-like spin fluctuation and emergent global coherence. GGG has minimal disorder and primarily originates from subtle off-stoichiometry, yet this is sufficient to pin the collective degrees of freedom, akin to weak random-field effects in nematic liquid crystals, which pose challenges to the traditional spin-glass where randomness and disorder are essential. GGG, a rare example of a unique interplay of geometrical frustration, hidden order, and nontrivial thermodynamics, which emphasizes the necessity to transcend beyond simple dipolar or Ising paradigms toward embracing the rich, multipolar landscape of glassy quantum magnets.
High spin trillium lattice KSrFe$_2$(PO$_4$)$_3$, with $S = 5/2$, shows a frozen ground state~\cite{boya2022signatures}. Its low-temperature magnetic specific heat follows a non-trivial power law. The unconventional behavior of magnetic specific heat, $C_m \propto T^{2.33}$ can be traced to a quantum treatment of thermal fluctuations, where the fluctuation stiffness $\kappa_{T}(q)$ exhibits a non-analytic temperature dependence at low energies~\cite{bergman2007order}. This generates soft nodal-line excitations that effectively reduce the dimensionality imposed by the lattice geometry~\cite{bergman2007order, plumb2019continuum}. Another example of a trillium lattice with Fe$^{3+}$ moments is K$_2$FeSn(PO$_4$)$_3$, which exhibits an anomaly in magnetic susceptibility consistent with antiferromagnetic ordering near 2~K~\cite{lmsf-73hn}. However, the specific heat follows a nearly field-independent sub-quadratic power-law dependence on temperature, without showing a clear $\lambda$-type anomaly, indicating the absence of conventional long-range magnetic order. Muon spin spectroscopy provides further support for this picture: the zero-field relaxation rate $\lambda_{\mathrm{ZF}}$ shows a crossover at $T^{\ast}\!\approx 11$~K associated with the slowing down of spin fluctuations, while at low temperatures the spectra are dominated by a Gaussian relaxation component that remains largely undecoupled even in longitudinal fields up to several tesla. This ``undecouplable Gaussian" line shape is commonly regarded as indicative of a distribution of local fields arising from short-lived singlet correlations and persistent spin dynamics~\cite{PhysRevLett.84.2957,PhysRevLett.127.157204}. The coexistence of quasistatic and dynamic local moments, along with field-independent specific heat, which indicates the formation of an unconventional frozen state. While ZFC–FC splitting in magnetic susceptibility has been linked to a weak ferromagnetic component below 2 K~\cite{lmsf-73hn}, we suggest that these characteristics are more naturally interpreted as indicators of topological spin freezing, rather than conventional magnetic ordering. 
\\

\section{Hierarchical Vs. Non-hierarchical Energy landscape}

Understanding the complex physics of hierarchical and non-hierarchical energy landscapes in disordered magnetic systems requires exploring the complex energy landscape configurations. The disordered magnetic system (here spin glass and spin jam state) prevents a unique ground state due to competing interactions and random disorder, instead trapping it in a multitude of metastable configurations that are separated by varying energy barriers. In hierarchical energy landscapes, observed in the spin glass state, the energy states are decorated in a nested, ultrametric fashion and rugged funnel shape~\cite{PhysRevLett.43.1754,mezard1987spin,young1998spin,parisi1980order}. The Sherrington-Kirkpatrick model provides a mathematical foundation to describe such a rugged energy landscape. Later, Parisi's replica symmetry breaking solution of the SK model elucidates an appropriate mathematical framework within the mean field theory, where the energy landscape is organized in a nested, tree-like hierarchy visualization as ``valleys within valleys".Each valley represents one particular spin orientation, while deeper subvalleys produced small fluctuations of spin configuration. These tree-like hierarchical landscapes induce the spin configuration to move slowly towards the equilibrium state. As a result, memory and rejuvenation effects were observed in the canonical spin glass CuMn, AgMn system, which is typically probed through the thermoremanent magnetization (TRM) experimentally. These slow spin dynamics and memory effects can described through time-dependent spin correlation functions $C(t, t_w) = \frac{1}{N} \sum_{i=1}^N \langle s_i(t_w)s_i(t_w + t) \rangle$, where $t_w$ is the waiting time before measurement. The violation of time-translation invariance in $C(t,t_w)$ clearly encodes aging, where the spin correlations decay more slowly as $t_w$ increases, implying the system progressively confines to deeper minima of the energy landscape~\cite{bouchaud1992weak}.

\begin{figure*}[t]
\includegraphics[height=294.42596pt, width=510.0pt]{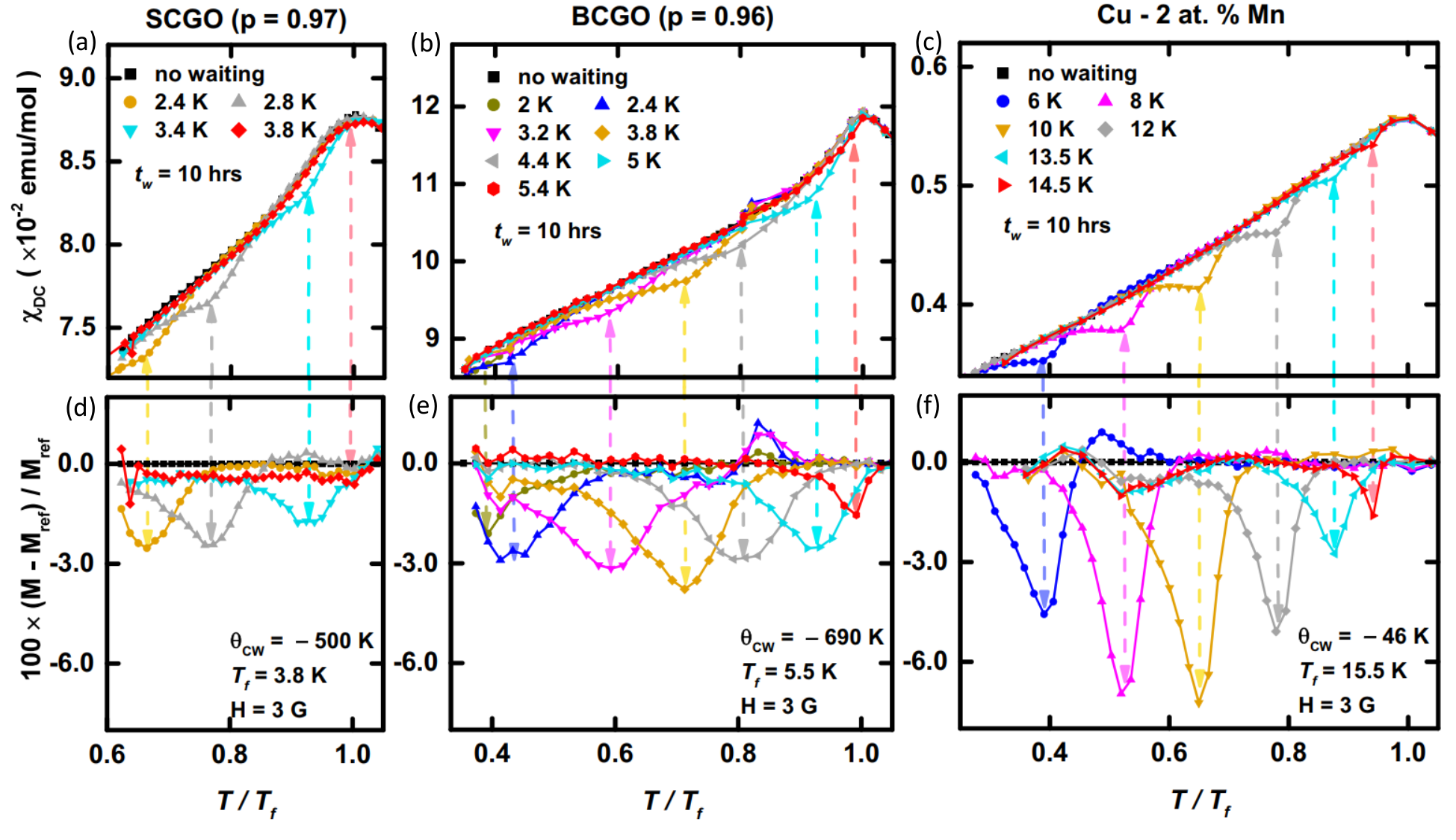}
  \caption{\textbf{Aging and memory effect in unconventional and conventional spin glass materials}. (a-c) DC magnetization (d-f) thermo-remanent
magnetization. Source: adapted from refs.\cite{samarakoon2016aging} with permission from PNAS.}
  \label{landscape}
\end{figure*}
In contrast, in non-hierarchical energy landscapes, observed in  spin jam state, the energy landscape exhibits decorated and shallow minima but a rough or corrugated bottom (Fig. \ref{figure1} (e)). The energy shape in spin jam state breaks an exactly flat landscape due to classical and quantum fluctuation, and a constrained driven mechanism arises from intrinsic spin texture. The non-hierarchical energy landscapes can be elaborated through a feature of broad basins, within which numerous microstates with  local minima are separated by energy barriers and exhibit no clear hierarchical organization~\cite{klich2014glassiness,yang2015spin,samarakoon2016aging,samarakoon2017scaling}. A qualitative and quantitative distinction between these two energy landscapes is described below in the context of aging and memory effects.

Memory effects can be observed in the response function
$ \chi(t, t_w) = \frac{\delta \langle s_i(t + t_w) \rangle}{\delta h(t_w)}$,
which depends on the history of the applied field $h(t_w)$.
The TRM method remains the most effective technique to study memory and aging effects, justifying the qualitative distinction between these two energy landscapes. A system is cooled with a waiting time $t_{w}$ using zero magnetic field at an intermediate temperature $T_w$ below the freezing temperature $T_{g}$, allowing it to relax into lower energy states. The longer the wait, the deeper the relaxation, which is termed ``aging." After reaching base temperature, to measure TRM a weak magnetic field is applied. Upon reheating, the system revisits these states near $T_w$, demonstrating memory of the aging process before transitioning to higher energy states at elevated temperatures. The TRM measurements on SCGO (p = 0.97), BCGO (p = 0.96), and CuMn(2\%) were conducted as reported in the reference. In SCGO and BCGO, which are spin jam prototypes with a highly frustrating quasi-2D triangular network, aging and memory effects are weaker compared to CuMn(2\%). TRM data at waiting times ($t_w$) from 6 minutes to 100 hours at $T_w/T_g \sim 0.7$ show intensifying aging and memory effects with increasing $t_w$. Significant aging is observed in CuMn(2\%) even for $t_w = 6$ minutes, while SCGO and BCGO show minimal aging at this $t_w$. As $t_w$ increases in CuMn(2\%), a memory dip develops at $T_w$ for $t_w > 3$ hours, whereas SCGO only exhibits a weak memory effect even at $t_w = 100$ hours Fig. \ref{landscape}(a,d,b,e). How the memory effect behaves for both states can be studied by the aging-induced relative change in magnetization $(M_{\text{aging}} - M_{\text{ref}})/M_{\text{ref}}$, where $M_{\text{aging}}$ and $M_{\text{ref}}$  are the magnetization with and without aging. This quantity increases in CuMn(2\%) from 3.4\% to 8.2\% with an increase in $t_w$, while in SCGO and BCGO it rises from 0.6\% to 2.4\% and 0.7\% to 3.1\%, respectively, from 6 minutes to 10 hours, with a diminished rate of increase for $t_w > 10$ hours, reaching 2.7\% at $t_w = 100$ hours for SCGO. While both spin glasses and spin jams display aging and memory effects, canonical spin glasses exhibit pronounced memory effects even after short waiting times, indicative of a highly rugged, hierarchical energy landscape (Fig. \ref{landscape}(c, f)). In contrast, spin jamm state displays weaker memory effects and slower aging, consistent with a comparatively flatter and less hierarchical energy landscape. Another preferential distinction between these two states is that the spin jam state looks like a shoulder-like feature, while spin glass shows a substantial dip in DC magnetization with aging~\cite{samarakoon2016aging,samarakoon2017scaling}.In these systems, the barriers between metastable configurations are more uniformly distributed, leading to smoother relaxation dynamics and reduced sensitivity to thermal or temporal perturbations.

\begin{figure*}[t]
\includegraphics[height=290.28474pt, width=433.50311pt]{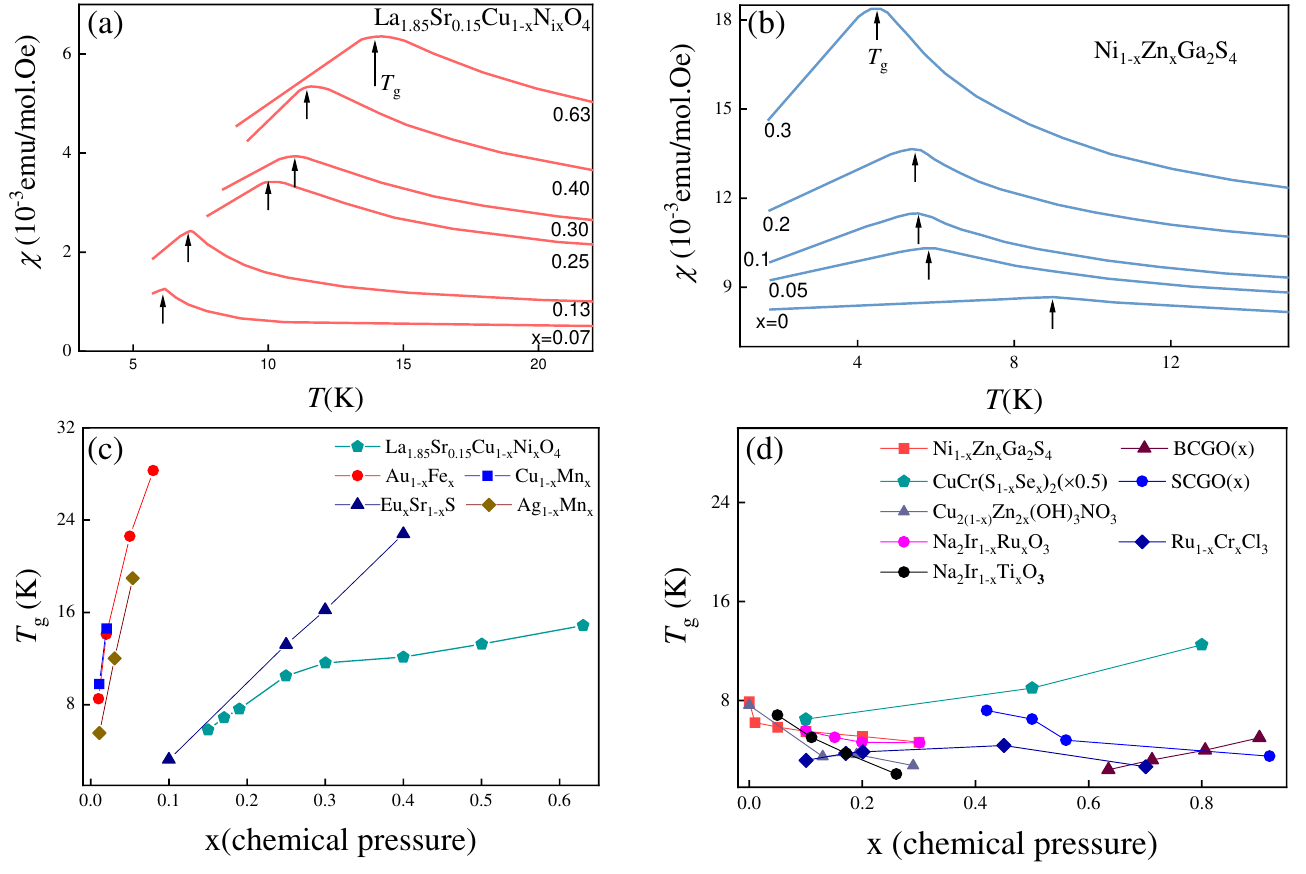}
  \caption{\textbf{ Conventional Spin Glass vs. Topological Spin Glass}. (a) and (b) represent the temperature-dependent magnetic susceptibility for the canonical and the topological spin glasses, respectively. The data highlight the contrasting magnetic responses between the two systems. (c) and (d) depict the dependence of the freezing temperature \( T_g \) on doping concentration \( x \) for canonical spin glass systems as well as topological glasses. Source: (a-d) adapted from refs.\cite{PhysRevB.84.024409,PhysRevB.19.1633,PhysRevB.20.1245,PhysRevB.6.4220,wu2011spin,PhysRevB.99.214410,PhysRevB.92.134412,manni2014effect,yang2016glassy,tewari2024signature,nambu2006coherent} with permission.}
  \label{Compare}
\end{figure*}

\section{Effect of Chemical Pressure on Spin Freezing}
Sometimes, even a subtle perturbation can trigger unexpectedly rich and complex physical phenomena in many-body system. Particularly frustrated magnets are quite sensitive to external perturbation in this context; their nearly degenerate ground states can be tipped from one phase to another by the slightest modification of lattice geometry or exchange interactions. A classic approach to tune the magnetic interaction is through chemical pressure, which is achieved by substituting the cations with different ionic radii. Such substitutions subtly modify the bond lengths and bond angles, therefore influencing the local geometry. Since the superexchange interactions sensitively depend on orbital overlap and bond configurations, these structural distortions can alter the balance among antiferromagnetic, ferromagnetic and anisotropic (e.g. Kitaev-type) couplings. In a wider context, such external tuning has been instrumental in realizing several emergent phases such as Bose–Einstein condensation~\cite{giamarchi2008bose,PhysRevLett.103.207203}, unconventional superconductivity~\cite{drozdov2015conventional,uji2001magnetic,PhysRevLett.101.057006,PhysRevX.5.041018} and exotic spin liquid states~\cite{mirebeau2002pressure,PhysRevLett.112.087204,PhysRevLett.93.257201,PhysRevLett.119.037201,PhysRevLett.127.157204}. In this section, we discuss the role of chemical pressure on the magnetic and topological behavior of the spin glass systems. In particular, we compare the influence of chemical pressure on conventional spin glasses, which are disorder-driven systems, and on systems that show topological spin freezing.

In conventional spin glasses,  such as La\(_{1.85}\)Sr\(_{0.15}\)Cu\(_{1-x}\)Ni\(_{x}\)O\(_4\)~\cite{PhysRevB.84.024409} and dilute CuMn alloys~ \cite{PhysRevB.23.1384}, the magnetic impurities are doped into a non-magnetic host lattice, which gives rise to random exchange interactions that involve ferro and antiferromagnetic correlations. With increasing the magnetic impurity concentration, the system approaches a percolation threshold and the glass transition temperature $T_g$ progressively increases with further doping. As evident from Fig.~\ref{Compare}(a), the rightward shift of peak in $\chi (T)$ suggests an enhancement of Tg with Ni doping at Cu site. The increased concentration of the interacting spins strengthens the collective freezing, highlighting the disorder driven origin of the glassy phase. Unlike in conventional spin glasses, the introduction of non-magnetic impurity at magnetic site~\cite{yang2016glassy,nambu2006coherent} not only introduces randomness in the exchange interactions but also perturbs the frustrated, constraint governed manifold. Interestingly, the freezing temperature in such systems often remains unchanged or even decreases slightly with doping (Fig.~\ref{Compare}(b, d)), indicating a robustness of the frozen state to weak compositional changes. This insensitivity underscores the fundamentally distinct origin of spin freezing. In this context the dynamics are arrested not by random distribution of local fields but by proliferation and entrapment of topological defects in the geometrical framework. These opposing trends show that the freezing in conventional spin glasses is driven by enhanced randomness and local interaction density, while the topologically frustrated systems achieve their glassy behavior from global constraints embedded in their lattice geometry. In the latter case, the chemical pressure subtly perturbs the balance of frustration and interactions. However, the topological nature of the constraints accounts for an exceptional resistance of these system to freezing temperatures. This finding adds to our understanding of spin glass systems and reveals the fundamental role of topology and emergent gauge structure in governing the collective behavior in quantum materials.

The freezing temperature $T_g$ depends both on the energy barrier required for collective spin flipping and on the spin correlation length~\cite{klich2014glassiness, syzranov2022eminuscent}. Particularly controlled chemical pressure can tune the degrees of frustration and reconfigure the energy landscape which leads to modifications of freezing temperature or the characteristics of the frozen spin state. In certain instances, the chemical pressure may even drive the system towards a different phase and modify the mechanism of spin freezing thereby affecting magnetic and topological properties. Topologically frustrated spin freezing can withstand up to a certian threshold. For instance, in the spin-\(\frac{1}{2}\) square lattice system Sr\(_2\)CuTe\(_{1-x}\)W\(_x\)O\(_6\), the Goldstone mode is observed, as indicated by the broad maxima in specific heat data, which remain field independent for an applied magnetic field range upto 9 T for doping concentrations of \(x = 0.1\), \(0.2\), and \(0.5\)~\cite{PhysRevB.98.054422, PhysRevLett.127.017201}. However, at higher doping levels (\(x = 0.7\), \(0.8\), \(0.9\), and \(1.0\)), the system transitions to a canted antiferromagnetic ordering, indicating a breakdown of the frustrated spin freezing beyond this doping threshold~\cite{PhysRevB.98.054422}.

To differentiate between the topological spin glass and the canonical two-level (TL) spin glass upon doping, one can analyze their distinct thermodynamic behaviors, such as specific heat and entropy. Experimentally, as Ga$^{3+}$ non-magnetic doping density(p) increases in the triangular pyramids $\text{Ba}_2\text{Sn}_2\text{ZnCr}_{7p}\text{Ga}_{10-7p}\text{O}_{22}$ (BSZCZO), topological glassiness wanes, leading to a crossover from a spin-jam state to a mixed or TL spin-glass state, with a corresponding linear decrease in the fractional population \( f \) of the spin-jam state~\cite{PhysRevB.109.104420}. At higher \(p\) (\( p \geq 0.93 \)) the system mostly follows a robust spin stiffness and high \( E_{\text{JAM}} \) whereas at lower \(p\) (\( p < 0.67 \)), TL spin-glass clusters dominate, indicated by a linear \( C_{\text{mag}} \) dependence.  The variation of zero point entropy $S_0$ with \(p\) provides additional insight into the distinction between these states. The total entropy, expressed as  \( S_{\text{tot}}(p) = f S_{\text{JAM}}(p) + [1 - f] S_{\text{TL}}(p) \) transitions from spin-jam to spin-glass dominance as \( p \) decreases~\cite{PhysRevB.109.104420}. This crossover elucidates a redistribution of entropy: when the non-magnetic disorder grows, extended spin wave-like modes are suppressed and the entropy shifts towards the localized tunneling-type excitations associated with strongly interacting Cr sites ~\cite{PhysRevLett.64.2070}. This transition is observed in  SrCr$_{9p}$Ga$_{12-9p}$O$_{19}$, where the dynamic susceptibility is $\chi''$ measured using INS, fitted with a sum of both the  tangent and Lorentzian contributions~\cite{yang2015spin}. This transition marks a crossover from the spin jam state mostly dominated by HS mode to a conventional frozen state in which dynamics are governed by localized spin clusters. In the higher $p$ region, the low-energy spin fluctuations exhibit progressive broadening and linear behavior, which indicates the emergence of Goldstone modes without long-range magnetic order~\cite{yang2015spin}. With an increase in the doping parameter, the paradigm of unconventioanal spin freezing shift to canonical spin freezing but it can also result in long-range magnetic ordering, as observed in systems such as Sr$_2$CuTe$_{1-x}$W$_x$O$_6$ (for $x$ = 0.7, 0.8, 0.9 and 1) ~\cite{PhysRevB.98.054422} and (D$_3$O)Fe$_{3-y}$Al$_y$(SO$_4$)$_2$(OD)$_6$~\cite{PhysRevB.61.6156}.

From this study, we observed that controlled doping can potentially affect the magnetic properties and spin dynamics, thereby opening up a fresh way to look at quantum magnetism. In this context, the existence of magnetic analogue of bosonic peak in spin glasses remains an open question since direct experimental evidence has yet to be obtained. Nonetheless, the recent studies offer a promising evidence suggesting the emerging magnetic bosonic peak (MBP) in conventional spin glasses. For example, the classical spin glass system $\text{Cu}_{1-x}\text{Mn}_{x}$ exhibits a localized bosonic peak, displaying an intriguing interplay among disorder, frustration and spin dynamics~\cite{PhysRevResearch.6.013006}. INS experiments demonstrate a broad low-energy magnetic excitations that scale with the Mn concentration. This indicates that as disorder increases, the energy spectrum of these excitations broadens and the peak energy shifts to higher values.  A characteristic peak is observed around 2 meV  below spin glass temperature \(T_g\) in the magnetic susceptibility $\chi''(Q, \omega)$ with a tail extending to higher energies. Such behavior is indicative of localized spin excitations that arises within the short range correlation region of spin glass state ~\cite{PhysRevResearch.6.013006}. The $Q$-independence of these excitations exhibits a pronounced enhancement at specific wave vectors indicating the presence of short range ferromagnetic interactions. When the temperature rises above \(T_g\), a notable changes in the spectrum are observed, signaling the onset of relaxation process. However, bolow \(T_g\) the excitations retain their Bose-scaled intensity, indicating that the MBP arrises from thermally populated localized excitations confined within the metastable regions of the spin glass energy landscape. The increasing Mn concentration leads to the broadening of excitation spectrum, highlighting the significance of disorder in creating these localized bosonic modes. The INS results suggest that the MBP arises from the collective response of the spins entrapped in the rugged energy landscape of the spin glass, where each metastable state contributes to the overall excitation spectrum.

\section{The Magnetic Trichotomy: Order, Glass, and Liquid}
\begin{figure}[t]
		\begin{center}
			\includegraphics[height=239.6986pt, width=249.90117pt]{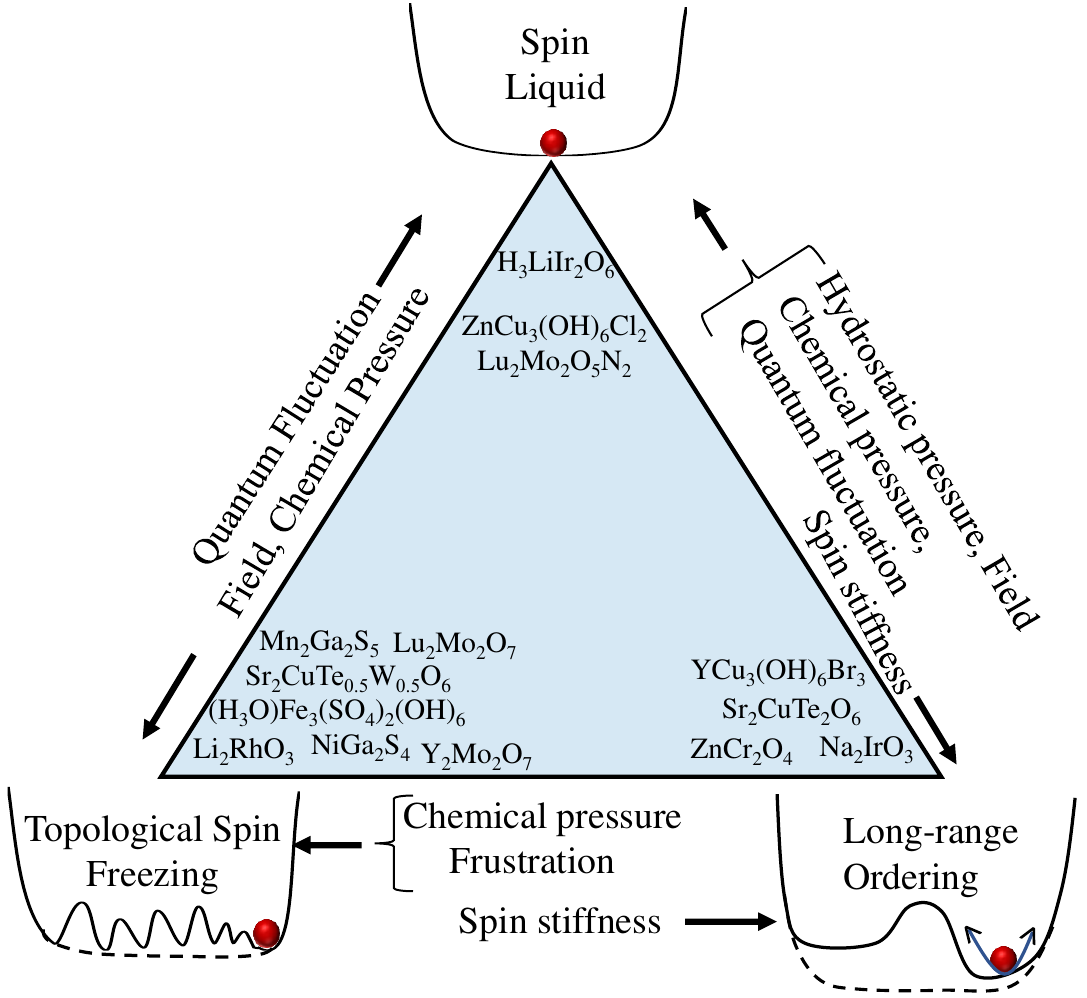}
			\caption{\textbf{Schematics of Novel States in Frustrated Quantum Magnets.} This highlights distinct regimes in selected frustrated materials, categorized by their low-energy excitations and underlying energy landscapes. Conventional zig-zag and spiral orders correspond to well-defined minima in the energy landscape, characteristic of classical magnetic order. In contrast, quantum spin liquids exhibit a nearly flat landscape with extensively degenerate ground states, while topological spin freezing arises from rugged landscapes with metastable states separated by finite energy barriers.
}
			\label{TriangularPhaseDiagram}
		\end{center}
	\end{figure}

The frustrated magnets exhibit a variety of ground states that are broadly categorized into three groups namely: long range magnetically ordered state, spin glass state and highly fluctuating entangled quantum spin liquid state. Fig.~\ref{TriangularPhaseDiagram} schematically depicts the contrasting energy landscapes of ordered, frozen, and liquid phases under the influence of key control parameters such as chemical substitution, pressure, and magnetic field in the frustrated magnets, along with some representative candidate materials. In frustrated magnetic ordered state, the onset of long-range magnetic order occurs below a critical temperature $T_\text{c}$ which is typically suppressed compared to characteristic exchange coupling strength $J$ ~\cite{moessner2006geometrical}. When the spin stiffness is large, and quantum fluctuations or thermal fluctuations act as a perturbation, the system lifts the degeneracy of frustrated manifold and selects an ordered configurations by breaking the rotational and possibly lattice symmetries. This phenomenon is known as order-by-disorder mechanism, which stabilizes a particular ordered state entropically or quantum mechanically ~\cite{villain1980order}. Depending on the exchange couplings and anisotropies the resulting magnetic order may be collinear (N\'eel), coplanar ($120^\circ$ structure) or non coplanar (multiple-$\mathbf{Q}$ textures)~\cite{takagi2018multiple,PhysRevB.103.054422}.These systems are characterized by a $\lambda $-type peak in specific heat akin to the $\lambda$ transition observed in the superfluid helium~\cite{keesom1935new}. This anomaly signals the establishment of low-energy collective modes, namely Goldstone modes which arise from the spontaneous symmetry breaking of continuous spin rotational symmetry. Such fluctuation-induced selection plays a crucial role on the ground state of frustrated magnetic systems. For instance, in classical spinels like ZnCr$_2$O$_4$, magnetic ordering is accompanied by a spin-lattice coupling-driven structural distortion that relieves frustration, a form of spin-Peierls-like transition that lowers the total free energy~\cite{PhysRevLett.134.086702}.  Despite the dominance of classical order in this regime, proximity to frustration often leaves clear fingerprints. These manifest as soft modes, roton-like minima in the magnon spectrum, or sub-extensive degeneracies, which can be revealed by tuning the system toward the other two regimes. In the QSL regime, the spin degrees of freedom do not order down to the lowest temperatures despite strong exchange interactions~\cite{balents2010spin}. The preservation of spin rotational symmetry, even at low temperatures, occurs due to the quantum superposition among macroscopically many spin configurations that lead to a state of long-range quantum entanglement ~\cite{wen2019experimental,broholm2020quantum}. The excitations in such systems are often fractionalized, such as spinons, Majorana fermions, vison and emergent gauge bosons whose statistics and dynamics are governed by an emergent gauge structure. The defining feature of this phase is the collapse of spin stiffness ($\rho_{\mathrm{eff}} \to 0$) driven by strong zero-point motion, implying that twisting the spin configuration requires vanishingly small energy. As a result, the low-energy excitations exhibit Dirac-like dispersions or nearly flat bands, which originate from destructive interference among the hopping amplitudes of the fractionalized quasiparticles.

In some frustrated magnetic lattices, symmetry and topology combine to protect gapless points, whereas in others the topological band gaps emerge, which give rise to quantized responses such as thermal Hall conductance. This regime emerges when the frustration is strong enough to suppress any magnetic order state, resulting in exotic quantum states with intriguing properties. A dynamically arrested regime arises between the quantum spin liquid and classical ordered phases, wherein frustration and topology together entrap the system within a manifold of metastable states. In contrast to conventional spin glasses, where freezing arises from quenched disorder, the constraints in these frustrated systems emerge intrinsically from the topology of their configuration space. Examples include closed flux loops in kagome lattices, Dirac strings in pyrochlores (Fig. \ref{figure1}(d)), and nontrivial winding numbers in triangular magnets. In such systems, the free energy landscape is highly rugged with multiple hierarchical valleys separated by barriers that originate not only from exchange energies but also from topological invariants.

As a result, spin dynamics slow down dramatically, and the spin correlation function decays with a timescale that is much longer than the microscopic spin precession periods, but the system never becomes truly static. The experimental probes uncover this ``frozen yet fluctuating" nature: the spin lattice relaxation rate remains finite down to the lowest measurable temperature and higher order susceptibilities show non-Gaussian fluctuations.

\section{Operational Definition and Role of Topology in Topological Spin Glass} In this review, we have discussed the unconventional spin glasses and their properties in detail. The terminology ``topological spin freezing'' encompasses a diverse class of frustrated magnets in which the system's equilibrium is hindered by a topological constraint, whether arising from real-space structure or emergent gauge symmetries. So, for all the above systems, topology is not just a static, invariant property (like a hole in a donut) that can be easily measured; instead, it acts as a dynamical constraint on spin dynamics. Here, we provide a technical framework that distinguishes the various roles of topology in spin dynamical arrest. To make it clearer, we describe three distinct physical scenarios where non-local properties govern the dynamics:

1. \textbf{Real-Space Topological Defects (Classical/Semi-Classical):} In systems like 2D antiferromagnetic kagomes, the ground state degeneracy allows continuous zero-energy modes known as ``spin folds" or ``weathervane modes"~\cite{chandra1993anisotropic}. In the presence of XY anisotropy, the system undergoes a KT transition that involves the binding of non-Abelian defects. These defects, which are either closed or open domain walls, separate regions of uniform and staggered chirality, and cost zero energy but are non-commuting under propagation. Here, the glassiness arises from these non-Abelian topological processes rather than thermal activation. For triangular lattices with noncollinear spin configurations, the pairing of associated $Z_2$ vortices is thought to be the reason behind the glassiness observed in some of the Heisenberg triangular lattices even in the clean limit~\cite{PhysRevB.78.180404}. In the pyrochlore fluoride CsNiCrF$_6$, the weak interloop couplings induce a spin-nematic state that is associated with the slowing down of the relaxation process during the rearrangements of spin loops, which is a possible mechanism for spin freezing~\cite{xvxt-whns}. Similarly, the multipolar reorganization of the emergent loop excitations in gadolinium gallium garnet GGG leads to spin freezing in the sub-Kelvin region~\cite{paddison2015hidden}. A similar scenario may explain the anomalous spin freezing observed in the low dilution limit of ZnCr$_{2(1-x)}$Ga$_{2x}$O$_4$~\cite{PhysRevLett.110.017203} and Zn$_{1-x}$Cd$_x$Cr$_2$O$_4$~\cite{PhysRevB.64.024408}. The parent compound ZnCr$_2$O$_4$, as discussed previously, forms a composite structure of hexagonal spin loops of Cr moments, which may undergo a slow relaxation process in the rearrangements of spin loops due to site dilution. However, in the large dilution limit, such as in MgCrGaO$_4$~\cite{w6j5-ljfj} and ZnCrGaO$_4$ (unpublished data), where there is nearly 50\% vacancy in the pyrochlore site of the Cr sublattice, the spins remain dynamic down to a temperature three orders of magnitude lower than $\theta_\text{CW}$. This could be due to magnetic loops of varying lengths (open- or closed-ended), which give rise to algebraic spin correlations.

2. \textbf{Emergent Gauge Structure:} A topological spin glass in a weakly nonmagnetically diluted spin ice emerges from the interplay between quenched disorder and the underlying topological spin liquid~\cite{PhysRevLett.114.247207}. The microscopic starting point is the spin-ice rule on a pyrochlore lattice, with the ``2-in–2-out” ice rule setting along local ⟨111⟩ axes~\cite{PhysRevLett.83.1854}. Projecting the 3D pyrochlore into 2D, we can represent each spin as a dumbbell shape with both ends carrying a pair of opposite magnetic charges $\pm Q$ and dipole moments $\mu$~\cite{castelnovo2008magnetic}. The key energy scales are the dipolar interaction energy $D$ and the monopole excitation gap $\Delta$, which controls the onset of the Coulombic phase that is characterized by gauge fields and algebraic spin correlations. Upon nonmagnetic dilution, the vacancy could be interpreted as a "ghost spin'' creating a pair of opposite magnetic charges on adjacent tetrahedra. This mapping simplifies the calculation: instead of considering dense interacting spin systems, we can track a dilute gas of ghost dipoles embedded in a fluctuating Coulomb background~\cite{PhysRevLett.114.247207}. Although in the new framework, the effective Hamiltonian is dipolar energy, it should be renormalized to account for the fluctuating spin-ice background. As a result, the effective interaction becomes $D'=D+\frac{3T}{\sqrt{2\pi}}$ where $T$ is the temperature scale, demonstrating that the entropic fluctuations mediate the long-range interaction between the defects. In this framework, the Coulomb phase effectively ``dresses'' the impurities and encoded the topological correlations into their interactions. Another important aspect is that the impurities are charge neutral, so they do not disturb Coulomb phase correlations; instead, they coexist with the gauge structure. The freezing of this ghost spin occurs at a critical temperature $T_g\propto x$, where $x$ is the dilution amount. Monte Carlo simulation confirms a glass transition with an Edward-Anderson type order parameter defined as: $q_\text{EA}^{\alpha\beta}(k)=\frac{1}{N}\sum_i\mu_i^{\alpha(1)}\mu_i^{\beta(2)} \exp{(ik\cdot r_i)}$, with $\alpha,\beta=x,y,z$ are spin components and $N$ is the number of ghost spins~\cite{PhysRevLett.114.247207}. A diverging susceptibility occurs below $T_g$ and $q_\text{EA}$ grows continuously. Importantly, the freezing occurs only in the impurity sector, and the bulk remains in the Coulombic phase. This is fundamentally different from the conventional spin glass, where the freezing destroys such underlying correlations. Here, the glassiness is built on top of the topological structure of the spin ice state rather than competing with it.

\begin{figure}[t]
		\begin{center}
			\includegraphics[height=120.6986pt, width=249.90117pt]{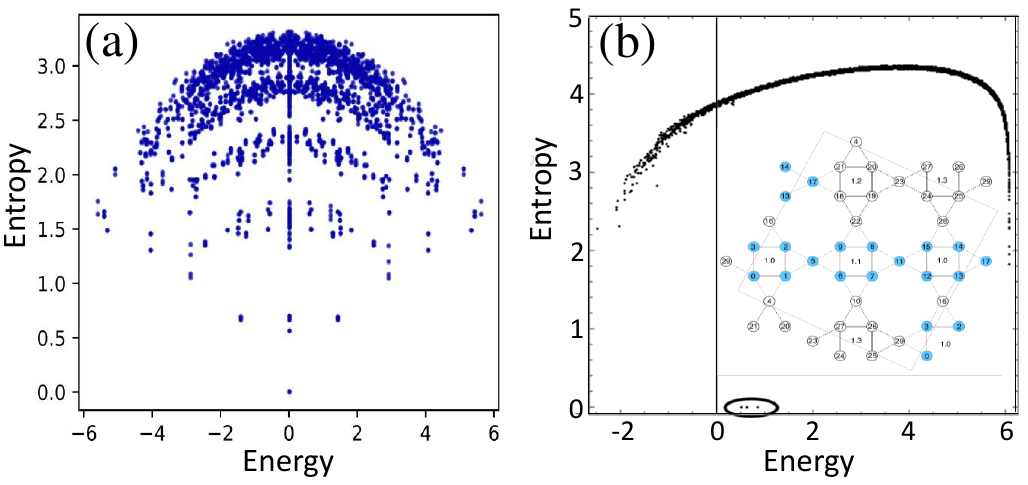}
			\caption{\textbf{Weak broken ergodicity in geometically frustrated magnets.} Entangle entropy 
for (a) square-kagome (a) and (b) kagome magnet. The circled in (a) is localized magnons in square-kagome. Sources: (a) adapted from the ref.~\cite{PhysRevB.101.241111} and (b) adapted from ref.~\cite{PhysRevB.102.224303}.
}
			\label{integrability}
		\end{center}
	\end{figure}

3. \textbf{Glassiness due to Hilbert Space Fragmentation:} In frustrated magnets, the existence of spin liquids and associated fractionalized excitations is well defined close to the ground state or in thermal equilibrium. However, the geometry of the lattice and quantum mechanical effects can conspire together to fragment the Hilbert space, where the quantum many-body scar emerges as a nonthermal eigenstate, violating the eigenstate thermalization hypothesis (ETH)~\cite{PhysRevB.103.235133,PhysRevA.110.023301}. These scarred states obstruct ergodic relaxation and, as a result, can give rise to spin-glass behavior in disorder-free systems~\cite{PhysRevX.11.011011,PhysRevB.102.224303}. The convergence of frustration, scar theory, and fragmentation opens a new frontier for the study of quantum glassiness in clean many-body systems~\cite{PhysRevB.102.224303,PhysRevB.104.L121103,PhysRevB.102.241115}. In many-body ergodic systems, the initial state will evolve to the thermal state under the unitary evolution of the underlying Hamiltonian. Since thermalization erases memory of the initial state, it is essential to identify non-ergodic quantum systems that preserve this information.

For a spin$-1/2$ kagome lattice constructed from XYZ Hamiltonian with $J_x=J_y$ and $J_z/J_x=-1/2$, exhibits emergent Hilbert space fragmentation~\cite{PhysRevB.103.235133, PhysRevB.101.241111}. This setting allows the exact construction of an exponentially large set of eigenstates of three distinct types. The key point is that geometric frustration imposes a local constraint: no two neighboring sites can share the same state. As a result, the accessible Hilbert space is strongly reduced. Importantly, these states are not trivially related; instead, they belong to different topological sectors that cannot be connected by local spin operations. Figure~\ref{integrability}(a) illustrates the entanglement entropy (EE) of all eigenstates of the staggered XXZ Hamiltonian. As the Hilbert space is turned into disconnected sectors, it exhibits distinct EE values despite sharing the same energy~\cite{PhysRevB.101.241111}. Thus, the EE value is not a single-valued function of energy as expected for systems obeying the ETH rule. Figure~\ref{integrability}(b) shows how a square-kagome lattice can host quantum scar states due to the geometric frustration and localized excitations~\cite{PhysRevB.102.224303}. Here, each square plaquette is connected by triangular units, with different exchange couplings assigned to the bonds within the square and to the connecting triangular bonds. Localized magnons arise from spin-flip excitations confined to the square plaquette. Due to destructive interference arising from geometric frustration, these magnons cannot propagate through the lattice and remain spatially localized. So the frustration dynamically projects out hopping processes, which generate a constrained subspace that remains decoupled from the thermal continuum, thereby embedding low-entanglement eigenstates within the many-body spectrum~\cite{PhysRevB.102.224303}. As discussed previously, in exactly solvable models like the Kitaev system, the sheer abundance of conserved quantities (local $Z_2$
 gauge fluxes or plaquette operators) arising from the problem's integrability hinders the exploration of the full Hilbert space, potentially leading to glassiness.

In the above-discussed systems, geometrical frustration or interference mechanisms partition the Hilbert space into dynamically disconnected subspaces. Thus, the resulting glassiness is not governed by energetic barriers but by topological obstructions in configuration space, leading to dynamical arrest even in the absence of disorder. In this sense, the system realizes a topological spin glass, in which glassiness emerges from the interplay of frustration, emergent constraints, and nontrivial topology.

\begin{figure*}[t]
\includegraphics[height=334.5181pt, width=433.50311pt]{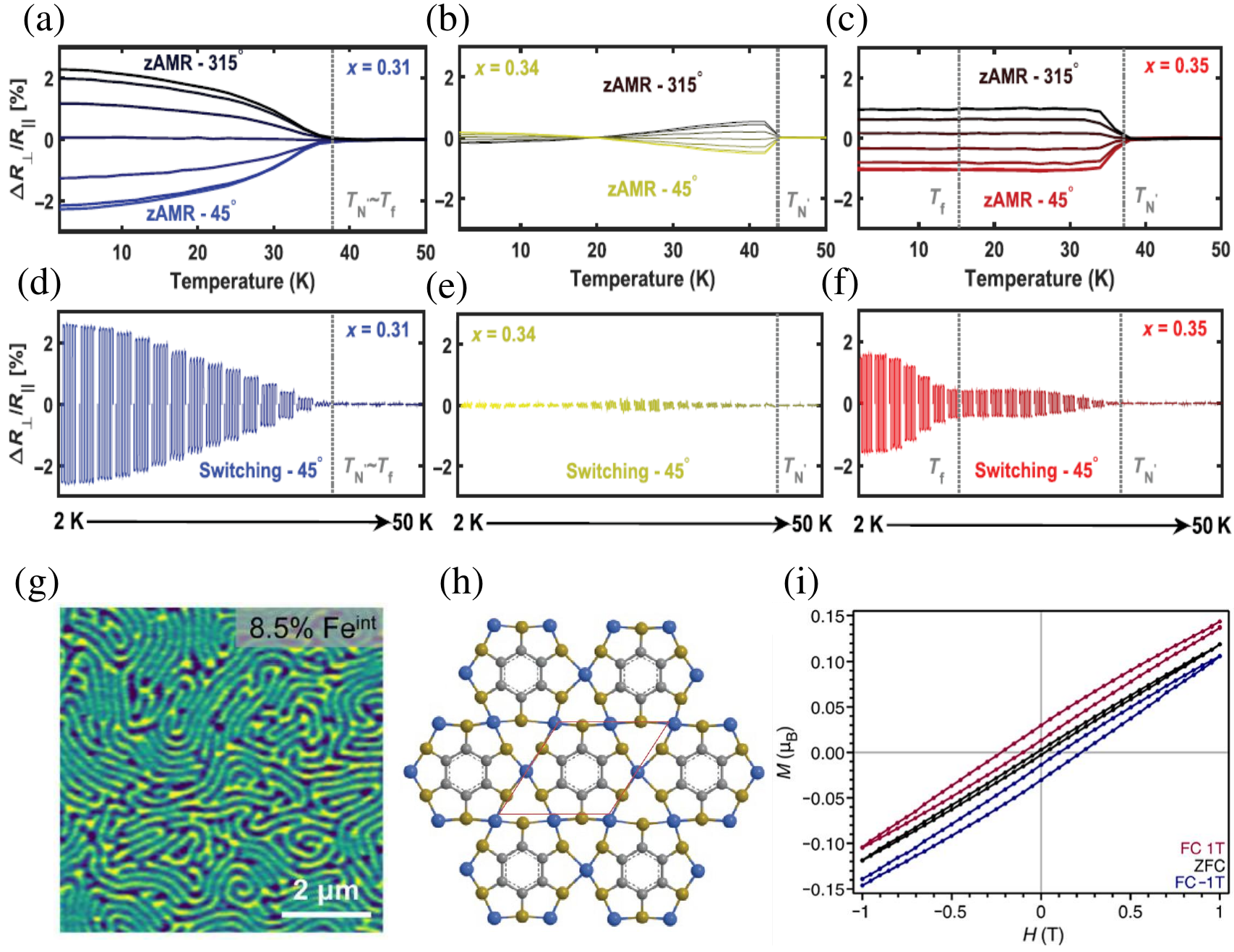}
  \caption{\textbf{ Technological prospects of topological spin freezing}. Temperature-dependent zero-field anisotropic magnetoresistance (zAMR) and electrical switching in Fe$_{x}$NbS$_{2}$ for compositions near the commensurate value $x = 1/3$. Panels (a--c) show zAMR curves for $x = 0.31$, 0.34, and 0.35, marking the onset of antiferromagnetic order at $T_{N}$. Panels (d--f) display the corresponding current-induced switching amplitudes, revealing a striking enhancement in the Fe-rich sample ($x = 0.35$). The switching maximum coincides with the spin-glass freezing temperature $T_{g}$, establishing that the frozen spin component actively mediates torque transfer to the AFM order. This result demonstrates how small deviations from stoichiometry ($\delta \neq 0$) transform Fe$_{1/3+\delta}$NbS$_{2}$ into a unique system where topological spin freezing boosts electrical control of antiferromagnetism. (g) Magnetic force microscopy image of Fe$_3$GaTe$_2$ nanoflakes with 8.5\% Fe$_{\text{int}}$ concentration, measured at room temperature under zero-field-cooled (ZFC) conditions. (g) Schematic structure of kagome Mn$_3$(C$_6$S$_6$). (i) Field-dependent magnetization of Mn$_3$(C$_6$S$_6$), recorded after cooling in +1~T (red), --1~T (blue), and zero field (black). The shifted hysteresis loops demonstrate symmetric displacements relative to the ZFC loop.  Adapted from the refs~\cite{maniv2021antiferromagnetic, zhang2024spin,wang2025electronic,murphy2021exchange}. }
  \label{application}
\end{figure*}

\section{Technological Relevance and Prospects of Frustrated Spin Glasses}
Spin glasses were originally explored in disordered magnetic alloys; however, the implications of spin glass theory across disciplines make it a model system in many settings. Quenched disorder, magnetic frustration, and a highly rugged energy landscape are the defining features of spin glasses. These characteristics have provided deep conceptual insights that have influenced diverse fields, including optimization, computer science, neuroscience, and information theory~\cite{binder1986spin, mezard1987spin, parisi2006spin}. For example, the mapping between Ising spin-glass Hamiltonian and NP-hard combinatorial optimization problems. It is also extended in the study of optimization algorithms, simulated annealing, and more recently, quantum annealing~\cite{kirkpatrick1983optimization, PhysRevE.58.5355}. The glassy energy landscape acts as a computational resource in quantum optimization devices. Similarly, spin-glass models laid the foundation for Hopfield neural networks. It is one of the earliest theoretical models of associative memory in neuroscience~\cite{hopfield1982neural}.

 In Hopfield networks, stored memories align with energy minima, enabling the system to retrieve a complete pattern from incomplete or noisy inputs. The principles of spin glass physics can be applied to coding theory and complex neural networks~\cite{mezard2009information,nishimori2001statistical}. Spin glasses can be useful in improving the design of disordered magnets and magnetic storage devices~\cite{mydosh2015spin}.

Topological spin glasses can have much more profound applications, as they exhibit the characteristic frustration and disorder, along with topological constraints. The glassy behavior in topological spin glasses arises not just from random local interactions, but from the complex interplay of defects, gauge fluxes, and emergent strings that are constrained by the system’s underlying topology~\cite{castelnovo2008magnetic, RevModPhys.89.025003, knolle2019field}. 
Qualitatively, this suggests the possibility of a new paradigm in spin glass physics, in which the emergent collective state exhibits an intrinsic robustness analogous to the "quantum protectorate" state~\cite{laughlin2000theory,anderson2000sources}. Here, the collective excitations are merely affected by the microscopic perturbations.
For example, spin freezing in Kitaev magnets and pyrochlore spin ice could provide a route to realize a topologically stable quantum memory where the information is encoded in the nonlocal degree of freedom that remains robust against local perturbations. This form of topological freezing inherently possesses aspects of fault tolerance, which can be important in quantum information science~\cite{kitaev2003fault,kitaev2006anyons}. The interplay between glassiness and topology can have dual aspects. In one way, it guides us on how to suppress and control freezing by suitable material design and control parameters, in order to stabilize more exotic states, such as spin liquids and high \(T_c\) superconductors. On the other hand, it suggests methods to harness topologically protected metastable states for robust memory or spintronic devices~\cite{lee2007high, vojta2018frustration}. Thus the topological extension of spin freezing offers exciting possibilities for quantum-functional materials that combine complexity, stability, and emergent protective properties.

Electrical control of antiferromagnetic order has become a central pursuit in spintronics, motivated by the absence of stray fields, ultrafast dynamics, and robustness against magnetic perturbations~\cite{jungwirth2016antiferromagnetic,RevModPhys.90.015005}. Unlike ferromagnets, however, the vanishing net magnetization of antiferromagnets makes it difficult to couple external currents directly to the Néel vector. The spin–orbit torque (SOT) mechanism provides one route, where current-induced spin accumulation transfers angular momentum to staggered sublattices. Yet such mechanisms often require large current densities and produce domain-fragmented switching, limiting their reproducibility and scalability. These limitations have driven the search for alternative routes in which collective magnetic backgrounds, rather than rigid domain walls, mediate torque transfer. A particularly striking experimental observation linking spin freezing to functionality arises from electrical switching measurements in magnetically intercalated transition-metal dichalcogenide, Fe$_{1/3+\delta}$NbS$_2$, where the Fe ions form a periodic sublattice within the van der Waals (vdW) gaps of the NbS$_2$ host, stabilizing long-range antiferromagnetic order below about 42 K~\cite{nair2020electrical}. As shown in Fig.~\ref{application}(a-f), the amplitude of current-induced switching exhibits a sharp maximum precisely at the spin-glass freezing temperature, $T_g$, even though the long-range antiferromagnetic order occurs at a distinctly higher temperature, $T_N'$\cite{maniv2021antiferromagnetic}. This behavior provides unambiguous evidence that the frozen spin-glass component is not an inert background but actively facilitates torque transfer to the antiferromagnetic order parameter. From a theoretical perspective, this enhancement can be understood within the framework of collective spin hydrodynamics: below $T_g$, the Edwards–Anderson order parameter acquires a finite value, leading to the emergence of low-energy collective modes akin to HS modes in glassy magnets~\cite{PhysRevB.16.2154}, which allow the disordered but rigid spin background to sustain long-wavelength twists acting as a topological channel for redistributing angular momentum injected by electrical currents~\cite{maniv2021antiferromagnetic}. Consequently, the torque transfer does not rely solely on conventional spin–orbit processes acting on the rigid N\'eel vector, but instead is mediated by the winding dynamics of the frozen spin texture. The sharp resonance-like peak in the switching amplitude at $T_g$ reflects this dynamic interplay: as the spin glass forms, the stiffness of the frozen texture is maximized, enabling efficient coupling to the antiferromagnetic vector. The critical role of the spin glass becomes especially clear when comparing different intercalation levels. Near-stoichiometric compositions such as for $x=0.33$ and $x=0.34$, the material displays sharp antiferromagnetic transitions but no glassiness in these systems; the electrical switching is weak and unstable, with pulse-to-pulse variations that underscore the limitations of a purely ordered background. By contrast, off-stoichiometric compounds with $x=0.31$ (Fe-deficient) and $x=0.35$ (Fe-excess) both host a coexisting spin glass that couples strongly to the canted antiferromagnetic order. For $x=0.31$, the freezing temperature coincides with $T_{N}$, leading to switching that tracks the antiferromagnetic order parameter but with markedly enhanced stability. While for $x=0.35$, the situation is even more striking: although the antiferromagnetic canting transition occurs near 37 K, the switching amplitude is dramatically amplified only at the lower spin-glass freezing temperature around 15 K. The NMR reveals that the frozen glass exerts an exchange field on the antiferromagnetic sublattices, producing asymmetric line splittings and demonstrating that the glass texture biases the antiferromagnet. Remarkably, the sense of Néel-vector rotation under current is opposite for $x=0.31$ and $x=0.35$ compositions, indicating that the microscopic spin-glass textures associated with vacancies and interstitials encode opposite topological winding biases. These observations firmly establish that the glass is not merely a disordered state but a topologically constrained collective state whose dynamics open a robust channel for torque transfer. In light of in another vdW material NbFeTe$_2$, the presence of a spin-glass state plays a pivotal role in shaping its low-field magnetic and transport behavior~\cite{PhysRevResearch.7.023219}. The spin-glass behavior appears in the low-field regime ($B < 0.7$ T) below the ferromagnetic ordering temperature. The Nb deficiencies and the resulting local inversion-symmetry breaking induce canting of Fe spins, giving rise to finite scalar spin chirality within spin clusters that generate the observed topological Hall effect~\cite{PhysRevResearch.7.023219}. 

An illuminating case is the ferromagnetically coupled triangular vdW Fe$_3$GaTe$_2$ lattice. Here, the partial intercalation of Fe ions into octahedral sites of the van der Waals gap gives rise to a random distribution of local moments~\cite{zhang2024spin}. As a result of these intercalations, there appears a random distribution of the exchange field pertaining to random anisotropy. This manifests as macroscopic spin freezing, as evident from the ZFC-FC bifurcation of magnetization. The glassiness doesn't degrade the ferromagnetic order rather it produces a heterogeneous magnetic landscape. Regions with strong and nearly uniform ferromagnetic alignment coexist with "Ferri" regions where the antiferromagnetically coupled disordered spins dilute the net moment. Cross-correlated measurements using magnetic force microscopy, XMCD in photoemission electron microscopy, and 4D Lorentz scanning transmission electron microscopy combine to reveal this mesoscale phase separation. At room temperature, the induction field is about $\sim$0.3 T in the Ferro phase and around $\sim$0.15 T in the Ferri phase. The disordered magnetic background has a strong influence on the stability and organization of skyrmions. In the Ferri regions, skyrmions form a triangular lattice with high bond-orientational order. Whereas in the Ferro regions, which contain many dislocations, the triangular lattice melts into a liquid-like spin arrangement. It is further revealed that the skyrmions possess a curling profiles that suggest the emergence of three-dimensionally twisted textures pinned by the glassy environment. These findings suggest that the quenched disorder not only controls the degree of positional order but also enriches the topology of the individual solitons. With a moderately applied magnetic field it is observed that the ring-shaped dislocations collapse into skyrmions with $ Q=0$.

Another interesting outcome of topological spin freezing is its effect on exchange bias. In conventional systems, exchange bias manifests as a horizontal shift of the hysteresis loop in layered ferromagnet/antiferromagnet materials that are promising candidates for magneto-electronic devices~\cite{nayak2015design}. This is attributed to the pinning of magnetic moments at the interface between the two layers~\cite{PhysRev.102.1413}. However, recent studies have shown that exchange bias can also appear in frustrated spin glasses without the need for heterostructures. For example, in Mn$_3$(C$_6$S$_6$) (Fig.~\ref{application}(h)), which is a kagome lattice made from Mn$^{2+}$ ions and benzenehexathiol (BHT)
ligands~\cite{murphy2021exchange}. This $S=5/2$ kagome compound is a topological spin glass transition with a glass transition temperature of 12 K and a Curie-Weiss temperature of $-253$ K~\cite{murphy2021exchange,wang2025electronic}. The topological glassy background produces a substantial shift in the magnetic hysteresis loop at low temperatures. When this material is cooled in a $\pm$1 T field, the loop shifts sideways by about 1625 Oe at 2 K, and its coercivity changes by around 395 Oe (see Fig.~\ref{application}(i)). This effect is not lost after waiting an hour at low temperature without a field, and both bias and coercivity level off after repeating the loop several times. Notably, it is possible to create the bias even after cooling in zero field, simply by applying a field at low temperature. This behavior is not what one would expect from minor-loop effects, nor does it match what is typically seen in conventional ferromagnet/antiferromagnet systems. A similar phenomenon is also observed in other topological spin glass hydronium iron jarosite (H$_3$O)Fe$_3$(SO$_4$)$_2$(OH)$_6$, which shows an exchange bias of about 1340 Oe after field cooling under 4 T. Frustrated magnets like Mn$_3$(C$_6$S$_6$) represent tunable platforms for hosting different quantum ground states, where chemical substitutions can switch between glassy, superconducting, or topologically insulating behavior. For instance, copper-based variants Cu$_3$(BHT) become superconducting below 0.25 K~\cite{huang2015two,takenaka2021strongly} , while Ni$_3$(BHT)$_2$ turns out to be a topological insulator~\cite{wang2013prediction}. These varied properties arise from the same type of chemical framework and underscore how tweaking the lattice or introducing controlled disorder can stabilize new ground states or magnetic textures. Even in ferrimagnetic alloys like TbCo, spin freezing can exist alongside long-range magnetic order, influencing how spins transport or switch under a field~\cite{park2022observation}. These examples illustrate emerging ways to harness spin glassiness or frustration for future device applications, particularly where magnetic memory or switching is crucial.

Finally, we give a brief view of self-induced spin glasses owing to specific crystalline lattice structures that generate competing exchange interactions. These self induced spin-freezing effects can have complex magnetic states and may lead to promising applications in future technologies~\cite{kamber2020self,verlhac2022thermally}. At low temperatures, elemental crystalline Nd shows a self-induced spin glass  state~\cite{kamber2020self}. Spin-polarized STM images show that the system is not randomly disordered. Instead, it is characterized by multi-$Q$ local order with nanoscale patches having distinct periodicities, each defined by multiple nearly degenerate wavev ectors. In reciprocal space, this appears as broad “$Q$ pockets,” reminiscent of flat valleys in the magnetic free-energy surface, where no single ordering vector dominates. The result is a frozen state with strong local correlations but without global coherence. Crucially, this state exhibits aging and dynamic heterogeneity, with relaxation times depending on the underlying $Q$ vector. Such behavior parallels the heterogeneous dynamics of structural glasses and underscores the rugged multiwell character of the spin-$Q$ energy landscape~\cite{ediger2000spatially}. A counterintuitive twist emerges upon heating. Instead of disorder growing with temperature, Nd undergoes a transition into a long-range multi-$Q$ ordered state above the glass transition temperature ($\approx 8$ K)~\cite{verlhac2022thermally}. Here, frustration is partially lifted because the weaker cubic–cubic sublattice correlations collapse, while the stronger hexagonal–hexagonal correlations stabilize order. Atomistic spin dynamics simulations confirm this order-from-disorder mechanism, predicting a sequence: low-$T$ spin glass$\rightarrow$intermediate multi-$Q$ order$\rightarrow$high-$T$ paramagnet. This strange thermal pathway shows how sublattice competition works as a topological constraint on the magnetic manifold, determining whether the system freezes or orders. Putting neodymium in the bigger picture of frustrated and topological spin glasses shows important differences. In triangular and kagome frustrated magnets, defect-induced textures (e.g., $Z_2$ vortices or spin folds) act as topological barriers to long-range order~\cite{kawamura1985phase, chandra1993anisotropic}, whereas in pyrochlores, Dirac strings and monopole excitations result in constrained freezing~\cite{snyder2001spin,PhysRevLett.114.247207}. Nd attains glassiness via $Q$-space degeneracy, wherein the "topology" is situated not in defects of real-space spins but in the flat valleys of reciprocal space. Nd shows that disorder is not necessary; frustration built into crystal symmetry can be enough to freeze.

\section{Summary and Outlook}

Understanding the complex, many-body non-equilibrium phenomena cross different length scales and disciplines leads to numerous fundamental discoveries that open new avenues for potential applications. In this review, we investigated promising frustrated spin lattices that show anomalous spin freezing behavior. The study reveals that the interplay between competing interactions, emergent degrees of freedom, unconventional electronic band structure, acting together with quenched disorder, results in the realization of a non-trivial spin glass state in a broad range of frustrated magnets. Early studies on canonical spin glasses have raised a fundamental query about the presence of Goldstone modes in the frozen state; however, experimental findings do not confirm this. Recent observations in highly frustrated spin glasses have revealed the existence of Goldstone modes in the frozen state. The tuning parameters, such as the applied magnetic field, chemical pressure, and hydrostatic pressure, indicate that we can transition from a liquid-like dynamic state and a spin solid state to a short-range driven spin glass state. Overall, the frustrated magnets show a paradigm shift from canonical spin glass to quantum fluctuation-driven topological spin glass in the limit of clean, densely populated magnetic systems. 

The combined study of bulk thermodynamic properties, along with microscopic signatures via  NMR, $\mu$SR, and INS experiments, performed on topologically frustrated spin glasses supports the presence of a large ground-state degeneracy, the onset of short-range spin correlations, and the emergence of gapless excitations. In summary, our research underscores the importance of topological effects in the study of frustrated magnets. The ubiquitous nature of glassiness of topological origin in several classes of frustrated magnets offers an ideal platform to understand many emergent quantum manybody phenomena through the prism of topology, underlying
symmetry, and external stimuli. Topological defects in spin glasses, including vortices, domain walls, or frustrated spin clusters, can act as pinning centers in inducing glassy dynamics and metastable states essential for topological spin textures, including skyrmions.
Skyrmions hold great potential for technological applications in spintronics and neuromorphic computing, where controlled skyrmion motion is required. In frustrated quantum materials, non-trivial fractional excitations, spin correlations, and competition between emergent degrees of freedom play a vital role in inducing exotic states such as Kitaev QSL with non-Abelian statistics, and classical spin liquids with fractions that hold tremendous potential to address some of the fundamental questions in physics and materials science. In this context, experimental detection and detailed insights into these exotic quasiparticle excitations by microscopic experimental techniques in complementary time scales are highly essential. Suitable strategies to manipulate and control exotic excitations in the topological quantum states of frustrated quantum materials are other uncharted areas to be pursued, as this may aid in the design and growth of novel quantum materials with enhanced functionalities to address challenges in computing, energy, and storage. Furthermore, insights into the role of external stimuli such as pressure and magnetic fields are crucial to realizing emergent quantum phenomena in the vicinity of a quantum critical point that go beyond the standard paradigm. The exploration of topological spin freezing is just beginning, and a growing eagerness to uncover the potential discoveries that lie ahead, promising to unravel numerous mysteries across various fields. Moreover, the implications of our work extend beyond the realm of fundamental physics. The suitable choice of a model frustrated magnet is crucial for achieving topological spin freezing with exotic low-energy quasi-particle excitations. However, unavoidable disorder and defects in real materials pose a strong constraint. Frustrated spin glasses are known for their complex and highly degenerate energy landscape with a large number of metastable states, which makes it challenging to control, predict, manipulate, and exploit their magnetic behavior. In this vein, efficient routes to design frustrated spin glass materials are desired wherein one can control frustration, disorder, exchange couplings, and anisotropies for the faithful realization of topological spin freezing and associated excitations, leading to ultimate functionalities as per demand. The insights gained from studying topological spin freezing may aid in the design and development of promising quantum materials for next-generation technologies in spintronics and computing Given the prevalence of complex energy landscapes in frustrated magnets, complementary techniques aimed at mapping the complete topography of frustrated magnets, ideally single crystals and heterostructures, are highly essential for the unambiguous identification of exotic low energy states in the topological spin glass state. In addition, advanced theoretical efforts are required to elucidate the complex spin freezing mechanism in frustrated magnets and to guide experiments, and vice versa. This approach may delineate generic glass features in a large class of frustrated quantum magnets from conventional spin glass materials, which possibly shed deep insights into the complex hierarchy of states in frustrated quantum materials. As we move towards a future where quantum technologies become increasingly prevalent, the ability to control and manipulate topological states will be crucial. Frustrated spin glass represents an ideal class of material for its fundamental appeal and potential technological applications.  Frustrated magnets offer a paradigmatic model for verifying complex disorder, emergence, and topological order in quantum materials. In view of the experimental constraints in developing structurally perfect ideal frustrated materials,  efforts may be directed towards the deployment of quantum simulators, which are synthetic lattices made of cold atoms, qubits, or nanomagnets, for the best realization of topological spin freezing in frustrated quantum magnets and exotic low-energy excitations.  This approach will validate long-standing theoretical predictions regarding the quantum behavior of spin glasses and the presence of exotic states such as quantum spin liquids~\cite{semeghini2021probing,samajdar2021quantum,wang2006artificial}.

From a technological perspective, the inherent complexity of spin glasses in frustrated quantum magnets represents a fundamental characteristic. The highly degenerate, disordered landscape of metastable states of frustrated spin glass materials offers a viable ground for neuromorphic computing. Frustrated spin glass networks act like inherent annealers that can solve complex, real-life problems, including protein folding and complex traffic networks efficiently. Frustrated spin glass materials are promising candidates to host high exchange bias, the topological Hall effect, anisotropic magnetoresistance, and topological spin textures, such as skyrmions, which are key ingredients for high-speed and high-density spintronic devices. Notably, exotic characteristic features,  including topological order, complex energy landscape, quantum memory, and highly entangled states of topological spin glass materials, are proposed to provide a basis for robust quantum information processing technology~\cite{placke2024topological}. As Philip W. Anderson famously quoted~\cite{anderson1988spin}, “The history of spin glass may be the best example I know of the dictum that a real scientific mystery is worth pursuing to the ends of the Earth for its own sake, independently of any obvious practical importance or intellectual glamour.” This perspective continues to resonate today: spin glasses remain a fertile arena where subtle quantum mechanical effects, once thought esoteric, can guide the design of materials and devices for quantum technologies yet to be imagined.

\section*{Acknowledgement}
P.K. acknowledges the funding by the Anusandhan National Research Foundation (ANRF), Department of Science and Technology, India through Research Grants.

\bibliography{topo}

@article{PhysRevLett.64.2070,
  title = {{Strong frustration and dilution-enhanced order in a quasi-2D spin glass}},
  author = {Ramirez, A. P. and Espinosa, G. P. and Cooper, A. S.},
  journal = {Physical Review Letters},
  volume = {64},
  issue = {17},
  pages = {2070-2073},
  numpages = {0},
  year = {1990},
  month = {Apr},
  publisher = {American Physical Society},
  doi = {10.1103/PhysRevLett.64.2070},
  url = {https://link.aps.org/doi/10.1103/PhysRevLett.64.2070}
}

@article{PhysRevLett.86.894,
  title = {{Geometric Magnetic Frustration in ${\mathrm{Ba}}_{2}{\mathrm{Sn}}_{2}{\mathrm{Ga}}_{3}{\mathrm{ZnCr}}_{7}{O}_{22}$: A Two-Dimensional Spinel Based Kagom\'e Lattice}},
  author = {Hagemann, I. S. and Huang, Q. and Gao, X. P. A. and Ramirez, A. P. and Cava, R. J.},
  journal = {Physical Review Letters},
  volume = {86},
  issue = {5},
  pages = {894-897},
  numpages = {0},
  year = {2001},
  month = {Jan},
  publisher = {American Physical Society},
  doi = {10.1103/PhysRevLett.86.894},
  url = {https://link.aps.org/doi/10.1103/PhysRevLett.86.894}
}

@article{PhysRev.79.357,
  title = {{Antiferromagnetism. The Triangular Ising Net}},
  author = {Wannier, G. H.},
  journal = {Physical Review},
  volume = {79},
  issue = {2},
  pages = {357-364},
  numpages = {0},
  year = {1950},
  month = {Jul},
  publisher = {American Physical Society},
  doi = {10.1103/PhysRev.79.357},
  url = {https://link.aps.org/doi/10.1103/PhysRev.79.357}
}

@article{PhysRevB.47.15342,
  title = {Spin folding in the two-dimensional Heisenberg kagom\'e antiferromagnet},
  author = {Ritchey, I. and Chandra, P. and Coleman, P.},
  journal = {Physical Review B},
  volume = {47},
  issue = {22},
  pages = {15342-15345},
  numpages = {0},
  year = {1993},
  month = {Jun},
  publisher = {American Physical Society},
  doi = {10.1103/PhysRevB.47.15342},
  url = {https://link.aps.org/doi/10.1103/PhysRevB.47.15342}
}

@article{VILLAIN1980105,
title = {{Dynamics of spin glasses}},
journal = {Journal of Magnetism and Magnetic Materials},
volume = {15-18},
pages = {105-107},
year = {1980},
issn = {0304-8853},
doi = {10.1016/0304-8853(80)90971-3},
url = {https://www.sciencedirect.com/science/article/pii/0304885380909713},
author = {J. Villain}
}

@article{holleis2021anomalous,
  title={{Anomalous and anisotropic nonlinear susceptibility in the proximate Kitaev magnet $\alpha$-RuCl$_3$}},
  author={Holleis, Ludwig and Prestigiacomo, Joseph C and Fan, Zhijie and Nishimoto, Satoshi and Osofsky, Michael and Chern, Gia-Wei and van den Brink, Jeroen and Shivaram, BS},
doi={https://doi.org/10.1038/s41535-021-00364-z},
url= {https://www.nature.com/articles/s41535-021-00364-z#further-reading},
  journal={npj Quantum Materials},
  volume={6},
  number={1},
  pages={66},
  year={2021},
  publisher={Nature Publishing Group UK London}
  
}

@article{PhysRevB.65.144413,
  title = {{Spin-glass behavior in the $S=1/2$ fcc ordered perovskite ${\mathrm{Sr}}_{2}{\mathrm{CaReO}}_{6}$}},
  author = {Wiebe, C. R. and Greedan, J. E. and Luke, G. M. and Gardner, J. S.},
  journal = {Physical Review B},
  volume = {65},
  issue = {14},
  pages = {144413},
  numpages = {9},
  year = {2002},
  month = {Mar},
  publisher = {American Physical Society},
  doi = {10.1103/PhysRevB.65.144413},
  url = {https://link.aps.org/doi/10.1103/PhysRevB.65.144413}
}

@article{PhysRevD.15.2929,
  title = {Fate of the false vacuum: Semiclassical theory},
  author = {Coleman, Sidney},
  journal = {Physical Review D},
  volume = {15},
  issue = {10},
  pages = {2929-2936},
  numpages = {0},
  year = {1977},
  month = {May},
  publisher = {American Physical Society},
  doi = {10.1103/PhysRevD.15.2929},
  url = {https://link.aps.org/doi/10.1103/PhysRevD.15.2929}
}

@article{de2021glassy,
  title={Glassy gravity},
  author={De Giuli, Eric and Zee, A},
  journal={Europhysics Letters},
  volume={133},
  number={2},
  pages={20008},
  year={2021},
  publisher={IOP Publishing},
doi = {10.1209/0295-5075/133/20008},
url = {https://iopscience.iop.org/article/10.1209/0295-5075/133/20008/meta?casa_token=3EJAKpfyYccAAAAA:SLw2nTdwVanmzmPp0Kukk4qVBYR-tI0NQHPPtX7dA6r1kE6OvQCQHO4UJ4XlrvtMGIseg2LjovRd93FsuLN08VNYBDCH](https://iopscience.iop.org/article/10.1209/0295-5075/133/20008/meta?casa_token=3EJAKpfyYccAAAAA:SLw2nTdwVanmzmPp0Kukk4qVBYR-tI0NQHPPtX7dA6r1kE6OvQCQHO4UJ4XlrvtMGIseg2LjovRd93FsuLN08VNYBDCH}

}

@article{lazicki2021metastability,
  title={Metastability of diamond ramp-compressed to 2 terapascals},
  author={Lazicki, Amy and McGonegle, David and Rygg, JR and Braun, DG and Swift, DC and Gorman, MG and Smith, RF and Heighway, PG and Higginbotham, Andrew and Suggit, Matthew J and others},
  journal={Nature},
  volume={589},
  number={7843},
  pages={532-535},
  year={2021},
  publisher={Nature Publishing Group UK London},
doi = {https://doi.org/10.1038/s41586-020-03140-4},
url = {https://www.nature.com/articles/s41586-020-03140-4#citeas}
}

@article{hawking1984cosmological,
  title={The cosmological constant is probably zero},
  author={Hawking, Stephen William},
  journal={Physics Letters B},
  volume={134},
  number={6},
  pages={403-404},
  year={1984},
  publisher={Elsevier},
doi = {https://doi.org/10.1016/0370-2693(84)91370-4},
url = {https://www.sciencedirect.com/science/article/abs/pii/0370269384913704}
}

@article{PhysRevD.16.1248,
  title = {Erratum: Fate of the false vacuum: semiclassical theory},
  author = {Coleman, Sidney},
  journal = {Physical Review D},
  volume = {16},
  issue = {4},
  pages = {1248-1248},
  numpages = {0},
  year = {1977},
  month = {Aug},
  publisher = {American Physical Society},
  doi = {10.1103/PhysRevD.16.1248},
  url = {https://link.aps.org/doi/10.1103/PhysRevD.16.1248}
}

@article{nakatsuji2005spin,
  title={{Spin Disorder on a Triangular Lattice}},
  author={Nakatsuji, Satoru and Nambu, Yusuke and Tonomura, Hiroshi and Sakai, Osamu and Jonas, Seth and Broholm, Collin and Tsunetsugu, Hirokazu and Qiu, Yiming and Maeno, Yoshiteru},
  journal={Science},
  volume={309},
  number={5741},
  pages={1697-1700},
  year={2005},
  publisher={American Association for the Advancement of Science},
doi = {10.1126/science.1114727},
url = {https://www.science.org/doi/full/10.1126/science.1114727}
}

@article{PhysRevLett.99.137207,
  title = {{Spin-Liquid State in the $S=1/2$ Hyperkagome Antiferromagnet ${\mathrm{Na}}_{4}{\mathrm{Ir}}_{3}{\mathrm{O}}_{8}$}},
  author = {Okamoto, Yoshihiko and Nohara, Minoru and Aruga-Katori, Hiroko and Takagi, Hidenori},
  journal = {Physical Review Letters},
  volume = {99},
  issue = {13},
  pages = {137207},
  numpages = {4},
  year = {2007},
  month = {Sep},
  publisher = {American Physical Society},
  doi = {10.1103/PhysRevLett.99.137207},
  url = {https://link.aps.org/doi/10.1103/PhysRevLett.99.137207}
}

@article{ramirez1990strong,
  title={{Strong frustration and dilution-enhanced order in a quasi-2D spin glass}},
  author={Ramirez, AP and Espinosa, GP and Cooper, AS},
  journal={Physical Review Letters},
  volume={64},
  number={17},
  pages={2070},
  year={1990},
  publisher={APS},
doi = {10.1103/PhysRevLett.64.2070},
  url = {https://link.aps.org/doi/10.1103/PhysRevLett.64.2070}
}

@article{PhysRevB.88.174415,
  title = {{Experimental evidence of a collinear antiferromagnetic ordering in the frustrated CoAl${}_{2}$O${}_{4}$ spinel}},
  author = {Roy, B. and Pandey, Abhishek and Zhang, Q. and Heitmann, T. W. and Vaknin, D. and Johnston, D. C. and Furukawa, Y.},
  journal = {Physical Review B},
  volume = {88},
  issue = {17},
  pages = {174415},
  numpages = {12},
  year = {2013},
  month = {Nov},
  publisher = {American Physical Society},
  doi = {10.1103/PhysRevB.88.174415},
  url = {https://link.aps.org/doi/10.1103/PhysRevB.88.174415}
}

@article{khatua2023experimental,
  title={Experimental signatures of quantum and topological states in frustrated magnetism},
  author={Khatua, J and Sana, B and Zorko, A and Gomil{\v{s}}ek, M and Sethupathi, K and Rao, MS Ramachandra and Baenitz, M and Schmidt, B and Khuntia, P},
  journal={Physics Reports},
  volume={1041},
  pages={1-60},
  year={2023},
  publisher={Elsevier},
doi = {https://doi.org/10.1016/j.physrep.2023.09.008},
url = {https://www.sciencedirect.com/science/article/pii/S037015732300306X},
}

@article{khuntia2020gapless,
  title={{Gapless ground state in the archetypal quantum kagome antiferromagnet ZnCu$_3$(OH)$_6$Cl$_2$}},
  author={Khuntia, P and Velazquez, Matias and Barth{\'e}lemy, Quentin and Bert, Fabrice and Kermarrec, Edwin and Legros, A and Bernu, Bernard and Messio, L and Zorko, Andrej and Mendels, P},
  journal={Nature Physics},
  volume={16},
  number={4},
  pages={469-474},
  year={2020},
  publisher={Nature Publishing Group UK London},
doi = {https://doi.org/10.1038/s41567-020-0792-1},
url = {https://www.nature.com/articles/s41567-020-0792-1#citeas}
}

@article{PhysRevB.16.2154,
  title = {Hydrodynamic theory of spin waves in spin glasses and other systems with noncollinear spin orientations},
  author = {Halperin, B. I. and Saslow, W. M.},
  journal = {Physical Review B},
  volume = {16},
  issue = {5},
  pages = {2154-2162},
  numpages = {0},
  year = {1977},
  month = {Sep},
  publisher = {American Physical Society},
  doi = {10.1103/PhysRevB.16.2154},
  url = {https://link.aps.org/doi/10.1103/PhysRevB.16.2154}
}

@article{PhysRev.188.898,
  title = {{Hydrodynamic Theory of Spin Waves}},
  author = {Halperin, B. I. and Hohenberg, P. C.},
  journal = {Physical Review},
  volume = {188},
  issue = {2},
  pages = {898-918},
  numpages = {0},
  year = {1969},
  month = {Dec},
  publisher = {American Physical Society},
  doi = {10.1103/PhysRev.188.898},
  url = {https://link.aps.org/doi/10.1103/PhysRev.188.898}
}

@article{okamoto2007spin,
  title={{Spin-Liquid State in the S= 1/2 Hyperkagome Antiferromagnet Na$_4$Ir$_3$O$_8$}},
  author={Okamoto, Yoshihiko and Nohara, Minoru and Aruga-Katori, Hiroko and Takagi, Hidenori},
  journal={Physical Review Letters},
  volume={99},
  number={13},
  pages={137207},
  year={2007},
  publisher={APS},
doi = {10.1103/PhysRevLett.99.137207},
  url = {https://link.aps.org/doi/10.1103/PhysRevLett.99.137207}
}

@article{kawamura1985phase,
  title={Phase transition of the Heisenberg antiferromagnet on the triangular lattice in a magnetic field},
  author={Kawamura, Hikaru and Miyashita, Seiji},
  journal={Journal of the Physical Society of Japan},
  volume={54},
  number={12},
  pages={4530-4538},
  year={1985},
  publisher={The Physical Society of Japan},
doi = {https://doi.org/10.1143/JPSJ.54.4530},
url = {https://journals.jps.jp/doi/10.1143/JPSJ.54.4530}
}

@book{weinberg1995quantum,
  title={The quantum theory of fields},
  author={Weinberg, Steven},
  volume={2},
  year={1995},
  publisher={Cambridge university press}
}

@article{PhysRevLett.29.1698,
  title = {{Approximate Symmetries and Pseudo-Goldstone Bosons}},
  author = {Weinberg, Steven},
  journal = {Physical Review Letters},
  volume = {29},
  issue = {25},
  pages = {1698-1701},
  numpages = {0},
  year = {1972},
  month = {Dec},
  publisher = {American Physical Society},
  doi = {10.1103/PhysRevLett.29.1698},
  url = {https://link.aps.org/doi/10.1103/PhysRevLett.29.1698}
}

@article{PhysRevLett.128.141601,
  title = {{Damping of Pseudo-Goldstone Fields}},
  author = {Delacr\'etaz, Luca V. and Gout\'eraux, Blaise and Ziogas, Vaios},
  journal = {Physical Review Letters},
  volume = {128},
  issue = {14},
  pages = {141601},
  numpages = {8},
  year = {2022},
  month = {Apr},
  publisher = {American Physical Society},
  doi = {10.1103/PhysRevLett.128.141601},
  url = {https://link.aps.org/doi/10.1103/PhysRevLett.128.141601}
}

@article{PhysRevLett.121.237201,
  title = {{Pseudo-Goldstone Gaps and Order-by-Quantum Disorder in Frustrated Magnets}},
  author = {Rau, Jeffrey G. and McClarty, Paul A. and Moessner, Roderich},
  journal = {Physical Review Letters},
  volume = {121},
  issue = {23},
  pages = {237201},
  numpages = {7},
  year = {2018},
  month = {Dec},
  publisher = {American Physical Society},
  doi = {10.1103/PhysRevLett.121.237201},
  url = {https://link.aps.org/doi/10.1103/PhysRevLett.121.237201}
}

@article{PhysRevResearch.2.043023,
  title = {{Emergence of nematic paramagnet via quantum order-by-disorder and pseudo-Goldstone modes in Kitaev magnets}},
  author = {Gohlke, Matthias and Chern, Li Ern and Kee, Hae-Young and Kim, Yong Baek},
  journal = {Physical Review Research},
  volume = {2},
  issue = {4},
  pages = {043023},
  numpages = {13},
  year = {2020},
  month = {Oct},
  publisher = {American Physical Society},
  doi = {10.1103/PhysRevResearch.2.043023},
  url = {https://link.aps.org/doi/10.1103/PhysRevResearch.2.043023}
}

@article{PhysRevLett.4.380,
  title = {{Axial Vector Current Conservation in Weak Interactions}},
  author = {Nambu, Yoichiro},
  journal = {Physical Review Letters},
  volume = {4},
  issue = {7},
  pages = {380-382},
  numpages = {0},
  year = {1960},
  month = {Apr},
  publisher = {American Physical Society},
  doi = {10.1103/PhysRevLett.4.380},
  url = {https://link.aps.org/doi/10.1103/PhysRevLett.4.380}
}

@article{PhysRevB.77.054429,
  title = {{Spin dynamics and spin freezing behavior in the two-dimensional antiferromagnet $\mathrm{Ni}{\mathrm{Ga}}_{2}{\mathrm{S}}_{4}$ revealed by Ga-NMR, NQR and $\ensuremath{\mu}\mathrm{SR}$ measurements}},
  author = {Takeya, Hideo and Ishida, Kenji and Kitagawa, Kentaro and Ihara, Yoshihiko and Onuma, Keisuke and Maeno, Yoshiteru and Nambu, Yusuke and Nakatsuji, Satoru and MacLaughlin, Douglas E. and Koda, Akihiko and Kadono, Ryosuke},
  journal = {Physical Review B},
  volume = {77},
  issue = {5},
  pages = {054429},
  numpages = {13},
  year = {2008},
  month = {Feb},
  publisher = {American Physical Society},
  doi = {10.1103/PhysRevB.77.054429},
  url = {https://link.aps.org/doi/10.1103/PhysRevB.77.054429}
}

@article{PhysRevLett.124.087201,
  title = {{Spin-Orbital Glass Transition in a Model of a Frustrated Pyrochlore Magnet without Quenched Disorder}},
  author = {Mitsumoto, Kota and Hotta, Chisa and Yoshino, Hajime},
  journal = {Physical Review Letters},
  volume = {124},
  issue = {8},
  pages = {087201},
  numpages = {6},
  year = {2020},
  month = {Feb},
  publisher = {American Physical Society},
  doi = {10.1103/PhysRevLett.124.087201},
  url = {https://link.aps.org/doi/10.1103/PhysRevLett.124.087201}
}

@article{PhysRevLett.118.067201,
  title = {{Orbital Dimer Model for the Spin-Glass State in ${\mathrm{Y}}_{2}{\mathrm{Mo}}_{2}{\mathrm{O}}_{7}$}},
  author = {Thygesen, Peter M. M. and Paddison, Joseph A. M. and Zhang, Ronghuan and Beyer, Kevin A. and Chapman, Karena W. and Playford, Helen Y. and Tucker, Matthew G. and Keen, David A. and Hayward, Michael A. and Goodwin, Andrew L.},
  journal = {Physical Review Letters},
  volume = {118},
  issue = {6},
  pages = {067201},
  numpages = {6},
  year = {2017},
  month = {Feb},
  publisher = {American Physical Society},
  doi = {10.1103/PhysRevLett.118.067201},
  url = {https://link.aps.org/doi/10.1103/PhysRevLett.118.067201}
}

@article{PhysRevLett.84.2957,
  title = {{Entropy Balance and Evidence for Local Spin Singlets in a Kagom\'e-Like Magnet}},
  author = {Ramirez, A. P. and Hessen, B. and Winklemann, M.},
  journal = {Physical Review Letters},
  volume = {84},
  issue = {13},
  pages = {2957-2960},
  numpages = {0},
  year = {2000},
  month = {Mar},
  publisher = {American Physical Society},
  doi = {10.1103/PhysRevLett.84.2957},
  url = {https://link.aps.org/doi/10.1103/PhysRevLett.84.2957}
}

@article{PhysRevLett.127.157204,
  title = {{Magnetic Field Induced Quantum Spin Liquid in the Two Coupled Trillium Lattices of ${\mathrm{K}}_{2}{\mathrm{Ni}}_{2}({\mathrm{SO}}_{4}{)}_{3}$}},
  author = {\ifmmode \check{Z}\else \v{Z}\fi{}ivkovi\ifmmode \acute{c}\else \'{c}\fi{}, Ivica and Favre, Virgile and Salazar Mejia, Catalina and Jeschke, Harald O. and Magrez, Arnaud and Dabholkar, Bhupen and Noculak, Vincent and Freitas, Rafael S. and Jeong, Minki and Hegde, Nagabhushan G. and Testa, Luc and Babkevich, Peter and Su, Yixi and Manuel, Pascal and Luetkens, Hubertus and Baines, Christopher and Baker, Peter J. and Wosnitza, Jochen and Zaharko, Oksana and Iqbal, Yasir and Reuther, Johannes and R\o{}nnow, Henrik M.},
  journal = {Physical Review Letters},
  volume = {127},
  issue = {15},
  pages = {157204},
  numpages = {7},
  year = {2021},
  month = {Oct},
  publisher = {American Physical Society},
  doi = {10.1103/PhysRevLett.127.157204},
  url = {https://link.aps.org/doi/10.1103/PhysRevLett.127.157204}
}

@article{boya2022signatures,
  title={{Signatures of spin-liquid state in a 3D frustrated lattice compound KSrFe2 (PO4) 3 with S= 5/2}},
  author={Boya, K and Nam, K and Kargeti, K and Jain, A and Kumar, R and Panda, SK and Yusuf, SM and Paulose, PL and Voma, UK and Kermarrec, Edwin and others},
  journal= {APL Materials},
  volume={10},
  number={10},
  year={2022},
  publisher={AIP Publishing},
doi = {10.1063/5.0096942},
url = {https://pubs.aip.org/aip/apm/article/10/10/101103/2834990}
}

@article{palmer1982broken,
  title={Broken ergodicity},
  author={Palmer, Richard G},
  journal={Advances in Physics},
  volume={31},
  number={6},
  pages={669-735},
  year={1982},
  publisher={Taylor \& Francis},
doi = {10.1080/00018738200101438},
url = {https://www.tandfonline.com/doi/abs/10.1080/00018738200101438}
}

@article{bergman2007order,
  title={Order-by-disorder and spiral spin-liquid in frustrated diamond-lattice antiferromagnets},
  author={Bergman, Doron and Alicea, Jason and Gull, Emanuel and Trebst, Simon and Balents, Leon},
  journal={Nature Physics},
  volume={3},
  number={7},
  pages={487-491},
  year={2007},
  publisher={Nature Publishing Group UK London},
doi = {10.1038/nphys622},
url = {https://doi.org/10.1038/nphys622}
}

@article{majzlan2004thermodynamic,
  title={{Thermodynamic properties, low-temperature heat-capacity anomalies, and single-crystal X-ray refinement of hydronium jarosite,(H$_3$O)Fe$_3$(SO$_4$)$_2$(OH)$_6$}},
  author={Majzlan, J and Stevens, R and Boerio-Goates, J and Woodfield, BF and Navrotsky, A and Burns, PC and Crawford, MK and Amos, TG},
  journal={Physics and Chemistry of Minerals},
  volume={31},
  pages={518-531},
  year={2004},
  publisher={Springer},
  doi={10.1007/s00269-004-0405-z},
  url={https://link.springer.com/article/10.1007/s00269-004-0405-z}
}

@article{wills1998magnetic,
  title={{Magnetic correlations in deuteronium jarosite, a model S= 5/2 Kagom{\'e} antiferromagnet}},
  author={Wills, AS and Harrison, A and Mentink, SAM and Mason, TE and Tun, Z},
  journal={Europhysics Letters},
  volume={42},
  number={3},
  pages={325},
  year={1998},
  publisher={IOP Publishing},
  doi={10.1209/epl/i1998-00250-2},
  url={https://iopscience.iop.org/article/10.1209/epl/i1998-00250-2/meta}
}

@article{fujihala2014unconventional,
  title={{Unconventional spin freezing in the highly two-dimensional spin-1 2 kagome antiferromagnet Cd$_2$Cu$_3$(OH)$_6$(SO$_4$)$_2$4H$_2$O: Evidence of partial order and coexisting spin singlet state on a distorted kagome lattice}},
  author={Fujihala, Masayoshi and Zheng, Xu-Guang and Morodomi, Hiroki and Kawae, Tatsuya and Matsuo, Akira and Kindo, Koichi and Watanabe, Isao},
  journal={Physical Review B},
  volume={89},
  number={10},
  pages={100401},
  year={2014},
  publisher={APS},
  doi={10.1103/PhysRevB.89.100401},
  url={https://journals.aps.org/prb/abstract/10.1103/PhysRevB.89.100401}
}

@article{yang2016glassy,
  title={{Glassy Behavior and Isolated Spin Dimers in a New Frustrated Magnet BaCr$_9p$Ga$_{12-9p}$O$_{19}$}},
  author={Yang, Junjie and Samarakoon, Anjana M and Hong, Kyun Woo and Copley, John RD and Huang, Qingzhen and Tennant, Alan and Sato, Taku J and Lee, Seung-Hun},
  journal={Journal of the Physical Society of Japan},
  volume={85},
  number={9},
  pages={094712},
  year={2016},
  publisher={The Physical Society of Japan},
  doi={10.7566/JPSJ.85.094712},
  url={https://journals.jps.jp/doi/full/10.7566/JPSJ.85.094712}
}

@article{hagemann2001geometric,
  title={{Geometric magnetic frustration in Ba$_2$Sn$_2$Ga$_3$ZnCr$_7$O$_{22}$: A two-dimensional spinel based Kagom{\'e} lattice}},
  author={Hagemann, IS and Huang, Q and Gao, XPA and Ramirez, AP and Cava, RJ},
  journal={Physical review letters},
  volume={86},
  number={5},
  pages={894},
  year={2001},
  publisher={APS},
  doi={10.1103/PhysRevLett.86.894},
  url={https://journals.aps.org/prl/abstract/10.1103/PhysRevLett.86.894}
}

@article{shen2022spin,
  title={{Spin freezing in the van der Waals material Mn$_2$Ga$_2$S$_5$}},
  author={Shen, Jie and Xu, Xitong and He, Miao and Liu, Yonglai and Han, Yuyan and Qu, Zhe},
  journal={Chinese Physics B},
  volume={31},
  number={6},
  pages={067105},
  year={2022},
  publisher={IOP Publishing},
  doi={10.1088/1674-1056/ac657c},
  url={https://iopscience.iop.org/article/10.1088/1674-1056/ac657c/meta}
}

@article{nambu2006coherent,
  title={{Coherent behavior and nonmagnetic impurity effects of spin disordered state in NiGa$_2$S$_4$}},
  author={Nambu, Yusuke and Nakatsuji, Satoru and Maeno, Yoshiteru},
  journal={Journal of the Physical Society of Japan},
  volume={75},
  number={4},
  pages={043711},
  year={2006},
  publisher={The Physical Society of Japan},
  doi={10.1143/JPSJ.75.043711},
  url={https://journals.jps.jp/doi/abs/10.1143/JPSJ.75.043711}
}

@article{girctu2000glassiness,
  title={{Glassiness and canted antiferromagnetism in three geometrically frustrated triangular quantum Heisenberg antiferromagnets with additional Dzyaloshinskii-Moriya interaction}},
  author={G{\^\i}r{\c{t}}u, Mihai A and Wynn, Charles M and Fujita, Wataru and Awaga, Kunio and Epstein, Arthur J},
  journal={Physical Review B},
  volume={61},
  number={6},
  pages={4117},
  year={2000},
  publisher={APS},
  doi={10.1103/PhysRevB.61.4117},
  url={https://journals.aps.org/prb/abstract/10.1103/PhysRevB.61.4117}
}

@article{tewari2024signature,
  title={{Signature of elementary excitations in Se-substituted layered triangular-lattice antiferromagnet CuCrS$_2$}},
  author={Tewari, Girish C and Kousar, H Sajida and Srivastava, Divya and Karppinen, Maarit},
  journal={Journal of Magnetism and Magnetic Materials},
  volume={590},
  pages={171688},
  year={2024},
  publisher={Elsevier},
  doi={10.1016/j.jmmm.2023.171688},
  url={https://www.sciencedirect.com/science/article/pii/S0304885323013380}
}

@article{samarakoon2016aging,
  title={Aging, memory, and nonhierarchical energy landscape of spin jam},
  author={Samarakoon, Anjana and Sato, Taku J and Chen, Tianran and Chern, Gai-Wei and Yang, Junjie and Klich, Israel and Sinclair, Ryan and Zhou, Haidong and Lee, Seung-Hun},
  journal={Proceedings of the National Academy of Sciences},
  volume={113},
  number={42},
  pages={11806-11810},
  year={2016},
  publisher={National Acad Sciences},
doi={https://doi.org/10.1073/pnas.1608057113},
url={https://www.pnas.org/doi/abs/10.1073/pnas.1608057113}
}

@article{baskaran2023metastable,
  title={{Metastable Kitaev Spin Liquids in Isotropic Quantum Heisenberg Magnets}},
  author={Baskaran, Ganapathy},
  journal={arXiv preprint arXiv:2309.07119},
  year={2023},
doi={https://doi.org/10.48550/arXiv.2309.07119},
url={https://arxiv.org/abs/2309.07119v1}
}

@article{PhysRev.65.117,
  title = {{Crystal Statistics. I. A Two-Dimensional Model with an Order-Disorder Transition}},
  author = {Onsager, Lars},
  journal = {Physical Review},
  volume = {65},
  issue = {3-4},
  pages = {117-149},
  numpages = {0},
  year = {1944},
  month = {Feb},
  publisher = {American Physical Society},
  doi = {10.1103/PhysRev.65.117},
  url = {https://link.aps.org/doi/10.1103/PhysRev.65.117}
}

@article{zenesini2024false,
  title={False vacuum decay via bubble formation in ferromagnetic superfluids},
  author={Zenesini, Alessandro and Berti, Anna and Cominotti, Riccardo and Rogora, Chiara and Moss, Ian G and Billam, Thomas P and Carusotto, Iacopo and Lamporesi, Giacomo and Recati, Alessio and Ferrari, Gabriele},
  journal={Nature Physics},
  volume={20},
  number={4},
  pages={558-563},
  year={2024},
  publisher={Nature Publishing Group UK London},
doi={https://doi.org/10.1038/s41567-023-02345-4},
url={10.1038/s41567-023-02345-4}
}

@article{PhysRevD.35.1747,
  title = {Dynamics of false-vacuum bubbles},
  author = {Blau, Steven K. and Guendelman, E. I. and Guth, Alan H.},
  journal = {Physical Review D},
  volume = {35},
  issue = {6},
  pages = {1747-1766},
  numpages = {0},
  year = {1987},
  month = {Mar},
  publisher = {American Physical Society},
  doi = {10.1103/PhysRevD.35.1747},
  url = {https://link.aps.org/doi/10.1103/PhysRevD.35.1747}
}

@article{PhysRevD.49.6410,
  title = {{False vacuum inflation with Einstein gravity}},
  author = {Copeland, Edmund J. and Liddle, Andrew R. and Lyth, David H. and Stewart, Ewan D. and Wands, David},
  journal = {Physical Review D},
  volume = {49},
  issue = {12},
  pages = {6410-6433},
  numpages = {0},
  year = {1994},
  month = {Jun},
  publisher = {American Physical Society},
  doi = {10.1103/PhysRevD.49.6410},
  url = {https://link.aps.org/doi/10.1103/PhysRevD.49.6410}
}

@article{bovzovic2021unstable,
  title={An unstable pathway to room temperature superconductivity?},
  author={Bo{\v{z}}ovi{\'c}, Ivan},
  journal={Proceedings of the National Academy of Sciences},
  volume={118},
  number={33},
  pages={e2111471118},
  year={2021},
  publisher={National Acad Sciences},
doi = {10.1073/pnas.2111471118},
url={https://doi.org/10.1073/pnas.2111471118}
}

@article{PhysRevB.96.094432,
  title = {{Local magnetism and spin dynamics of the frustrated honeycomb rhodate ${\mathrm{Li}}_{2}{\mathrm{RhO}}_{3}$}},
  author = {Khuntia, P. and Manni, S. and Foronda, F. R. and Lancaster, T. and Blundell, S. J. and Gegenwart, P. and Baenitz, M.},
  journal = {Physical Review B},
  volume = {96},
  issue = {9},
  pages = {094432},
  numpages = {6},
  year = {2017},
  month = {Sep},
  publisher = {American Physical Society},
  doi = {10.1103/PhysRevB.96.094432},
  url = {https://link.aps.org/doi/10.1103/PhysRevB.96.094432}
}

@article{yang2015spin,
  title={Spin jam induced by quantum fluctuations in a frustrated magnet},
  author={Yang, Junjie and Samarakoon, Anjana and Dissanayake, Sachith and Ueda, Hiroaki and Klich, Israel and Iida, Kazuki and Pajerowski, Daniel and Butch, Nicholas P and Huang, Q and Copley, John RD and others},
  journal={Proceedings of the National Academy of Sciences},
  volume={112},
  number={37},
  pages={11519-11523},
  year={2015},
  publisher={National Acad Sciences},
doi={https://doi.org/10.1073/pnas.1503126112},
url={https://www.pnas.org/doi/abs/10.1073/pnas.1503126112}
}

@article{PhysRevB.89.054433,
  title = {{Liquidlike correlations in single-crystalline Y${}_{2}$Mo${}_{2}$O${}_{7}$: An unconventional spin glass}},
  author = {Silverstein, H. J. and Fritsch, K. and Flicker, F. and Hallas, A. M. and Gardner, J. S. and Qiu, Y. and Ehlers, G. and Savici, A. T. and Yamani, Z. and Ross, K. A. and Gaulin, B. D. and Gingras, M. J. P. and Paddison, J. A. M. and Foyevtsova, K. and Valenti, R. and Hawthorne, F. and Wiebe, C. R. and Zhou, H. D.},
  journal = {Physical Review B},
  volume = {89},
  issue = {5},
  pages = {054433},
  numpages = {16},
  year = {2014},
  month = {Feb},
  publisher = {American Physical Society},
  doi = {10.1103/PhysRevB.89.054433},
  url = {https://link.aps.org/doi/10.1103/PhysRevB.89.054433}
}

@article{PhysRevB.104.L201106,
  title = {False vacuum decay in quantum spin chains},
  author = {Lagnese, Gianluca and Surace, Federica Maria and Kormos, M\'arton and Calabrese, Pasquale},
  journal = {Physical Review B},
  volume = {104},
  issue = {20},
  pages = {L201106},
  numpages = {6},
  year = {2021},
  month = {Nov},
  publisher = {American Physical Society},
  doi = {10.1103/PhysRevB.104.L201106},
  url = {https://link.aps.org/doi/10.1103/PhysRevB.104.L201106}
}

@article{braden2018towards,
  title={Towards the cold atom analog false vacuum},
  author={Braden, Jonathan and Johnson, Matthew C and Peiris, Hiranya V and Weinfurtner, Silke},
  journal={Journal of High Energy Physics},
  volume={2018},
  number={7},
  pages={1-39},
  year={2018},
  publisher={Springer},
doi = {https://doi.org/10.1007/JHEP07%282018%29014},
url={https://link.springer.com/article/10.1007/JHEP07(2018)014}
}

@article{anderson1973resonating,
  title={Resonating valence bonds: A new kind of insulator?},
  author={Anderson, Philip W},
  journal={Materials Research Bulletin},
  volume={8},
  number={2},
  pages={153-160},
  year={1973},
  publisher={Elsevier},
url={https://doi.org/10.1016/0025-5408(73)90167-0},
doi={https://doi.org/10.1016/0025-5408(73)90167-0}
}

@article{balents2010spin,
  title={Spin liquids in frustrated magnets},
  author={Balents, Leon},
  journal={Nature},
  volume={464},
  number={7286},
  pages={199-208},
  year={2010},
  publisher={Nature Publishing Group UK London},
doi={10.1038/nature08917},
url={https://doi.org/10.1038/nature08917}
}

@article{10.1063/1.2186278,
    author = {Moessner, Roderich and Ramirez, Arthur P.},
    title = "{Geometrical frustration}",
    journal = {Physics Today},
    volume = {59},
    number = {2},
    pages = {24-29},
    year = {2006},
    month = {02},
    issn = {0031-9228},
    doi = {10.1063/1.2186278},
    url = {https://doi.org/10.1063/1.2186278},
    
}

@article{KHATUA20231,
title = {Experimental signatures of quantum and topological states in frustrated magnetism},
journal = {Physics Reports},
volume = {1041},
pages = {1-60},
year = {2023},

issn = {0370-1573},
doi = {https://doi.org/10.1016/j.physrep.2023.09.008},
url = {https://www.sciencedirect.com/science/article/pii/S037015732300306X},
author = {J. Khatua and B. Sana and A. Zorko and M. Gomilšek and K. Sethupathi and M.S. Ramachandra Rao and M. Baenitz and B. Schmidt and P. Khuntia}
}

@article{ramirez1999zero,
  title={Zero-point entropy in ‘spin ice’},
  author={Ramirez, Arthur P and Hayashi, A and Cava, Robert Joseph and Siddharthan, R and Shastry, BS},
  journal={Nature},
  volume={399},
  number={6734},
  pages={333-335},
  year={1999},
  publisher={Nature Publishing Group UK London},
url={https://doi.org/10.1038/20619},
doi={10.1038/20619}
}

@article{snyder2001spin,
  title={How ‘spin ice’ freezes},
  author={Snyder, J and Slusky, JS and Cava, RJ and Schiffer, P},
  journal={Nature},
  volume={413},
  number={6851},
  pages={48-51},
  year={2001},
  publisher={Nature Publishing Group UK London},
url={https://doi.org/10.1038/35092516},
doi={10.1038/35092516}
}

@article{RevModPhys.58.801,
  title = {Spin glasses: Experimental facts, theoretical concepts, and open questions},
  author = {Binder, K. and Young, A. P.},
  journal = {Reviews of Modern Physics},
  volume = {58},
  issue = {4},
  pages = {801-976},
  numpages = {0},
  year = {1986},
  month = {Oct},
  publisher = {American Physical Society},
  doi = {10.1103/RevModPhys.58.801},
  url = {https://link.aps.org/doi/10.1103/RevModPhys.58.801}
}

@book{mydosh1993spin,
  title={Spin glasses: an experimental introduction},
  author={Mydosh, John A},
  year={1993},
  publisher={CRC Press},
url = {https://doi.org/10.1201/9781482295191}, 
doi={10.1201/9781482295191}
}

@article{RevModPhys.56.755,
  title = {Heavy-fermion systems},
  author = {Stewart, G. R.},
  journal = {Reviews of Modern Physics},
  volume = {56},
  issue = {4},
  pages = {755-787},
  numpages = {0},
  year = {1984},
  month = {Oct},
  publisher = {American Physical Society},
  doi = {10.1103/RevModPhys.56.755},
  url = {https://link.aps.org/doi/10.1103/RevModPhys.56.755}
}

@article{PhysRevLett.58.2790,
  title = {{Resonating-valence-bond theory of phase transitions and superconductivity in ${\mathrm{La}}_{2}$${\mathrm{CuO}}_{4}$-based compounds}},
  author = {Anderson, P. W. and Baskaran, G. and Zou, Z. and Hsu, T.},
  journal = {Physical Review Letters},
  volume = {58},
  issue = {26},
  pages = {2790-2793},
  numpages = {0},
  year = {1987},
  month = {Jun},
  publisher = {American Physical Society},
  doi = {10.1103/PhysRevLett.58.2790},
  url = {https://link.aps.org/doi/10.1103/PhysRevLett.58.2790}
}

@article{edwards1975theory,
  title={Theory of spin glasses},
  author={Edwards, Samuel Frederick and Anderson, Phil W},
  journal={Journal of Physics F: Metal Physics},
  volume={5},
  number={5},
  pages={965},
  year={1975},
  publisher={IOP Publishing},
url={https://dx.doi.org/10.1088/0305-4608/5/5/017},
doi={10.1088/0305-4608/5/5/017}
}

@article{PhysRevB.79.140402,
  title = {{Halperin-Saslow modes as the origin of the low-temperature anomaly in ${\text{NiGa}}_{2}{\text{S}}_{4}$}},
  author = {Podolsky, Daniel and Kim, Yong Baek},
  journal = {Physical Review B},
  volume = {79},
  issue = {14},
  pages = {140402},
  numpages = {4},
  year = {2009},
  month = {Apr},
  publisher = {American Physical Society},
  doi = {10.1103/PhysRevB.79.140402},
  url = {https://link.aps.org/doi/10.1103/PhysRevB.79.140402}
}

@book{stein2013spin,
  title={Spin glasses and complexity},
  author={Stein, Daniel L and Newman, Charles M},
  volume={4},
  year={2013},
  publisher={Princeton University Press}
}

@article{anderson1988spin,
  title={Spin glass i: A scaling law rescued},
  author={Anderson, Philip W},
  journal={Physics Today},
  volume={41},
  number={1},
  pages={9-11},
  year={1988},
  publisher={American Institute of Physics}
}

@article{PhysRevB.64.094436,
  title = {{Short-range order in the topological spin glass $({\mathrm{D}}_{3}\mathrm{O}){\mathrm{Fe}}_{3}({\mathrm{SO}}_{4}{)}_{2}(\mathrm{OD}{)}_{6}$ using polarized neutron diffraction}},
  author = {Wills, A. S. and Oakley, G. S. and Visser, D. and Frunzke, J. and Harrison, A. and Andersen, K. H.},
  journal = {Physical Review B},
  volume = {64},
  issue = {9},
  pages = {094436},
  numpages = {8},
  year = {2001},
  month = {Aug},
  publisher = {American Physical Society},
  doi = {10.1103/PhysRevB.64.094436},
  url = {https://link.aps.org/doi/10.1103/PhysRevB.64.094436}
}

@article{chandra1993anisotropic,
  title={The anisotropic kagome antiferromagnet: a topological spin glass?},
  author={Chandra, P and Coleman, P and Ritchey, I},
  journal={Journal de Physique I},
  volume={3},
  number={2},
  pages={591-610},
  year={1993},
  publisher={EDP Sciences},
url={https://doi.org/10.1051/jp1:1993104},
doi={10.1051/jp1:1993104}
}

@article{PhysRevB.101.024413,
  title = {Goldstone modes in the emergent gauge fields of a frustrated magnet},
  author = {Garratt, S. J. and Chalker, J. T.},
  journal = {Physical Review B},
  volume = {101},
  issue = {2},
  pages = {024413},
  numpages = {6},
  year = {2020},
  month = {Jan},
  publisher = {American Physical Society},
  doi = {10.1103/PhysRevB.101.024413},
  url = {https://link.aps.org/doi/10.1103/PhysRevB.101.024413}
}

@article{PhysRevLett.127.017201,
  title = {Freezing of a Disorder Induced Spin Liquid with Strong Quantum Fluctuations},
  author = {Hu, Xiao and Pajerowski, Daniel M. and Zhang, Depei and Podlesnyak, Andrey A. and Qiu, Yiming and Huang, Qing and Zhou, Haidong and Klich, Israel and Kolesnikov, Alexander I. and Stone, Matthew B. and Lee, Seung-Hun},
  journal = {Physical Review Letters},
  volume = {127},
  issue = {1},
  pages = {017201},
  numpages = {6},
  year = {2021},
  month = {Jun},
  publisher = {American Physical Society},
  doi = {10.1103/PhysRevLett.127.017201},
  url = {https://link.aps.org/doi/10.1103/PhysRevLett.127.017201}
}

@article{el2010electron,
  title={Electron spin resonance in S= 1 2 antiferromagnets at high temperature},
  author={El Shawish, S and Cepas, Olivier and Miyashita, S},
  journal={Physical Review B—Condensed Matter and Materials Physics},
  volume={81},
  number={22},
  pages={224421},
  year={2010},
  publisher={APS},
url={https://journals.aps.org/prb/abstract/10.1103/PhysRevB.81.224421},
doi={10.1103/PhysRevB.81.224421}
}

@article{zorko2008dzyaloshinsky,
  title={Dzyaloshinsky-Moriya anisotropy in the spin-1/2 kagome compound ZnCu 3 (OH) 6 Cl 2},
  author={Zorko, Andei and Nellutla, S and Van Tol, J and Brunel, LC and Bert, F and Duc, F and Trombe, J-C and De Vries, MA and Harrison, <? format?> A and Mendels, P},
  journal={Physical review letters},
  volume={101},
  number={2},
  pages={026405},
  year={2008},
  publisher={APS},
url={https://journals.aps.org/prl/abstract/10.1103/PhysRevLett.101.026405},
doi={10.1103/PhysRevLett.101.026405}
}

@article{sahasrabudhe2024chiral,
  title={Chiral excitations and the intermediate-field regime in the Kitaev magnet $\alpha$- RuCl 3},
  author={Sahasrabudhe, Anuja and Prosnikov, Mikhail A and Koethe, Thomas C and Stein, Philipp and Tsurkan, Vladimir and Loidl, Alois and Gr{\"u}ninger, Markus and Hedayat, Hamoon and van Loosdrecht, Paul HM},
  journal={Physical Review Research},
  volume={6},
  number={2},
  pages={L022005},
  year={2024},
  publisher={APS},
url={https://journals.aps.org/prresearch/abstract/10.1103/PhysRevResearch.6.L022005},
doi={10.1103/PhysRevResearch.6.L022005}
}

@article{sahasrabudhe2020high,
  title={High-field quantum disordered state in $\alpha$-RuCl 3: Spin flips, bound states, and multiparticle continuum},
  author={Sahasrabudhe, Anuja and Kaib, DAS and Reschke, Stephan and German, R and Koethe, TC and Buhot, J and Kamenskyi, Dmytro and Hickey, C and Becker, P and Tsurkan, Vladimir and others},
  journal={Physical Review B},
  volume={101},
  number={14},
  pages={140410},
  year={2020},
  publisher={APS},
url={https://journals.aps.org/prb/abstract/10.1103/PhysRevB.101.140410},
dio={10.1103/PhysRevB.101.140410}
}

@article{PhysRevLett.80.2929,
  title = {{Properties of a Classical Spin Liquid: The Heisenberg Pyrochlore Antiferromagnet}},
  author = {Moessner, R. and Chalker, J. T.},
  journal = {Physical Review Letters},
  volume = {80},
  issue = {13},
  pages = {2929-2932},
  numpages = {0},
  year = {1998},
  month = {Mar},
  publisher = {American Physical Society},
  doi = {10.1103/PhysRevLett.80.2929},
  url = {https://link.aps.org/doi/10.1103/PhysRevLett.80.2929}
}

@article{samarakoon2017scaling,
  title={{Scaling of Memories and Crossover in Glassy Magnets}},
  author={Samarakoon, AM and Takahashi, M and Zhang, D and Yang, J and Katayama, N and Sinclair, R and Zhou, HD and Diallo, SO and Ehlers, G and Tennant, DA and others},
  journal={Scientific reports},
  volume={7},
  number={1},
  pages={12053},
  year={2017},
  publisher={Nature Publishing Group UK London},
url={https://doi.org/10.1038/s41598-017-12187-9},
doi={10.1038/s41598-017-12187-9}
}

@article{klich2014glassiness,
  title={Glassiness and exotic entropy scaling induced by quantum fluctuations in a disorder-free frustrated magnet},
  author={Klich, I and Lee, S-H and Iida, K},
  journal={Nature communications},
  volume={5},
  number={1},
  pages={3497},
  year={2014},
  publisher={Nature Publishing Group UK London},
url={https://doi.org/10.1038/ncomms4497},
doi={10.1038/ncomms4497}
}

@article{syzranov2022eminuscent,
  title={Eminuscent phase in frustrated magnets: a challenge to quantum spin liquids},
  author={Syzranov, SV and Ramirez, AP},
  journal={Nature Communications},
  volume={13},
  number={1},
  pages={2993},
  year={2022},
  publisher={Nature Publishing Group UK London},
url={https://doi.org/10.1038/s41467-022-30739-0},
doi={10.1038/s41467-022-30739-0}
}

@article{fischer1980ferromagnetic,
  title={Ferromagnetic modes in spin glasses and dilute ferromagnets},
  author={Fischer, KH},
  journal={Zeitschrift f{\"u}r Physik B Condensed Matter},
  volume={39},
  number={1},
  pages={37-46},
  year={1980},
  publisher={Springer},
url={https://link.springer.com/article/10.1007/BF01292636},
doi={10.1007/BF01292636}
}

@article{PhysRevLett.117.237203,
  title = {{Magnetic Excitations and Electronic Interactions in ${\mathrm{Sr}}_{2}{\mathrm{CuTeO}}_{6}$: A Spin-$1/2$ Square Lattice Heisenberg Antiferromagnet}},
  author = {Babkevich, P. and Katukuri, Vamshi M. and F\aa{}k, B. and Rols, S. and Fennell, T. and Paji\ifmmode \acute{c}\else \'{c}\fi{}, D. and Tanaka, H. and Pardini, T. and Singh, R. R. P. and Mitrushchenkov, A. and Yazyev, O. V. and R\o{}nnow, H. M.},
  journal = {Physical Review Letters},
  volume = {117},
  issue = {23},
  pages = {237203},
  numpages = {6},
  year = {2016},
  month = {Dec},
  publisher = {American Physical Society},
  doi = {10.1103/PhysRevLett.117.237203},
  url = {https://link.aps.org/doi/10.1103/PhysRevLett.117.237203}
}

@article{PhysRevB.94.064411,
  title = {{Spin wave excitations in the tetragonal double perovskite ${\mathrm{Sr}}_{2}{\mathrm{CuWO}}_{6}$}},
  author = {Walker, H. C. and Mustonen, O. and Vasala, S. and Voneshen, D. J. and Le, M. D. and Adroja, D. T. and Karppinen, M.},
  journal = {Physical Review B},
  volume = {94},
  issue = {6},
  pages = {064411},
  numpages = {8},
  year = {2016},
  month = {Aug},
  publisher = {American Physical Society},
  doi = {10.1103/PhysRevB.94.064411},
  url = {https://link.aps.org/doi/10.1103/PhysRevB.94.064411}
}

@article{vasala2014magnetic,
  title={{Magnetic structure of Sr$_2$CuWO$_6$}},
  author={Vasala, S and Avdeev, M and Danilkin, S and Chmaissem, O and Karppinen, M},
  journal={Journal of Physics: Condensed Matter},
  volume={26},
  number={49},
  pages={496001},
  year={2014},
  publisher={IOP Publishing},
url={https://iopscience.iop.org/article/10.1088/0953-8984/26/49/496001/meta},
doi={10.1088/0953-8984/26/49/496001}
}

@article{mustonen2018spin,
  title={{Spin-liquid-like state in a spin-1/2 square-lattice antiferromagnet perovskite induced by $d^{10}$-$d^0$ cation mixing}},
  author={Mustonen, O and Vasala, S and Sadrollahi, E and Schmidt, KP and Baines, C and Walker, HC and Terasaki, I and Litterst, FJ and Baggio-Saitovitch, E and Karppinen, M},
  journal={Nature Communications},
  volume={9},
  number={1},
  pages={1085},
  year={2018},
  publisher={Nature Publishing Group UK London},
url={https://doi.org/10.1038/s41467-018-03435-1},
doi={10.1038/s41467-018-03435-1}
}

@article{PhysRevB.98.054422,
  title = {{Valence-bond-glass state with a singlet gap in the spin-$\frac{1}{2}$ square-lattice random ${J}_{1}\text{\ensuremath{-}}{J}_{2}$ Heisenberg antiferromagnet ${\mathrm{Sr}}_{2}{\mathrm{CuTe}}_{1\ensuremath{-}x}{\mathrm{W}}_{x}{\mathrm{O}}_{6}$}},
  author = {Watanabe, Masari and Kurita, Nobuyuki and Tanaka, Hidekazu and Ueno, Wataru and Matsui, Kazuki and Goto, Takayuki},
  journal = {Physical Review B},
  volume = {98},
  issue = {5},
  pages = {054422},
  numpages = {6},
  year = {2018},
  month = {Aug},
  publisher = {American Physical Society},
  doi = {10.1103/PhysRevB.98.054422},
  url = {https://link.aps.org/doi/10.1103/PhysRevB.98.054422}
}

@article{PhysRevB.89.241102,
  title = {{Effect of nonmagnetic dilution in the honeycomb-lattice iridates ${\mathrm{Na}}_{2}$${\mathrm{IrO}}_{3}$ and ${\mathrm{Li}}_{2}$${\mathrm{IrO}}_{3}$}},
  author = {Manni, S. and Tokiwa, Y. and Gegenwart, P.},
  journal = {Physical Review B},
  volume = {89},
  issue = {24},
  pages = {241102},
  numpages = {5},
  year = {2014},
  month = {Jun},
  publisher = {American Physical Society},
  doi = {10.1103/PhysRevB.89.241102},
  url = {https://link.aps.org/doi/10.1103/PhysRevB.89.241102}
}

@article{PhysRevB.99.214410,
  title = {{Spin-glass state and reversed magnetic anisotropy induced by Cr doping in the Kitaev magnet $\ensuremath{\alpha}\text{\ensuremath{-}}{\mathrm{RuCl}}_{3}$}},
  author = {Bastien, G. and Roslova, M. and Haghighi, M. H. and Mehlawat, K. and Hunger, J. and Isaeva, A. and Doert, T. and Vojta, M. and B\"uchner, B. and Wolter, A. U. B.},
  journal = {Physical Review B},
  volume = {99},
  issue = {21},
  pages = {214410},
  numpages = {11},
  year = {2019},
  month = {Jun},
  publisher = {American Physical Society},
  doi = {10.1103/PhysRevB.99.214410},
  url = {https://link.aps.org/doi/10.1103/PhysRevB.99.214410}
}

@article{PhysRevB.92.134412,
  title = {{Fragile magnetic order in the honeycomb lattice Iridate ${\mathrm{Na}}_{2}{\mathrm{IrO}}_{3}$ revealed by magnetic impurity doping}},
  author = {Mehlawat, Kavita and Sharma, G. and Singh, Yogesh},
  journal = {Physical Review B},
  volume = {92},
  issue = {13},
  pages = {134412},
  numpages = {7},
  year = {2015},
  month = {Oct},
  publisher = {American Physical Society},
  doi = {10.1103/PhysRevB.92.134412},
  url = {https://link.aps.org/doi/10.1103/PhysRevB.92.134412}
}

@article{takagi2019concept,
  title={{Concept and realization of Kitaev quantum spin liquids}},
  author={Takagi, Hidenori and Takayama, Tomohiro and Jackeli, George and Khaliullin, Giniyat and Nagler, Stephen E},
  journal={Nature Reviews Physics},
  volume={1},
  number={4},
  pages={264-280},
  year={2019},
  publisher={Nature Publishing Group UK London},
url={https://doi.org/10.1038/s42254-019-0038-2},
doi={10.1038/s42254-019-0038-2}
}

@article{kitaev2006anyons,
  title={Anyons in an exactly solved model and beyond},
  author={Kitaev, Alexei},
  journal={Annals of Physics},
  volume={321},
  number={1},
  pages={2-111},
  year={2006},
  publisher={Elsevier},
url={https://doi.org/10.1016/j.aop.2005.10.005},
doi={10.1016/j.aop.2005.10.005}
}

@article{PhysRevB.108.165118,
  title = {{Emergent glassiness in the disorder-free Kitaev model: Density matrix renormalization group study on a one-dimensional ladder setting}},
  author = {Yogendra, K. B. and Das, Tanmoy and Baskaran, G.},
  journal = {Physical Review B},
  volume = {108},
  issue = {16},
  pages = {165118},
  numpages = {11},
  year = {2023},
  month = {Oct},
  publisher = {American Physical Society},
  doi = {10.1103/PhysRevB.108.165118},
  url = {https://link.aps.org/doi/10.1103/PhysRevB.108.165118}
}

@article{PhysRevResearch.4.013103,
  title = {{Quantum many-body scars in spin-1 Kitaev chains}},
  author = {You, Wen-Long and Zhao, Zhuan and Ren, Jie and Sun, Gaoyong and Li, Liangsheng and Ole\ifmmode \acute{s}\else \'{s}\fi{}, Andrzej M.},
  journal = {Physical Review Research},
  volume = {4},
  issue = {1},
  pages = {013103},
  numpages = {12},
  year = {2022},
  month = {Feb},
  publisher = {American Physical Society},
  doi = {10.1103/PhysRevResearch.4.013103},
  url = {https://link.aps.org/doi/10.1103/PhysRevResearch.4.013103}
}

@article{serbyn2021quantum,
  title={Quantum many-body scars and weak breaking of ergodicity},
  author={Serbyn, Maksym and Abanin, Dmitry A and Papi{\'c}, Zlatko},
  journal={Nature Physics},
  volume={17},
  number={6},
  pages={675-685},
  year={2021},
  publisher={Nature Publishing Group UK London},
url={https://doi.org/10.1038/s41567-021-01230-2},
doi={10.1038/s41567-021-01230-2}
}

@article{PhysRevB.99.054421,
  title = {{Double-peak specific heat and spin freezing in the spin-2 triangular lattice antiferromagnet ${\mathrm{FeAl}}_{2}{\mathrm{Se}}_{4}$}},
  author = {Li, Kunkun and Jin, Shifeng and Guo, Jiangang and Xu, Yanping and Su, Yixi and Feng, Erxi and Liu, Yu and Zhou, Shengqiang and Ying, Tianping and Li, Shiyan and Wang, Ziqiang and Chen, Gang and Chen, Xiaolong},
  journal = {Physical Review B},
  volume = {99},
  issue = {5},
  pages = {054421},
  numpages = {8},
  year = {2019},
  month = {Feb},
  publisher = {American Physical Society},
  doi = {10.1103/PhysRevB.99.054421},
  url = {https://link.aps.org/doi/10.1103/PhysRevB.99.054421}
}

@article{PhysRevB.109.104420,
  title = {Zero-point entropies of spin-jam and spin-glass states in a frustrated magnet},
  author = {Piyakulworawat, C. and Thennakoon, A. and Yang, J. and Yoshizawa, H. and Ueta, D. and Sato, T. J. and Sheng, K. and Chen, W.-T. and Pai, W.-W. and Matan, K. and Lee, S.-H.},
  journal = {Physical Review B},
  volume = {109},
  issue = {10},
  pages = {104420},
  numpages = {7},
  year = {2024},
  month = {Mar},
  publisher = {American Physical Society},
  doi = {10.1103/PhysRevB.109.104420},
  url = {https://link.aps.org/doi/10.1103/PhysRevB.109.104420}
}

@article{anderson1972anomalous,
  title={Anomalous low-temperature thermal properties of glasses and spin glasses},
  author={Anderson, P W and Halperin, Bertrand I and Varma, C M},
  journal={Philosophical Magazine},
  volume={25},
  number={1},
  pages={1-9},
  year={1972},
  publisher={Taylor \& Francis},
url={https://doi.org/10.1080/14786437208229210},
doi={10.1080/14786437208229210}
}

@article{PhysRevB.61.6156,
  title = {{Magnetic properties of pure and diamagnetically doped jarosites: Model kagom\'e antiferromagnets with variable coverage of the magnetic lattice}},
  author = {Wills, A. S. and Harrison, A. and Ritter, C. and Smith, R. I.},
  journal = {Physical Review B},
  volume = {61},
  issue = {9},
  pages = {6156-6169},
  numpages = {0},
  year = {2000},
  month = {Mar},
  publisher = {American Physical Society},
  doi = {10.1103/PhysRevB.61.6156},
  url = {https://link.aps.org/doi/10.1103/PhysRevB.61.6156}
}

@article{nandkishore2015many,
  title={{Many-Body Localization and Thermalization in Quantum Statistical Mechanics}},
  author={Nandkishore, Rahul and Huse, David A},
  journal={Annu. Rev. Condens. Matter Phys.},
  volume={6},
  number={1},
  pages={15-38},
  year={2015},
  publisher={Annual Reviews},
url={https://doi.org/10.1146/annurev-conmatphys-031214-014726},
doi={10.1146/annurev-conmatphys-031214-014726}
}

@article{PhysRevLett.117.037209,
  title = {{Low-Energy Spin Dynamics of the Honeycomb Spin Liquid Beyond the Kitaev Limit}},
  author = {Song, Xue-Yang and You, Yi-Zhuang and Balents, Leon},
  journal = {Physical Review Letters},
  volume = {117},
  issue = {3},
  pages = {037209},
  numpages = {6},
  year = {2016},
  month = {Jul},
  publisher = {American Physical Society},
  doi = {10.1103/PhysRevLett.117.037209},
  url = {https://link.aps.org/doi/10.1103/PhysRevLett.117.037209}
}

@article{PhysRevB.84.024409,
  title = {{Spin-glass behavior in Ni-doped La${}_{1.85}$Sr${}_{0.15}$CuO${}_{4}$}},
  author = {Malinowski, A. and Bezusyy, V. L. and Minikayev, R. and Dziawa, P. and Syryanyy, Y. and Sawicki, M.},
  journal = {Physical Review B},
  volume = {84},
  issue = {2},
  pages = {024409},
  numpages = {17},
  year = {2011},
  month = {Jul},
  publisher = {American Physical Society},
  doi = {10.1103/PhysRevB.84.024409},
  url = {https://link.aps.org/doi/10.1103/PhysRevB.84.024409}
}

@article{kofu2021spin,
  title={Spin glass behavior and magnetic boson peak in a structural glass of a magnetic ionic liquid},
  author={Kofu, Maiko and Watanuki, Ryuta and Sakakibara, Toshiro and Ohira-Kawamura, Seiko and Nakajima, Kenji and Matsuura, Masato and Ueki, Takeshi and Akutsu, Kazuhiro and Yamamuro, Osamu},
  journal={Scientific Reports},
  volume={11},
  number={1},
  pages={12098},
  year={2021},
  publisher={Nature Publishing Group UK London},
url={https://doi.org/10.1038/s41598-021-91619-z},
doi={10.1038/s41598-021-91619-z}
}

@article{PhysRevResearch.6.013006,
  title = {Magnetic boson peak in classical spin glasses},
  author = {Kofu, Maiko and Ohira-Kawamura, Seiko and Murai, Naoki and Ishii, Rieko and Hirai, Daigorou and Arima, Hiroshi and Funakoshi, Kenichi},
  journal = {Physical Review Research},
  volume = {6},
  issue = {1},
  pages = {013006},
  numpages = {9},
  year = {2024},
  month = {Jan},
  publisher = {American Physical Society},
  doi = {10.1103/PhysRevResearch.6.013006},
  url = {https://link.aps.org/doi/10.1103/PhysRevResearch.6.013006}
}

@article{mekata2003kagome,
  title={{Kagome: The Story of the Basketweave Lattice}},
  author={Mekata, Mamoru},
  journal={Physics Today},
  volume={56},
  number={2},
  pages={12-13},
  year={2003},
  publisher={AIP Publishing},
url={https://doi.org/10.1063/1.1564329},
doi={10.1063/1.1564329}
}

@article{husimi1950statistics,
  title={{The Statistics of Honeycomb and Triangular Lattice. I}},
  author={Husimi, Kodi and Sy{\^o}zi, Itiro},
  journal={Progress of Theoretical Physics},
  volume={5},
  number={2},
  pages={177-186},
  year={1950},
  publisher={Oxford University Press},
url={https://doi.org/10.1143/ptp/5.2.177},
doi={10.1143/ptp/5.2.177}
}

@article{PhysRevLett.65.3173,
  title = {{Antiferromagnetic fluctuations and short-range order in a Kagom\'e lattice}},
  author = {Broholm, C. and Aeppli, G. and Espinosa, G. P. and Cooper, A. S.},
  journal = {Physical Review Letters},
  volume = {65},
  issue = {25},
  pages = {3173-3176},
  numpages = {0},
  year = {1990},
  month = {Dec},
  publisher = {American Physical Society},
  doi = {10.1103/PhysRevLett.65.3173},
  url = {https://link.aps.org/doi/10.1103/PhysRevLett.65.3173}
}

@article{syozi1951statistics,
  title={{Statistics of Kagom{\'e} Lattice}},
  author={Sy{\^o}zi, Itiro},
  journal={Progress of Theoretical Physics},
  volume={6},
  number={3},
  pages={306-308},
  year={1951},
  publisher={Oxford University Press},
doi={10.1143/ptp/6.3.306},
url={https://doi.org/10.1143/ptp/6.3.306}
}

@article{yan2011spin,
  title={{Spin-Liquid Ground State of the S= 1/2 Kagome Heisenberg Antiferromagnet}},
  author={Yan, Simeng and Huse, David A and White, Steven R},
  journal={Science},
  volume={332},
  number={6034},
  pages={1173-1176},
  year={2011},
  publisher={American Association for the Advancement of Science},
url={https://doi.org/10.1126/science.1201080},
doi={10.1126/science.1201080}
}

@article{PhysRevLett.98.107204,
  title = {{Spin Dynamics of the Spin-$1/2$ Kagome Lattice Antiferromagnet ${\mathrm{ZnCu}}_{3}(\mathrm{OH}{)}_{6}{\mathrm{Cl}}_{2}$}},
  author = {Helton, J. S. and Matan, K. and Shores, M. P. and Nytko, E. A. and Bartlett, B. M. and Yoshida, Y. and Takano, Y. and Suslov, A. and Qiu, Y. and Chung, J.-H. and Nocera, D. G. and Lee, Y. S.},
  journal = {Physical Review Letters},
  volume = {98},
  issue = {10},
  pages = {107204},
  numpages = {4},
  year = {2007},
  month = {Mar},
  publisher = {American Physical Society},
  doi = {10.1103/PhysRevLett.98.107204},
  url = {https://link.aps.org/doi/10.1103/PhysRevLett.98.107204}
}

@article{mielke1991ferromagnetic,
  title={{Ferromagnetic ground states for the Hubbard model on line graphs}},
  author={Mielke, Andreas},
  journal={Journal of Physics A: Mathematical and General},
  volume={24},
  number={2},
  pages={L73},
  year={1991},
  publisher={IOP Publishing},
url={https://iopscience.iop.org/article/10.1088/0305-4470/24/2/005/meta},
doi={10.1088/0305-4470/24/2/005
}
}

@article{leykam2018artificial,
  title={Artificial flat band systems: from lattice models to experiments},
  author={Leykam, Daniel and Andreanov, Alexei and Flach, Sergej},
  journal={Advances in Physics: X},
  volume={3},
  number={1},
  pages={1473052},
  year={2018},
  publisher={Taylor \& Francis},
url={https://doi.org/10.1080/23746149.2018.1473052},
doi={10.1080/23746149.2018.1473052}
}

@article{yin2022topological,
  title={{Topological kagome magnets and superconductors}},
  author={Yin, Jia-Xin and Lian, Biao and Hasan, M Zahid},
  journal={Nature},
  volume={612},
  number={7941},
  pages={647-657},
  year={2022},
  publisher={Nature Publishing Group UK London},
url={https://doi.org/10.1038/s41586-022-05516-0},
doi={10.1038/s41586-022-05516-0}
}

@article{kang2020dirac,
  title={{Dirac fermions and flat bands in the ideal kagome metal FeSn}},
  author={Kang, Mingu and Ye, Linda and Fang, Shiang and You, Jhih-Shih and Levitan, Abe and Han, Minyong and Facio, Jorge I and Jozwiak, Chris and Bostwick, Aaron and Rotenberg, Eli and others},
  journal={Nature materials},
  volume={19},
  number={2},
  pages={163-169},
  year={2020},
  publisher={Nature Publishing Group UK London},
url={https://doi.org/10.1038/s41563-019-0531-0},
doi={10.1038/s41563-019-0531-0}
}

@article{PhysRevLett.120.087201,
  title = {{Spin-Glass Ground State in a Triangular-Lattice Compound ${\mathrm{YbZnGaO}}_{4}$}},
  author = {Ma, Zhen and Wang, Jinghui and Dong, Zhao-Yang and Zhang, Jun and Li, Shichao and Zheng, Shu-Han and Yu, Yunjie and Wang, Wei and Che, Liqiang and Ran, Kejing and Bao, Song and Cai, Zhengwei and \ifmmode \check{C}\else \v{C}\fi{}erm\'ak, P. and Schneidewind, A. and Yano, S. and Gardner, J. S. and Lu, Xin and Yu, Shun-Li and Liu, Jun-Ming and Li, Shiyan and Li, Jian-Xin and Wen, Jinsheng},
  journal = {Physical Review Letters},
  volume = {120},
  issue = {8},
  pages = {087201},
  numpages = {5},
  year = {2018},
  month = {Feb},
  publisher = {American Physical Society},
  doi = {10.1103/PhysRevLett.120.087201},
  url = {https://link.aps.org/doi/10.1103/PhysRevLett.120.087201}
}

@article{PhysRevLett.122.137201,
  title = {{Rearrangement of Uncorrelated Valence Bonds Evidenced by Low-Energy Spin Excitations in ${\mathrm{YbMgGaO}}_{4}$}},
  author = {Li, Yuesheng and Bachus, Sebastian and Liu, Benqiong and Radelytskyi, Igor and Bertin, Alexandre and Schneidewind, Astrid and Tokiwa, Yoshifumi and Tsirlin, Alexander A. and Gegenwart, Philipp},
  journal = {Physical Review Letters},
  volume = {122},
  issue = {13},
  pages = {137201},
  numpages = {6},
  year = {2019},
  month = {Apr},
  publisher = {American Physical Society},
  doi = {10.1103/PhysRevLett.122.137201},
  url = {https://link.aps.org/doi/10.1103/PhysRevLett.122.137201}
}

@article{huang2024emergent,
  title={{Emergent Halperin-Saslow mode, gauge glass and quenched disorders in quantum Ising magnet TmMgGaO4}},
  author={Huang, Chun-Jiong and Wang, Xiaoqun and Wang, Ziqiang and Chen, Gang},
  journal={International Journal of Modern Physics B},
  volume={38},
  number={03},
  pages={2450040},
  year={2024},
  publisher={World Scientific},
url={https://doi.org/10.1142/S0217979224500401},
doi={10.1142/S0217979224500401}
}

@article{li2020kosterlitz,
  title={{Kosterlitz-Thouless melting of magnetic order in the triangular quantum Ising material TmMgGaO4}},
  author={Li, Han and Liao, Yuan Da and Chen, Bin-Bin and Zeng, Xu-Tao and Sheng, Xian-Lei and Qi, Yang and Meng, Zi Yang and Li, Wei},
  journal={Nature Communications},
  volume={11},
  number={1},
  pages={1111},
  year={2020},
  publisher={Nature Publishing Group UK London},
url={https://doi.org/10.1038/s41467-020-14907-8},
doi={10.1038/s41467-020-14907-8}
}

@article{wu2023topology,
  title={Topology of vibrational modes predicts plastic events in glasses},
  author={Wu, Zhen Wei and Chen, Yixiao and Wang, Wei-Hua and Kob, Walter and Xu, Limei},
  journal={Nature Communications},
  volume={14},
  number={1},
  pages={2955},
  year={2023},
  publisher={Nature Publishing Group UK London},
doi={10.1038/s41467-023-38547-w},
url={https://doi.org/10.1038/s41467-023-38547-w}
}

@article{baggioli2023topological,
  title={Topological defects reveal the plasticity of glasses},
  author={Baggioli, Matteo},
  journal={Nature Communications},
  volume={14},
  number={1},
  pages={2956},
  year={2023},
  publisher={Nature Publishing Group UK London},
url={https://doi.org/10.1038/s41467-023-38549-8},
doi={10.1038/s41467-023-38549-8}
}

@article{PhysRevB.76.054452,
  title = {{Structure, magnetization, and NMR studies of the spin-glass compound ${({\mathrm{Li}}_{x}{\mathrm{V}}_{1\ensuremath{-}x})}_{3}\mathrm{B}{\mathrm{O}}_{5}$ ($x\ensuremath{\approx}0.40$ and 0.33)}},
  author = {Zong, X. and Niazi, A. and Borsa, F. and Ma, X. and Johnston, D. C.},
  journal = {Physical Review B},
  volume = {76},
  issue = {5},
  pages = {054452},
  numpages = {15},
  year = {2007},
  month = {Aug},
  publisher = {American Physical Society},
  doi = {10.1103/PhysRevB.76.054452},
  url = {https://link.aps.org/doi/10.1103/PhysRevB.76.054452}
}

@article{broholm2020quantum,
  title={Quantum spin liquids},
  author={Broholm, C and Cava, RJ and Kivelson, SA and Nocera, DG and Norman, MR and Senthil, T},
  journal={Science},
  volume={367},
  number={6475},
  pages={eaay0668},
  year={2020},
  publisher={American Association for the Advancement of Science},
url={https://doi.org/10.1126/science.aay0668},
doi={10.1126/science.aay0668}
}

@incollection{ecker2007chiral,
  title={Chiral symmetry},
  author={Ecker, Gerhard},
  booktitle={Broken Symmetries: Proceedings of the 37. Internationale Universit{\"a}tswochen f{\"u}r Kern-und Teilchenphysik, Schladming, Austria, February 28-March 7, 1998},
  pages={83-129},
  year={2007},
  publisher={Springer},
url={https://link.springer.com/chapter/10.1007/bfb0105525},
doi={10.1007/bfb0105525}
}

@article{kocic1993universal,
  title={Universal properties of chiral symmetry breaking},
  author={Koci{\'c}, Aleksandar and Kogut, John B and Lombardo, Maria-Paola},
  journal={Nuclear Physics B},
  volume={398},
  number={2},
  pages={376-402},
  year={1993},
  publisher={Elsevier},
url={https://doi.org/10.1016/0550-3213(93)90115-6},
doi={10.1016/0550-3213(93)90115-6}
}

@book{nakahara2018geometry,
  title={Geometry, topology and physics},
  author={Nakahara, Mikio},
  year={2018},
  publisher={CRC press}
}

@article{bramwell1998frustration,
  title={{Frustration in Ising-type spin models on the pyrochlore lattice}},
  author={Bramwell, ST and Harris, MJ},
  journal={Journal of Physics: Condensed Matter},
  volume={10},
  number={14},
  pages={L215},
  year={1998},
  publisher={IOP Publishing},
url={https://iopscience.iop.org/article/10.1088/0953-8984/10/14/002/meta},
doi={10.1088/0953-8984/10/14/002}
}

@article{fennell2009magnetic,
  title={{Magnetic Coulomb phase in the spin ice Ho$_2$Ti$_2$O$_7$}},
  author={Fennell, Tom and Deen, PP and Wildes, AR and Schmalzl, K and Prabhakaran, D and Boothroyd, AT and Aldus, RJ and McMorrow, DF and Bramwell, ST},
  journal={Science},
  volume={326},
  number={5951},
  pages={415-417},
  year={2009},
  publisher={American Association for the Advancement of Science},
url={https://doi.org/10.1126/science.1177582}, 
doi={10.1126/science.1177582}
}

@article{morris2009dirac,
  title={{Dirac strings and magnetic monopoles in the spin ice Dy$_2$Ti$_2$O$_7$}},
  author={Morris, David Jonathan Pryce and Tennant, David Alan and Grigera, Santiago Andres and Klemke, B and Castelnovo, C and Moessner, R and Czternasty, C and Meissner, M and Rule, KC and Hoffmann, J-U and others},
  journal={Science},
  volume={326},
  number={5951},
  pages={411-414},
  year={2009},
  publisher={American Association for the Advancement of Science},
url={https://doi.org/10.1126/science.1178868},
doi={10.1126/science.1178868}
}

@article{pauling1935structure,
  title={{The structure and Entropy of Ice and of Other Crystals with some Randomness of Atomic Arrangement}},
  author={Pauling, Linus},
  journal={Journal of the American Chemical Society},
  volume={57},
  number={12},
  pages={2680-2684},
  year={1935},
  publisher={ACS Publications},
url={https://pubs.acs.org/doi/pdf/10.1021/ja01315a102},
doi={10.1021/ja01315a102}
}

@misc{wen2003quantum,
  title={Quantum Field Theory of Many-body Systems-from the Origin of Sound to an Origin of Light and Fermions},
  author={Wen, Xiao-Gang},
  year={2003},
  publisher={Oxford University Press}
}

@article{PhysRevLett.95.217201,
  title = {{Why Spin Ice Obeys the Ice Rules}},
  author = {Isakov, S. V. and Moessner, R. and Sondhi, S. L.},
  journal = {Physical Review Letters},
  volume = {95},
  issue = {21},
  pages = {217201},
  numpages = {4},
  year = {2005},
  month = {Nov},
  publisher = {American Physical Society},
  doi = {10.1103/PhysRevLett.95.217201},
  url = {https://link.aps.org/doi/10.1103/PhysRevLett.95.217201}
}

@article{wen2019choreographed,
  title={{Choreographed entanglement dances: Topological states of quantum matter}},
  author={Wen, Xiao-Gang},
  journal={Science},
  volume={363},
  number={6429},
  pages={eaal3099},
  year={2019},
  publisher={American Association for the Advancement of Science},
url={https://doi.org/10.1126/science.aal3099},
doi={10.1126/science.aal3099}
}

@article{PhysRevB.74.092406,
  title = {{Possible ferro-spin nematic order in $\mathrm{Ni}{\mathrm{Ga}}_{2}{\mathrm{S}}_{4}$}},
  author = {Bhattacharjee, Subhro and Shenoy, Vijay B. and Senthil, T.},
  journal = {Physical Review B},
  volume = {74},
  issue = {9},
  pages = {092406},
  numpages = {4},
  year = {2006},
  month = {Sep},
  publisher = {American Physical Society},
  doi = {10.1103/PhysRevB.74.092406},
  url = {https://link.aps.org/doi/10.1103/PhysRevB.74.092406}
}

@article{PhysRevLett.114.247207,
  title = {{Topological Spin Glass in Diluted Spin Ice}},
  author = {Sen, Arnab and Moessner, R.},
  journal = {Physical Review Letters},
  volume = {114},
  issue = {24},
  pages = {247207},
  numpages = {5},
  year = {2015},
  month = {Jun},
  publisher = {American Physical Society},
  doi = {10.1103/PhysRevLett.114.247207},
  url = {https://link.aps.org/doi/10.1103/PhysRevLett.114.247207}
}

@article{castelnovo2012spin,
  title={{Spin ice, Fractionalization, and Topological Order}},
  author={Castelnovo, C and Moessner, R and Sondhi, Shivaji Lal},
  journal={Annu. Rev. Condens. Matter Phys.},
  volume={3},
  number={1},
  pages={35-55},
  year={2012},
  publisher={Annual Reviews},
url={https://doi.org/10.1146/annurev-conmatphys-020911-125058},
doi={10.1146/annurev-conmatphys-020911-125058}
}

@article{PhysRevB.92.014406,
  title = {{$\mathrm{NaCaN}{\mathrm{i}}_{2}{\mathrm{F}}_{7}$: A frustrated high-temperature pyrochlore antiferromagnet with $S=1\phantom{\rule{0.16em}{0ex}}\mathrm{N}{\mathrm{i}}^{2+}$}},
  author = {Krizan, J. W. and Cava, R. J.},
  journal = {Physical Review B},
  volume = {92},
  issue = {1},
  pages = {014406},
  numpages = {6},
  year = {2015},
  month = {Jul},
  publisher = {American Physical Society},
  doi = {10.1103/PhysRevB.92.014406},
  url = {https://link.aps.org/doi/10.1103/PhysRevB.92.014406}
}

@article{PhysRevLett.113.117201,
  title = {{From Spin Glass to Quantum Spin Liquid Ground States in Molybdate Pyrochlores}},
  author = {Clark, L. and Nilsen, G. J. and Kermarrec, E. and Ehlers, G. and Knight, K. S. and Harrison, A. and Attfield, J. P. and Gaulin, B. D.},
  journal = {Physical Review Letters},
  volume = {113},
  issue = {11},
  pages = {117201},
  numpages = {5},
  year = {2014},
  month = {Sep},
  publisher = {American Physical Society},
  doi = {10.1103/PhysRevLett.113.117201},
  url = {https://link.aps.org/doi/10.1103/PhysRevLett.113.117201}
}

@article{greedan1986spin,
  title={{Spin-glass-like behavior in Y$_2$Mo$_2$O$_7$, a concentrated, crystalline system with negligible apparent disorder}},
  author={Greedan, JE and Sato, M and Yan, Xu and Razavi, FS},
  journal={Solid State Communications},
  volume={59},
  number={12},
  pages={895-897},
  year={1986},
  publisher={Elsevier},
url={https://doi.org/10.1016/0038-1098(86)90652-6},
doi={10.1016/0038-1098(86)90652-6}
}

@article{clark2013oxygen,
  title={{Oxygen miscibility gap and spin glass formation in the pyrochlore Lu$_2$Mo$_2$O$_7$}},
  author={Clark, Lucy and Ritter, C and Harrison, A and Attfield, JP},
  journal={Journal of Solid State Chemistry},
  volume={203},
  pages={199-203},
  year={2013},
  publisher={Elsevier},
url={https://doi.org/10.1016/j.jssc.2013.04.012},
doi={10.1016/j.jssc.2013.04.012}
}

@article{PhysRevB.89.214401,
  title = {{${\text{NaCaCo}}_{2}{\mathrm{F}}_{7}$: A single-crystal high-temperature pyrochlore antiferromagnet}},
  author = {Krizan, J. W. and Cava, R. J.},
  journal = {Physical Review B},
  volume = {89},
  issue = {21},
  pages = {214401},
  numpages = {5},
  year = {2014},
  month = {Jun},
  publisher = {American Physical Society},
  doi = {10.1103/PhysRevB.89.214401},
  url = {https://link.aps.org/doi/10.1103/PhysRevB.89.214401}
}

@article{cai2018musr,
  title={{$\mu$SR study of spin freezing and persistent spin dynamics in NaCaNi$_2$F$_7$}},
  author={Cai, Y and Wilson, MN and Hallas, AM and Liu, L and Frandsen, BA and Dunsiger, SR and Krizan, JW and Cava, RJ and Rubel, O and Uemura, YJ and others},
  journal={Journal of Physics: Condensed Matter},
  volume={30},
  number={38},
  pages={385802},
  year={2018},
  publisher={IOP Publishing},
url={https://iopscience.iop.org/article/10.1088/1361-648X/aad91c/meta},
doi={10.1088/1361-648X/aad91c}
}

@article{PhysRevB.107.024407,
  title = {{Local spin dynamics in the geometrically frustrated Mo pyrochlore antiferromagnet ${\mathrm{Lu}}_{2}{\mathrm{Mo}}_{2}{\mathrm{O}}_{5\ensuremath{-}y}{\mathrm{N}}_{2}$}},
  author = {Dey, S. K. and Ishida, K. and Okabe, H. and Hiraishi, M. and Koda, A. and Honda, T. and Yamaura, J. and Kageyama, H. and Kadono, R.},
  journal = {Physical Review B},
  volume = {107},
  issue = {2},
  pages = {024407},
  numpages = {11},
  year = {2023},
  month = {Jan},
  publisher = {American Physical Society},
  doi = {10.1103/PhysRevB.107.024407},
  url = {https://link.aps.org/doi/10.1103/PhysRevB.107.024407}
}

@article{PhysRevB.93.014433,
  title = {{Static and dynamic $XY$-like short-range order in a frustrated magnet with exchange disorder}},
  author = {Ross, K. A. and Krizan, J. W. and Rodriguez-Rivera, J. A. and Cava, R. J. and Broholm, C. L.},
  journal = {Physical Review B},
  volume = {93},
  issue = {1},
  pages = {014433},
  numpages = {8},
  year = {2016},
  month = {Jan},
  publisher = {American Physical Society},
  doi = {10.1103/PhysRevB.93.014433},
  url = {https://link.aps.org/doi/10.1103/PhysRevB.93.014433}
}

@article{cox1998exotic,
  title={{Exotic Kondo effects in metals: magnetic ions in a crystalline electric field and tunnelling centres}},
  author={Cox, DL and Zawadowski, Alfred},
  journal={Advances in Physics},
  volume={47},
  number={5},
  pages={599-942},
  year={1998},
  publisher={Taylor \& Francis},
doi={https://doi.org/10.1080/000187398243500},
url={https://www.tandfonline.com/doi/abs/10.1080/000187398243500}
}

@article{field1997rediscovering,
  title={{Rediscovering the Archimedean Polyhedra: Piero della Francesca, Luca Pacioli, Leonardo da Vinci, Albrecht D{\"u}rer, Daniele Barbaro, and Johannes Kepler}},
  author={Field, Judith V},
  journal={Archive for History of Exact Sciences},
  volume={50},
  number={3/4},
  pages={241-289},
  year={1997},
  publisher={JSTOR},
doi={https://doi.org/10.1007/BF00374595},
url={https://www.jstor.org/stable/41134110}
}

@article{conrad1971essential,
  title={{The essential closure of an Archimedean lattice-ordered group}},
  author={Conrad, Paul},
journal={Duke Math. J.},
  year={1971},
doi={10.1215/S0012-7094-71-03819-1},
url={https://projecteuclid.org/journals/duke-mathematical-journal/volume-38/issue-1/The-essential-closure-of-an-Archimedean-lattice-ordered-group/10.1215/S0012-7094-71-03819-1.short}
}

@article{martinez1973archimedean,
  title={Archimedean lattices},
  author={Martinez, Jorge},
  journal={Algebra Universalis},
  volume={3},
  pages={247-260},
  year={1973},
  publisher={Springer},
doi={10.1007/BF02945124},
url={https://link.springer.com/article/10.1007/BF02945124}
}

@article{Yu2015IsingAO,
  title={{Ising antiferromagnet on the Archimedean lattices.}},
  author={Unjong Yu},
  journal={Physical review. E, Statistical, nonlinear, and soft matter physics},
  year={2015},
  volume={91 6},
  pages={062121},
doi={https://doi.org/10.1103/PhysRevE.91.062121},
url={https://journals.aps.org/pre/abstract/10.1103/PhysRevE.91.062121}
}

@article{shrock1997lower,
  title={{Lower bounds and series for the ground-state entropy of the Potts antiferromagnet on Archimedean lattices and their duals}},
  author={Shrock, Robert and Tsai, Shan-Ho},
  journal={Physical Review E},
  volume={56},
  number={4},
  pages={4111},
  year={1997},
  publisher={APS},
doi={https://doi.org/10.1103/PhysRevE.56.4111},
url={https://journals.aps.org/pre/abstract/10.1103/PhysRevE.56.4111}
}

@article{farnell2018interplay,
  title={{Interplay between lattice topology, frustration, and spin quantum number in quantum antiferromagnets on Archimedean lattices}},
  author={Farnell, DJJ and G{\"o}tze, O and Schulenburg, J and Zinke, R and Bishop, RF and Li, PHY},
  journal={Physical Review B},
  volume={98},
  number={22},
  pages={224402},
  year={2018},
  publisher={APS},
doi={https://doi.org/10.1103/PhysRevB.98.224402},
url={https://journals.aps.org/prb/abstract/10.1103/PhysRevB.98.224402}
}

@article{richter2004quantum,
      title={{Quantum Magnetism in Two Dimensions: From Semi-classical N{\'e}el Order to Magnetic Disorder}},
  author={Richter, Johannes and Schulenburg, J{\"o}rg and Honecker, Andreas},
  journal={Quantum Magnetism},
  pages={85-153},
  year={2004},
  publisher={Springer},
url={https://link.springer.com/chapter/10.1007/BFb0119592}
}

@article{basko2006metal,
  title={Metal-insulator transition in a weakly interacting many-electron system with localized single-particle states},
  author={Basko, Denis M and Aleiner, Igor L and Altshuler, Boris L},
  journal={Annals of physics},
  volume={321},
  number={5},
  pages={1126-1205},
  year={2006},
  publisher={Elsevier},
doi= {10.1016/j.aop.2005.11.014},
url={https://doi.org/10.1016/j.aop.2005.11.014}
}

@book{mezard1987spin,
  title={Spin glass theory and beyond: An Introduction to the Replica Method and Its Applications},
  author={M{\'e}zard, Marc and Parisi, Giorgio and Virasoro, Miguel Angel},
  volume={9},
  year={1987},
  publisher={World Scientific Publishing Company}
}

@article{PhysRevB.38.373,
  title = {Nonequilibrium dynamics of spin glasses},
  author = {Fisher, Daniel S. and Huse, David A.},
  journal = {Physical Review B},
  volume = {38},
  issue = {1},
  pages = {373-385},
  numpages = {0},
  year = {1988},
  month = {Jul},
  publisher = {American Physical Society},
  doi = {10.1103/PhysRevB.38.373},
  url = {https://link.aps.org/doi/10.1103/PhysRevB.38.373}
}

@article{moessner2006geometrical,
  title={Geometrical frustration},
  author={Moessner, Roderich and Ramirez, Arthur P},
  journal={Physics Today},
  volume={59},
  number={2},
  pages={24-29},
  year={2006},
  publisher={AIP Publishing},
doi={10.1063/1.2186278},
url={https://doi.org/10.1063/1.2186278}
}

@article{PhysRevB.89.134419,
  title = {{Characterization of magnetic properties of ${\mathrm{Sr}}_{2}$${\mathrm{CuWO}}_{6}$ and ${\mathrm{Sr}}_{2}$${\mathrm{CuMoO}}_{6}$}},
  author = {Vasala, Sami and Saadaoui, Hassan and Morenzoni, Elvezio and Chmaissem, Omar and Chan, Ting-Shan and Chen, Jin-Ming and Hsu, Ying-Ya and Yamauchi, Hisao and Karppinen, Maarit},
  journal = {Physical Review B},
  volume = {89},
  issue = {13},
  pages = {134419},
  numpages = {9},
  year = {2014},
  month = {Apr},
  publisher = {American Physical Society},
  doi = {10.1103/PhysRevB.89.134419},
  url = {https://link.aps.org/doi/10.1103/PhysRevB.89.134419}
}

@book{le2011muon,
  title={Muon spin rotation, relaxation, and resonance: applications to condensed matter},
  author={Le Yaouanc, Alain and De Reotier, Pierre Dalmas},
  number={147},
  year={2011},
  publisher={OUP Oxford}
}

@article{nakatsuji2010novel,
  title={{Novel Geometrical Frustration Effects in the Two-Dimensional Triangular-Lattice Antiferromagnet NiGa$_2$S$_4$ and Related Compounds}},
  author={Nakatsuji, Satoru and Nambu, Yusuke and Onoda, Shigeki},
  journal={Journal of the Physical Society of Japan},
  volume={79},
  number={1},
  pages={011003},
  year={2010},
  publisher={The Physical Society of Japan},
doi={10.1143/JPSJ.79.011003},
url={https://journals.jps.jp/doi/abs/10.1143/JPSJ.79.011003}
}

@article{kobe1995frustration,
  title={{Frustration: How it can be measured}},
  author={Kobe, S and Klotz, T},
  journal={Physical Review E},
  volume={52},
  number={5},
  pages={5660},
  year={1995},
  publisher={APS},
doi={10.1103/PhysRevE.52.5660},
url={https://journals.aps.org/pre/abstract/10.1103/PhysRevE.52.5660}
}

@article{zhang2005thermodynamic,
  title={{Thermodynamic and magnetic properties of the Archimedean lattices}},
  author={Zhang, QL},
  journal={International Journal of Modern Physics B},
  volume={19},
  number={28},
  pages={4259-4267},
  year={2005},
  publisher={World Scientific},
doi={10.1142/S0217979205032681},
url={https://www.worldscientific.com/doi/abs/10.1142/S0217979205032681}
}

@article{witczak2014correlated,
  title={{Correlated Quantum Phenomena in the Strong Spin-Orbit Regime}},
  author={Witczak-Krempa, William and Chen, Gang and Kim, Yong Baek and Balents, Leon},
  journal={Annu. Rev. Condens. Matter Phys.},
  volume={5},
  number={1},
  pages={57-82},
  year={2014},
  publisher={Annual Reviews},
doi={10.1146/annurev-conmatphys-020911-125138},
url={https://doi.org/10.1146/annurev-conmatphys-020911-125138}
}

@article{PhysRevLett.101.076402,
  title = {{Novel ${J}_{\mathrm{eff}}=1/2$ Mott State Induced by Relativistic Spin-Orbit Coupling in ${\mathrm{Sr}}_{2}{\mathrm{IrO}}_{4}$}},
  author = {Kim, B. J. and Jin, Hosub and Moon, S. J. and Kim, J.-Y. and Park, B.-G. and Leem, C. S. and Yu, Jaejun and Noh, T. W. and Kim, C. and Oh, S.-J. and Park, J.-H. and Durairaj, V. and Cao, G. and Rotenberg, E.},
  journal = {Physical Review Letters},
  volume = {101},
  issue = {7},
  pages = {076402},
  numpages = {4},
  year = {2008},
  month = {Aug},
  publisher = {American Physical Society},
  doi = {10.1103/PhysRevLett.101.076402},
  url = {https://link.aps.org/doi/10.1103/PhysRevLett.101.076402}
}

@book{cao2021physics,
  title={Physics of Spin-orbit-coupled Oxides},
  author={Cao, Gang and DeLong, Lance},
  year={2021},
  publisher={Oxford University Press}
}

@article{RevModPhys.70.1039,
  title = {Metal-insulator transitions},
  author = {Imada, Masatoshi and Fujimori, Atsushi and Tokura, Yoshinori},
  journal = {Reviews of Modern Physics},
  volume = {70},
  issue = {4},
  pages = {1039-1263},
  numpages = {0},
  year = {1998},
  month = {Oct},
  publisher = {American Physical Society},
  doi = {10.1103/RevModPhys.70.1039},
  url = {https://link.aps.org/doi/10.1103/RevModPhys.70.1039}
}

@book{cao2013frontiers,
  title={Frontiers of 4D-and 5D-transition Metal Oxides},
  author={Cao, Gang and Delong, Lance E},
  year={2013},
  publisher={World Scientific}
}

@article{landau1957theory,
  title={{THE THEORY OF A FERMI LIQUID}},
  author={Landau, Lev Davidovich},
  journal={Soviet Physics Jetp-Ussr},
  volume={3},
  number={6},
  pages={920-925},
  year={1957},
  publisher={AMER INST PHYSICS 1305 WALT WHITMAN RD, STE 300, MELVILLE, NY 11747-4501 USA},
url={https://doi.org/10.1016/b978-0-08-010586-4.50095-x},
doi={10.1016/b978-0-08-010586-4.50095-x}
}

@article{abrikosov1959theory,
  title={The theory of a fermi liquid (the properties of liquid 3He at low temperatures)},
  author={Abrikosov, AA and Khalatnikov, IM},
  journal={Reports on Progress in Physics},
  volume={22},
  number={1},
  pages={329},
  year={1959},
  publisher={IOP Publishing},
doi={10.1088/0034-4885/22/1/310},
url={https://iopscience.iop.org/article/10.1088/0034-4885/22/1/310/meta}
}

@article{PhysRevLett.35.1779,
  title = {{$4f$-Virtual-Bound-State Formation in Ce${\mathrm{Al}}_{3}$ at Low Temperatures}},
  author = {Andres, K. and Graebner, J. E. and Ott, H. R.},
  journal = {Physical Review Letters},
  volume = {35},
  issue = {26},
  pages = {1779-1782},
  numpages = {0},
  year = {1975},
  month = {Dec},
  publisher = {American Physical Society},
  doi = {10.1103/PhysRevLett.35.1779},
  url = {https://link.aps.org/doi/10.1103/PhysRevLett.35.1779}
}

@article{lee1986theories,
  title={{Theories of heavy-electron systems}},
  author={Lee, PA and Rice, TM and Serene, JW and Sham, LJ and Wilkins, JW},
  journal={Comments Condens. Matter Phys.;(United Kingdom)},
  volume={12},
  number={3},
  year={1986},

}

@article{coleman2006heavy,
  title={Heavy fermions: Electrons at the edge of magnetism},
  author={Coleman, Piers},
  journal={arXiv preprint cond-mat/0612006},
  year={2006},
url={
https://doi.org/10.48550/arXiv.cond-mat/0612006},
doi={10.48550/arXiv.cond-mat/0612006}
}

@article{coleman2001fermiliquids,
  title={How do Fermiliquids get heavy and die?},
  author={Coleman, Piers and P{\'e}pin, C and Si, Qimiao and Ramazashvili, Revaz},
  journal={Journal of Physics: Condensed Matter},
  volume={13},
  number={35},
  pages={R723},
  year={2001},
  publisher={IOP Publishing},
url={https://iopscience.iop.org/article/10.1088/0953-8984/13/35/202/meta},
doi={10.1088/0953-8984/13/35/202}
}

@article{von202040,
  title={40 years of the quantum Hall effect},
  author={von Klitzing, Klaus and Chakraborty, Tapash and Kim, Philip and Madhavan, Vidya and Dai, Xi and McIver, James and Tokura, Yoshinori and Savary, Lucile and Smirnova, Daria and Rey, Ana Maria and others},
  journal={Nature Reviews Physics},
  volume={2},
  number={8},
  pages={397-401},
  year={2020},
  publisher={Nature Publishing Group UK London},
url={https://doi.org/10.1038/s42254-020-0209-1},
doi={10.1038/s42254-020-0209-1}
}

@article{PhysRevLett.50.1395,
  title = {{Anomalous Quantum Hall Effect: An Incompressible Quantum Fluid with Fractionally Charged Excitations}},
  author = {Laughlin, R. B.},
  journal = {Physical Review Letters},
  volume = {50},
  issue = {18},
  pages = {1395-1398},
  numpages = {0},
  year = {1983},
  month = {May},
  publisher = {American Physical Society},
  doi = {10.1103/PhysRevLett.50.1395},
  url = {https://link.aps.org/doi/10.1103/PhysRevLett.50.1395}
}

@article{harrison2000musr,
  title={{$\mu$SR studies of the kagom{\'e} antiferromagnet (H$_3$O)Fe$_3$(OH)$_6$(SO$_4$)$_2$}},
  author={Harrison, A and Kojima, KM and Wills, AS and Fudamato, Y and Larkin, MI and Luke, GM and Nachumi, B and Uemura, YJ and Visser, D and Lord, James S},
  journal={Physica B: Condensed Matter},
  volume={289},
  pages={217-220},
  year={2000},
  publisher={Elsevier},
url={https://doi.org/10.1016/S0921-4526(00)00371-9},
doi={10.1016/S0921-4526(00)00371-9}
}

@article{PhysRevLett.113.247601,
  title = {{Short-Range Correlations in the Magnetic Ground State of ${\mathrm{Na}}_{4}{\mathrm{Ir}}_{3}{\mathrm{O}}_{8}$}},
  author = {Dally, Rebecca and Hogan, Tom and Amato, Alex and Luetkens, Hubertus and Baines, Chris and Rodriguez-Rivera, Jose and Graf, Michael J. and Wilson, Stephen D.},
  journal = {Physical Review Letters},
  volume = {113},
  issue = {24},
  pages = {247601},
  numpages = {5},
  year = {2014},
  month = {Dec},
  publisher = {American Physical Society},
  doi = {10.1103/PhysRevLett.113.247601},
  url = {https://link.aps.org/doi/10.1103/PhysRevLett.113.247601}
}

@article{PhysRevLett.115.047201,
  title = {{Frozen State and Spin Liquid Physics in ${\mathrm{Na}}_{4}{\mathrm{Ir}}_{3}{\mathrm{O}}_{8}$: An NMR Study}},
  author = {Shockley, A. C. and Bert, F. and Orain, J-C. and Okamoto, Y. and Mendels, P.},
  journal = {Physical Review Letters},
  volume = {115},
  issue = {4},
  pages = {047201},
  numpages = {5},
  year = {2015},
  month = {Jul},
  publisher = {American Physical Society},
  doi = {10.1103/PhysRevLett.115.047201},
  url = {https://link.aps.org/doi/10.1103/PhysRevLett.115.047201}
}

@book{aleksandr1949mathematical,
  title={Mathematical foundations of statistical mechanics},
  author={Aleksandr, I and Khinchin, Akovlevich},
  year={1949},
  publisher={Courier Corporation}
}

@article{turner2018weak,
  title={Weak ergodicity breaking from quantum many-body scars},
  author={Turner, Christopher J and Michailidis, Alexios A and Abanin, Dmitry A and Serbyn, Maksym and Papi{\'c}, Zlatko},
  journal={Nature Physics},
  volume={14},
  number={7},
  pages={745-749},
  year={2018},
  publisher={Nature Publishing Group UK London},
doi={10.1038/s41567-018-0137-5},
url={https://doi.org/10.1038/s41567-018-0137-5}
}

@article{PhysRevLett.116.250401,
  title = {{Phase Structure of Driven Quantum Systems}},
  author = {Khemani, Vedika and Lazarides, Achilleas and Moessner, Roderich and Sondhi, S. L.},
  journal = {Physical Review Letters},
  volume = {116},
  issue = {25},
  pages = {250401},
  numpages = {6},
  year = {2016},
  month = {Jun},
  publisher = {American Physical Society},
  doi = {10.1103/PhysRevLett.116.250401},
  url = {https://link.aps.org/doi/10.1103/PhysRevLett.116.250401}
}

@book{anderson2018basic,
  title={Basic notions of condensed matter physics},
  author={Anderson, Philip W},
  year={2018},
  publisher={CRC press}
}

@article{PhysRev.127.965,
  title = {{Broken Symmetries}},
  author = {Goldstone, Jeffrey and Salam, Abdus and Weinberg, Steven},
  journal = {Physical Review},
  volume = {127},
  issue = {3},
  pages = {965-970},
  numpages = {0},
  year = {1962},
  month = {Aug},
  publisher = {American Physical Society},
  doi = {10.1103/PhysRev.127.965},
  url = {https://link.aps.org/doi/10.1103/PhysRev.127.965}
}

@article{PhysRevLett.13.508,
  title = {{Broken Symmetries and the Masses of Gauge Bosons}},
  author = {Higgs, Peter W.},
  journal = {Physical Review Letters},
  volume = {13},
  issue = {16},
  pages = {508-509},
  numpages = {0},
  year = {1964},
  month = {Oct},
  publisher = {American Physical Society},
  doi = {10.1103/PhysRevLett.13.508},
  url = {https://link.aps.org/doi/10.1103/PhysRevLett.13.508}
}

@article{PhysRev.130.439,
  title = {{Plasmons, Gauge Invariance, and Mass}},
  author = {Anderson, P. W.},
  journal = {Physical Review},
  volume = {130},
  issue = {1},
  pages = {439-442},
  numpages = {0},
  year = {1963},
  month = {Apr},
  publisher = {American Physical Society},
  doi = {10.1103/PhysRev.130.439},
  url = {https://link.aps.org/doi/10.1103/PhysRev.130.439}
}

@article{jena2025nature,
  title={{The nature of low-temperature spin-freezing in frustrated Kitaev magnets}},
  author={Jena, U and Khuntia, P},
  journal={Communications Materials},
  volume={6},
  number={1},
  pages={63},
  year={2025},
  publisher={Nature Publishing Group UK London},
url={https://doi.org/10.1038/s43246-025-00765-8},
doi={10.1038/s43246-025-00765-8}
}

@article{mattis1976solvable,
  title={Solvable spin systems with random interactions},
  author={Mattis, DC},
  journal={Physics Letters A},
  volume={56},
  number={5},
  pages={421-422},
  year={1976},
  publisher={Elsevier},
url={https://doi.org/10.1016/0375-9601(76)90396-0},
doi={10.1016/0375-9601(76)90396-0}
}

@article{RevModPhys.40.677,
  title = {{Metal-Insulator Transition}},
  author = {MOTT, N. F.},
  journal = {Reviews of Modern Physics},
  volume = {40},
  issue = {4},
  pages = {677-683},
  numpages = {0},
  year = {1968},
  month = {Oct},
  publisher = {American Physical Society},
  doi = {10.1103/RevModPhys.40.677},
  url = {https://link.aps.org/doi/10.1103/RevModPhys.40.677}
}

@article{bednorz1986possible,
  title={{Possible high {$T_\text{c}$} superconductivity in the Ba-La-Cu-O system}},
  author={Bednorz, J George and M{\"u}ller, K Alex},
  journal={Zeitschrift f{\"u}r Physik B Condensed Matter},
  volume={64},
  number={2},
  pages={189-193},
  year={1986},
  publisher={Springer},
url={https://doi.org/10.1007/BF01303701},
doi={10.1007/BF01303701}
}

@article{RevModPhys.78.17,
  title = {Doping a Mott insulator: Physics of high-temperature superconductivity},
  author = {Lee, Patrick A. and Nagaosa, Naoto and Wen, Xiao-Gang},
  journal = {Reviews of Modern Physics},
  volume = {78},
  issue = {1},
  pages = {17-85},
  numpages = {0},
  year = {2006},
  month = {Jan},
  publisher = {American Physical Society},
  doi = {10.1103/RevModPhys.78.17},
  url = {https://link.aps.org/doi/10.1103/RevModPhys.78.17}
}

@book{mott2004metal,
  title={Metal-insulator transitions},
  author={Mott, Nevill},
  year={2004},
  publisher={CRC Press}
}

@article{PhysRevLett.94.040402,
  title = {{Quantum Glassiness in Strongly Correlated Clean Systems: An Example of Topological Overprotection}},
  author = {Chamon, Claudio},
  journal = {Physical Review Letters},
  volume = {94},
  issue = {4},
  pages = {040402},
  numpages = {4},
  year = {2005},
  month = {Jan},
  publisher = {American Physical Society},
  doi = {10.1103/PhysRevLett.94.040402},
  url = {https://link.aps.org/doi/10.1103/PhysRevLett.94.040402}
}

@article{PhysRevLett.118.266601,
  title = {{Disorder-Free Localization}},
  author = {Smith, A. and Knolle, J. and Kovrizhin, D. L. and Moessner, R.},
  journal = {Physical Review Letters},
  volume = {118},
  issue = {26},
  pages = {266601},
  numpages = {5},
  year = {2017},
  month = {Jun},
  publisher = {American Physical Society},
  doi = {10.1103/PhysRevLett.118.266601},
  url = {https://link.aps.org/doi/10.1103/PhysRevLett.118.266601}
}

@article{PhysRevB.95.155133,
  title = {Glassy quantum dynamics in translation invariant fracton models},
  author = {Prem, Abhinav and Haah, Jeongwan and Nandkishore, Rahul},
  journal = {Physical Review B},
  volume = {95},
  issue = {15},
  pages = {155133},
  numpages = {14},
  year = {2017},
  month = {Apr},
  publisher = {American Physical Society},
  doi = {10.1103/PhysRevB.95.155133},
  url = {https://link.aps.org/doi/10.1103/PhysRevB.95.155133}
}

@article{PhysRevLett.18.1049,
  title = {{Infrared Catastrophe in Fermi Gases with Local Scattering Potentials}},
  author = {Anderson, P. W.},
  journal = {Physical Review Letters},
  volume = {18},
  issue = {24},
  pages = {1049-1051},
  numpages = {0},
  year = {1967},
  month = {Jun},
  publisher = {American Physical Society},
  doi = {10.1103/PhysRevLett.18.1049},
  url = {https://link.aps.org/doi/10.1103/PhysRevLett.18.1049}
}

@article{PhysRevB.101.174204,
  title = {{Localization from Hilbert space shattering: From theory to physical realizations}},
  author = {Khemani, Vedika and Hermele, Michael and Nandkishore, Rahul},
  journal = {Physical Review B},
  volume = {101},
  issue = {17},
  pages = {174204},
  numpages = {17},
  year = {2020},
  month = {May},
  publisher = {American Physical Society},
  doi = {10.1103/PhysRevB.101.174204},
  url = {https://link.aps.org/doi/10.1103/PhysRevB.101.174204}
}

@article{PhysRevB.103.235133,
  title = {{Frustration-induced emergent Hilbert space fragmentation}},
  author = {Lee, Kyungmin and Pal, Arijeet and Changlani, Hitesh J.},
  journal = {Physical Review B},
  volume = {103},
  issue = {23},
  pages = {235133},
  numpages = {13},
  year = {2021},
  month = {Jun},
  publisher = {American Physical Society},
  doi = {10.1103/PhysRevB.103.235133},
  url = {https://link.aps.org/doi/10.1103/PhysRevB.103.235133}
}

@article{grover2014quantum,
  title={Quantum disentangled liquids},
  author={Grover, Tarun and Fisher, Matthew PA},
  journal={Journal of Statistical Mechanics: Theory and Experiment},
  volume={2014},
  number={10},
  pages={P10010},
  year={2014},
  publisher={IOP Publishing},
doi={10.1088/1742-5468/2014/10/P10010},
url={https://iopscience.iop.org/article/10.1088/1742-5468/2014/10/P10010/meta}
}

@book{abragam1961principles,
  title={The principles of nuclear magnetism},
  author={Abragam, Anatole},
  number={32},
  year={1961},
  publisher={Oxford university press}
}

@article{darwin1859origins,
      title={{On the origins of species by means of natural selection}},
  author={Darwin, Charles},
  journal={London: Murray},
  volume={247},
  pages={1859},
  year={1859},
url ={https://doi.org/10.1098/rsnr.2018.0015},
doi={10.1098/rsnr.2018.0015}
}

@article{takayama1978contribution,
  title={A contribution to the theory of spin waves in a metallic spin glass},
  author={Takayama, H},
  journal={Journal of Physics F: Metal Physics},
  volume={8},
  number={11},
  pages={2417},
  year={1978},
  publisher={IOP Publishing},
url ={https://iopscience.iop.org/article/10.1088/0305-4608/8/11/025/meta},
doi={10.1088/0305-4608/8/11/025}
}

@article{PhysRev.109.1492,
  title = {{Absence of Diffusion in Certain Random Lattices}},
  author = {Anderson, P. W.},
  journal = {Physical Review},
  volume = {109},
  issue = {5},
  pages = {1492-1505},
  numpages = {0},
  year = {1958},
  month = {Mar},
  publisher = {American Physical Society},
  doi = {10.1103/PhysRev.109.1492},
  url = {https://link.aps.org/doi/10.1103/PhysRev.109.1492}
}

@article{dzyaloshinsk1978concept,
  title={On the concept of local invariance in the theory of spin glasses},
  author={Dzyaloshinsk, IE and Volovik, GE},
  journal={Journal de Physique},
  volume={39},
  number={6},
  pages={693-700},
  year={1978},
  publisher={Soci{\'e}t{\'e} Fran{\c{c}}aise de Physique},
url={https://doi.org/10.1051/jphys:01978003906069300},
doi={10.1051/jphys:01978003906069300}
}

@book{lovesey1984theory,
  title={Theory of neutron scattering from condensed matter},
  author={Lovesey, Stephen W},
  year={1984},
publisher={Oxford science publication}
}

@article{lee1996spin,
  title={Spin-glass and non-spin-glass features of a geometrically frustratedmagnet},
  author={Lee, S-H and Broholm, C and Aeppli, G and Ramirez, AP and Perring, TG and Carlile, CJ and Adams, M and Jones, TJL and Hessen, B},
  journal={Europhysics Letters},
  volume={35},
  number={2},
  pages={127},
  year={1996},
  publisher={IOP Publishing},
url={https://iopscience.iop.org/article/10.1209/epl/i1996-00543-x/meta},
doi={10.1209/epl/i1996-00543-x}
}

@book{anderson1997concepts,
  title={Concepts in solids: lectures on the theory of solids},
  author={Anderson, Philip W},
  volume={58},
  year={1997},
  publisher={World Scientific}
}

@article{ginzburg1978microscopic,
  title={Microscopic theory of spin waves in spin glasses},
  author={Ginzburg, SL},
  journal={Sov. Phys. JETP,},
  volume={48},
  number={4},
  pages={756-759},
  year={1978},
url={},
doi={}
}

@article{andreev1976macroscopic,
  title={Macroscopic theory of spin waves},
  author={Andreev, AF and Marchenko, VI},
  journal={Zh. Eksp. Teor. Fiz},
  volume={70},
  pages={1522-1538},
  year={1976}
}

@article{de2019topological,
  title={{Topological flat band, Dirac fermions and quantum spin Hall phase in 2D Archimedean lattices}},
  author={De Lima, F Crasto and Ferreira, Gerson J and Miwa, RH},
  journal={Physical Chemistry Chemical Physics},
  volume={21},
  number={40},
  pages={22344-22350},
  year={2019},
  publisher={Royal Society of Chemistry},
url={https://pubs.rsc.org/en/content/articlelanding/2019/cp/c9cp04760c},
doi={10.1039/C9CP04760C}
}

@article{farnell2014quantum,
  title={Quantum s= 1 2 antiferromagnets on Archimedean lattices: The route from semiclassical magnetic order to nonmagnetic quantum states},
  author={Farnell, Damian JJ and G{\"o}tze, O and Richter, J and Bishop, RF and Li, PHY},
  journal={Physical Review B},
  volume={89},
  number={18},
  pages={184407},
  year={2014},
  publisher={APS},
url={https://journals.aps.org/prb/abstract/10.1103/PhysRevB.89.184407},
doi={10.1103/PhysRevB.89.184407}
}

@article{PhysRevB.19.1633,
  title = {{Low-dc-field susceptibility of $\mathrm{Cu}\mathrm{Mn}$ spin glass}},
  author = {Nagata, Shoichi and Keesom, P. H. and Harrison, H. R.},
  journal = { Physical Review B},
  volume = {19},
  issue = {3},
  pages = {1633-1638},
  numpages = {0},
  year = {1979},
  month = {Feb},
  publisher = {American Physical Society},
  doi = {10.1103/PhysRevB.19.1633},
  url = {https://link.aps.org/doi/10.1103/PhysRevB.19.1633}
}

@article{PhysRevB.6.4220,
  title = {{Magnetic Ordering in Gold-Iron Alloys}},
  author = {Cannella, V. and Mydosh, J. A.},
  journal = { Physical Review B},
  volume = {6},
  issue = {11},
  pages = {4220-4237},
  numpages = {0},
  year = {1972},
  month = {Dec},
  publisher = {American Physical Society},
  doi = {10.1103/PhysRevB.6.4220},
  url = {https://link.aps.org/doi/10.1103/PhysRevB.6.4220}
}

@article{wu2011spin,
  title={{Spin glassiness and power law scaling in anisotropic triangular spin-1/2 antiferromagnets}},
  author={Wu, Jian and Wildeboer, Julia S and Werner, Fletcher and Seidel, Alexander and Nussinov, Z and Solin, SA},
  journal={Europhysics Letters},
  volume={93},
  number={6},
  pages={67001},
  year={2011},
  publisher={IOP Publishing},
url={https://iopscience.iop.org/article/10.1209/0295-5075/93/67001/meta},
doi={10.1209/0295-5075/93/67001}
}

@article{PhysRevB.20.1245,
  title = {{Insulating spin-glass system ${\mathrm{Eu}}_{x}{\mathrm{Sr}}_{1\ensuremath{-}x}\mathrm{S}$}},
  author = {Maletta, H. and Felsch, W.},
  journal = { Physical Review B},
  volume = {20},
  issue = {3},
  pages = {1245-1260},
  numpages = {0},
  year = {1979},
  month = {Aug},
  publisher = {American Physical Society},
  doi = {10.1103/PhysRevB.20.1245},
  url = {https://link.aps.org/doi/10.1103/PhysRevB.20.1245}
}

@article{manni2014effect,
  title={{Effect of nonmagnetic dilution in the honeycomb-lattice iridates Na$_2$IrO$_3$ and Li$_2$IrO$_3$}},
  author={Manni, S and Tokiwa, Yoshifumi and Gegenwart, Philipp},
  journal={Physical Review B},
  volume={89},
  number={24},
  pages={241102},
  year={2014},
  publisher={APS},
url={https://journals.aps.org/prb/abstract/10.1103/PhysRevB.89.241102},
doi={10.1103/PhysRevB.89.241102}
}

@book{nagaoka2012anderson,
  title={{Anderson Localization: Proceedings of the Fourth Taniguchi International Symposium, Sanda-shi, Japan, November 3-8, 1981}},
  author={Nagaoka, Yosuke and Fukuyama, Hidetoshi},
  volume={39},
  year={2012},
  publisher={Springer Science \& Business Media}
}

@article{lounasmaa1999vortices,
  title={Vortices in rotating superfluid 3He},
  author={Lounasmaa, Olli V and Thuneberg, Erkki},
  journal={Proceedings of the National Academy of Sciences},
  volume={96},
  number={14},
  pages={7760-7767},
  year={1999},
  publisher={The National Academy of Sciences},
url={https://www.pnas.org/doi/abs/10.1073/pnas.96.14.7760},
doi={10.1073/pnas.96.14.7760}
}

@article{fischer1979electrical,
  title={On the electrical resistivity of spin glasses},
  author={Fischer, KH},
  journal={Zeitschrift f{\"u}r Physik B Condensed Matter},
  volume={34},
  pages={45-53},
  year={1979},
  publisher={Springer},
url={https://doi.org/10.1007/BF01362778},
doi={10.1007/BF01362778}
}

@book{fukushima2018experimental,
  title={Experimental pulse NMR: a nuts and bolts approach},
  author={Fukushima, Eiichi},
  year={2018},
  publisher={CRC press}
}

@article{PhysRevB.78.094403,
  title = {{Spin-orbit effects in ${\text{Na}}_{4}{\text{Ir}}_{3}{\text{O}}_{8}$: A hyper-kagome lattice antiferromagnet}},
  author = {Chen, Gang and Balents, Leon},
  journal = {Physical Review B},
  volume = {78},
  issue = {9},
  pages = {094403},
  numpages = {21},
  year = {2008},
  month = {Sep},
  publisher = {American Physical Society},
  doi = {10.1103/PhysRevB.78.094403},
  url = {https://link.aps.org/doi/10.1103/PhysRevB.78.094403}
}

@article{RevModPhys.88.041002,
  title = {Colloquium: Herbertsmithite and the search for the quantum spin liquid},
  author = {Norman, M. R.},
  journal = {Reviews of Modern Physics},
  volume = {88},
  issue = {4},
  pages = {041002},
  numpages = {14},
  year = {2016},
  month = {Dec},
  publisher = {American Physical Society},
  doi = {10.1103/RevModPhys.88.041002},
  url = {https://link.aps.org/doi/10.1103/RevModPhys.88.041002}
}

@article{PhysRevB.99.054416,
  title = {Dynamics and energy landscape of the jammed spin liquid},
  author = {Bilitewski, Thomas and Zhitomirsky, Mike E. and Moessner, Roderich},
  journal = {Physical Review B},
  volume = {99},
  issue = {5},
  pages = {054416},
  numpages = {15},
  year = {2019},
  month = {Feb},
  publisher = {American Physical Society},
  doi = {10.1103/PhysRevB.99.054416},
  url = {https://link.aps.org/doi/10.1103/PhysRevB.99.054416}
}

@article{PhysRevLett.56.1601,
  title = {{Ordered Phase of Short-Range Ising Spin-Glasses}},
  author = {Fisher, Daniel S. and Huse, David A.},
  journal = {Physical Review Letters},
  volume = {56},
  issue = {15},
  pages = {1601-1604},
  numpages = {0},
  year = {1986},
  month = {Apr},
  publisher = {American Physical Society},
  doi = {10.1103/PhysRevLett.56.1601},
  url = {https://link.aps.org/doi/10.1103/PhysRevLett.56.1601}
}

@book{venkataraman2012beyond,
  title={{Beyond the crystalline state: an emerging perspective}},
  author={Venkataraman, Ganesan and Sahoo, Debendranath and Balakrishnan, Venkataraman},
  volume={84},
  year={2012},
  publisher={Springer Science \& Business Media}
}

@article{moudgalya2022quantum,
  title={{Quantum many-body scars and Hilbert space fragmentation: A review of exact results}},
  author={Moudgalya, Sanjay and Bernevig, B Andrei and Regnault, Nicolas},
  journal={Reports on Progress in Physics},
  volume={85},
  number={8},
  pages={086501},
  year={2022},
  publisher={IOP Publishing},
doi={10.1088/1361-6633/ac73a0},
url={https://iopscience.iop.org/article/10.1088/1361-6633/ac73a0/meta}
}

@article{bluvstein2021controlling,
  title={{Controlling quantum many-body dynamics in driven Rydberg atom arrays}},
  author={Bluvstein, Dolev and Omran, Ahmed and Levine, Harry and Keesling, Alexander and Semeghini, Giulia and Ebadi, Sepehr and Wang, Tout T and Michailidis, Alexios A and Maskara, Nishad and Ho, Wen Wei and others},
  journal={Science},
  volume={371},
  number={6536},
  pages={1355-1359},
  year={2021},
  publisher={American Association for the Advancement of Science},
url={https://doi.org/10.1126/science.abg2530},
doi={10.1126/science.abg2530}
}

@article{trebst2022kitaev,
  title={Kitaev materials},
  author={Trebst, Simon and Hickey, Ciar{\'a}n},
  journal={Physics Reports},
  volume={950},
  pages={1-37},
  year={2022},
  publisher={Elsevier},
url={https://www.sciencedirect.com/science/article/pii/S0370157321004051},
doi={10.1016/j.physrep.2021.11.003}
}

@article{hu2021freezing,
  title={Freezing of a disorder induced spin liquid with strong quantum fluctuations},
  author={Hu, Xiao and Pajerowski, Daniel M and Zhang, Depei and Podlesnyak, Andrey A and Qiu, Yiming and Huang, Qing and Zhou, Haidong and Klich, Israel and Kolesnikov, Alexander I and Stone, Matthew B and others},
  journal={Physical review letters},
  volume={127},
  number={1},
  pages={017201},
  year={2021},
  publisher={APS},
url={https://journals.aps.org/prl/abstract/10.1103/PhysRevLett.127.017201},
doi={10.1103/PhysRevLett.127.017201}
}

@book{dotsenko1995introduction,
  title={An introduction to the theory of spin glasses and neural networks},
  author={Dotsenko, Viktor},
  volume={54},
  year={1995},
  publisher={World Scientific}
}

@article{PhysRev.106.893,
  title = {{Magnetic Properties of Cu-Mn Alloys}},
  author = {Yosida, Kei},
  journal = {Physical Review},
  volume = {106},
  issue = {5},
  pages = {893-898},
  numpages = {0},
  year = {1957},
  month = {Jun},
  publisher = {American Physical Society},
  doi = {10.1103/PhysRev.106.893},
  url = {https://link.aps.org/doi/10.1103/PhysRev.106.893}
}

@article{arrott1965neutron,
  title={{Neutron Diffraction Studies of Pd, Ni, FeMn, and Cu(Mn) Single Crystals}},
  author={Arrott, Anthony},
  journal={Journal of Applied Physics},
  volume={36},
  number={3},
  pages={1093-1093},
  year={1965},
  publisher={American Institute of Physics},
url={https://doi.org/10.1063/1.1714112},
doi={10.1063/1.1714112}
}

@article{gonser1965magnetic,
  title={{Magnetic Transitions in Dilute Solutions of Iron in Gold and Copper}},
  author={Gonser, U and Grant, RW and Meechan, CJ and Muir Jr, AH and Wiedersich, H},
  journal={Journal of Applied Physics},
  volume={36},
  number={7},
  pages={2124-2131},
  year={1965},
  publisher={American Institute of Physics},
url={https://doi.org/10.1063/1.1714431},
doi={10.1063/1.1714431}
}

@article{PhysRevB.1.349,
  title = {{Descriptive Model for the Magnetic Behavior of Au-Fe Alloys}},
  author = {Borg, R. J.},
  journal = {Physical Review B},
  volume = {1},
  issue = {1},
  pages = {349-351},
  numpages = {0},
  year = {1970},
  month = {Jan},
  publisher = {American Physical Society},
  doi = {10.1103/PhysRevB.1.349},
  url = {https://link.aps.org/doi/10.1103/PhysRevB.1.349}
}

@article{kondo1964resistance,
  title={Resistance Minimum in Dilute Magnetic Alloys},
  author={Kondo, Jun},
  journal={Progress of theoretical physics},
  volume={32},
  number={1},
  pages={37-49},
  year={1964},
  publisher={Oxford University Press},
url={https://doi.org/10.1143/PTP.32.37},
doi={10.1143/PTP.32.37}
}

@article{maiorano2018support,
  title={{Support for the value 5/2 for the spin glass lower critical dimension at zero magnetic field}},
  author={Maiorano, Andrea and Parisi, Giorgio},
  journal={Proceedings of the National Academy of Sciences},
  volume={115},
  number={20},
  pages={5129-5134},
  year={2018},
  publisher={National Academy of Sciences},
url={https://doi.org/10.1073/pnas.1720832115},
doi={10.1073/pnas.1720832115}
}

@book{boothroyd2020principles,
  title={{Principles of Neutron Scattering from Condensed Matter}},
  author={Boothroyd, Andrew T},
  year={2020},
  publisher={Oxford University Press}
}

@book{squires1996introduction,
  title={{Introduction to the Theory of Thermal Neutron Scattering}},
  author={Squires, Gordon Leslie},
  year={1996},
  publisher={Courier Corporation}
}

@article{PhysRevMaterials.6.L021401,
  title = {{Quantum paramagnetism in the hyperhoneycomb Kitaev magnet $\ensuremath{\beta}\text{\ensuremath{-}}{\mathrm{ZnIrO}}_{3}$}},
  author = {Haraguchi, Yuya and Matsuo, Akira and Kindo, Koichi and Katori, Hiroko Aruga},
  journal = {Physcial Review Matererials},
  volume = {6},
  issue = {2},
  pages = {L021401},
  numpages = {6},
  year = {2022},
  month = {Feb},
  publisher = {American Physical Society},
  doi = {10.1103/PhysRevMaterials.6.L021401},
  url = {https://link.aps.org/doi/10.1103/PhysRevMaterials.6.L021401}
}

@book{zubtsovskii2022topotactic,
  title={Topotactic synthesis and characterization of new Kitaev iridates},
  author={Zubtsovskii, Aleksandr},
publisher={Universität Augsburg, 2022},
  year={2022}
}

@article{haraguchi2023monoclinic,
  title={{Monoclinic distortion in hyperhoneycomb Kitaev material $\beta$-ZnIrO$_3$ revealed by improved sample quality}},
  author={Haraguchi, Yuya and Katori, Hiroko Aruga},
  journal={Chemistry Letters},
  volume={52},
  number={5},
  pages={404-407},
  year={2023},
  publisher={Oxford University Press},
url={https://doi.org/10.1246/cl.230108},
doi={10.1246/cl.230108}
}

@article{PhysRevB.89.045117,
  title = {{Heisenberg-Kitaev model on the hyperhoneycomb lattice}},
  author = {Lee, Eric Kin-Ho and Schaffer, Robert and Bhattacharjee, Subhro and Kim, Yong Baek},
  journal = { Physical Review B},
  volume = {89},
  issue = {4},
  pages = {045117},
  numpages = {13},
  year = {2014},
  month = {Jan},
  publisher = {American Physical Society},
  doi = {10.1103/PhysRevB.89.045117},
  url = {https://link.aps.org/doi/10.1103/PhysRevB.89.045117}
}

@article{ito2001nature,
  title={{Nature of spin freezing transition of geometrically frustrated pyrochlore system R$_2$Ru$_2$O$_7$ (R= rare earth elements and Y)}},
  author={Ito, M and Yasui, Y and Kanada, M and Harashina, H and Yoshii, S and Murata, K and Sato, M and Okumura, H and Kakurai, K},
  journal={Journal of Physics and Chemistry of Solids},
  volume={62},
  number={1-2},
  pages={337-341},
  year={2001},
  publisher={Elsevier},
url={https://doi.org/10.1016/S0022-3697(00)00159-1},
doi={10.1016/S0022-3697(00)00159-1}
}

@article{kasuya1956theory,
  title={{A theory of metallic ferro-and antiferromagnetism on Zener's model}},
  author={Kasuya, Tadao},
  journal={Progress of theoretical physics},
  volume={16},
  number={1},
  pages={45-57},
  year={1956},
  publisher={Oxford University Press},
url={https://doi.org/10.1143/PTP.16.45},
doi={10.1143/PTP.16.45}
}

@article{PhysRev.96.99,
  title = {{Indirect Exchange Coupling of Nuclear Magnetic Moments by Conduction Electrons}},
  author = {Ruderman, M. A. and Kittel, C.},
  journal = {Physocal Review},
  volume = {96},
  issue = {1},
  pages = {99-102},
  numpages = {0},
  year = {1954},
  month = {Oct},
  publisher = {American Physical Society},
  doi = {10.1103/PhysRev.96.99},
  url = {https://link.aps.org/doi/10.1103/PhysRev.96.99}
}

@article{tholence1979spin,
  title={{Spin-glass versus’’blocking’’in dilute Eu$_x$Sr$_{1- x}$S}},
  author={Tholence, JL},
  journal={Journal of Applied Physics},
  volume={50},
  number={B11},
  pages={7369-7371},
  year={1979},
  publisher={American Institute of Physics},
doi={10.1063/1.326945},
url={https://doi.org/10.1063/1.326945}
}

@article{maletta1982magnetic,
  title={{Magnetic ordering in Eu$_x$Sr$_{1- x}$S, a diluted Heisenberg system with competing interactions}},
  author={Maletta, H},
  journal={Journal of Applied physics},
  volume={53},
  number={3},
  pages={2185-2190},
  year={1982},
  publisher={American Institute of Physics},
url={https://doi.org/10.1063/1.330773},
doi={10.1063/1.330773}
}

@article{kobler2001impact,
  title={{The impact of fourth-order exchange interactions on the critical temperatures of Eu$_x$Sr$_{1-x}$S and Eu$_x$Sr$_{1-x}$Te}},
  author={K{\"o}bler, U and Fischer, K},
  journal={Journal of Physics: Condensed Matter},
  volume={13},
  number={1},
  pages={123},
  year={2001},
  publisher={IOP Publishing},
url={https://iopscience.iop.org/article/10.1088/0953-8984/13/1/313/meta},
doi={10.1088/0953-8984/13/1/313}
}

@book{grunbaum1987tilings,
  title={Tilings and patterns},
  author={Gr{\"u}nbaum, Branko and Shephard, Geoffrey Colin},
  year={1987},
  publisher={Courier Dover Publications}
}

@article{PhysRevB.79.214436,
  title = {{Quadrupolar correlations and spin freezing in $S=1$ triangular lattice antiferromagnets}},
  author = {Stoudenmire, E. M. and Trebst, Simon and Balents, Leon},
  journal = {Physical Review B},
  volume = {79},
  issue = {21},
  pages = {214436},
  numpages = {13},
  year = {2009},
  month = {Jun},
  publisher = {American Physical Society},
  doi = {10.1103/PhysRevB.79.214436},
  url = {https://link.aps.org/doi/10.1103/PhysRevB.79.214436}
}

@article{bastien2019spin,
  title={Spin-glass state and reversed magnetic anisotropy induced by Cr doping in the Kitaev magnet $\alpha$-RuCl 3},
  author={Bastien, G and Roslova, Maria and Haghighi, MH and Mehlawat, K and Hunger, J and Isaeva, A and Doert, T and Vojta, M and B{\"u}chner, B and Wolter, AUB},
  journal={Physical Review B},
  volume={99},
  number={21},
  pages={214410},
  year={2019},
  publisher={APS},
doi = {10.1103/PhysRevB.99.214410},
  url = {https://journals.aps.org/prb/abstract/10.1103/PhysRevB.99.214410}
}

@article{mehlawat2015fragile,
  title={Fragile magnetic order in the honeycomb lattice Iridate Na 2 IrO 3 revealed by magnetic impurity doping},
  author={Mehlawat, Kavita and Sharma, G and Singh, Yogesh},
  journal={Physical Review B},
  volume={92},
  number={13},
  pages={134412},
  year={2015},
  publisher={APS},
doi = {10.1103/PhysRevB.92.134412},
  url = {https://journals.aps.org/prb/abstract/10.1103/PhysRevB.92.134412}
}

@article{edwards1980ground,
  title={The ground state of a spin glass},
  author={Edwards, SF and Tanaka, F},
  journal={Journal of Physics F: Metal Physics},
  volume={10},
  number={11},
  pages={2471},
  year={1980},
  publisher={IOP Publishing},
url={https://iopscience.iop.org/article/10.1088/0305-4608/10/11/019/meta},
doi={10.1088/0305-4608/10/11/019}
}

@article{tanaka1980ground,
  title={{Ground-state entropy of the infinite-range model of a spin glass}},
  author={Tanaka, F},
  journal={Journal of Physics C: Solid State Physics},
  volume={13},
  number={32},
  pages={L951},
  year={1980},
  publisher={IOP Publishing},
url={https://iopscience.iop.org/article/10.1088/0022-3719/13/32/006/meta},
doi={10.1088/0022-3719/13/32/006}
}

@article{PhysRevLett.98.037203,
  title = {{Specific Heat of the Dilute Ising Magnet ${\mathrm{LiHo}}_{x}{\mathrm{Y}}_{1\ensuremath{-}x}{\mathrm{F}}_{4}$}},
  author = {Quilliam, J. A. and Mugford, C. G. A. and Gomez, A. and Kycia, S. W. and Kycia, J. B.},
  journal = {Physical Review Letters},
  volume = {98},
  issue = {3},
  pages = {037203},
  numpages = {4},
  year = {2007},
  month = {Jan},
  publisher = {American Physical Society},
  doi = {10.1103/PhysRevLett.98.037203},
  url = {https://link.aps.org/doi/10.1103/PhysRevLett.98.037203}
}

@article{PhysRevB.72.184401,
  title = {{Slow spin-glass and fast spin-liquid components in quasi-two-dimensional ${\mathrm{La}}_{2}(\mathrm{Cu},\mathrm{Li}){\mathrm{O}}_{4}$}},
  author = {Chen, Y. and Bao, Wei and Qiu, Y. and Lorenzo, J. E. and Sarrao, J. L. and Ho, D. L. and Lin, Min Y.},
  journal = {Physical Review B},
  volume = {72},
  issue = {18},
  pages = {184401},
  numpages = {6},
  year = {2005},
  month = {Nov},
  publisher = {American Physical Society},
  doi = {10.1103/PhysRevB.72.184401},
  url = {https://link.aps.org/doi/10.1103/PhysRevB.72.184401}
}

@article{PhysRevB.62.9148,
  title = {{Static and dynamic spin correlations in the spin-glass phase of slightly doped ${\mathrm{La}}_{2\ensuremath{-}x}{\mathrm{Sr}}_{x}{\mathrm{CuO}}_{4}$}},
  author = {Matsuda, M. and Fujita, M. and Yamada, K. and Birgeneau, R. J. and Kastner, M. A. and Hiraka, H. and Endoh, Y. and Wakimoto, S. and Shirane, G.},
  journal = {Physical Review B},
  volume = {62},
  issue = {13},
  pages = {9148-9154},
  numpages = {0},
  year = {2000},
  month = {Oct},
  publisher = {American Physical Society},
  doi = {10.1103/PhysRevB.62.9148},
  url = {https://link.aps.org/doi/10.1103/PhysRevB.62.9148}
}

@article{PhysRevB.69.014424,
  title = {Spin-glass phase of cuprates},
  author = {Hasselmann, N. and Castro Neto, A. H. and Morais Smith, C.},
  journal = {Physical Review B},
  volume = {69},
  issue = {1},
  pages = {014424},
  numpages = {21},
  year = {2004},
  month = {Jan},
  publisher = {American Physical Society},
  doi = {10.1103/PhysRevB.69.014424},
  url = {https://link.aps.org/doi/10.1103/PhysRevB.69.014424}
}

@article{PhysRevLett.83.604,
  title = {{Charge Segregation, Cluster Spin Glass, and Superconductivity in ${\mathrm{La}}_{1.94}{\mathrm{Sr}}_{0.06}{\mathrm{CuO}}_{4}$}},
  author = {Julien, M.-H. and Borsa, F. and Carretta, P. and Horvati\ifmmode \acute{c}\else \'{c}\fi{}, M. and Berthier, C. and Lin, C. T.},
  journal = {Physics Review Letters},
  volume = {83},
  issue = {3},
  pages = {604-607},
  numpages = {0},
  year = {1999},
  month = {Jul},
  publisher = {American Physical Society},
  doi = {10.1103/PhysRevLett.83.604},
  url = {https://link.aps.org/doi/10.1103/PhysRevLett.83.604}
}

@article{PhysRevLett.104.107201,
  title = {{Thermal Quenches in Spin Ice}},
  author = {Castelnovo, C. and Moessner, R. and Sondhi, S. L.},
  journal = {Physical Review Letters},
  volume = {104},
  issue = {10},
  pages = {107201},
  numpages = {4},
  year = {2010},
  month = {Mar},
  publisher = {American Physical Society},
  doi = {10.1103/PhysRevLett.104.107201},
  url = {https://link.aps.org/doi/10.1103/PhysRevLett.104.107201}
}

@article{mirebeau2002pressure,
  title={{Pressure-induced crystallization of a spin liquid}},
  author={Mirebeau, I and Goncharenko, IN and Cadavez-Peres, P and Bramwell, ST and Gingras, MJP and Gardner, JS},
  journal={Nature},
  volume={420},
  number={6911},
  pages={54-57},
  year={2002},
  publisher={Nature Publishing Group UK London},
url={https://doi.org/10.1038/nature01157},
doi={10.1038/nature01157}
}

@article{PhysRevLett.112.087204,
  title = {{Pressure-Induced Quantum Critical and Multicritical Points in a Frustrated Spin Liquid}},
  author = {Thede, M. and Mannig, A. and M\aa{}nsson, M. and H\"uvonen, D. and Khasanov, R. and Morenzoni, E. and Zheludev, A.},
  journal = {Physical Review Letters},
  volume = {112},
  issue = {8},
  pages = {087204},
  numpages = {5},
  year = {2014},
  month = {Feb},
  publisher = {American Physical Society},
  doi = {10.1103/PhysRevLett.112.087204},
  url = {https://link.aps.org/doi/10.1103/PhysRevLett.112.087204}
}

@article{PhysRevLett.93.257201,
  title = {{Pressure-Induced Quantum Phase Transition in the Spin-Liquid $\mathrm{T}\mathrm{l}\mathrm{C}\mathrm{u}{\mathrm{C}\mathrm{l}}_{3}$}},
  author = {R\"uegg, Ch. and Furrer, A. and Sheptyakov, D. and Str\"assle, Th. and Kr\"amer, K. W. and G\"udel, H.-U. and M\'el\'esi, L.},
  journal = {Physical Review Letters},
  volume = {93},
  issue = {25},
  pages = {257201},
  numpages = {4},
  year = {2004},
  month = {Dec},
  publisher = {American Physical Society},
  doi = {10.1103/PhysRevLett.93.257201},
  url = {https://link.aps.org/doi/10.1103/PhysRevLett.93.257201}
}

@article{PhysRevLett.119.037201,
  title = {{Evidence for a Field-Induced Quantum Spin Liquid in $\ensuremath{\alpha}$-${\mathrm{RuCl}}_{3}$}},
  author = {Baek, S.-H. and Do, S.-H. and Choi, K.-Y. and Kwon, Y. S. and Wolter, A. U. B. and Nishimoto, S. and van den Brink, Jeroen and B\"uchner, B.},
  journal = {Physical Review Letters},
  volume = {119},
  issue = {3},
  pages = {037201},
  numpages = {5},
  year = {2017},
  month = {Jul},
  publisher = {American Physical Society},
  doi = {10.1103/PhysRevLett.119.037201},
  url = {https://link.aps.org/doi/10.1103/PhysRevLett.119.037201}
}

@article{giamarchi2008bose,
  title={{Bose-Einstein condensation in magnetic insulators}},
  author={Giamarchi, Thierry and R{\"u}egg, Christian and Tchernyshyov, Oleg},
  journal={Nature Physics},
  volume={4},
  number={3},
  pages={198-204},
  year={2008},
  publisher={Nature Publishing Group UK London},
url={https://doi.org/10.1038/nphys893},
doi={10.1038/nphys893}
}

@article{PhysRevLett.103.207203,
  title = {{Field-Induced Bose-Einstein Condensation of Triplons up to 8 K in ${\mathrm{Sr}}_{3}{\mathrm{Cr}}_{2}{\mathbf{O}}_{8}$}},
  author = {Aczel, A. A. and Kohama, Y. and Marcenat, C. and Weickert, F. and Jaime, M. and Ayala-Valenzuela, O. E. and McDonald, R. D. and Selesnic, S. D. and Dabkowska, H. A. and Luke, G. M.},
  journal = {Physical Review Letters},
  volume = {103},
  issue = {20},
  pages = {207203},
  numpages = {4},
  year = {2009},
  month = {Nov},
  publisher = {American Physical Society},
  doi = {10.1103/PhysRevLett.103.207203},
  url = {https://link.aps.org/doi/10.1103/PhysRevLett.103.207203}
}

@article{uji2001magnetic,
  title={{Magnetic-field-induced superconductivity in a two-dimensional organic conductor}},
  author={Uji, Shinya and Shinagawa, H and Terashima, T and Yakabe, T and Terai, Y and Tokumoto, M and Kobayashi, A and Tanaka, H and Kobayashi, H},
  journal={Nature},
  volume={410},
  number={6831},
  pages={908-910},
  year={2001},
  publisher={Nature Publishing Group UK London},
url={https://doi.org/10.1038/35073531},
doi={10.1038/35073531}
}

@article{PhysRevLett.101.057006,
  title = {{Pressure Induced Superconductivity in ${\mathrm{CaFe}}_{2}{\mathrm{As}}_{2}$}},
  author = {Torikachvili, Milton S. and Bud'ko, Sergey L. and Ni, Ni and Canfield, Paul C.},
  journal = {Physical Review Letters},
  volume = {101},
  issue = {5},
  pages = {057006},
  numpages = {4},
  year = {2008},
  month = {Jul},
  publisher = {American Physical Society},
  doi = {10.1103/PhysRevLett.101.057006},
  url = {https://link.aps.org/doi/10.1103/PhysRevLett.101.057006}
}

@article{drozdov2015conventional,
  title={{Conventional superconductivity at 203 kelvin at high pressures in the sulfur hydride system}},
  author={Drozdov, AP and Eremets, MI and Troyan, IA and Ksenofontov, Vadim and Shylin, Sergii I},
  journal={Nature},
  volume={525},
  number={7567},
  pages={73-76},
  year={2015},
  publisher={Nature Publishing Group UK London},
url={https://doi.org/10.1038/nature14964},
doi={10.1038/nature14964}
}

@article{PhysRevX.5.041018,
  title = {Electron-Doped ${\mathrm{Sr}}_{2}{\mathrm{IrO}}_{4}$: An Analogue of Hole-Doped Cuprate Superconductors Demonstrated by Scanning Tunneling Microscopy},
  author = {Yan, Y. J. and Ren, M. Q. and Xu, H. C. and Xie, B. P. and Tao, R. and Choi, H. Y. and Lee, N. and Choi, Y. J. and Zhang, T. and Feng, D. L.},
  journal = {Physical Review X},
  volume = {5},
  issue = {4},
  pages = {041018},
  numpages = {7},
  year = {2015},
  month = {Nov},
  publisher = {American Physical Society},
  doi = {10.1103/PhysRevX.5.041018},
  url = {https://link.aps.org/doi/10.1103/PhysRevX.5.041018}
}

@article{PhysRevB.23.1384,
  title = {{Susceptibility of the $\mathrm{Cu}\mathrm{Mn}$ spin-glass: Frequency and field dependences}},
  author = {Mulder, C. A. M. and van Duyneveldt, A. J. and Mydosh, J. A.},
  journal = {Physical Review B},
  volume = {23},
  issue = {3},
  pages = {1384-1396},
  numpages = {0},
  year = {1981},
  month = {Feb},
  publisher = {American Physical Society},
  doi = {10.1103/PhysRevB.23.1384},
  url = {https://link.aps.org/doi/10.1103/PhysRevB.23.1384}
}

@article{reed1978high,
  title={{High temperature series for the Heisenberg spin-glass}},
  author={Reed, P},
  journal={Physics Letters A},
  volume={68},
  number={5-6},
  pages={473-474},
  year={1978},
  publisher={Elsevier},
url={https://doi.org/10.1016/0375-9601(78)90633-3},
doi={10.1016/0375-9601(78)90633-3}
}

@article{PhysRevB.32.7384,
  title = {{Dynamics of three-dimensional Ising spin glasses in thermal equilibrium}},
  author = {Ogielski, Andrew T.},
  journal = {Physical Review B},
  volume = {32},
  issue = {11},
  pages = {7384-7398},
  numpages = {0},
  year = {1985},
  month = {Dec},
  publisher = {American Physical Society},
  doi = {10.1103/PhysRevB.32.7384},
  url = {https://link.aps.org/doi/10.1103/PhysRevB.32.7384}
}

@article{PhysRevB.53.R484,
  title = {{Phase transition in the three-dimensional $\ifmmode\pm\else\textpm\fi{}J$ Ising spin glass}},
  author = {Kawashima, N. and Young, A. P.},
  journal = {Physical Review B},
  volume = {53},
  issue = {2},
  pages = {R484-R487},
  numpages = {0},
  year = {1996},
  month = {Jan},
  publisher = {American Physical Society},
  doi = {10.1103/PhysRevB.53.R484},
  url = {https://link.aps.org/doi/10.1103/PhysRevB.53.R484}
}

@article{kawamura1984phase,
  title={{Phase transition of the two-dimensional Heisenberg antiferromagnet on the triangular lattice}},
  author={Kawamura, Hikaru and Miyashita, Seiji},
  journal={Journal of the Physical Society of Japan},
  volume={53},
  number={12},
  pages={4138-4154},
  year={1984},
  publisher={The Physical Society of Japan},
url={https://doi.org/10.1143/JPSJ.53.4138},
doi={10.1143/JPSJ.53.4138}
}

@article{guerra2018freezing,
  title={{Freezing on a sphere}},
  author={Guerra, Rodrigo E and Kelleher, Colm P and Hollingsworth, Andrew D and Chaikin, Paul M},
  journal={Nature},
  volume={554},
  number={7692},
  pages={346-350},
  year={2018},
  publisher={Nature Publishing Group UK London},
url={https://doi.org/10.1038/nature25468},
doi={10.1038/nature25468}
}

@article{PhysRevLett.69.832,
  title = {{Order from disorder in a kagom\'e antiferromagnet}},
  author = {Chubukov, Andrey},
  journal = {Physical Review Letters},
  volume = {69},
  issue = {5},
  pages = {832-835},
  numpages = {0},
  year = {1992},
  month = {Aug},
  publisher = {American Physical Society},
  doi = {10.1103/PhysRevLett.69.832},
  url = {https://link.aps.org/doi/10.1103/PhysRevLett.69.832}
}

@article{pokrovsky1979properties,
  title={Properties of ordered, continuously degenerate systems},
  author={Pokrovsky, VL},
  journal={Advances in Physics},
  volume={28},
  number={5},
  pages={595-656},
  year={1979},
  publisher={Taylor \& Francis},
url={https://doi.org/10.1080/00018737900101425},
doi={10.1080/00018737900101425}
}

@article{kosterlitz1973ordering,
  title={{Ordering, metastability and phase transitions in two-dimensional systems}},
  author={Kosterlitz, John Michael and Thouless, David James},
  journal={Journal of Physics C: Solid State Physics},
  volume={6},
  number={7},
  pages={1181},
  year={1973},
  publisher={IOP Publishing},
url={https://iopscience.iop.org/article/10.1088/0022-3719/6/7/010/meta},
doi={10.1088/0022-3719/6/7/010}
}

@article{sonier2002muon,
  title={Muon Spin Rotation/Relaxation},
  author={Sonier, JE},
  journal={Resonance brochure},
  year={2002}
}

@article{PhysRevLett.74.2379,
  title = {Frustration Induced Spin Freezing in a Site-Ordered Magnet: Gadolinium Gallium Garnet},
  author = {Schiffer, P. and Ramirez, A. P. and Huse, D. A. and Gammel, P. L. and Yaron, U. and Bishop, D. J. and Valentino, A. J.},
  journal = {Physical Review Letters},
  volume = {74},
  issue = {12},
  pages = {2379-2382},
  numpages = {0},
  year = {1995},
  month = {Mar},
  publisher = {American Physical Society},
  doi = {10.1103/PhysRevLett.74.2379},
  url = {https://link.aps.org/doi/10.1103/PhysRevLett.74.2379}
}

@article{hillier2022muon,
  title={Muon spin spectroscopy},
  author={Hillier, Adrian D and Blundell, Stephen J and McKenzie, Iain and Umegaki, Izumi and Shu, Lei and Wright, Joseph A and Prokscha, Thomas and Bert, Fabrice and Shimomura, Koichiro and Berlie, Adam and others},
  journal={Nature Reviews Methods Primers},
  volume={2},
  number={1},
  pages={4},
  year={2022},
  publisher={Nature Publishing Group UK London},
doi={10.1038/s43586-021-00089-0},
url={https://www.nature.com/articles/s43586-021-00089-0#article-info}
}

@book{blundell2021muon,
  title={Muon spectroscopy: an introduction},
  author={Blundell, Stephen J and Blundell, Stephen and De Renzi, Roberto and Lancaster, Tom and Pratt, Francis L},
  year={2021},
  publisher={Oxford University Press},
url={https://academic.oup.com/book/43671?login=true},
doi={10.1093/oso/9780198858959.001.0001}
}

@article{wolf1962magnetic,
  title={The magnetic properties of rare earth ions in garnets},
  author={Wolf, WP and Ball, M and Hutchings, MT and Leask, MJM and MJ, H and Wyatt, A},
  journal={J. Phys. Soc. Jpn.},
  volume={17},
  number={Supp. B-1},
  pages={443},
  year={1962},
url={https://www.jps.or.jp/books/jpsjs/17BI/jpsj.17sbI.443.pdf
    },
doi={ Not available }
}

@article{de1997muon,
  title={Muon spin rotation and relaxation in magnetic materials},
  author={De R{\'e}otier, P Dalmas and Yaouanc, A},
  journal={Journal of Physics: Condensed Matter},
  volume={9},
  number={43},
  pages={9113},
  year={1997},
  publisher={IOP Publishing},
doi={10.1088/0953-8984/9/43/002},
url={https://iopscience.iop.org/article/10.1088/0953-8984/9/43/002/meta}
}

@article{blundell1999spin,
  title={Spin-polarized muons in condensed matter physics},
  author={Blundell, SJ},
  journal={Contemporary Physics},
  volume={40},
  number={3},
  pages={175-192},
  year={1999},
  publisher={Taylor \& Francis},
doi={10.1080/001075199181521},
url={https://www.tandfonline.com/doi/abs/10.1080/001075199181521}
}

@book{slichter2013principles,
  title={Principles of magnetic resonance},
  author={Slichter, Charles P},
  volume={1},
  year={2013},
  publisher={Springer Science \& Business Media},
url={https://books.google.co.in/books?hl=en&lr=&id=jF3xCAAAQBAJ&oi=fnd&pg=PA1&dq=Principles+of+Magnetic+Resonance&ots=n-hMfYlbyN&sig=HscgtOoBbCNjMIDYB2BTjAtIQMI&redir_esc=y#v=onepage&q=Principles%20of%20Magnetic%20Resonance&f=false}
}

@article{paddison2015hidden,
  title={{Hidden order in spin-liquid Gd$_3$Ga$_5$O$_{12}$}},
  author={Paddison, Joseph AM and Jacobsen, Henrik and Petrenko, Oleg A and Fern{\'a}ndez-D{\'\i}az, Maria Teresa and Deen, Pascale P and Goodwin, Andrew L},
  journal={Science},
  volume={350},
  number={6257},
  pages={179-181},
  year={2015},
  publisher={American Association for the Advancement of Science},
url={https://doi.org/10.1126/science.aaa5326},
doi={10.1126/science.aaa5326}
}

@article{pretko2020fracton,
  title={Fracton phases of matter},
  author={Pretko, Michael and Chen, Xie and You, Yizhi},
  journal={International Journal of Modern Physics A},
  volume={35},
  number={06},
  pages={2030003},
  year={2020},
  publisher={World Scientific},
url={https://doi.org/10.1142/S0217751X20300033},
doi={10.1142/S0217751X20300033}
}

@article{parisi1980order,
  title={{The order parameter for spin glasses: a function on the interval 0-1}},
  author={Parisi, Giorgio},
  journal={Journal of Physics A: Mathematical and General},
  volume={13},
  number={3},
  pages={1101},
  year={1980},
  publisher={IOP Publishing},
url={https://iopscience.iop.org/article/10.1088/0305-4470/13/3/042/meta},
doi={10.1088/0305-4470/13/3/042}
}

@article{anderson1972more,
  title={{More is different: broken symmetry and the nature of the hierarchical structure of science.}},
  author={Anderson, Philip W},
  journal={Science},
  volume={177},
  number={4047},
  pages={393-396},
  year={1972},
  publisher={American Association for the Advancement of Science},
url={https://doi.org/10.1126/science.177.4047.393},
doi={10.1126/science.177.4047.393}
}

@article{garrahan2011kinetically,
  title={Kinetically constrained models},
  author={Garrahan, Juan P and Sollich, Peter and Toninelli, Cristina},
  journal={Dynamical heterogeneities in glasses, colloids, and granular media},
  volume={150},
  pages={111-137},
  year={2011},
  publisher={Oxford University Press Oxford}
}

@article{castelnovo2009topological,
  title={Topological order and quantum criticality},
  author={Castelnovo, Claudio and Trebst, Simon and Troyer, Matthias},
  journal={arXiv preprint arXiv:0912.3272},
  year={2009}
}

@article{arh2022ising,
  title={{The Ising triangular-lattice antiferromagnet neodymium heptatantalate as a quantum spin liquid candidate}},
  author={Arh, T and Sana, B and Pregelj, M and Khuntia, P and Jagli{\v{c}}i{\'c}, Z and Le, MD and Biswas, PK and Manuel, P and Mangin-Thro, L and Ozarowski, A and others},
  journal={Nature Materials},
  volume={21},
  number={4},
  pages={416-422},
  year={2022},
  publisher={Nature Publishing Group UK London},
url={https://doi.org/10.1038/s41563-021-01169-y},
doi={10.1038/s41563-021-01169-y}
}

@article{basov2017towards,
  title={Towards properties on demand in quantum materials},
  author={Basov, DN and Averitt, RD and Hsieh, D},
  journal={Nature Materials},
  volume={16},
  number={11},
  pages={1077-1088},
  year={2017},
  publisher={Nature Publishing Group UK London},
url={https://doi.org/10.1038/nmat5017},
doi={10.1038/nmat5017}
}

@article{tokura2017emergent,
  title={Emergent functions of quantum materials},
  author={Tokura, Yoshinori and Kawasaki, Masashi and Nagaosa, Naoto},
  journal={Nature Physics},
  volume={13},
  number={11},
  pages={1056-1068},
  year={2017},
  publisher={Nature Publishing Group UK London},
url={https://doi.org/10.1038/nphys4274},
doi={10.1038/nphys4274}
}

@book{leggett2006quantum,
  title={{Quantum liquids: Bose condensation and Cooper pairing in condensed-matter systems}},
  author={Leggett, Anthony J},
  year={2006},
  publisher={Oxford university press}
}

@book{peskin2018introduction,
  title={{An Introduction to quantum field theory}},
  author={Peskin, Michael E},
  year={2018},
  publisher={CRC press}
}

@article{binder1986spin,
  title={Spin glasses: Experimental facts, theoretical concepts, and open questions},
  author={Binder, Kurt and Young, A Peter},
  journal={Reviews of Modern physics},
  volume={58},
  number={4},
  pages={801},
  year={1986},
  publisher={APS},
url={https://journals.aps.org/rmp/abstract/10.1103/RevModPhys.58.801},
doi={10.1103/RevModPhys.58.801}
}

@incollection{ginzburg2009theory,
  title={On the theory of superconductivity},
  author={Ginzburg, Vitaly L and Landau, Lev D},
  booktitle={On superconductivity and superfluidity: a scientific autobiography},
  pages={113-137},
  year={2009},
  publisher={Springer}
}

@article{zurek1985cosmological,
  title={Cosmological experiments in superfluid helium?},
  author={Zurek, Wojciech H},
  journal={Nature},
  volume={317},
  number={6037},
  pages={505-508},
  year={1985},
  publisher={Nature Publishing Group UK London},
url={https://doi.org/10.1038/317505a0},
doi={10.1038/317505a0}
}

\end{document}